\documentclass[useAMS,usegraphicx,onecolumn,usenatbib]{mn2e}
\usepackage{bm}
\usepackage{graphicx}
\usepackage{latexsym,color}
\usepackage{hyperref}
\usepackage{amssymb}
\usepackage{natbib}

\title[On the  Polish doughnut accretion disk]
  {On the Polish doughnut accretion disk via the effective potential approach}
\author[D. Pugliese et al.]
  {D.~Pugliese$^{1}$\thanks{E-mail:
 giovanni.montani@frascati.enea.it (GM); daniela.pugliese@alice.it  (DP); grazia.bernardini@brera.inaf.it (MGB)},  G.~Montani$^{2,3}$, M.~G.~Bernardini$^{4}$
\\
$^1$ School of Mathematical Sciences, Queen Mary University of London, Mile
End Road, London E1 4NS, United Kingdom\\
$^2$ ENEA - C.R, UTFUS-MAG
Via Enrico Fermi 45, 00044
Frascati, Roma, Italy\\
$^3$ Dipartimento di Fisica, Universit\`a di Roma ``Sapienza", Piazzale Aldo Moro 5, I-00185 Roma, Italy\\
$^4$ INAF - Osservatorio Astronomico di Brera, via Emilio Bianchi 46, I-23807 Merate
(LC), Italy}
\date{\today}

\pagerange{\pageref{firstpage}--\pageref{lastpage}} \pubyear{0000}

\def\LaTeX{L\kern-.36em\raise.3ex\hbox{a}\kern-.15em
    T\kern-.1667em\lower.7ex\hbox{E}\kern-.125emX}

\begin{document}

\label{firstpage}

\maketitle

\begin{abstract}
We revisit  the  Polish doughnut model of accretion disks providing  a  comprehensive
analytical description of the Polish doughnut structure. We describe a perfect fluid
circularly orbiting around a Schwarzschild black hole, source of the gravitational
field, by the effective potential approach for the exact gravitational and centrifugal
effects. This analysis leads to a  detailed, analytical description of the accretion
disk, its toroidal surface, the thickness, the  distance from the source. We determine
the variation of these  features  with the effective potential and the fluid angular
momentum. Many analytical formulas are given. In particular it turns out that the
distance from the source of the inner surface of the torus increases with increasing
fluid angular momentum   but decreases with increasing energy function defined as the
value of the effective potential for that momentum. The location of torus  maximum
thickness moves towards the external regions of the surface   with increasing angular
momentum, until it reaches a maximum an then decreases. Assuming  a polytropic equation
of state we investigate some specific cases.
\end{abstract}


\newcommand{\ti}[1]{\mbox{\tiny{#1}}}
\newcommand{\im}{\mathop{\mathrm{Im}}}
\def\be{\begin{equation}}
\def\ee{\end{equation}}
\def\bea{\begin{eqnarray}}
\def\eea{\end{eqnarray}}
\newcommand{\tb}[1]{\textbf{\texttt{#1}}}
\newcommand{\ttb}[1]{\textbf{#1}}
\newcommand{\rtb}[1]{\textcolor[rgb]{1.00,0.00,0.00}{\tb{#1}}}
\newcommand{\btb}[1]{\textcolor[rgb]{0.00,0.00,1.00}{\tb{#1}}}

\newcommand{\il}{~}
\newcommand{\rc}{\rho_{\ti{C}}}
\newcommand{\dd}{\mathcal{D}}
\newcommand{\lie}{\mathcal{L}}
\newcommand{\Tem}{T^{\rm{em}}}

\section{Introduction}

 Accretion disks are one of the most intriguing issues in high energy Astrophysics. They
enter into very different contexts: from the proto-planetary disks to Gamma Ray Bursts
(GRB), from X-ray binaries to the Active Galactic Nuclei (AGN). Indeed, many aspects of
disk structure, its dynamics, the formation of ``jets", the exact mechanism behind the
accretion, its equilibrium, the confinement and stability under perturbations are still
uncertain.

Analytic and semi-analytic models for accretion onto a compact object are generally
stationary and axially symmetric. Thus, all physical quantities depend only on the
radial distance from the center $r$, and the vertical distance from the equatorial
symmetry plane $z$. Thick accretion disk ($z/r \gg 1$) models are basically designed
according to the following assumptions: first, a circular motion dynamic is prescribed.
Second, the matter being accreted is described by a perfect fluid energy-momentum
tensor. This requirement lies on the assumption that the time scale of dynamic
processes, involving the pressure, centrifugal and gravitational forces,  are smaller,
and much smaller in the case of vertically thin disks, than thermal ones, which in turn
are smaller or much smaller than the viscous time scale. This implies that, at variance
with many thin disk models, dissipative effects like viscosity or resistivity are
neglected, and accretion is a consequence of the strong gravitational field of the
attractor. Indeed this is a great advantage, since angular momentum transport in the
fluid is perhaps one of the most controversial aspects in thin disk models.

In this work we focus on the Polish doughnut model. The Polish doughnut is a fully
relativistic model of thick accretion disk with a toroidal shape, and is an example of
opaque (large optical depth) and super-Eddington (hight  matter accretion rates) disk
model. During the evolution of dynamic processes, the functional form of the angular
momentum and entropy distribution depends on the initial conditions of the system and on
the details of the dissipative processes. Paczy\'nski realized that it is physically
reasonable to assume ad hoc distributions. The Polish doughnut is characterized by a
constant angular momentum \citep{Abramowicz:2008bk}.

The development of this model was drawn up by Paczy\'nski and his collaborators in a
series of works
\citep{Pac-Wii,cc,Koz-Jar-Abr:1978:ASTRA:,Abr-Jar-Sik:1978:ASTRA:,Jaroszynski(1980),Abr-Cal-Nob:1980:ASTRJ2:,Abramowicz:1996ap,FisM76}. Using a perfect fluid
energy-momentum  tensor, \citet{Abr-Jar-Sik:1978:ASTRA:} wrote the equations of the  hydrodynamics
for  this model. \citet{Jaroszynski(1980)} discussed the case
of a  non barotropic fluid and showed an important result
concerning the  pressure  of the rotating fluid: for a perfect fluid matter
circularly rotating around a Schwarzschild black hole, the shapes and location of the
equipressure surfaces follow directly from the assumed angular momentum distribution
\citep{Jaroszynski(1980)}. More recently, \citet{Lei:2008ui} assumed an  angular momentum
distribution in the form that depends on three constant parameters, and different
configurations have been studied. A significant result concerning this
model, known as the \emph{von Zeipel condition}, has been extensively investigated in
\citet{Koz-Jar-Abr:1978:ASTRA:} and \citet{Jaroszynski(1980)}: the constant pressure surfaces
coincide with surfaces of constant density if and only if the surfaces at constant
angular momentum coincide with the surfaces at constant relativistic velocity. More
generally  we can say that in the static spacetimes the family of von Zeipel's surfaces
does not depend on the particular rotation law of the fluid, in the sense that it does
not depend on nothing but the background spacetime. An accurate study of  the von Zeipel
surfaces has been performed in \citet{M.A. Abramowicz} and \citet{Chakrabarti, Chakrabarti0}.

Paczy\'nski  realized  from the study of Roche lobe in the accretion disks of the binary
systems that the black hole Roche lobe overflow must induce the dynamical mass loss from
the disk, thus the accretion \citep{Boy:1965:PCPS:,Raine}. The accretion occurs at the
point of cusp of equipotential surfaces. This process is realized by the relativistic
Roche lobe overflow. This clearly is an explanation for the accretion that does not
involve other factors (as the dissipative ones)  than the strong gravitational field of
the attractor. However \citet{A1981} showed that it constitutes also an
important stabilizing mechanism against the thermal and viscous instabilities locally,
and against the so called Papaloizou and Pringle instability globally \citep{Blaes1987}.

The general relativistic effects  on  matter dynamics close to a Schwarzschild black
hole have been modeled in an approximate pseudo-Newtonian theory by Paczy\'nski and
Wiita introducing a properly chosen non-exact gravitational potential, known as
Paczy\'nski-Wiita (P-W) potential \citep{cc}. This potential  simulates the relativistic
effects of the gravitational field acting  on the fluid in the disk, in the
Schwarzschild spacetime. This is not the exact expression of the effective potential for
gravitational and centrifugal effects, and yet the P-W potential cannot properly be
considered a newtonian approximation, that is valid in the limit of weak gravitational
fields. The P-W potential differs from the exact relativistic one by a constant in  its
gravitational  part and it is a Newtonian way to write  some of the general relativistic
effects  characterizing the thick disks. With this approximation, the radius of a
marginally bound orbit, the last stable circular orbit radius  of  the Schwarzschild
spacetime and the form of the Keplerian angular momentum have been correctively
reproduced. A step-by-step derivation of the
P-W potential  and a detailed discussion of its main features can be found in \citet{Abramowicz:2009bh}. The agreement between
model predictions and simulations of accretion flows has been verified and found an
excellent outcome, in e.g. \citet{Igumenshchev} and \citet{Shafee}. The equipressure
surfaces for a Schwarzschild  black hole have been  compared with global
magnetohydrodynamic numerical simulations in \citet{Fragile:2007dk}, {(see also \citealt{DeVilliers,Hawley1987,Hawley1990,Hawley1991,Hawley1984}). } Recently, the study
of the Polish doughnut model has been developed for different attractors (see
\citealt{arXiv:0910.3184,astro-ph/0605094,Stu-Kov:2008:INTJMD:} for the Schwarzschild-de
Sitter and Kerr-de Sitter spacetimes).

In the present work we face the study of the Polish doughnut model in the Schwarzschild
background  using the approach of the general relativistic effective potential  in its
exact form. Gravitational and centrifugal forces carried out in the effective potential
and the pressure force  operate on a perfect fluid of the disk. When the latter
vanishes, the hydrodynamics of the fluid describes a geodesic disk whose equations are
formally resembling those of motion of a test particle orbiting in the same background.
We take advantage of this formal analogy using the  familiar  and well known results  on
the dynamics of the test particles to get a comparison between the Polish doughnut,
which is supported by the pressure, and the geodetic disk. In this way we can evaluate
the right weight of the pressure effects  on the dynamics of fluid and the shape of the
torus. Furthermore, we draw a complete and analytic description of the toroidal surface
of the disk, including the analysis of its extension in space, the distance from the
center attractor, its thickness etc, and understand how these features  are modified by
changing the angular momentum of the fluid and the effective potential. In particular we
find that the distance from the source of the inner surface of the torus increases with
increasing fluid angular momentum and decreases with increasing energy function defined
as the value of the effective potential for that momentum.

In Section \il\ref{Sec:first} we introduce the Polish doughnut model writing the
equations of the ideal hydrodynamics for a fluid circularly orbiting in the background
of the  Schwarzschild spacetime. In Section\il\ref{Sec:Gr} we detail the fluid pressure
gradients along the radial direction, and in Section \il\ref{Sec:ella} along the polar
angular direction to determine the regions of maximum and minimum pressure in the disk,
the regions of increasing pressure and the isobar surfaces.  In Section \ref{Sec:Poli} we trace the profile of the toroidal disk in the
Polish model by introducing and studying in detail the Boyer potential  for the
barotropic fluid. In Section \il\ref{Sec:polu} we investigate the case of polytropic
equation of state. In
Section\il\ref{Sec:proper} we analyze the proper fluid velocity, finding the regions of
the disks of maximum and minimum velocity. In Section\il\ref{Sec:KeplerANDangular} we
discuss the relativistic angular velocity examining the properties of the von Zeipel
surfaces. Conclusions follow.

\section{The Polish doughnut model}\label{Sec:first}

Consider a one-species particle perfect  fluid (simple fluid)
, where
\be\label{E:Tm}
T_{a b}=(\rho +p) U_{a} U_{b}+\  p g_{a b}
\ee
is the fluid energy momentum tensor,  $\rho$ and $p$ are  the total energy density and
pressure, respectively, as measured by an observer moving with the fluid, and $g_{a b}$
the metric tensor.  The time-like flow vector field  $U$  denotes the fluid
four-velocity\footnote{The fluid  four-velocity  satisfy $U^a U_a=-1$. We adopt the
geometrical  units $c=1=G$ and  the $(-,+,+,+)$ signature.  The radius $r$ has unit of
mass $[M]$, and the angular momentum  units of $[M]^2$, the velocities  $[U^t]=[U^r]=1$
and $[U^{\varphi}]=[U^{\vartheta}]=[M]^{-1}$ with $[U^{\varphi}/U^{t}]=[M]^{-1}$ and
$[U_{\varphi}/U_{t}]=[M]$. For the seek of convenience, we always consider the
dimensionless  energy and effective potential $[V_{sc}]=1$ and an angular momentum per
unit of mass $[L]/[M]=[M]$.}

The motion of the fluid  is described by the \emph{continuity  equation}:
\bea\label{E:1a0}
U^a\nabla_a\rho+(p+\rho)\nabla^aU_a=0\, ,
\eea
and the \emph{Euler equation}:
\bea
\label{Eulerif0}
(p+\rho)U^a\nabla_aU^c+ \ h^{bc}\nabla_b p=0\, ,
\eea
where $h_{ab}=g_{ab}+ U_a U_b$ \citep{Mis-Tho-Whe:1973:Gra:}.

Neglecting the fluid back reaction, we  consider the fluid motion  in  the Schwarzschild
spacetime background:
\begin{equation}\label{11metrica}
ds^2=-e^{\nu(r)}dt^2+e^{-\nu(r)}(r)dr^2
+r^2\left(d\vartheta^2+\sin^2\vartheta d\varphi^2\right),
\end{equation}
written in standard spherical coordinates, where $e^{\nu(r)}\equiv\left(1-2M/r\right)$.
We  define:
\be\label{Lafluido}
\Lambda\equiv U^r,\quad \Sigma\equiv U^t, \quad \Phi\equiv U^{\varphi},\quad
\Theta\equiv U^{\vartheta}\, ,
\ee
and we introduce the  set  of variables $\{E, V_{sc}, L, T\}$  by the following
relations:
\bea\label{La2fluido}
\Lambda =\sqrt{E^2-V_{sc}^2}\, ,\quad
\Sigma =\frac{E}{e^{\nu}}\, ,\quad
\Phi = \frac{L}{r^2\sin^2\vartheta}\, ,\quad
\Theta=\frac{T}{r^2}\, ,
\eea
where
\be\label{60}
V_{sc}\equiv\sqrt{e^{\nu(r)}\left(1+\frac{L^2}{r^2\sin^2\vartheta}+\frac{T^2}{r^2}\right)}
\ee
is the \emph{effective potential}\citep{Mis-Tho-Whe:1973:Gra:}.
 In fact,
from Eq.\il\ref{La2fluido} we obtain:
\be\label{rpunto12x}
\dot{r}^2=\left(E^2-V_{sc}^2\right)
\ee
(the dot represents differentiation with respect to the proper time):
Eq.\il\ref{rpunto12x} describes the motion inside the effective potential
${V_{{sc}}}$, defined as the energy at which the (radial) kinetic energy of the fluid
element vanishes.

Eq.\il\ref{rpunto12x} and the definitions in
Eqs.\il\ref{Lafluido},\ref{La2fluido},\ref{60} are formally the same as for the test
particle motion in the Schwarzschild spacetime. Obviously, for the particle motion,
$U^a$ in Eq.\il\ref{Lafluido} is the test particle four-velocity, and $(E, L)$  in
Eq.\il\ref{La2fluido} are two constants of motion, the particle energy and angular
momentum per unit of mass as seen by infinity, respectively.

Using  this similarity,  we can make a one-to-one comparison of the motion of the fluid,
under the action of the pressure forces balanced by the effective potential, with the
test particles dynamics  regulated by the gravitational and centrifugal forces as
described by the effective potential
\citep{Wald,Mis-Tho-Whe:1973:Gra:,Pugliese:2011xn,Pugliese:2011py,Pugliese:2010ps}. The
comparison with the case of dust disk through the effective potential enables us to
evaluate the relationship between the contribution of pressure and the gravitational and
centrifugal effects to the dynamics of the system, especially in relation to the
angular momentum of the fluid in rotation and  the  disk shape.

We consider the case of a fluid circular configuration, defined by the constraints
$\Lambda=0$ (i.e. ${V_{sc}}=E$), restricted to a fixed plane $\sin\vartheta=\sigma\neq0$. No
motion is assumed in the $\vartheta$ angular direction, which means $\Theta=0$. For the
symmetries of the problem, we always assume $\partial_t \mathbf{Q}=0$ and
$\partial_{\varphi} \mathbf{Q}=0$, being $\mathbf{Q}$ a generic tensor of the spacetime
(we can refer to  this assumption as the condition of  ideal hidrodynamics of
equilibrium).

Within our assumptions $(\Lambda=0, \Theta=0,  \partial_tp=\partial_{\varphi}p=0)$, from
the Euler equation \ref{Eulerif0} we derive the expressions for the \emph{radial
pressure gradient} $G_r$ and the \emph{angular pressure gradient} $G_{\vartheta}$:

\be\label{vaperr}
{G_r\equiv \frac{ \nabla_rp}{\rho+p}=-\left(\frac{\nu'e^{\nu}}{2}\Sigma^2-r\sigma^2\Phi^2\right)\,}
,
\ee
and
\be\label{vaperfi}
{G_{\vartheta}\equiv \frac{
\nabla_{\vartheta}p}{\rho+p}=+\sigma\sqrt{1-\sigma^2}r^2\Phi^2\, ,}
\ee

\section{The radial pressure gradient $G_r$}\label{Sec:Gr}
{The first part  of the  present  work is dedicated to the study of the fluid angular
momentum, in particular, we are interested especially in the
comparison between the geodetic disk case, and the case of a fluid subjected to a non-zero
pressure, for this purpose in this and following Sections we will study the radial and
angular pressure gradient. This study
allows us to evaluate the pressure contribution  to the disk dynamics  along the orbital
radius and the plans on which the accretion disk stretches.
This analysis  introduces the second part of the work in which we finally trace the
profile of the disk obtained from the analysis of the constant pressure surfaces.
}

Equation \ref{vaperr}  can be written as:

\be\label{olgettin}
{\frac{\nabla_rp}{\rho+p}=-\frac{e^{-\nu}}{2}\left(\frac{\partial
V_{{sc}}^2}{\partial r}\right)_{L}}
\ee
in terms of the  partial derivative of $ V_{sc}$ computed  keeping  $L=\mbox{constant}$.
Eq.\il\ref{olgettin} yields to:
\be\label{E:la}
\partial_rp=0\quad\mbox{for}\quad \left(\frac{\partial  V_{{sc}}^2}{\partial
r}\right)_{L}=0\, ,
\ee
and
\be
{\lim_{r\rightarrow\infty}G_r=0,\quad\lim_{r\rightarrow2M}G_r=-\infty\,.}
\ee
Assuming $\rho>0$ and $p>0$, from Eq.\il\ref{olgettin} it follows that  the pressure
increases (decreases) with the orbital radius $r$ as $V_{sc}$  decreases (increases),
and that the critical points of $p$ (as a function of r) are the same as of $ V_{{sc}}$
as a function of $r$ at $L$ constant. Thus, solving Eq.\il\ref{E:la} for the unknown
$L$ we find that these critical points are for:
\be\label{Eq:wEo}
L_{\ti{K}}=\pm\sqrt{\frac{\sigma^2 M r^2}{(r-3M)}}\, .
\ee
This function is defined in the range $r>r_{{lco}}$, where $r_{{lco}}\equiv3M$ is the
last circular orbit radius for a test particle in the Schwarzschild
spacetime\footnote{The   angular proper  velocity  of the fluid with $L=L_{\ti{K}}$ is
$\Phi_{\ti{K}}=L_{\ti{K}}/r^2\sigma^2$. This function has no critical point. It is defined
in $r\in]r_{lco},\infty[$, where it is a monotonically decreasing function of  the
orbital radius $r/M$, it increases approaching $r=r_{lco}$, and goes to zero at
infinity.}. Equation \ref{E:la} is therefore satisfied only in the range
$r>r_{lco}$, for fixed $\sigma$. In $r=r_{lco}$ it is   ${G_{r}=-1/(3 M)}$. The angular momentum
$L_{\ti{K}}$ describes the isobar fluid configurations: where the fluid is
characterized by $L=L_{\ti{K}}$, the pressure $p$ is constant and  the Euler equation
\ref{Eulerif0} reduces to $U^a\nabla_aU^b=0$, describing the motion of a
pressure-free fluid (dust). The  curves $L=L_{\ti{K}}$ represent the critical points of
the pressure $p$. From Eq.\il\ref{Eq:wEo} it follows, according to the physics of the
free (test) particle (and dust defined by $p=0$)
\citep{Mis-Tho-Whe:1973:Gra:,Pugliese:2011xn,Pugliese:2011py,Pugliese:2010ps} that no
critical point exists in the range $[2M,r_{lco}]$, where $\partial_rp<0$ (pressure
always decreasing)
\citep{Koz-Jar-Abr:1978:ASTRA:,Abr-Jar-Sik:1978:ASTRA:,Pac-Wii,Abramowicz:2009bh,Lei:2008ui,2011}.

The angular momentum $L_{\ti{K}}$ as a function of $r$ \footnote{We
restrict our analysis to $\sigma\in(0,1]$ (the function is even in $\sigma\in[-1,1]$) and
$L_{\ti{K}}\geq0$}  for  fixed  $\sigma$,
$L_{\ti{K}}$ tends to infinity as the orbital radius approaches $r=r_{{lco}}$, then
monotonically decreases until it reaches its minimum value for $r=r_{{lsco}}$
$\left(L_{\ti{K}}(r_{{lsco}})=2 \sqrt{3 \sigma^2} M\right.$ and ${\left.G_r(r_{{lsco}})=-\frac{
M^{-1}}{24}\left(1-\frac{L^2}{12 M^2 \sigma^2}\right)\right)}$ where $r_{{lsco}}=6M$ is the
last \emph{stable} circular orbit radius for a test particle in the Schwarzschild
geometry. Finally it increases for $r>r_{{lsco}}$. The angular momentum $L_{\ti{K}}$ is
a monotonically increasing function of $\sigma$, and in the boundary $\sigma=0$ it is
$L_{\ti{K}}=0$.

\subsection{Radial pressure gradient $G_r(L)$ vs angular momentum
$L$}\label{Sec:Neutral-L0}

\begin{figure}
\centering
\includegraphics[width=0.3\hsize,clip]{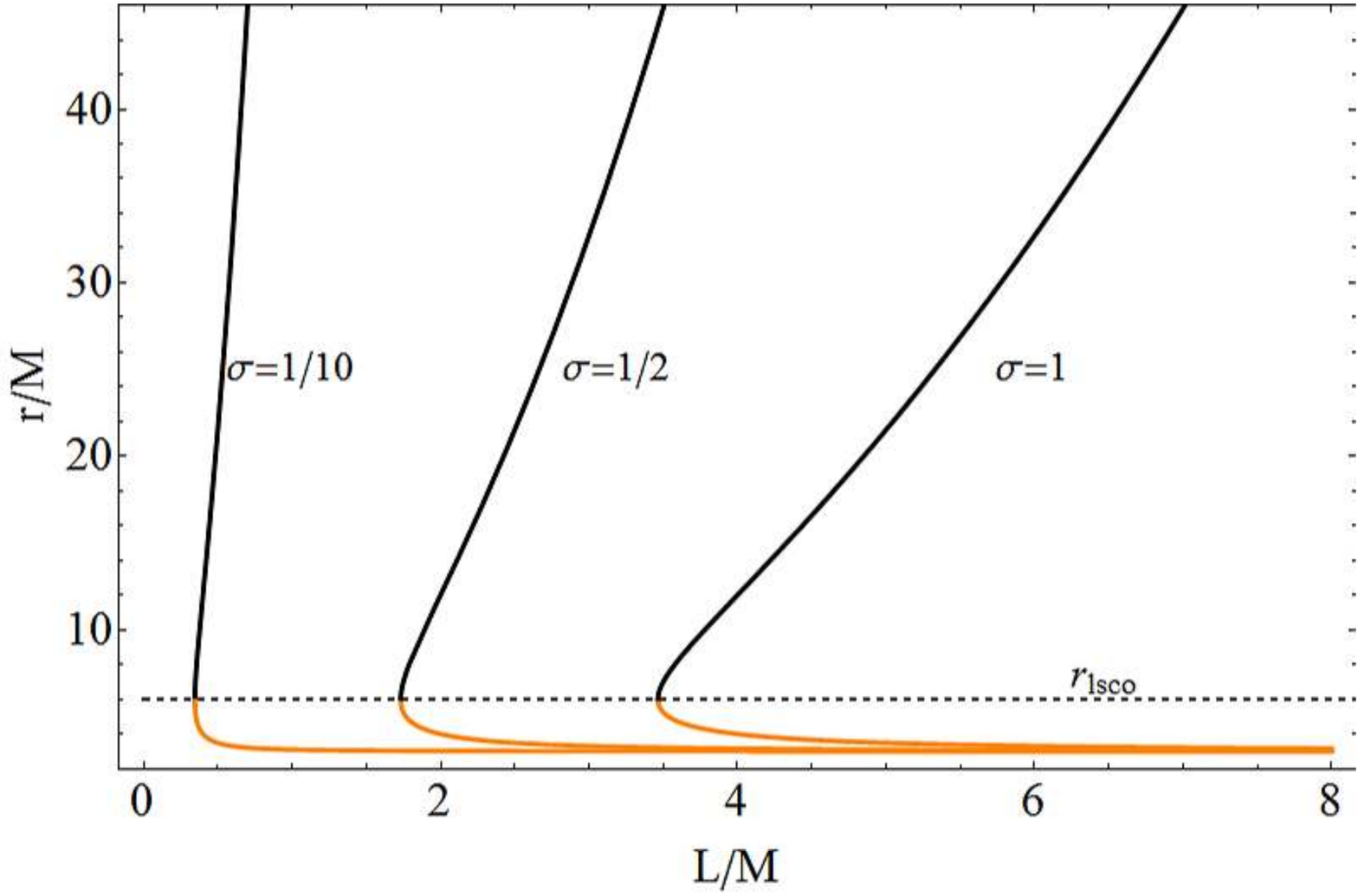}
\includegraphics[width=0.3\hsize,clip]{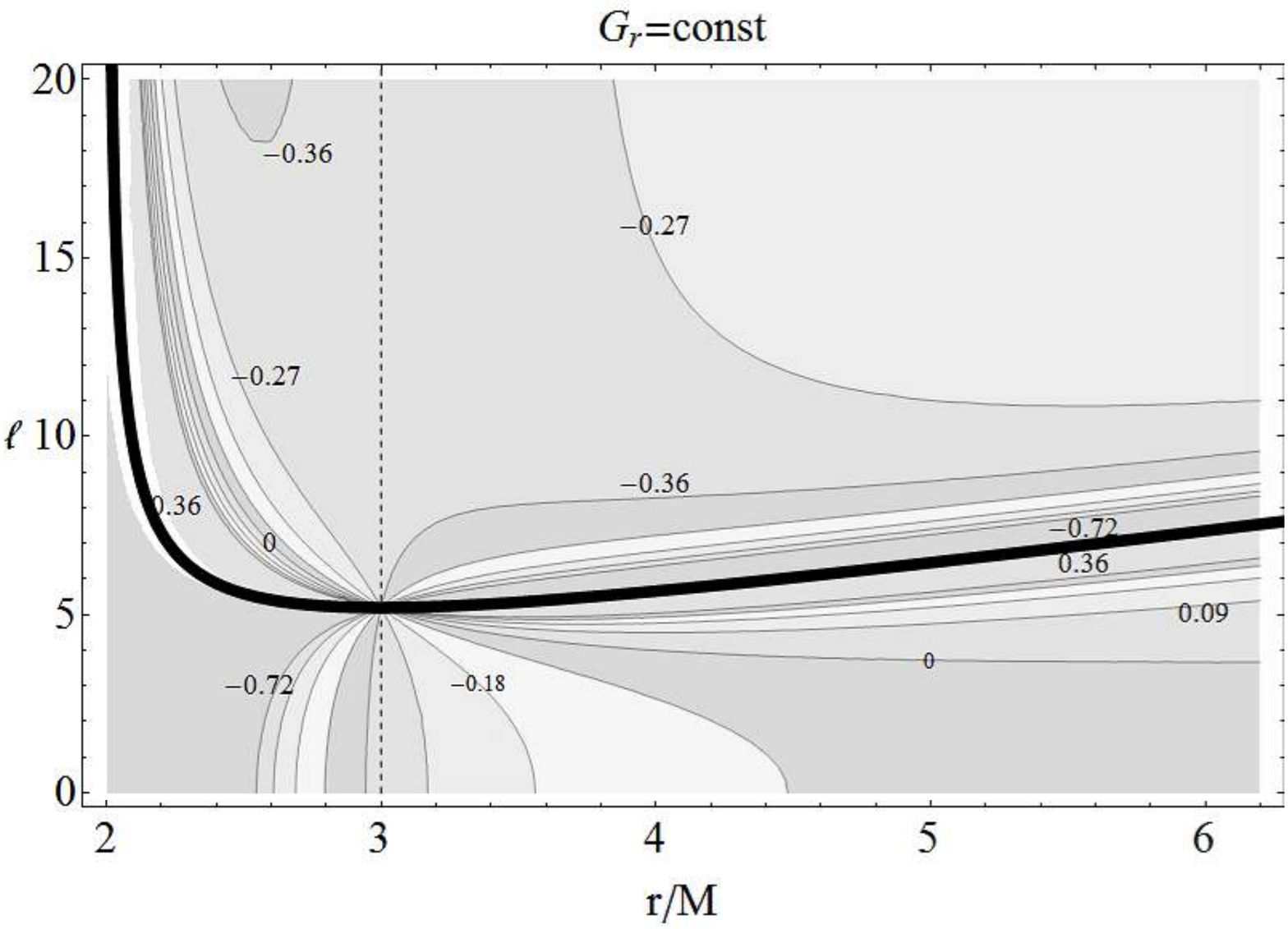}
\includegraphics[width=0.3\hsize,clip]{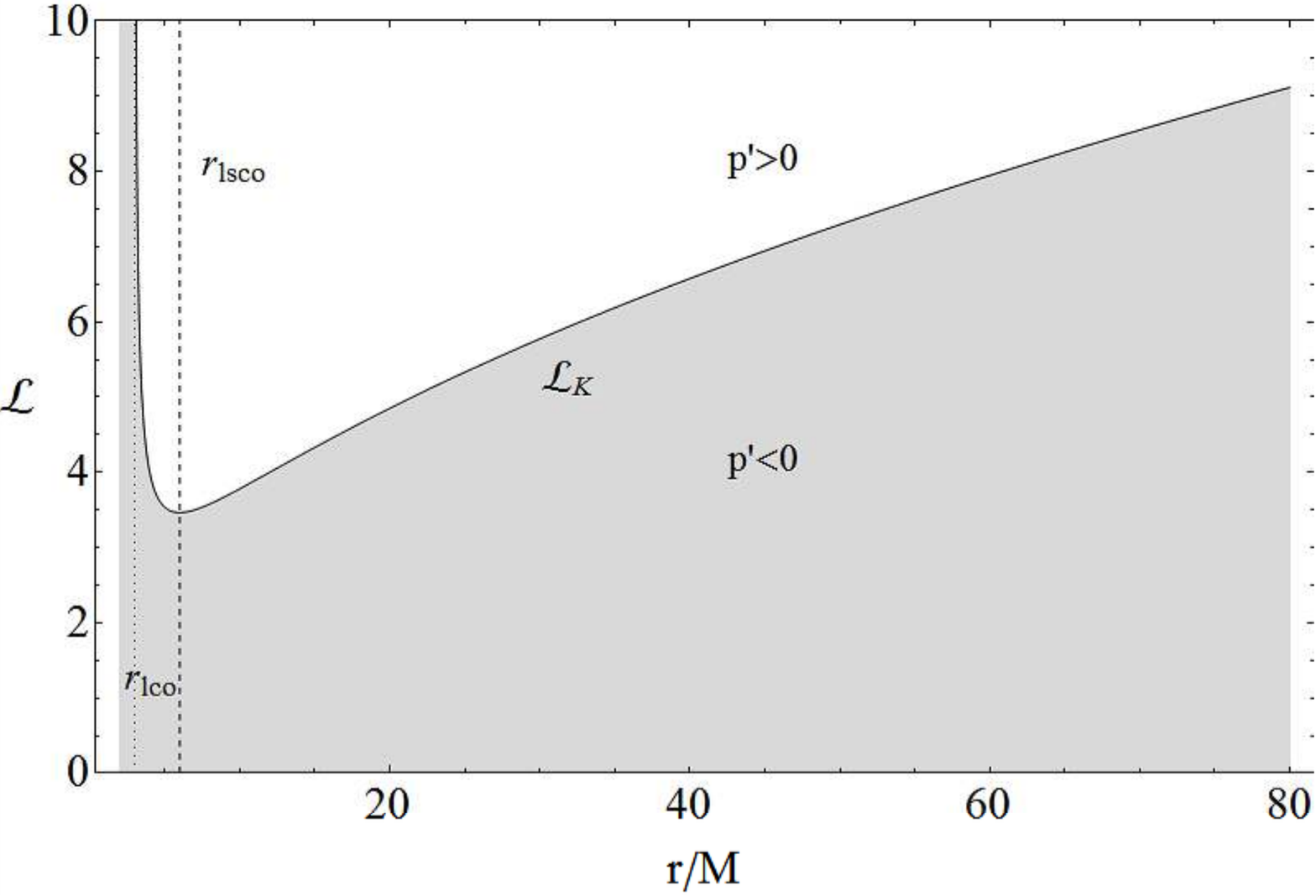}
\caption{{(Color online) \emph{Left panel}: $r_{\ti{L}}^{+}/M$ (black line) and
$r_{\ti{L}}^{-}/M$ (orange line)  as a function of $L/M$  for different  $\sigma$. The dashed
line marks $r_{\ti{L}}^{+}=r_{\ti{L}}^{-}=r_{lsco}$. This condition is fulfilled when
$\lie=2\sqrt{3}$. \emph{Center panel}: $G_r=\mbox{constant}$ as a function
of $\ell$ and $r/M$. The singularity at $\ell=\ell_r$ is marked with a thick black line.
Dashed line marks $r=r_{lco}$. \emph{Right panel}: {$\mathbf{\lie_{\ti{K}}=L_{\ti{K}}/(M\sigma)}$} as a
function of $r/M$.    $p'>0$ (and $p'<0$)  in the white (gray)
region. $G_{r}=0$ when $L=L_{\ti{K}}$ (solid line). Dashed line marks $r=r_{{lsco}}$,
dotted line $r=r_{{lco}}$}.
} \label{recognize}
\end{figure}

Figure\il\ref{recognize}\emph{-right panel} illustrates the sign of the radial pressure gradient $G_r$ as
a function of the dimensionless angular momentum {$\mathbf{\lie \equiv L/(M\sigma)}$}  and distance from
the attractor $r$. As we noticed in the previous Section, ${G_r<0}$ (pressure decreasing)
in the range $2M<r\leq r_{lco}$ for every value of the angular momentum. When
$r>r_{lco}$,  ${G_r<0}$  (pressure decreasing) for $0<L\leqslant L_{\ti{K}}$ while ${G_r>0}$
(pressure increasing) for $L> L_{\ti{K}}$.

We are now interested in finding explicitly the critical points for the pressure, i.e.
to find the solutions to Eq.\il\ref{E:la}. For a test particle within the effective
potential $V_{sc}$, the angular momentum $L$ is a constant of motion. The particle
motion is then described by $V'_{sc}=0$ (the prime $(')$ stays for the derivative with
respect to $r$), which is equivalent to Eq.\il\ref{E:la}. Therefore, the circular
orbit radii for a test particle in Schwarzschild spacetime:
\be
\label{testpartradii}
r^{\pm}_{\ti{L}}\equiv \frac{L^2}{2  \sigma^2 M^2}\left(M\pm \sqrt{L^2-12
\sigma^2M^2}\right)=\frac{\lie^2}{2}\left(M\pm \sigma M \sqrt{\lie^2-12}\right)
\ee
 are solutions to Eq.\il\ref{E:la}. $r_{\ti{L}}^{+}$ corresponds to the test particle
 stable orbit, and $r_{\ti{L}}^{-}$ to the unstable orbit. The isobar surfaces ($G_r=0$)
 are therefore located at $r=r_{\ti{L}}^{\pm}$. {Figure\il\ref{recognize}\emph{left panel}} describes
 the behavior of $r_{\ti{L}}^{\pm}$ as a function of $L$ and $\sigma$. We underline that the
 matter distribution around the accretor is not spherically symmetric, and hence it is
 not independent on $\sigma$. A complete characterization of the equipotential surfaces as a
 function of $\sigma$ can be found in Appendix\il(\ref{Sec:app}).

\subsection{Angular momentum $L$ vs. fluid angular momentum $l$}\label{Sec:NeutraLl}

It is also possible to describe the motion of the fluid orbiting in the Schwarzschild
background in terms of its angular momentum $l$, defined as:
\be\label{ldef}
l\equiv\frac{g_{\varphi\varphi}}{g_{tt}}\frac{\Phi}{\Sigma}=\frac{L}{V_{sc}}\, ,
\ee
In fact, equipotential surfaces define the marginally stable configurations with respect
to the axisymmetric perturbation \citep{Seguin}, characterized by $l$ constant. These
configurations have been detailed studied in \citet{Koz-Jar-Abr:1978:ASTRA:}. From
Eq.\il\ref{ldef} and the definition of the effective potential in Eq.\il\ref{60} the
following relation holds:
\be\label{EA}
L^2=\frac{r^2\sigma^2}{e^{-\nu}r^2 \sigma^2l^{-2}-1}\quad (l\neq0)\, .
\ee
The boundary case $l=0$ corresponds to $L=0$.

\begin{figure}
\centering
\includegraphics[width=0.3\hsize,clip]{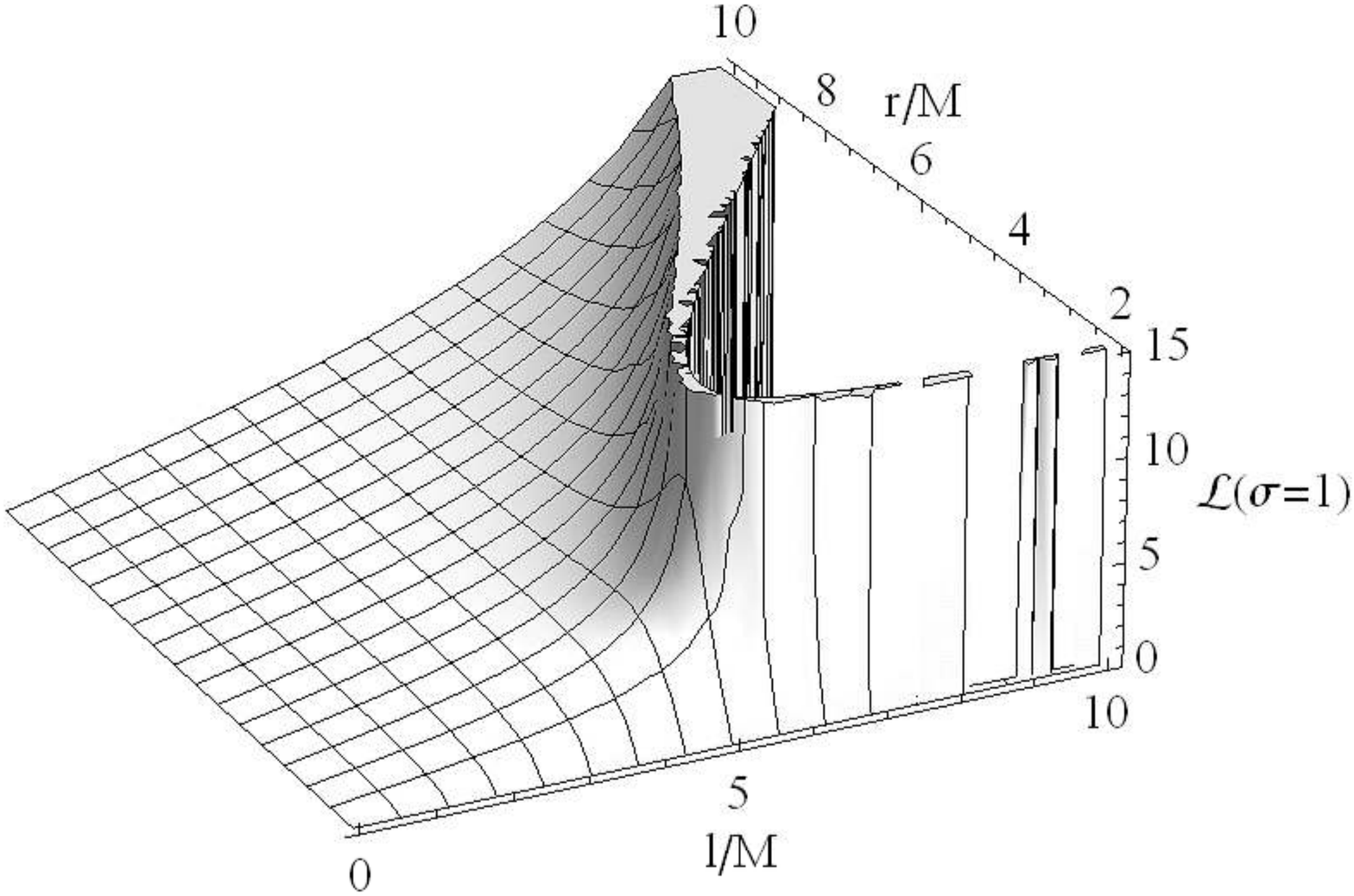}
\includegraphics[width=0.3\hsize,clip]{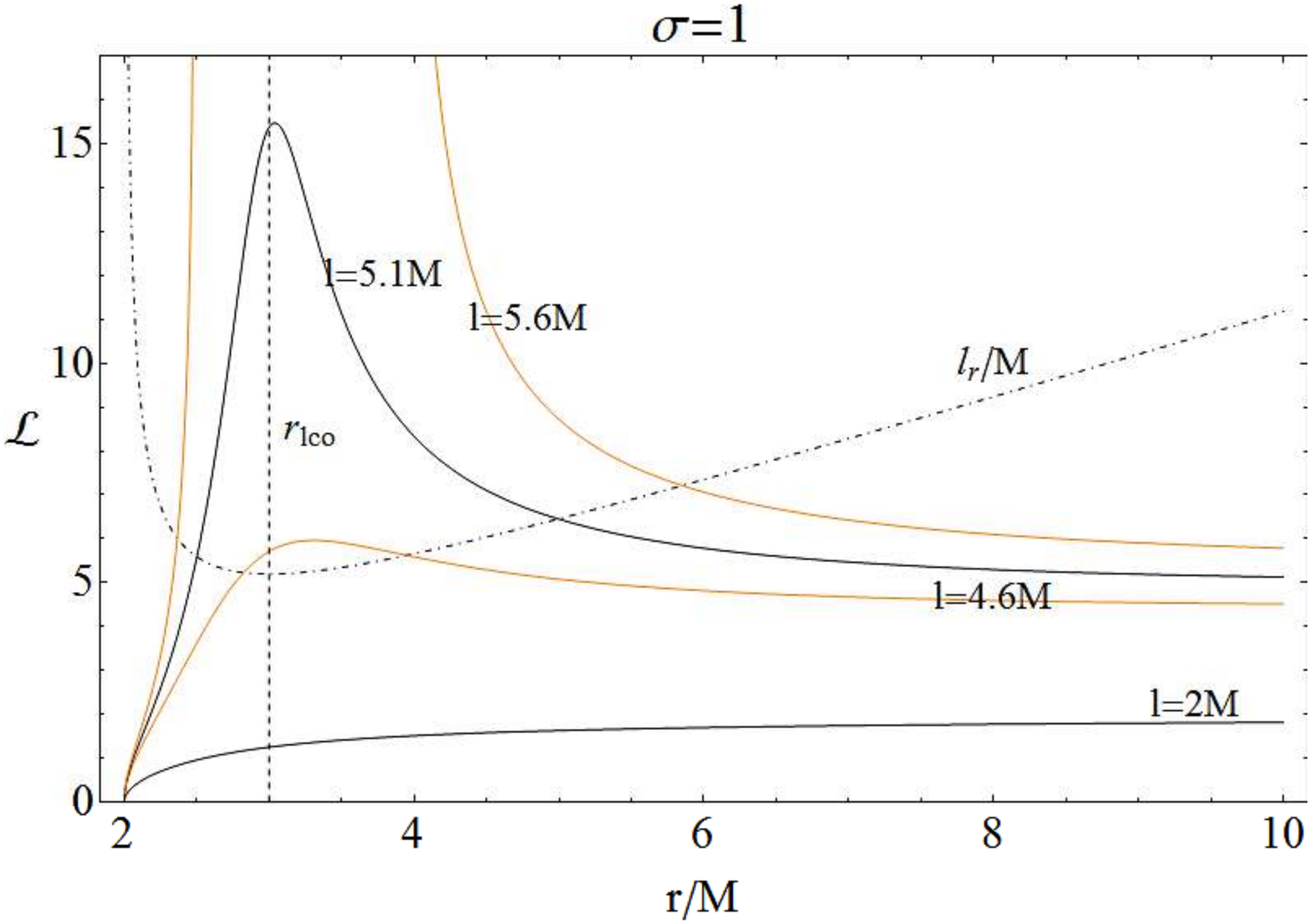}
\includegraphics[width=0.3\hsize,clip]{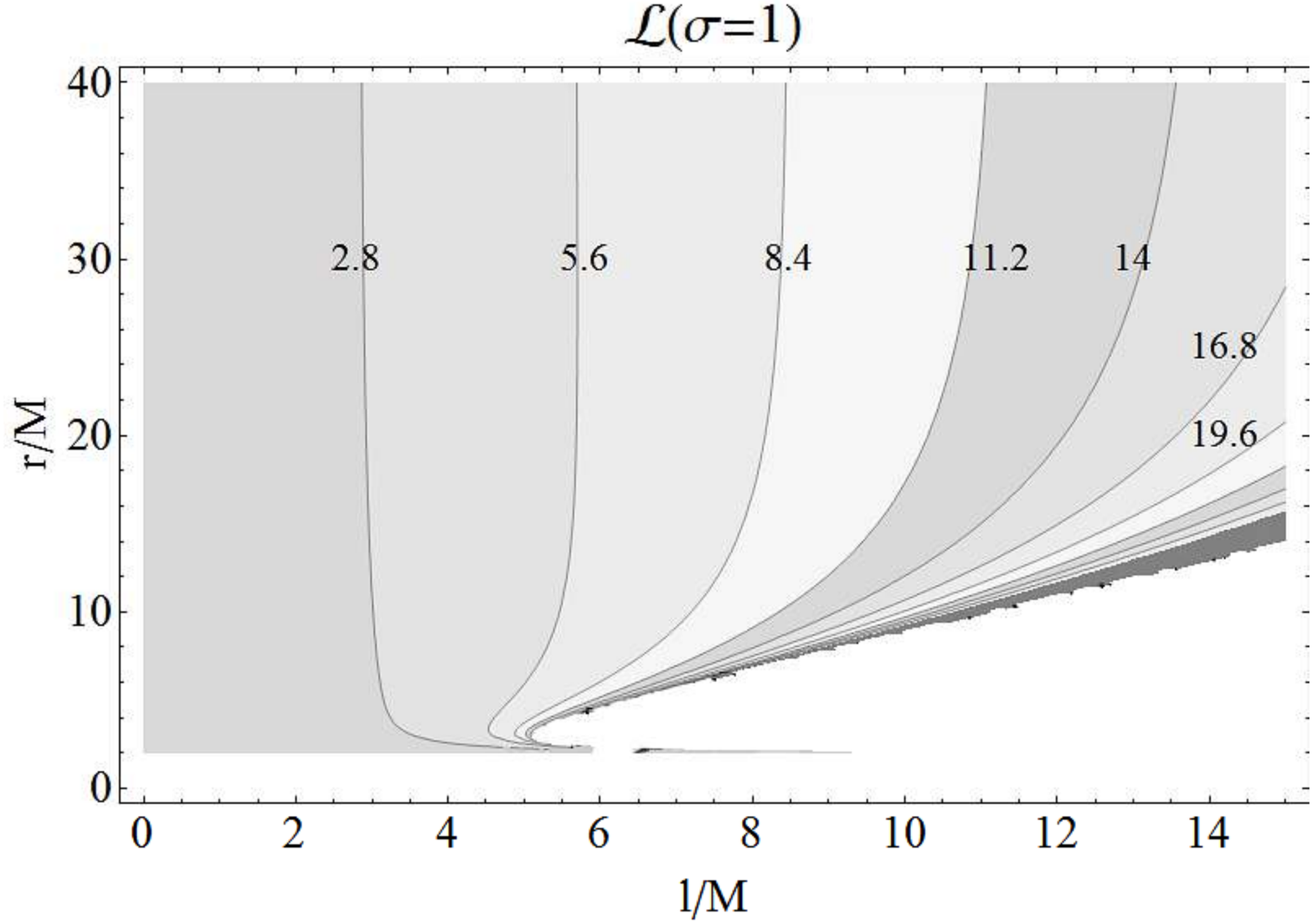}
\caption{{\emph{Left panel}: $\lie$ on the
equatorial plane $\sigma=1$ as a function of $l/M\in[0,10]$ and $r/M\in]2,10]$. \emph{Central
panel}: $\lie(\sigma=1)$ as a function of $r/M$, for different $l$. Asymptotically
($r\rightarrow\infty$), it is $L\rightarrow l$. The dotted-dashed curve marks $l=l_r$ as
a function of $r/M$: the angular momentum $L$ is defined for $l<l_r$. $r_{\ti{lco}}$ is
 a minimum of $l_r$. \emph{Right panel}: $\lie(\sigma=1)=\rm{constant}$
as a function of $r/M$ and $l/M$.}} \label{lampadabuia}
\end{figure}

We detail the angular momentum $L$ on the equatorial plane $\sigma=1$ as a function of
$l\neq0$ and $r$ in Figs.\il\ref{lampadabuia}. The two definitions of angular momentum
coincide in the asymptotic limit of flat spacetime ($r\rightarrow\infty$). It is
manifest from  Fig.\il\ref{lampadabuia}, \emph{central panel} that it is
$L=l=$constant for sufficiently large distances from the accretor. $L$ is not defined
everywhere in the plane $(r,l)$: from Eq.\il\ref{EA}, we notice that $L$ exists for
$0\leq l<l_{r}$, where:
\be
l_r\equiv\sqrt{\frac{\sigma^2 r^3}{(r-2M)}}\, ,
\ee
is defined for $r>2M$. In the boundary case $r=2M$, $L=0$. For fixed $\sigma$, $l_r$ is
increasing for $r>r_{lco}$, is minimum for $r=r_{lco}$, and is deceasing for
$2M<r<r_{lco}$. It is also progressively larger when approaching the equatorial plane
$\sigma=1$, where it is maximum.

\begin{figure}
\centering
\includegraphics[width=0.45\hsize,clip]{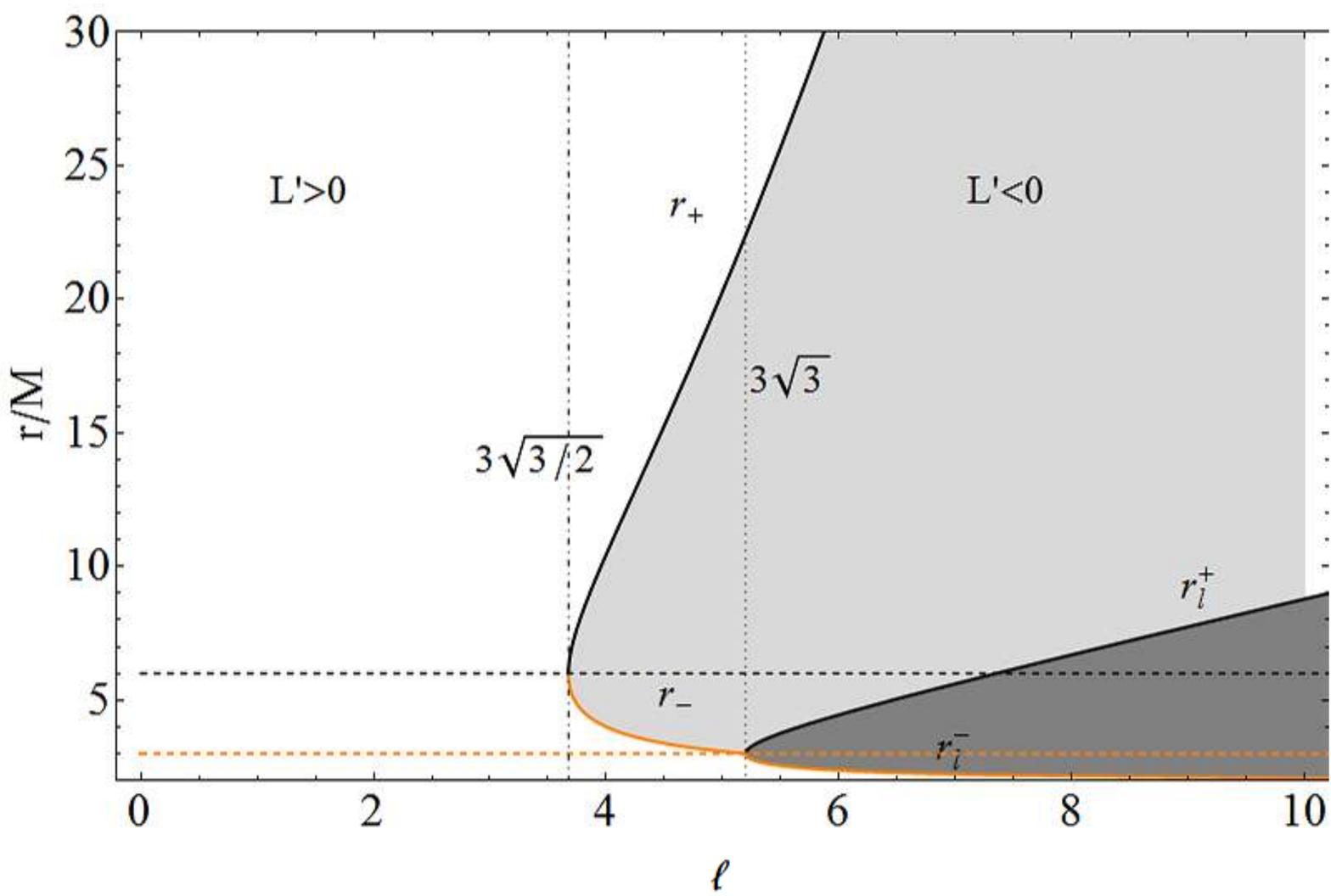}
\includegraphics[width=0.45\hsize,clip]{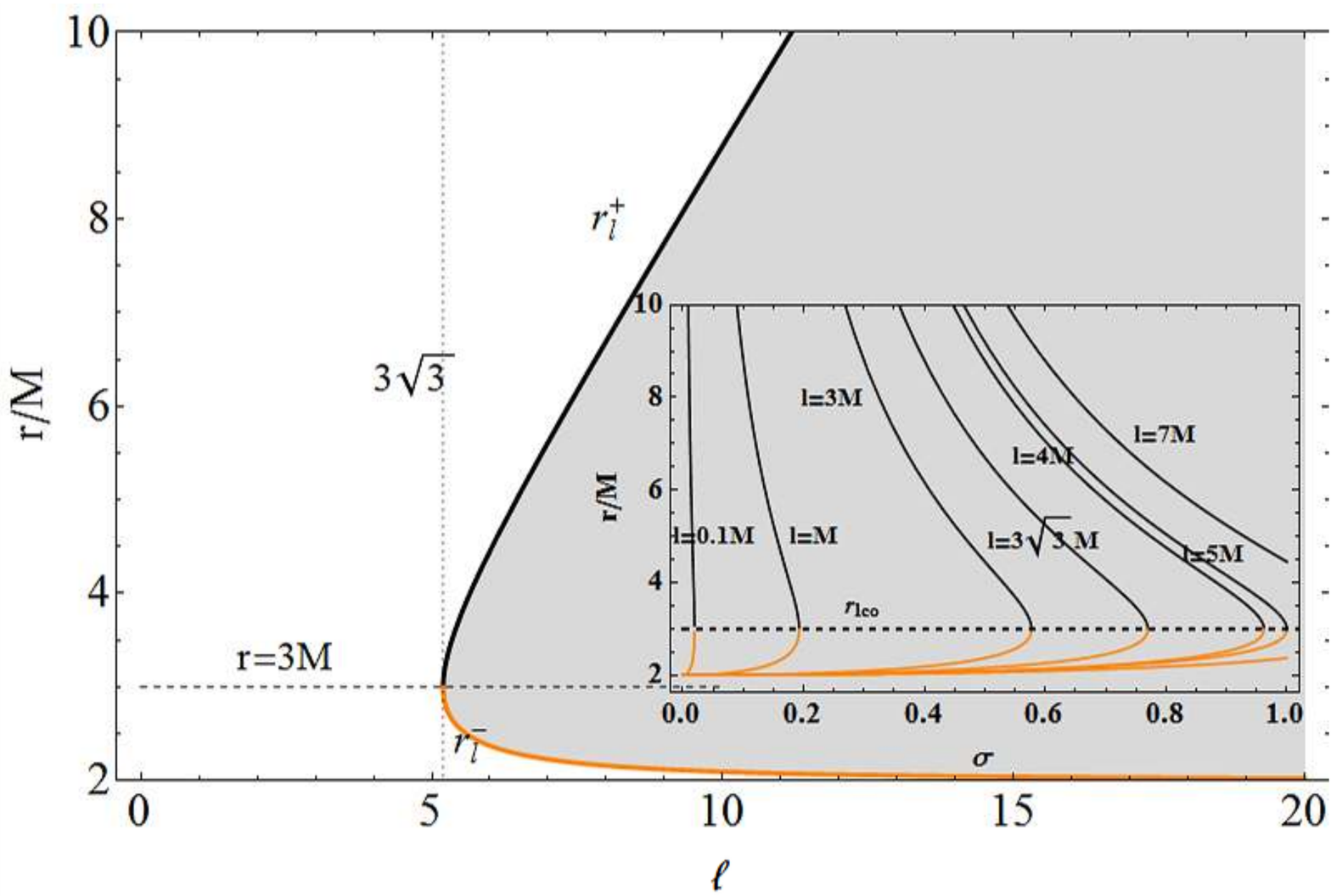}\\
\includegraphics[width=0.6\hsize,clip]{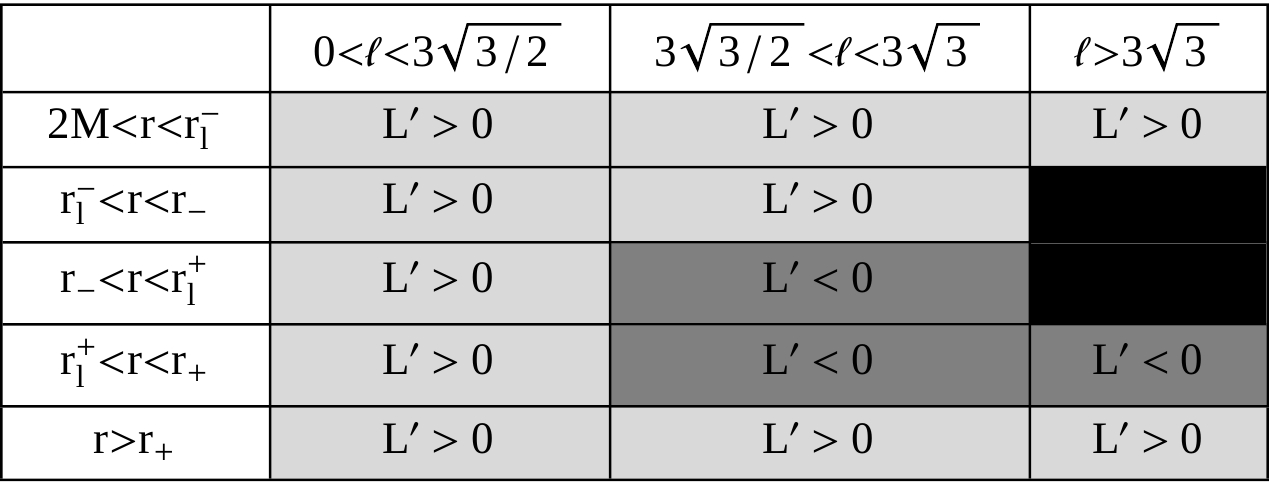}
\caption{{(Color online) {\emph{Upper left panel} $r_{\pm}$ and
$r^{\pm}_l$ as a function of $\ell$. $r_{+}=r_{-}=r_{lsco}=6M$  (dashed line) when $\ell=3
\sqrt{3/2}$ (dotted-dashed line). $r^{+}_l=r^{-}_l=r_{lco}=3M$ (dashed orange line) when
$\ell=3 \sqrt{3}$ (dotted line). White region corresponds to $L'>0$, light gray region
to $L'<0$. In the gray region the function $L$ is not defined. \emph{Upper right panel}
 $r_l^+$ (black curve) and
$r_l^-$ (orange curve) as a function of $\ell=l/(\sigma M)$. $r_l^+=r_l^-=3M$  (dashed line)
when $\ell=3 \sqrt{3}$ (dotted line). $L$ is not defined in the gray region.  \emph{Inset panel}  $r_l^+$ (black curve)
and $r_l^-$ (orange curve) as a function of $\sigma$ for different values of $l$. Dashed line
marks $r=r_{lco}$, where $r_l^+=r_l^-$. \emph{Lower panel} table
summarizing the intervals where $L'>0$ (light-gray) or $L'<0$ (gray). Black boxes
correspond to the interval $[r_{-},r_{+}]$, where the function $L$ is not defined.}}}
\label{tempo}
\end{figure}

We now face the problem of finding a relation to link $(r, \sigma)$ and $l$ to the condition
of the existence of $L$, and therefore of the velocity of the fluid $\Phi$. For this
purpose we re-define the radii in Eq.\il\ref{testpartradii}:
\bea\label{testpartradiil}
r^+_l/M&\equiv&\frac{2 \sqrt{\ell^2} \cos\left[\frac{1}{3} \arccos\left(-\frac{3
\sqrt{3}}{\sqrt{\ell^2}}\right)\right]}{\sqrt{3}},
\quad
r^-_l/M\equiv-\frac{2 {\sqrt{\ell^2}} \sin\left[\frac{\pi }{6}-\frac{1}{3}
\arccos\left(-\frac{3 \sqrt{3}}{{\sqrt{\ell^2}}}\right)\right]}{\sqrt{3}},
\eea
introducing the dimensionless quantity $\ell\equiv l/(\sigma M)>0$ ($\ell_r\equiv l_r/(\sigma M)$).
With this definition, $r^+_l=r^-_l=r_{lco}$ for $\ell=3\sqrt{3}$. $r^-_l$ approaches the
horizon $r=2M$ as the angular momentum  $\ell$ increases {(see Fig.\il\ref{tempo} \emph{upper right panel})}. The angular momentum $L$ and the velocity $\Phi$ are not defined
inside the region $[r^-_l,r^+_l]$ {(see Figs.\il\ref{tempo} \emph{bottom  panel} and \emph{upper left  panel})}. The
region $[r^-_l,r^+_l]$ increases with $\ell$. It varies also for different $\sigma$: its
behavior as a function of $\sigma$ is detailed in Appendix\il\ref{Sec:app}.

Now, we investigate the critical points of the angular momentum $L$ as function of $r$,
solutions of $L'=0$. For $r>r_{lco}$, $L$ is a constant with respect to the orbital
radius when $l=l_{\ti{K}}$, where
\be\label{lkkeple}
l_{\ti{K}}/M=\sqrt{\frac{\sigma^2 r^3}{M (r-2M)^2}}
\ee
is \emph{the Keplerian angular momentum} of the fluid. The critical points of $L$ are,
thus:
\bea\label{critrad}
r_-/M&\equiv&\frac{1}{3} \left(\ell^2-2 \sqrt{\ell^2 \left(\ell^2-12\right)}
\cos\left[\frac{1}{3} \left(\pi +\arccos\left[\frac{\ell^2 \left(54-18
\ell^2+\ell^4\right)}{\left(\ell^2
\left(\ell^2-12\right)\right)^{3/2}}\right]\right)\right]\right)\, ,\\
r_+/M&\equiv&\frac{1}{3} \left(\ell^2+2 \sqrt{\ell^2 \left(\ell^2-12\right)}
\cos\left[\frac{1}{3} \arccos\left(\frac{\ell^2 \left(54-18
\ell^2+\ell^4\right)}{\left[\ell^2
\left(\ell^2-12\right)\right]^{3/2}}\right)\right]\right)\, ,
\eea
where $r_-\leq r_+$, and $r_-=r_+=r_{lsco}$ when $\ell=3 \sqrt{{3}/{2}}$.

The angular momentum $L$ is a decreasing function of $r/M$ in all $r>r_{{lco}}$ with
$\ell_{\ti{K}}<\ell<\ell_r$, it increases with $r$ in $2M<r\leq r_{lco}$ with
$0<\ell<\ell_r$ and in $r>r_{lco}$ for $0<\ell<\ell_{\ti{K}}$ (see
Figs.\il\ref{tempo}).

\subsection{Radial pressure gradient $G_r(l)$ vs fluid angular momentum
$l$}\label{Sec:neutrall}

The radial gradient $G_r$, as a function of $\ell=l/(\sigma M)$, can be written as:

\be
{G_r =-\frac{M r^3-\ell^2 M^2(r-2M)^2}{r(r-2M)  \left[r^3-\ell^2 M^2 (r-2M)\right]}\, ,}
\ee

for $\sigma\neq0$ (see Figs.\il\ref{recognize}). It is not defined in
$\ell=\ell_r=\sqrt{r^3/M^2(r-2M)}$ and in $r=r^{\pm}_l$.

\begin{figure}
\centering
\includegraphics[width=0.45\hsize,clip]{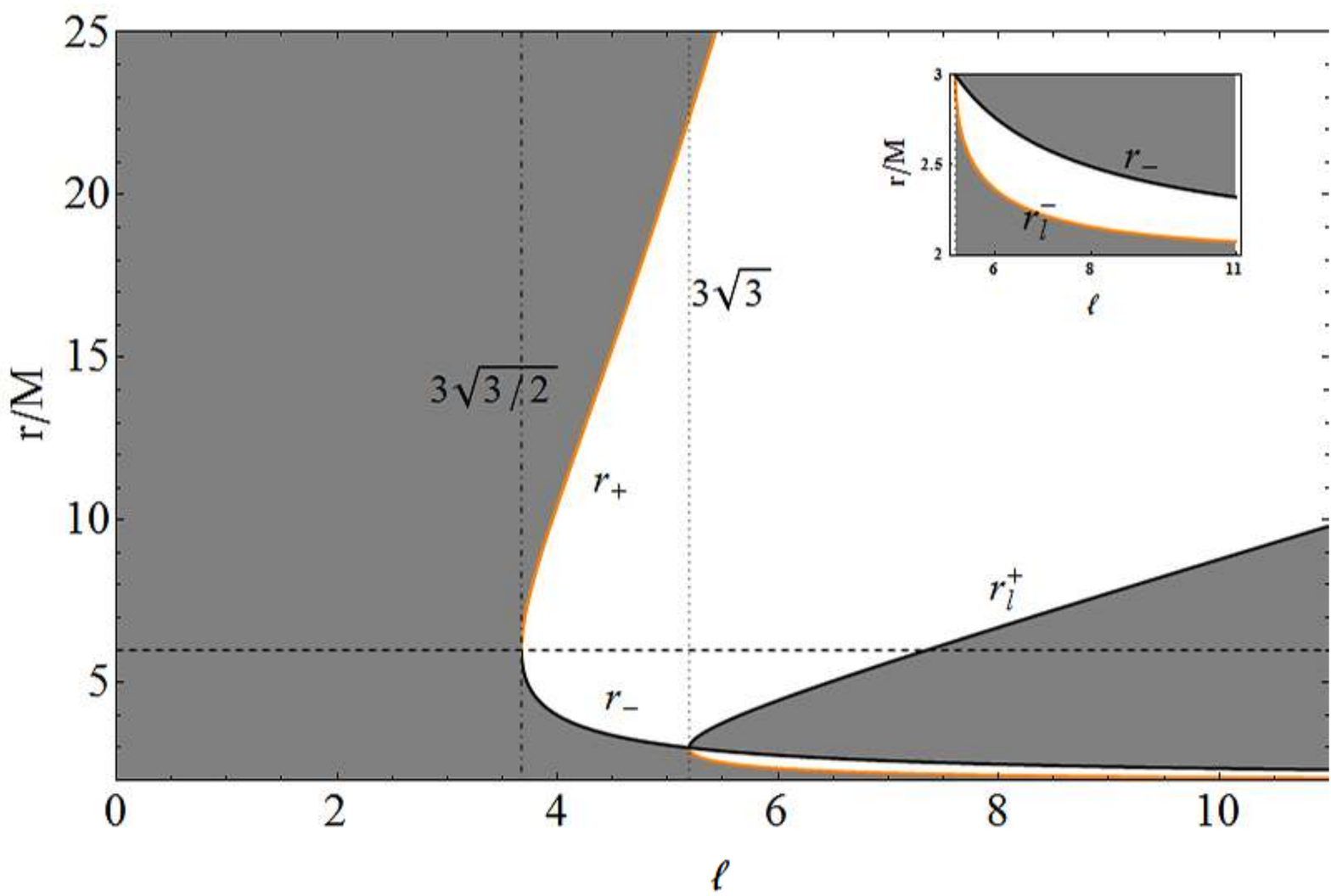}
\includegraphics[width=0.45\hsize,clip]{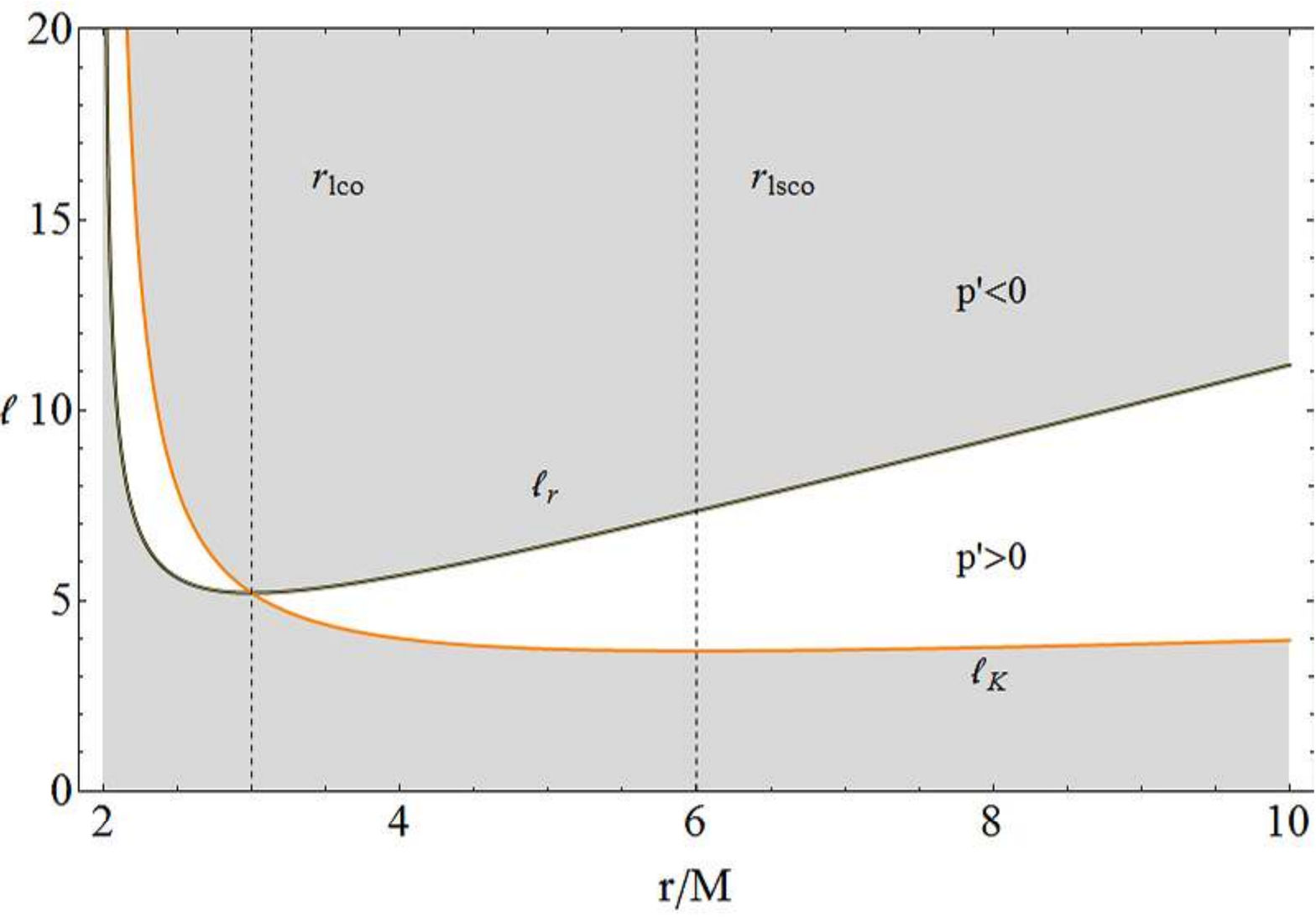}
\includegraphics[width=0.5\hsize,clip]{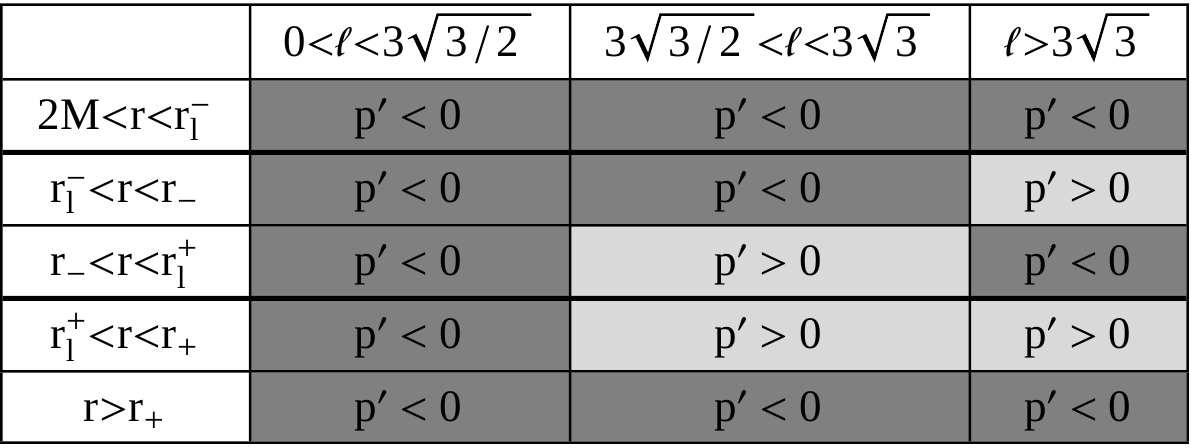}
\caption{{(Color online) {\emph{Upper left panel}: $r_{\pm}$ and
$r_l^{\pm}$ as a function of $\ell$.  White regions correspond to $p'>0$ and gray
regions to $p'<0$. Radii $r_l^{\pm}$ are defined for $\ell>3\sqrt{3}$: the gradient
$G_r$ is not defined in $r^{\pm}_l$. For $3\sqrt{3/2}<\ell<3\sqrt{3}$ the radius  $r_- $
is a minimum of the pressure, and  $r_+$ is a maximum. For $\ell>3\sqrt{3}$ the radius
$r_- $  and  $r_+ $ are both maximum of the pressure.  $r_{\pm}$, and
$r^{\pm}_l$ are plotted as a function of $\ell$, $r_{+}=r_{-}=r_{lsco}$  (dashed line) when $\ell=3 \sqrt{3/2}$ (dotted-dashed
line). $r^{+}_l=r^{-}_l=r_{lco}$ when $\ell=3 \sqrt{3}$ (dotted line). \emph{Inset:}
zoom of the region $r/M\in[2,3]$. \emph{Right panel}: $\ell_r$ (black line) and
$\ell_{\ti{K}}$ (orange line) as a function of $r/M$. $\ell_{r}=\ell_{\ti{K}}=3 \sqrt{3}$
in $r=r_{lco}$ (dashed line), where ${G_r=-(1/3) M^{-1}}$. The function is not defined in
$\ell=\ell_r$, while $G_r=0$ in $\ell=\ell_{\ti{K}}$.  In the gray regions $p'<0$, in
the white regions $p'>0$ . The angular momentum $\ell_{\ti{K}}$  has a {minimum} in
$r_{lsco}$ (dashed line). \emph{Lower panel}: table
summarizing the regions of increasing ($p'>0$--light-gray shaded regions) and decreasing
($p'<0$--gray shaded regions) pressure $p$.}}} \label{filodiff}
\end{figure}

We studied the sign of $G_r$ as a function of $L$ in Sect.\il\ref{Sec:Neutral-L0}: here
we face the problem for $G_r$ as a function of $\ell$, the fluid constant of motion. In
Appendix\il\ref{Sec:app}, we will discuss the $G_r$ sign in terms of $(l,\sigma,r)$
explicitly. According with the results found in Sect.\il\ref{Sec:NeutraLl}, $G_r =0$
(isobar surfaces) for the radii  $r_{\pm}$ with Keplerian angular momentum
$\ell_{\ti{K}}$. Thus, ${G_r<0}$ in:
\bea
2M<r\leq r_{lco}\quad\mbox{for}\quad 0\leq \ell<\ell_r\quad\mbox{and}\quad
\ell>\ell_{\ti{K}}\, ,
\nonumber \\
r>r_{lco}\quad\mbox{for}\quad 0\leq \ell<\ell_{\ti{K}}\quad\mbox{and}\quad \ell>\ell_r\,
.\nonumber
\eea
In terms of the radii $r_{\pm}$  and $r^{\pm}_l$,  ${G_r>0}$ in:
\bea
0\leq \ell<3 \sqrt{{3}/{2}}&\quad\mbox{in}\quad&r>2M\, ,
\nonumber \\
\ell=3 \sqrt{{3}/{2}}&\quad\mbox{in}\quad&r>2M, \quad r\neq r_{lsco}\, ,
\nonumber \\
3 \sqrt{{3}/{2}}<\ell\leq 3 \sqrt{3}&\quad\mbox{in}\quad& 2M<r<r_-,\quad r>r_+\, ,
\nonumber \\
\ell>3 \sqrt{3}&\quad\mbox{in}\quad& 2<r<r_l^-,\quad r_-<r<r_l^+,\quad r>r_+\,
.\nonumber
\eea
These intervals are portrayed in Figs.\il\ref{filodiff} \emph{upper panels}.

In Sect.\il\ref{Sec:NeutraLl}  we verified  that  $L'(l_{\ti{K}})=0$, where
$L(l_{\ti{K}})=L_{\ti{{K}}}$ for $r>r_{lco}$. Here  we showed  that  $l_{\ti{K}}$
satisfies the condition  $G_{r}(l_{\ti{K}})=0$ and therefore we can claim that
$l_{\ti{K}}$ is also a  critical  point  for   the pressure $p$. This is illustrated in
Figs.\il\ref{filodiff}.

\section{The polar angular pressure gradient $G_{\vartheta}$}\label{Sec:ella}

We now concentrate our attention on  the polar angular pressure gradient
$G_{\vartheta}$. From Eq.\il\ref{vaperfi}:

\be\label{E:possilea}
{G_{\vartheta}\equiv\frac{\nabla_{\vartheta}p}{\rho+p}\, }
\ee
On the equatorial plane $\sigma=1$, it is $\nabla_{\vartheta}p=0$: the angular gradient of
pressure on the equatorial plane is always zero. $\partial_{\vartheta}p=0$ also for
$L=0$. This implies that the case $L=0$ leads to a zero polar gradient of $p$, while
 ${G_r(L=0)=-{M}/((r-2M) r)<0}$, i.e. the pressure decreases as $1/r^2$  as approaching the
horizon $r=2M$.

\begin{figure}
\centering
\includegraphics[width=0.45\hsize,clip]{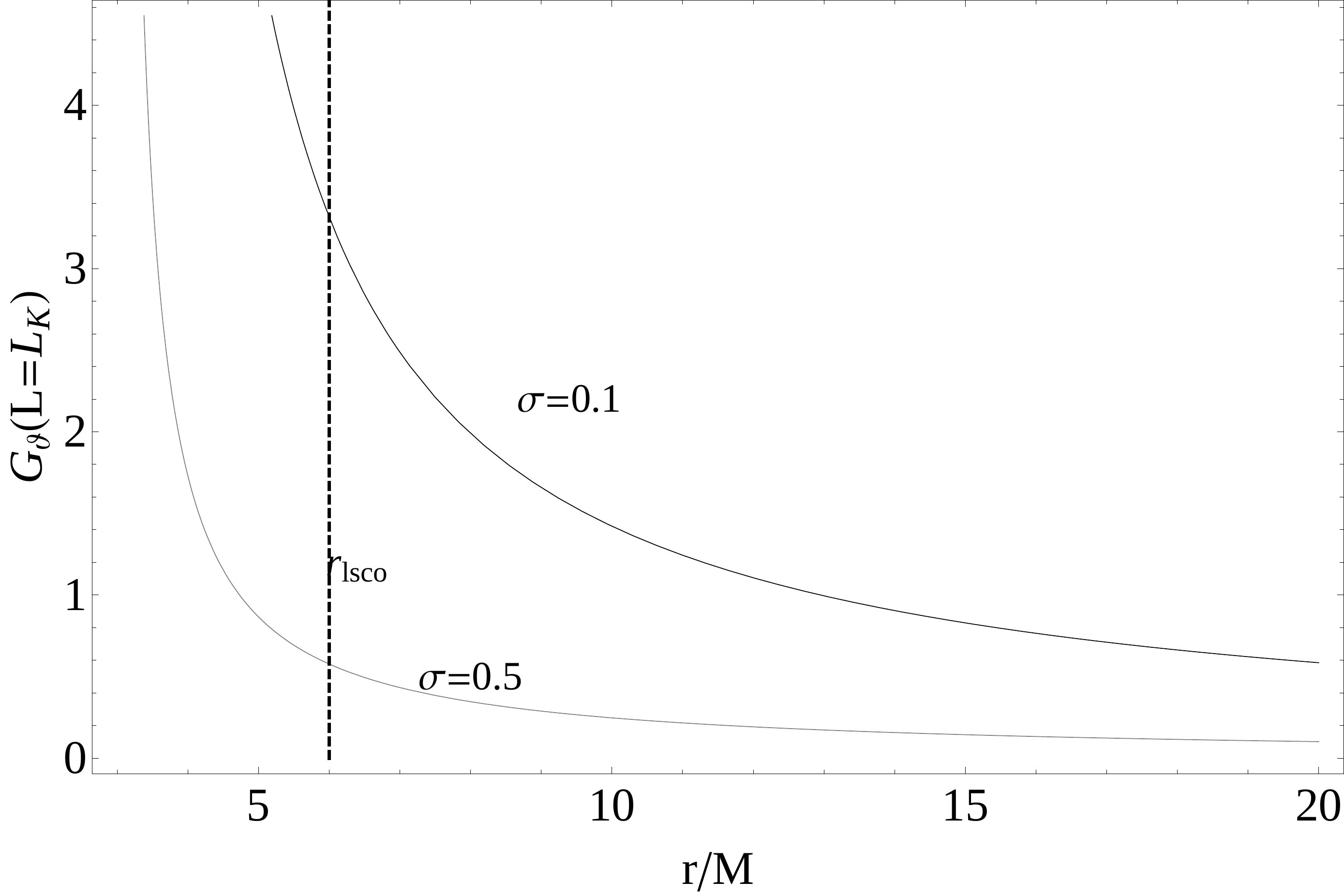}
\includegraphics[width=0.45\hsize,clip]{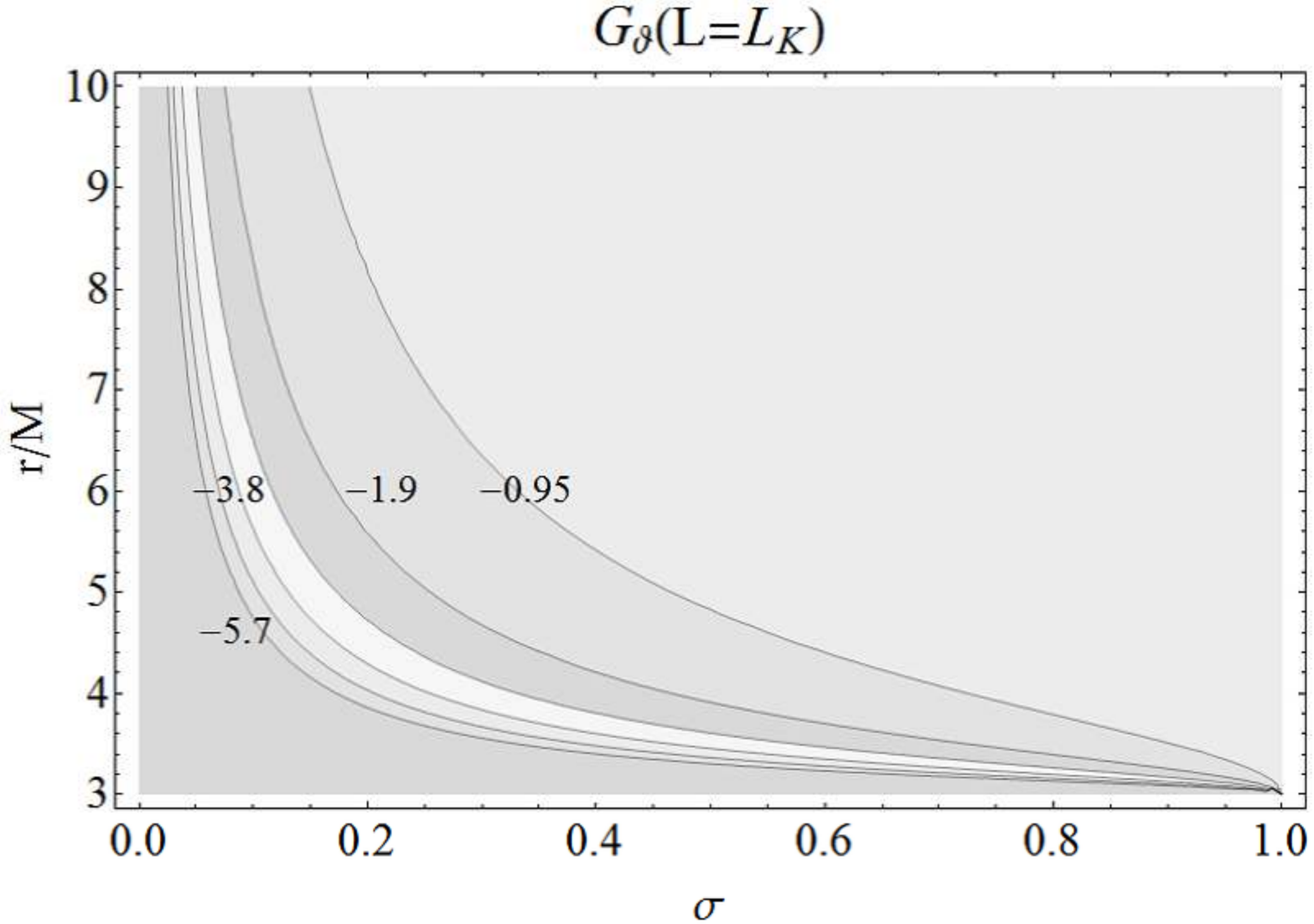}
\caption{{
{\emph{left
panel}: $G_{\vartheta}(L=L_{\ti{K}})$ as a function of $r/M\in]3,20]$ for different
values of $\sigma$. The black dashed line marks $r=r_{lsco}$. No solutions of
$\partial_\vartheta p=0$, and no test particle circular orbits exist for $r/M\in(2,3]$.
\emph{Right panel}: Curves $G_{\vartheta}(L=L_{\ti{K}})=$constant  as a function of
$r/M$ and $\sigma$.}}} \label{lampadat}
\end{figure}

For $L=L_{\ti{K}}$, $\partial_rp=0$ and therefore $p$ is a function of $\vartheta$ only.
In general, for $\sigma\neq1$ it is  $\nabla_{\vartheta}p(L=L_{\ti{K}})\neq0$. In fact it
results:
\be\label{E:gooo}
G_{\vartheta}(L=L_{\ti{K}})\equiv\frac{
\nabla_{\vartheta}p(L_{\ti{K}})}{\rho+p}=\frac{M\sqrt{1-\sigma^2}}{\sigma(r-3M)},
\ee
(see Figs.\il\ref{lampadat})\footnote{$G_{\vartheta}(L=L_{\ti{K}})$ is written as an
odd symmetric function of $\sigma$, but we are in an even symmetric theory in $\sigma$, as we have
not explicitly given any further constraints to the unknown functions $(\rho, p)$, we
implicitly take $\sigma>0$ and $\sqrt{1-\sigma^2}\geq0$ or $\vartheta\in (0,\pi/2]$.}. The
quantity in Eq.\il\ref{E:gooo} is not defined in $r=r_{{lco}}$ and, as for
$L_{\ti{K}}$, in the region $r\in(2M,r_{lco}]$. This  means that there is a fluid
configuration with pressure constant along the orbital radius  and a pressure variable
from plane to plane extended at radial distance $r>r_{lco}$.
$G_{\vartheta}(L=L_{\ti{K}})$ admits critical points for $\sigma=1$.

\subsection{Pressure gradient ratio $\Pi_{ r \vartheta}$ vs angular momentum $L$}

\begin{figure}
\centering
\includegraphics[width=0.45\hsize,clip]{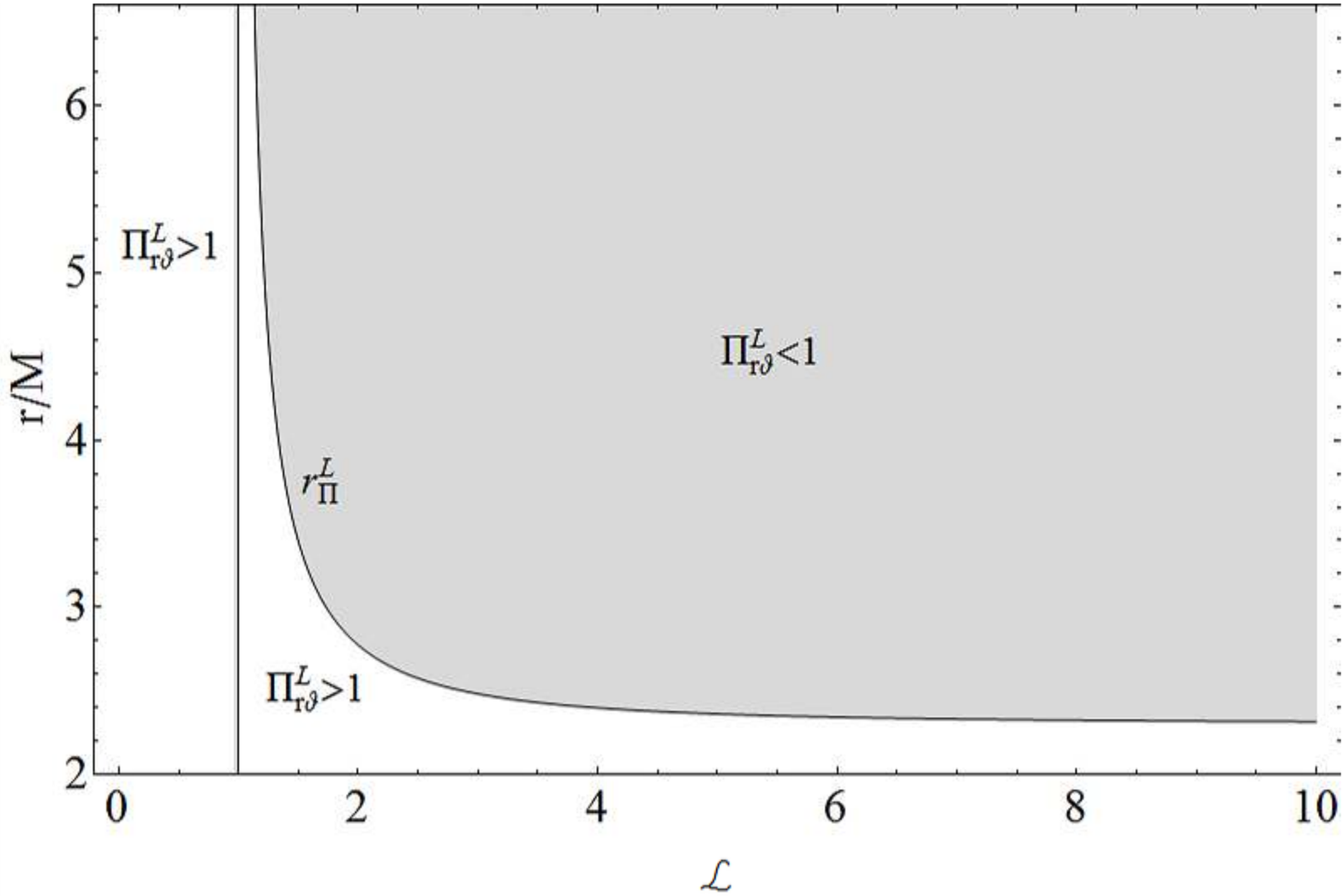}
\includegraphics[width=0.45\hsize,clip]{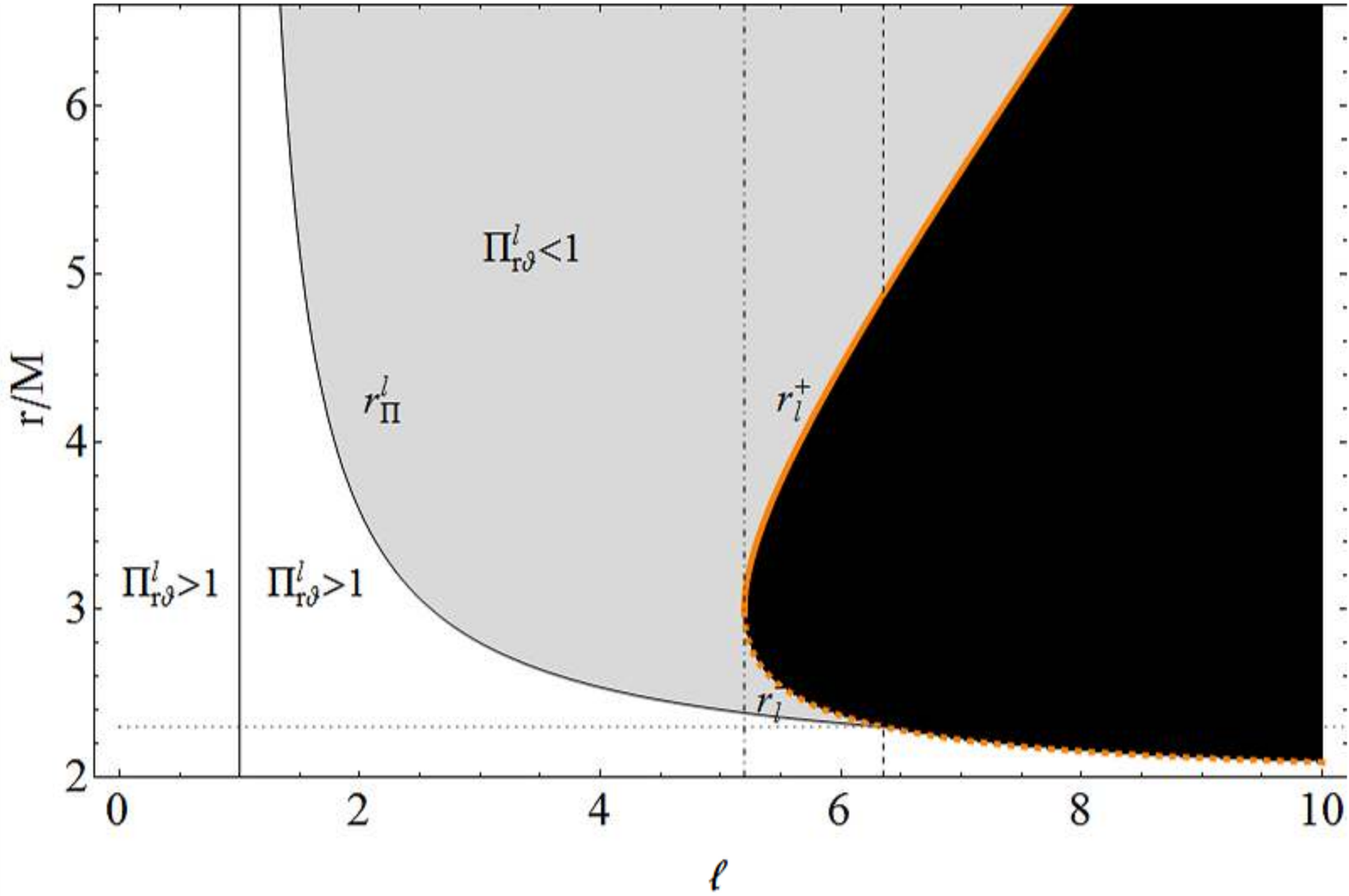}
\caption{(Color online) \emph{Left panel}: $r_\Pi^L$ as a function of $ \lie$.
Black line marks $ \lie=1$.  In the light gray region $\Pi^L_{r\vartheta}<1$,  in the
white region $\Pi^L_{r\vartheta}>1$. \emph{Right panel}: $r_\Pi^l$, $r^{\pm}_l$ as a
function of $\ell$. Black line marks the value $\ell=1$, dashed line $\ell\simeq6.35$,
dotted-dashed line $\ell=3\sqrt{3}$, dotted line $r=\frac{1}{2}
\left(1+\sqrt{13}\right)M$. In the black region $r^-_l<r<r^+_l$ the function $\Pi^l_{r
\vartheta}$ is not defined. In the gray region $\Pi^l_{r \vartheta}<1$, in the white
region $\Pi^l_{r \vartheta}>1$. In $r=r_\Pi^l$ (black line) it is  $\Pi^l_{r
\vartheta}=1$.
}
\label{dipinto}
\end{figure}

We define the pressure gradient ratio $\Pi_{ r \vartheta}=\Pi_{ r \vartheta}(L,r,\sigma)$ as
$\Pi_{r\vartheta}\equiv{\nabla_rp}/{\nabla_{\vartheta}p}$. It is clearly $\Pi_{ r
\vartheta}=0$ for $L=L_{\ti{K}}$ and it is not defined in $\sigma=1$. We rewrite this
function as:
\be
\Pi^L_{r\vartheta}\equiv\left|\cot\vartheta\Pi_{r \vartheta} M\right|=\frac{\left|  r^2-
M \lie^2 (r-3M)\right|}{ \lie^2  r (r-2M) },
\ee
This ratio is equal to one for:
\be
 \lie_\Pi\equiv\sqrt{\frac{r^2}{r(r-M)-3M^2}}\, ,\quad r_\Pi^L/M\equiv\frac{1}{2}
 \left[\frac{ \lie^2}{ \lie^2-1}+\sqrt{\frac{ \lie^2 \left(13  \lie^2-12\right)}{\left(
 \lie^2-1\right)^2}}\right]\, .
\ee
The ranges where $\Pi^L_{r\vartheta}$ is larger, lower or equal to one are summarized in
Fig.\il\ref{dipinto}, \emph{left panel}.

\subsection{Analysis of the pressure gradient ratio $\Pi_{ r \vartheta}$ vs angular
momentum $l$}

We consider now the ratio $\Pi_{r \vartheta}$ as function of $l$. We then define the
function $\Pi^l_{r \vartheta}=\Pi^l_{r \vartheta}(r, \ell)$:
\be
\Pi^l_{r \vartheta}\equiv \left|\cot\vartheta M \Pi_{r \vartheta}\right|=\frac{
\left|r^3-\ell^2 M(r-2M)^2\right|}{ \ell^2 r  (r-2M)^2 },
\ee
in the range of existence of $L(l)$. Then we exclude the range $r^-_l\leq r\leq r^+_l$.
The solution of $\Pi^l_{r \vartheta}=1$ are $r=r_\Pi^l>M(1+\sqrt{13})/2$ and
$\ell=\ell_\Pi\in(1,\ell_{(\Pi-)})$, where $\ell_{(\Pi-)}=\sqrt{ \left(41+11
\sqrt{13}\right)/2}=6.35063$ satisfies the condition
$r_\Pi^l(\ell_{(\Pi-)})=r_l^-(\ell_{(\Pi-)})\approx2.30278\,M$ (see
Fig.\il\ref{dipinto}, \emph{right panel}), with:
\bea
\ell_\Pi\equiv\sqrt{\frac{r^3}{(r+M) (r-2M)^2 }}\, ,
\eea
and
\bea
r_\Pi^l/M\equiv\frac{\ell^2}{\ell^2-1}+2 \ell^2\sqrt{\frac{1}{\left(\ell^2-1\right)^2}}
\cos\left(\frac{1}{3}
\arccos\left[\frac{1}{\ell^4}\sqrt{\frac{1}{\left(\ell^2-1\right)^2}}
\left(1-\ell^2\right) \left(\ell^4-4 \ell^2-2\right)\right]\right)\, .
\eea
The ranges where $\Pi^l_{r\vartheta}$ is larger, lower or equal to one are summarized in
Fig.\il\ref{dipinto}, \emph{right panel}.

\section{The Boyer potential}\label{Sec:Poli}

The disk fluid configuration in the Polish doughnut model has been widely studied by
many authors (see for example \citealt{Raine,Abramowicz:2011xu}). In particular, an
analytic theory of equilibrium configurations of rotating perfect fluid bodies was
initially developed by \citet{Boy:1965:PCPS:}. The ``Boyer's condition'' states
that the boundary of any stationary, barotropic, perfect fluid body has to be an
equipotential surface $W(l,\vartheta)=\mbox{constant}$. For a barotropic fluid the
surfaces of constant pressure are given by the equipotential surfaces of the potential
defined by the relation:
\be\label{boyereq}
\int_0^{p_{in}}\frac{dp}{\rho+p}=-(W-W_{in}),
\ee
where the subscript ``in''  refers to the inner edge of the disc. It is important to
notice here that, in the newtonian  limit,  the quantity $W$ is equal to the total
potential, i.e. to the sum of the gravitational and of the centrifugal effects.

As mentioned in Sect.\il\ref{Sec:NeutraLl}, all the main features of the equipotential
surfaces for a generic rotation law $\Omega=\Omega(l)$ are described by the
equipotential surface of the simplest configuration with uniform distribution of the
angular momentum density $l$, which are very important being marginally stable
\citep{Seguin}. At the same time, the equipotential surfaces of the marginally stable
configurations orbiting in a Schwarzschild spacetime are defined by the constant $l$. It
is therefore important to study the potential $W=W(l,\sigma)$ and to compare it with
$W=W(L,\sigma)$.

We can classify the equipotential surfaces in three classes: \emph{closed}, \emph{open},
and \emph{with a cusp} (self-crossing surfaces, which can be either closed or open). The
closed equipotential surfaces determine stationary equilibrium configurations: the fluid
can fill any closed surface. The open equipotential surfaces are important to model some
dynamical situations, for example the formation of  jets
\citep[see e.g.][]{2011,arXiv:0910.3184,Lei:2008ui,Abramowicz:1997sg,astro-ph/0605094,Abramowicz:2009bh,
Stu-Kov:2008:INTJMD:,Rez-Zan-Fon:2003:ASTRA:,Stu:2000:ACTPS2:,
Sla-Stu:2005:CLAQG:,Stu-Sla-Hle:2000:ASTRA:}.

The critical, self-crossing and closed equipotential surfaces are relevant in the theory
of thick accretion disks, since the accretion onto the black hole can occur through the
cusp of the equipotential surface. According to Paczy\'nski
\citep{Abr-Jar-Sik:1978:ASTRA:,Koz-Jar-Abr:1978:ASTRA:,Jaroszynski(1980),A1981}, the
accretion onto the source (black hole) is driven through the vicinity of the cusp due to
a little overcoming of the critical equipotential surface $W_{cusp}$ by the surface of
the disk. The accretion is thus driven by a violation of the hydrostatic equilibrium,
clearly  ruling  out the viscosity as a basis for accretion
\citep{Koz-Jar-Abr:1978:ASTRA:}. In the Paczy\'nski mechanism  the disk surface exceeds
the critical equipotential surface $W_{cusp}$ giving rise to a mechanical
non-equilibrium process that allows the  matter inflow  into the black hole. In this
accretion model the cusp of this equipotential surface corresponds to the inner edge of
the disk.

We calculate now the Boyer potential for our system integrating Eq.\il\ref{olgettin}:
\be\label{equazp}
\int^{p_{out}}_{p_{in}}\frac{dp}{\rho+p}=-\int_{r_{in}}^{r_{out}}\frac{e^{-\nu}}{2}\left(\frac{\partial
V_{sc}^2}{\partial r}\right)_{L}.
\ee
The integration range $[r_{in},r_{out}]\subset[2M, \infty]$ is the range of existence
for $V_{sc}(r,l)$, $(r-2M) \left[r^3-\ell^2 M^2(r-2M)\right]>0$. This condition implies
that we are excluding the range $[r_l^-, r_l^+]$. The general integral is:
\be
\int\frac{dp}{\rho+p}=-W,
\ee
where
\be\label{Edverdi}
W=\ln\left[\sqrt{\frac{(r-2M) r^2}{r^3-\ell^2 M^2 (r-2M)}}\right]=\ln V_{{sc}}.
\ee
In particular it results:
\be
\lim_{r\rightarrow\infty}W=0,\quad\lim_{r\rightarrow2M}W=+\infty
\ee
\citep[see also][]{Stu-Sla-Hle:2000:ASTRA:, Stu-Kov:2008:INTJMD:,
astro-ph/0411185,astro-ph/0605094, Sla-Stu:2005:CLAQG:,
Abr-Cal-Nob:1980:ASTRJ2:,Rez-Zan-Fon:2003:ASTRA:}. Clearly it is $W=0$ where
$V_{sc}=1$, and $W\gtrless0$ where $V_{sc}\gtrless1$. Its maxima and minima are the same
than the Schwarzschild effective potential, $V_{sc}=V_{sc}(l,r)$. In particular, we are
interested to study the equipotential surfaces, defined by the condition
$W=\rm{constant}$, that coincide with the surfaces $V_{sc}=\rm{K}>0$, being
$\rm{K}\equiv e^c$ the energy of a test particle circularly orbiting around the source.
In fact the surfaces of constant Boyer potential determine the shape of the torus
(disc). We study these surfaces as a function of $L=\mbox{constant}$ and
$l=\mbox{constant}$, respectively.

\begin{figure}
\centering
\includegraphics[width=0.45\hsize,clip]{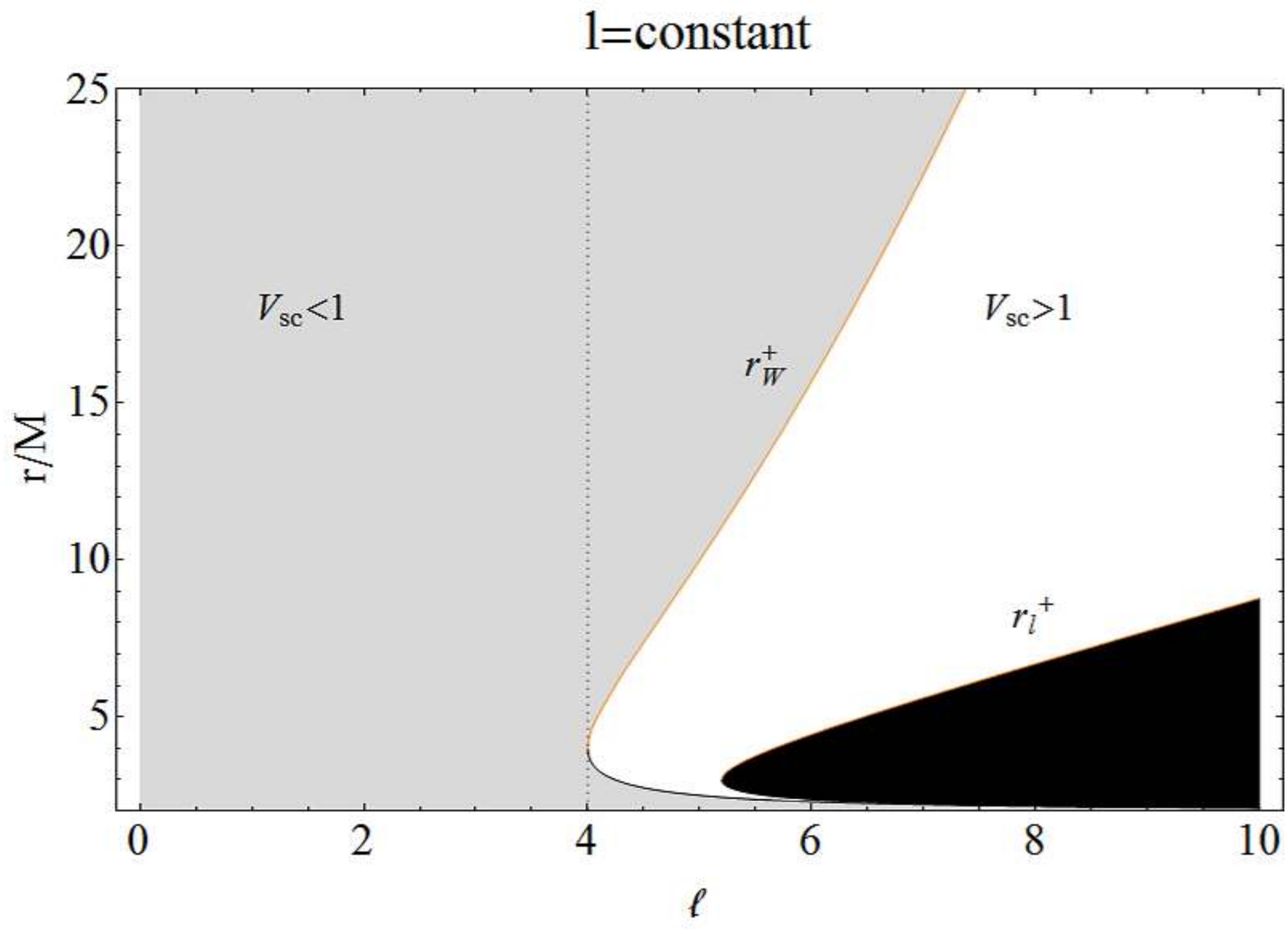}
\includegraphics[width=0.45\hsize,clip]{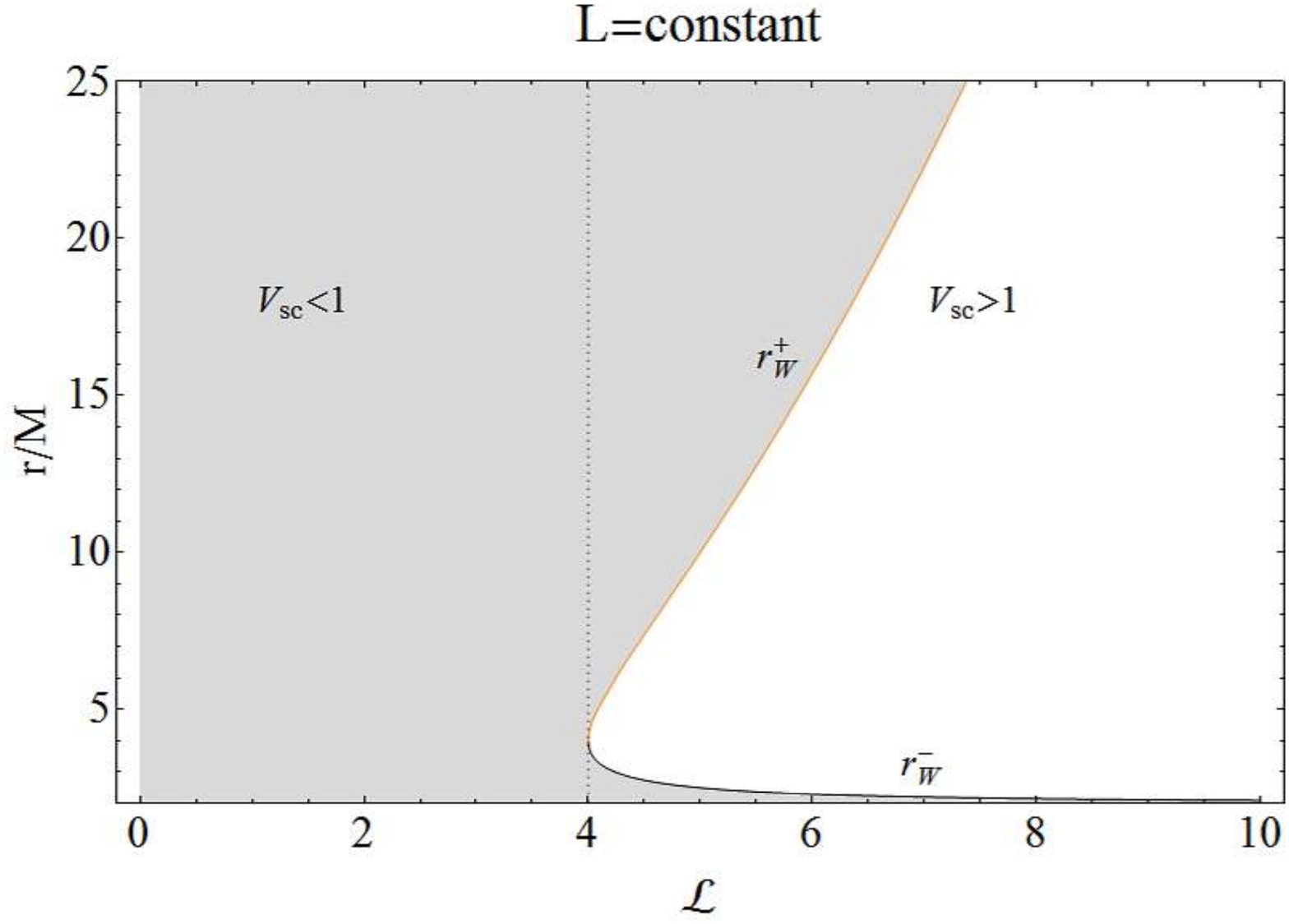}
\caption{(Color online) \emph{Left panel}: $r_W^+$ (orange curve) and $r_W^-$
(black curve) as a function of $\ell$. The black region $r^-_l<r<r_l^+$) is forbidden.
$V_{sc}>1$ in the white region, $V_{sc}<1$ in the gray region. Dotted line marks $
\ell=4$. \emph{Right panel}: $\tilde{r}_{W}^+$, (orange curve) and $\tilde{r}_{W}^-$
(black curve) as function of $ \lie$. $V_{sc}>1$ in the white region, $V_{sc}<1$ in the
gray region. Dotted line marks $ \lie=4$.}
\label{restyling}
\end{figure}

\subsection{Analysis  of the Boyer potential vs the angular momentum $L$}

We consider  the Boyer potential  $W(L,r)$ in  Eq.\il\ref{Edverdi} as function of  the
angular momentum $L$. The condition $W=0$ is satisfied in:
\bea
\tilde{r}_{W}^{\pm}&\equiv&\frac{M}{4}  \left(\lie^2\pm\sqrt{\lie^2(\lie^2-16)}\right)
\eea
(see Fig.\il\ref{restyling}, \emph{right panel}).

We notice that two relevant cases occur when the angular momentum is
${\lie^2=\lie_{k}^2}$, where ${{\lie}_{k}^2\equiv \frac{r^2
\left[2M+\left(\rm{K}^2-1\right) r\right]}{M^2(r-2M)}}$:
\textbf{\texttt{(I)}} when $0<\rm{K}<1$, $W<0$, in  $2M< r \leq \tilde{r}_{\rm{K}}$,
where $\tilde{r}_{\rm{K}}\equiv\frac{2M}{1-\rm{K}^2}$; \texttt{\textbf{(II)}} when
$\rm{K}\geq1$, $W\geq0$  in $r>2M$.

\begin{figure}
\centering
\includegraphics[width=0.3\hsize,clip]{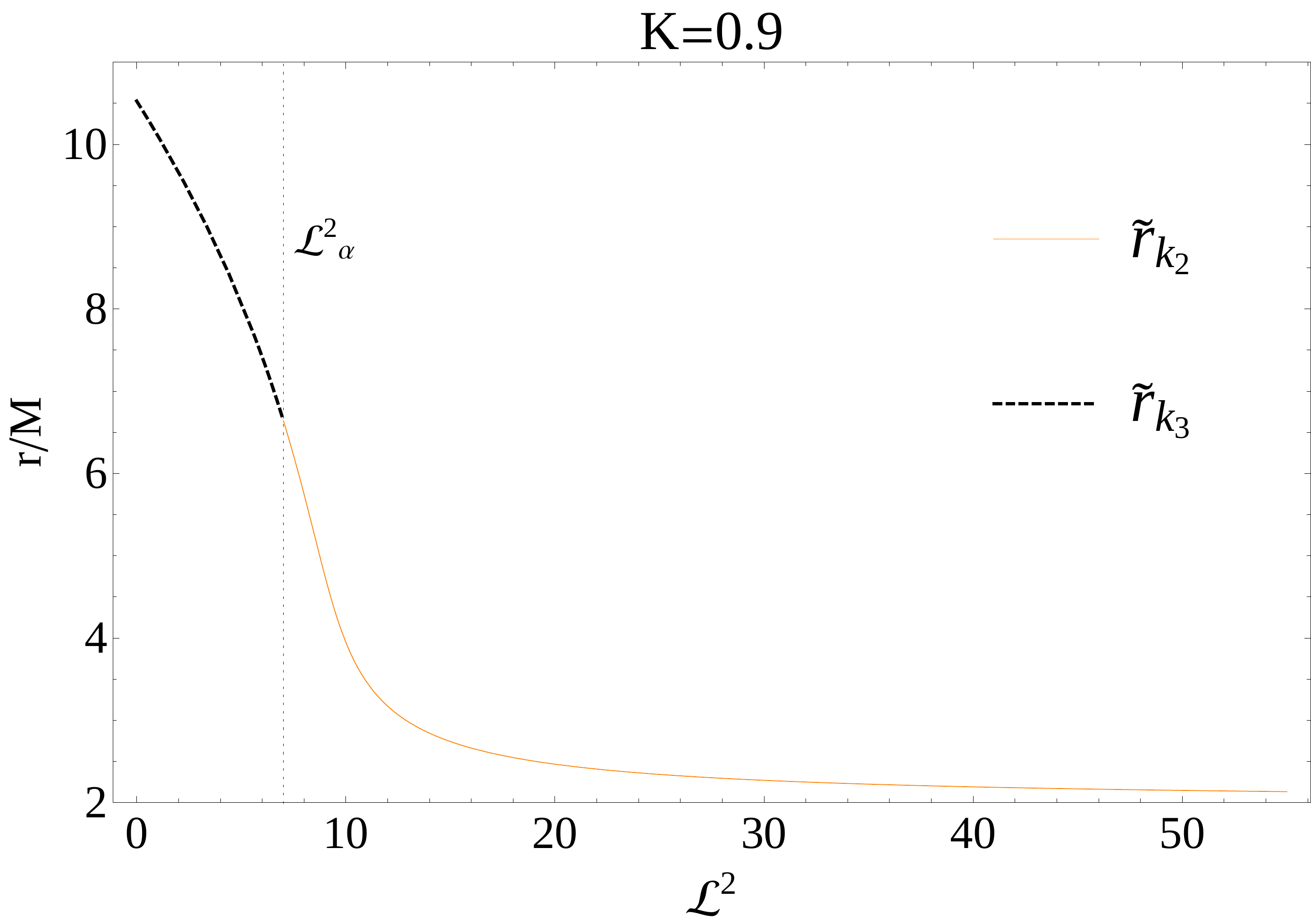}
\includegraphics[width=0.3\hsize,clip]{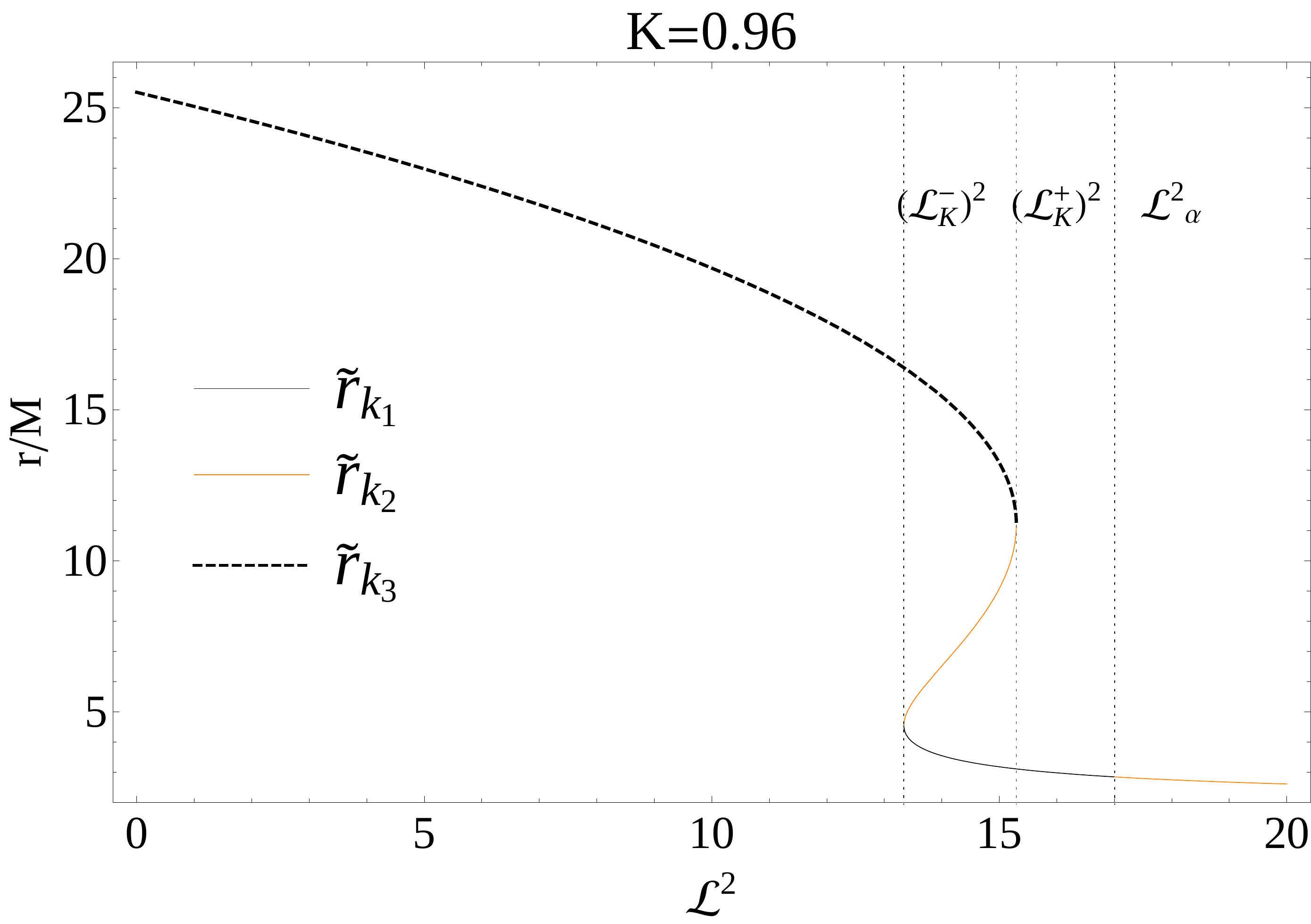}
\includegraphics[width=0.3\hsize,clip]{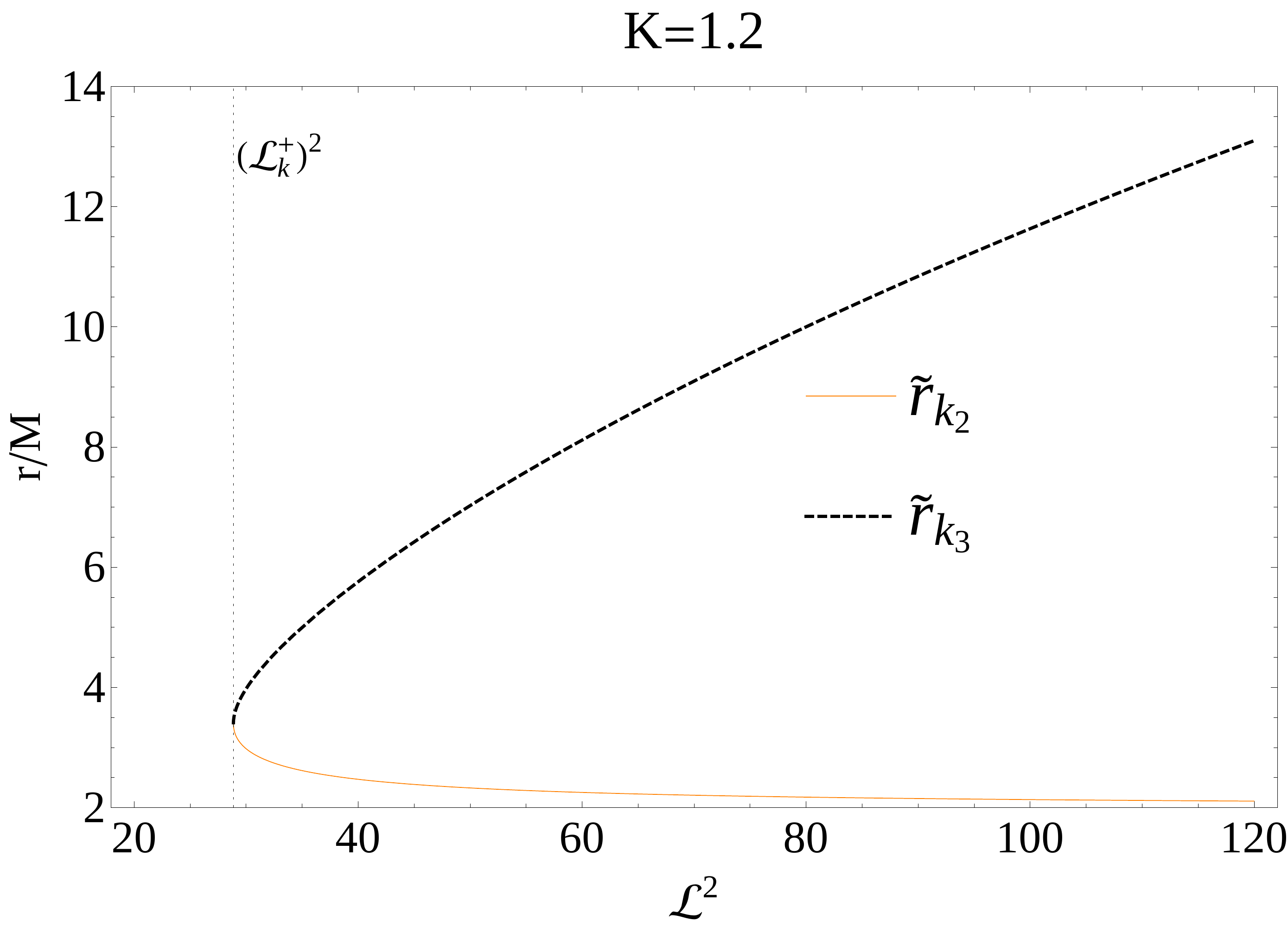}
\caption{(Color online) $\tilde{r}_{k_1}$, $\tilde{r}_{k_2}$, $\tilde{r}_{k_3}$
as a function of $\lie^2$ for different values of $\rm{K}$ in the ranges: $[0,
2\sqrt{2}/3]$ (left panel), $[ 2\sqrt{2}/3,1]$ (central panel), $\rm{K}>1$ (right
panel). Dotted lines are the momentum  $\lie^2_{\alpha}$ and $(\lie_K^{\pm})^2$.}
\label{eleonor}
\end{figure}

\begin{figure}
\centering
\includegraphics[width=0.45\hsize,clip]{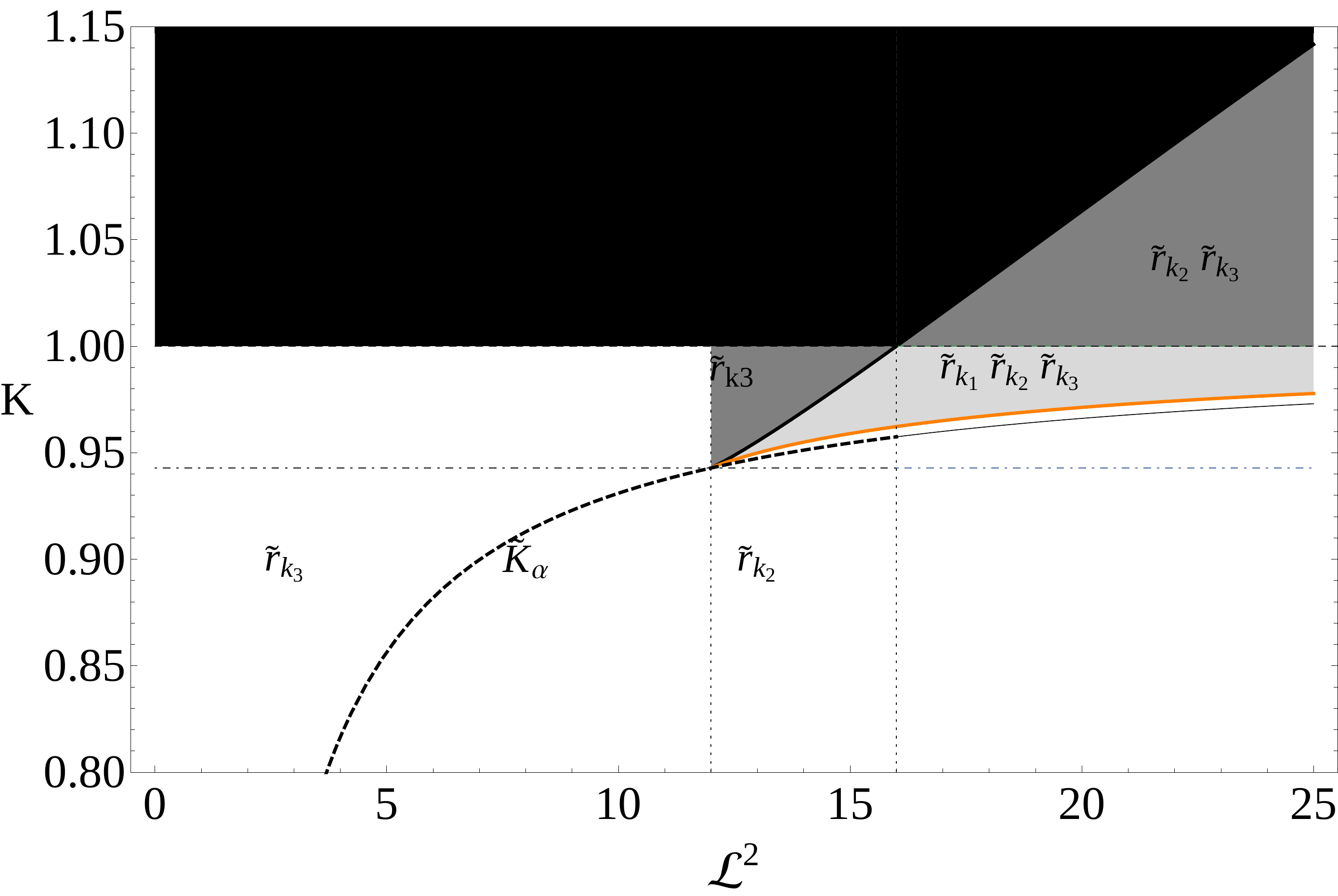}
\includegraphics[width=0.45\hsize,clip]{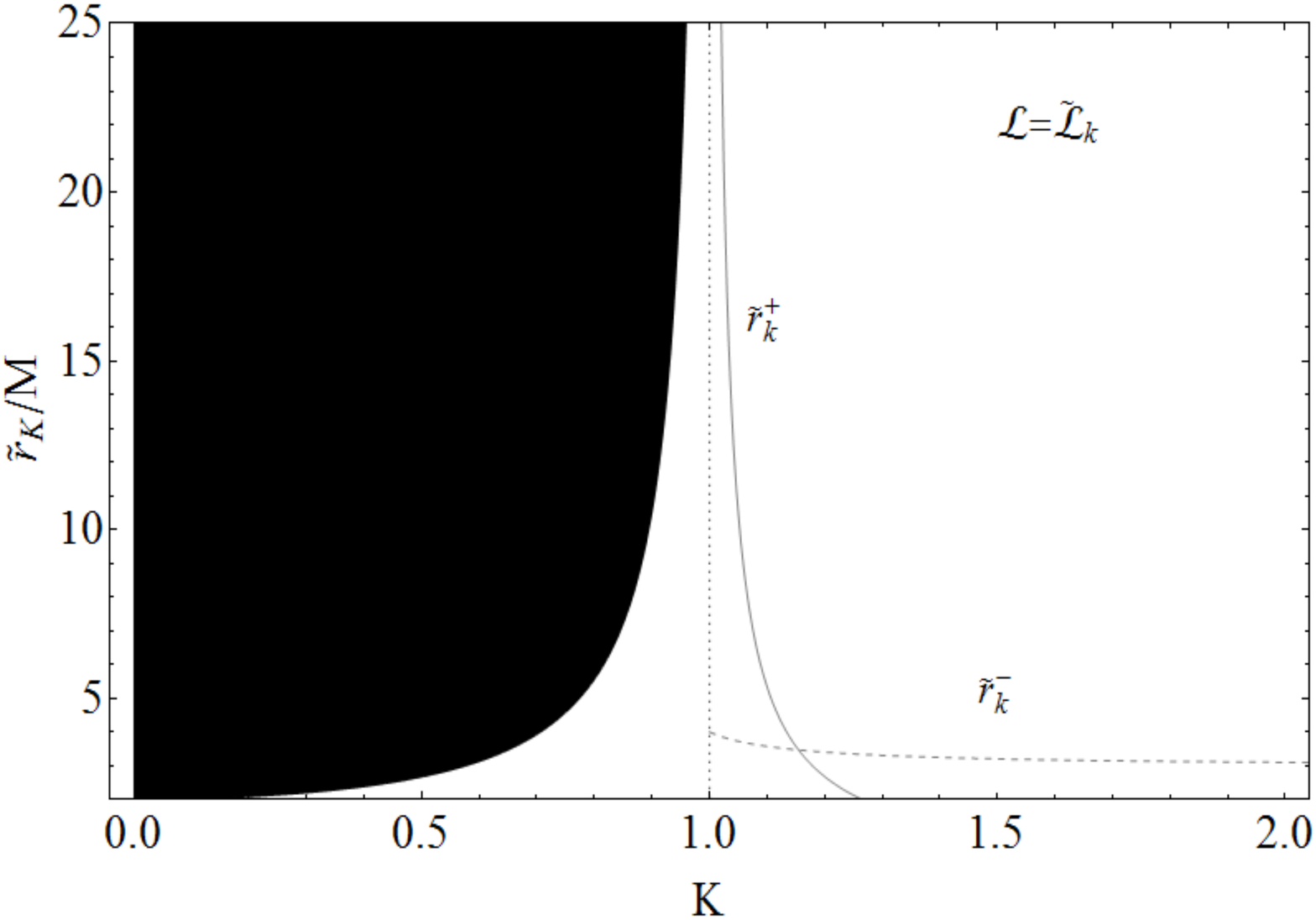}
\includegraphics[width=0.5\hsize,clip]{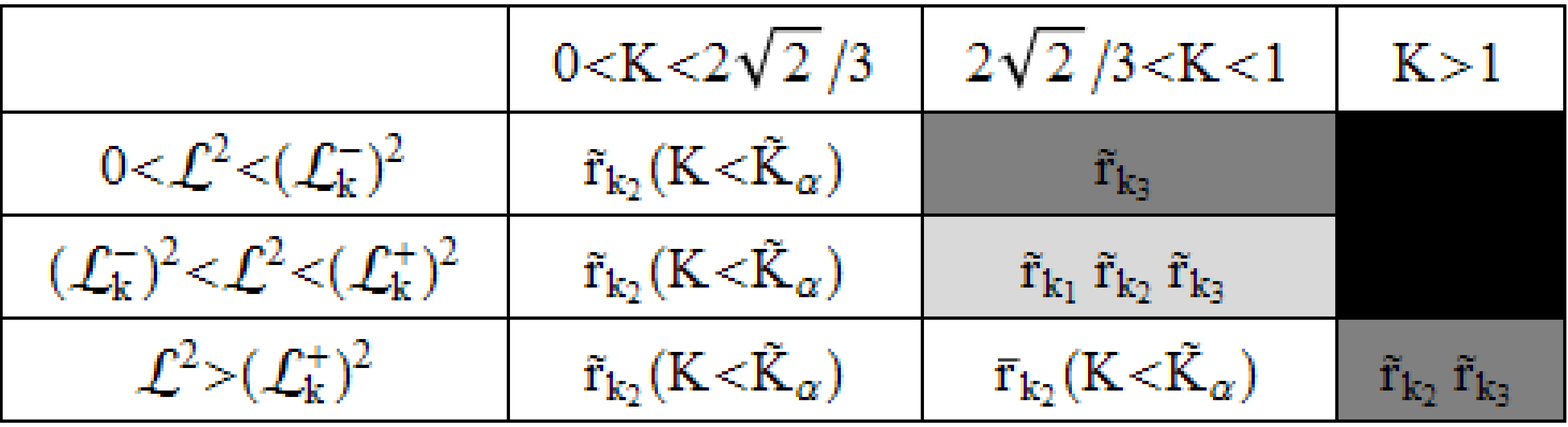}
\caption{(Color online) \emph{Upper left panel}: $\tilde{\rm{K}}^{+}$ (black
curve), $\tilde{\rm{K}}^{-}$ (orange curve) and $\tilde{\rm{K}}_{\alpha}$  (dashed thick
curve) as a function of $\lie^2$. The  regions of existence for the radii
$\tilde{r}_{k_1}, \tilde{r}_{k_2}, \tilde{r}_{k_3}$, solution of $W=\ln(\rm{K})$ are
marked in white, gray, light-gray. Black regions are forbidden. Dotted-dashed line marks
$\rm{K}=2\sqrt{2}/3$. \emph{Upper right panel}:  $\tilde{r}_{\rm{K}}$ as a function of
$\rm{K}$. Gray curve is $\tilde{r}_k^+$, dashed curve $\tilde{r}_k^-$. In the white
region $\lie^2={\lie^2}_{k}$ solution of $W=\ln(\rm{K})$. No solution exists in the
black region. \emph{Lower panel}: table summarizing the regions of existence of the
radii $\tilde{r}_{k_1}, \tilde{r}_{k_2}, \tilde{r}_{k_3}$. Black boxes are forbidden.}
\label{Angelica}
\end{figure}

The solutions of the equation $W=\ln(\rm{K})$ can be describer in terms of the
energies:
\bea
\label{ene_boyerL}
\tilde{\rm{K}}_{\pm}\equiv\sqrt{\frac{1}{54}
\left(36\pm\sqrt{\frac{(\lie^2-12)^3}{\lie^2}}+\lie^2\right)}, \quad
\tilde{\rm{K}}_{\alpha}\equiv\frac{1}{\sqrt{3}}\sqrt{\frac{3 \lie^2-4}{\lie^2}},
\eea
of the radii:
\bea
\label{r_boyerL}
\tilde{r}_{k_1}/M&\equiv&-\frac{2+2 \varpi  \sin\left[\frac{1}{6} \left(\pi +2
\arccos\psi\right)\right]}{3 \left(\rm{K}^2-1\right)},
\\
\tilde{r}_{k_2}/M&\equiv&-\frac{2+2\varpi  \cos\left[\frac{1}{3} \left(\pi
+\arccos\psi\right)\right]}{3 \left(\rm{K}^2-1\right)},
\\
\tilde{r}_{k_3}/M&\equiv&\frac{-2+2 \varpi  \cos\left(\frac{1}{3} \arccos\psi\right)}{3
\left(\rm{K}^2-1\right)},
\eea
(see Figs.\il\ref{eleonor}), where:
\be
\psi\equiv-\frac{8+9 \left(3 \rm{K}^4-5 \rm{K}^2+2\right) \lie^2}{\varpi ^3},\quad
\varpi \equiv \left(\rm{K}^2-1\right) \sqrt{\frac{4+3 \left(\rm{K}^2-1\right)
\lie^2}{\left(\rm{K}^2-1\right)^2}},
\ee
and of the angular momenta:
\be
\label{lie_boyerL}
\lie^2_{\alpha}\equiv \frac{4}{3 \left(\rm{K}^2-1\right)},\quad
{(\lie_{K}^{\pm})}^2\equiv{\frac{1}{2} \left[\frac{27 \rm{\rm{K}}^4-36
\rm{K}^2+8}{\rm{K}^2-1}\pm\sqrt{\frac{\rm{K}^2 \left(9
\rm{K}^2-8\right)^3}{\left(\rm{K}^2-1\right)^2}}\right]},
\ee
These solutions are summarized in Figs.\il\ref{Angelica}.

The critical points of the angular momentum $\lie^2_{k}$ are:
\be
\tilde{r}_k^{\pm}/M=\frac{1}{2} \left[\sqrt{\frac{\rm{K}^2 \left(9
\rm{K}^2-8\right)}{\left(\rm{K}^2-1\right)^2}}\pm\frac{3
\rm{K}^2-4}{\rm{K}^2-1}\right]\, .
\ee
When $2\sqrt{2}/{3}<\rm{K}<1$ there are two critical points $\tilde{r}_k^{\pm}$, while
when $\rm{K}>1$ there is only $r=\tilde{r}_k^+$ (see Fig.\il\ref{Angelica},
\emph{upper right panel}). For $\rm{K}=1$ it is $\tilde{r}_k^{\pm}=r_{mbo}$,  {where $r_{mbo}=4M$ is the
\emph{marginally bounded orbit} for a test particle in the Schwarzschild spacetime}, and in
$\rm{K}=2 \sqrt{2}/3$ it is $\tilde{r}_k^{\pm}=r_{lsco}$.

\medskip

\noindent

\begin{figure}
\centering
\includegraphics[width=0.45\hsize,clip]{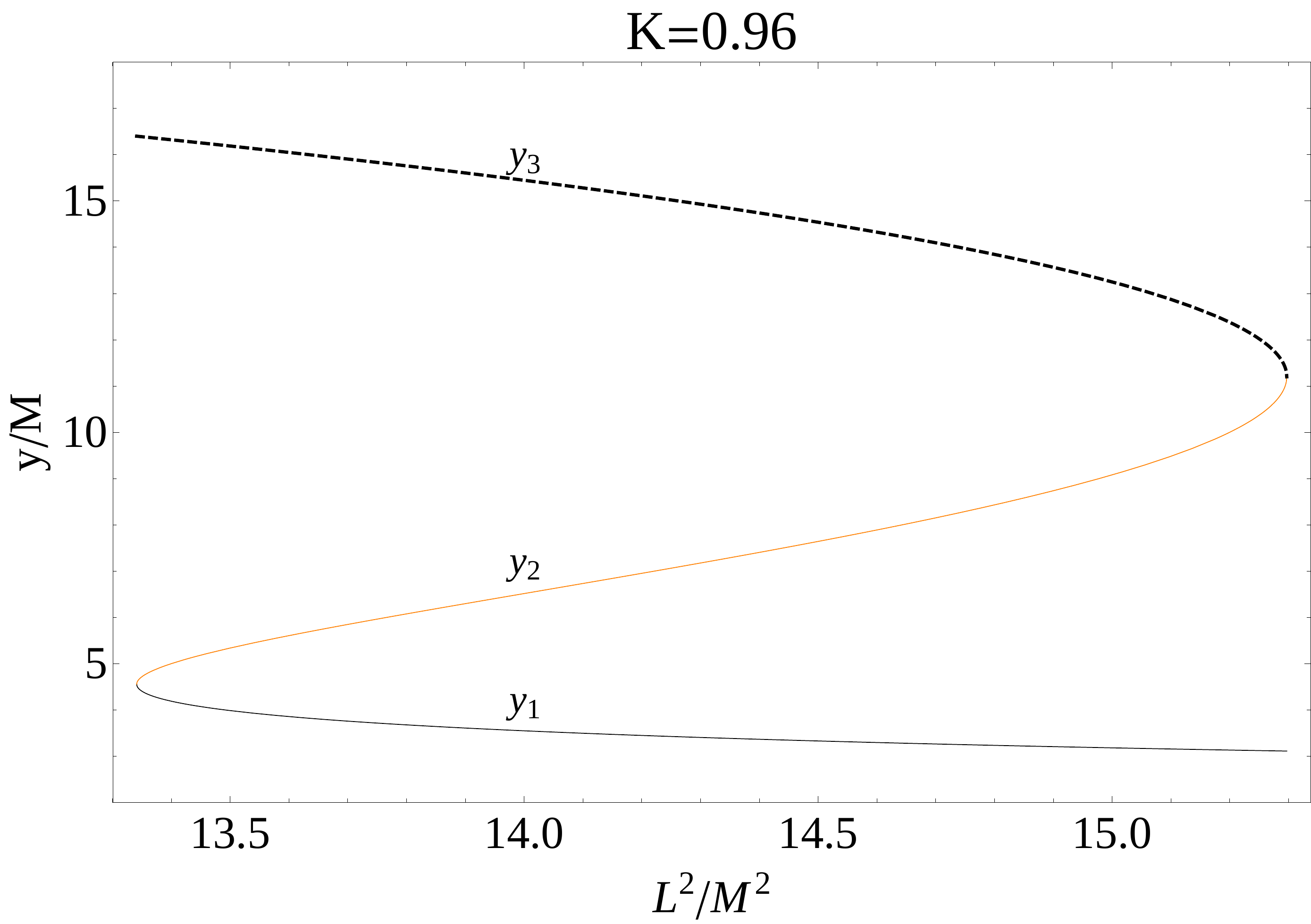}
\includegraphics[width=0.45\hsize,clip]{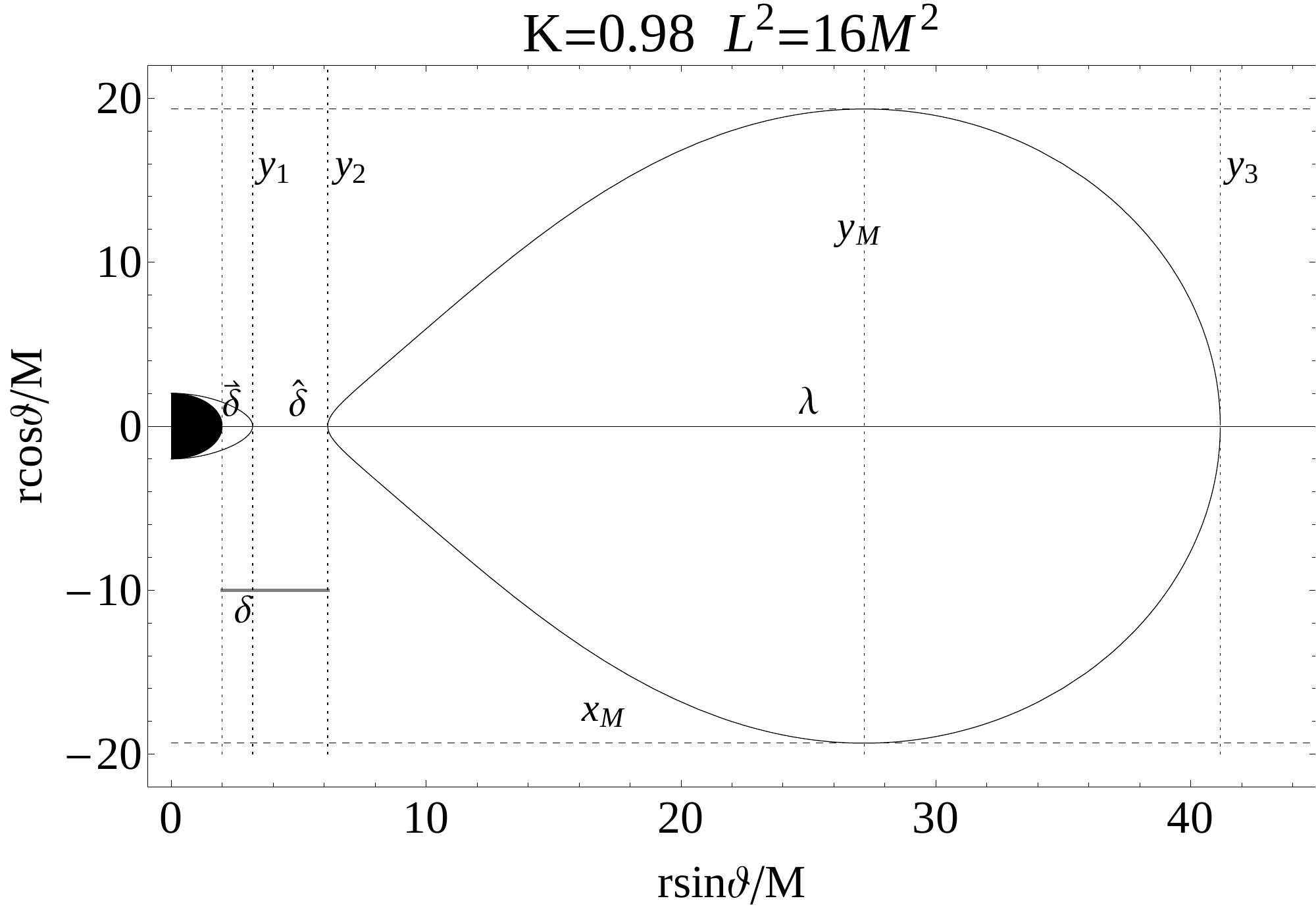}
\caption{{(Color online) \emph{Left panel}: $y_1$ (black curve), $y_2$ (orange
curve)  and $y_3$ (dashed curve) as a function of $L^2$ for $\rm{K}=0.96$. \emph{Right
panel}: the closed Boyer surface at $K=0.98$ and $L^2=16M^2$. $\lambda\equiv y_3-y_2$ is
the surface maximum diameter, $\delta\equiv y_2-2M$ is the distance from the source,
$\hat{\delta}\equiv y_2-y_1$ is the distance from the inner surface,
$\breve{\delta}\equiv y_1-2M$ is the distance of the inner surface from the horizon,
$h\equiv 2 x_M$ and $X_h\equiv y_M$.}}
\label{PKliod}
\end{figure}

\textbf{Closed surfaces:} the conditions for the existence of closed surfaces can be
obtained by noting that, in cartesian coordinate $(x,y)$, the closed surfaces should
satisfy the condition $V_{sc}(x=0)=\rm{K}$ with three solutions, say
$y=\{y_{1},y_{2},y_{3}\}$. The closed surfaces then exist when ${2
\sqrt{2}}/{3}<\rm{K}<1$ and $(\lie_{\rm{K}}^{-})^2<\lie^2<({\lie_{\rm{K}}^+})^2$ (see
Fig.\il\ref{Angelica}, \emph{lower panel}), with $(\lie_{\rm{K}}^{-})^2\geq16$. In
cartesian coordinate $(x,y)$ the surfaces $V_{sc}=\rm{K}$ are:
\be\label{rmpr}
x=\pm\sqrt{\frac{\left(2 L^2+2 y^2\right)^2}{\left(L^2+y^2-y^2\rm{K}^2 \right)^2}-y^2}.
\ee
The maximum diameter $(x=0)$ of the  closed Boyer surface lies between the points
$y=y_2$ and $y=y_3$, where
\bea\label{din}
{y_1}/M &\equiv&\sqrt{\frac{1}{9} \left(\varsigma -6 \theta  \sin\left[\frac{1}{6} (\pi
+2 \arccos\varepsilon )\right]\right)},
\\
{y_2}/M &\equiv&\sqrt{\frac{1}{9} \left(\varsigma -6 \theta  \cos\left[\frac{1}{3} (\pi
+\arccos\varepsilon)\right]\right)},
\\
{y_3}/M &\equiv&\sqrt{\frac{1}{9} \left[\varsigma +6\theta  \cos\left(\frac{1}{3}
\arccos\varepsilon \right)\right]},
\label{dan}
\eea
being,
\bea
\theta &\equiv&\sqrt{\frac{16+8 \left(3 \rm{K}^4-4 \rm{K}^2+1\right)
(L/M)^2+\left(\rm{K}^2-1\right)^2 (L/M)^4}{\left(\rm{K}^2-1\right)^4}},
\\\label{sudben}
\varepsilon  &\equiv&\frac{64+48 \left(3 \rm{K}^4-4 \rm{K}^2+1\right) (L/M)^2+6
\left(\rm{K}^2-1\right)^2 \left(9 \rm{K}^4-6 \rm{K}^2+2\right)
(L/M)^4-\left(\rm{K}^2-1\right)^3 (L/M)^6}{\left(\rm{K}^2-1\right)^6 \theta ^3},
\\
\varsigma &\equiv&\frac{6 \left[2+\left(\rm{K}^2-1\right)
(L/M)^2\right]}{\left(\rm{K}^2-1\right)^2}
\eea
(see Fig.\il\ref{PKliod}, \emph{left panel}).

\begin{figure}
\centering
\includegraphics[width=0.45\hsize,clip]{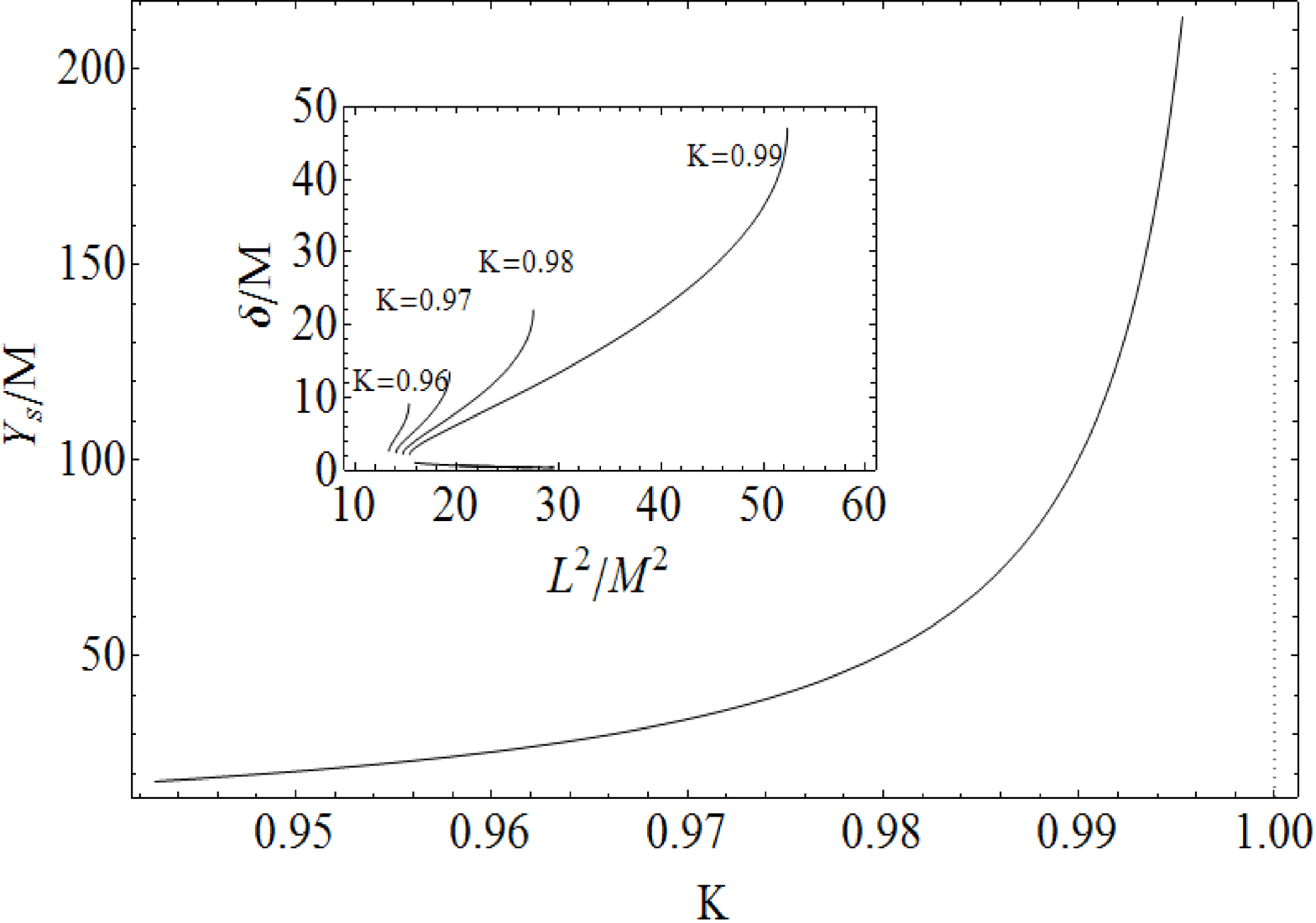}
\includegraphics[width=0.45\hsize,clip]{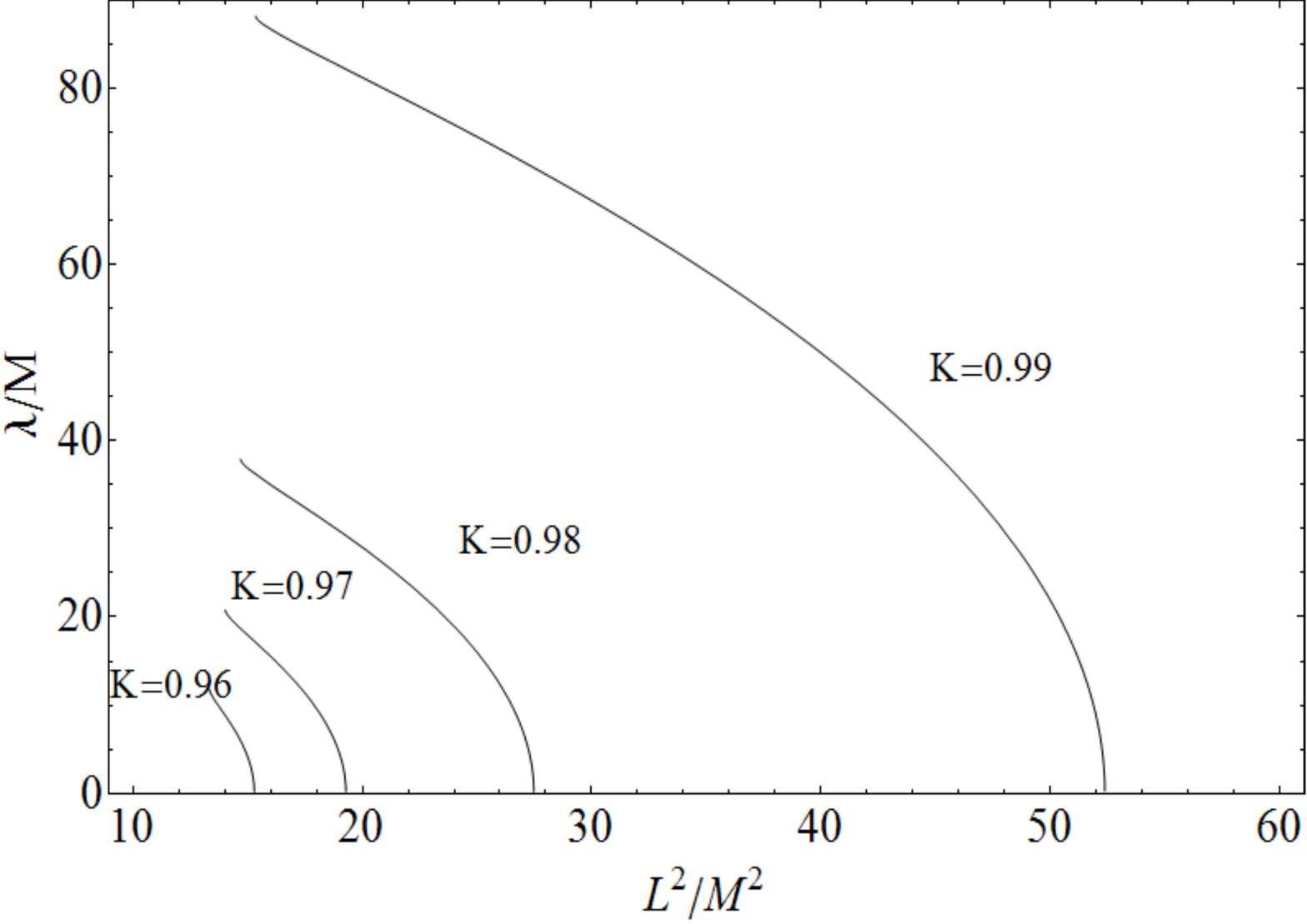}\\
\includegraphics[width=0.45\hsize,clip]{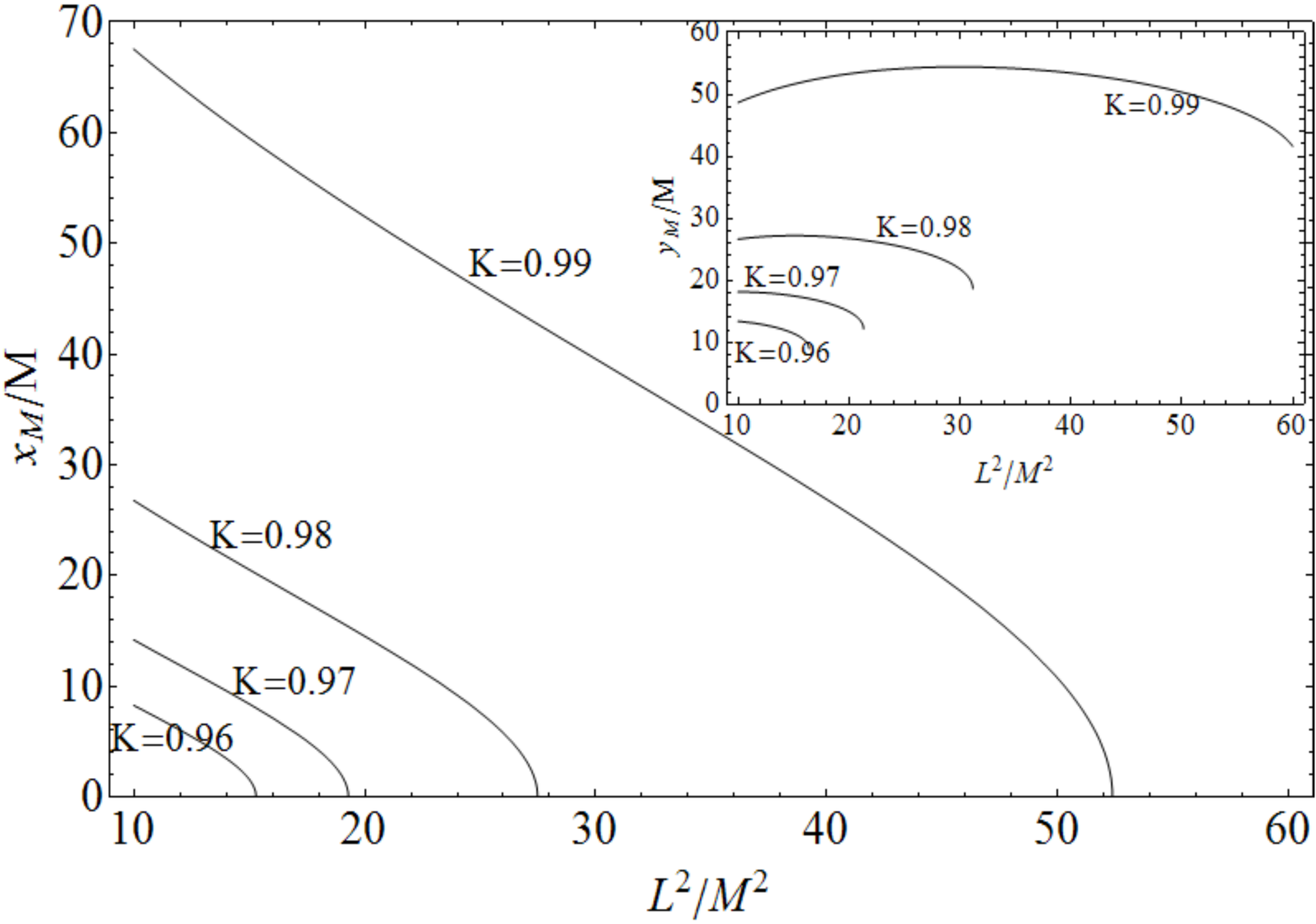}
\includegraphics[width=0.45\hsize,clip]{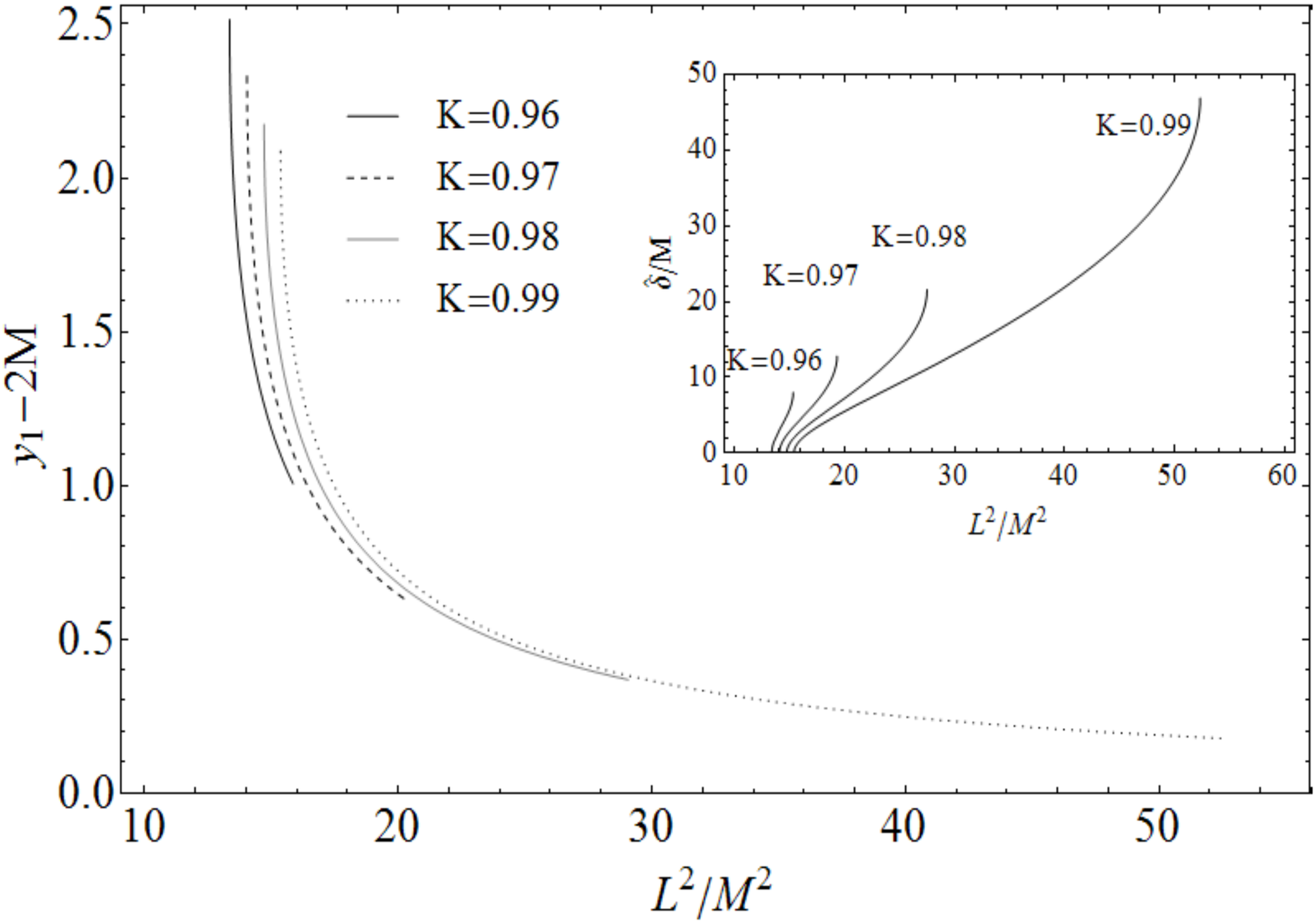}
\caption{\emph{Upper left panel}: $y_s$ as a function of the
energy $2\sqrt{2}/3<K<1$. \emph{Inset}: distance of the surface  from the $r=2M$
\(\delta\equiv y_2-2M\) as a function of $L^2$, for different values of $K$. It
increases with $L^2$ and decreases with $\rm{K}$. \emph{Upper right panel}: surface
maximum diameter \(\lambda\equiv y_3-y_2\) as a function of  $L^2$ for different values
of $K$. \emph{Lower left panel}: maximum $x_M=x(y_M)$ as a function of $L^2$ for
different energies $K$. \emph{Inset}: $y_M$ as a function of $L^2$, for different values
of $K$. \emph{Lower right panel}: distance $\breve{\delta}\equiv y_1-2M$ as a function
of $L^2$, for different values of $K$. \emph{Inset}: distance  $\hat{\delta}= y_2-y_1$
as a function of $L^2$, for different values of $K$.}
\label{Distrazione4elmot}
\end{figure}

Fig.\il\ref{PKliod}, \emph{right panel} portrays a closed Boyer surface. We can
characterize this surface introducing the following parameters:
\begin{enumerate}
\item the {surface maximum diameter}:
\(
\label{E:ert}
\lambda\equiv y_3-y_2,
\)
\item the {distance from the source} defined as:
\(
\delta\equiv y_2-2M,
\)
\item the {distance from the inner surface}:
\(
\hat{\delta}\equiv y_2-y_1,
\)
\item the {distance of the inner surface from the horizon}:
\(
\breve{\delta}\equiv y_1-2M,
\)
\item the surface {maximum height} defined as:
\(
h\equiv 2 x_M,
\)
\item the quantity
\(
\label{E:ertdon}
X_h\equiv y_M.
\)
\end{enumerate}
$(x_M,y_M)$ is the critical point of the surface in Eq.\il\ref{rmpr}. In what follows
we find the constraints for the set of parameters  $\{\lambda, \delta,\hat{\delta},
\breve{\delta}, h, X_h\}$. The point $y_3$ varies in the range $2M<y_3<y_s$, and $L^2$
lies on the surface $L^2_s$, which is a solution of $V_s=K$ on the plane $x=0$ for
$y>4M$ and $K\geq1$, and  for $4M<y<y_s$ for $K<1$:
\be
y_s\equiv\frac{2M}{\left|\rm{K}^2-1\right|},\quad L^2_s=2M \sqrt{\frac{\rm{K}^4
y^6}{(y^2-4M^2)^2}}+y^2 \left(\frac{\rm{K}^2  y^2}{y^2-4M^2}-1\right)
\ee
(see Figs.\il\ref{Distrazione4elmot}). $\lambda$ increases with the energy $\rm{K}$,
but decreases with the fluid angular momentum $L$. On the contrary, the distance from
the source $\delta=y_2-2M$ increases with $L$ and decreases with $\rm{K}$ (see
Figs.\il\ref{Distrazione4elmot}).

The maximum vertical distance of the closed surfaces is:
\be
y_M/M=\sqrt{\frac{3 (L/M)^2+4 \sqrt{6} \left(\rm{K}^2-1\right) \sqrt{-\frac{\rm{K}^2
(L/M)^2}{\left(\rm{K}^2-1\right)^3}} \cos\left[\frac{1}{3} \arccos\left[-\frac{3}{4}
\sqrt{\frac{3}{2}} \left(\rm{K}^2-1\right)^2 \sqrt{-\frac{\rm{K}^2
(L/M)^2}{\left(\rm{K}^2-1\right)^3}}\right]\right]}{3 \left(\rm{K}^2-1\right)}}.
\ee
As shown in Figs.\il\ref{Distrazione4elmot}, the maximum $x_M=x(y_M)$, and
consequently the height $h$, increases with $\rm{K}$ and decreases with $L^2$. On the
contrary $y_M$  increases with $\rm{K}$ and with $L^2$, until it reaches a maximum and
then decreases. The distance  $\hat{\delta}= y_2-y_1$ increases with $L^2$ and decreases
with $\rm{K}$. The distance  $\breve{\delta}\equiv y_1-2M$ increases with the energy
$\rm{K}$ and decreases with $L^2$.

\medskip
\noindent

\begin{figure}
\centering
\includegraphics[width=0.45\hsize,clip]{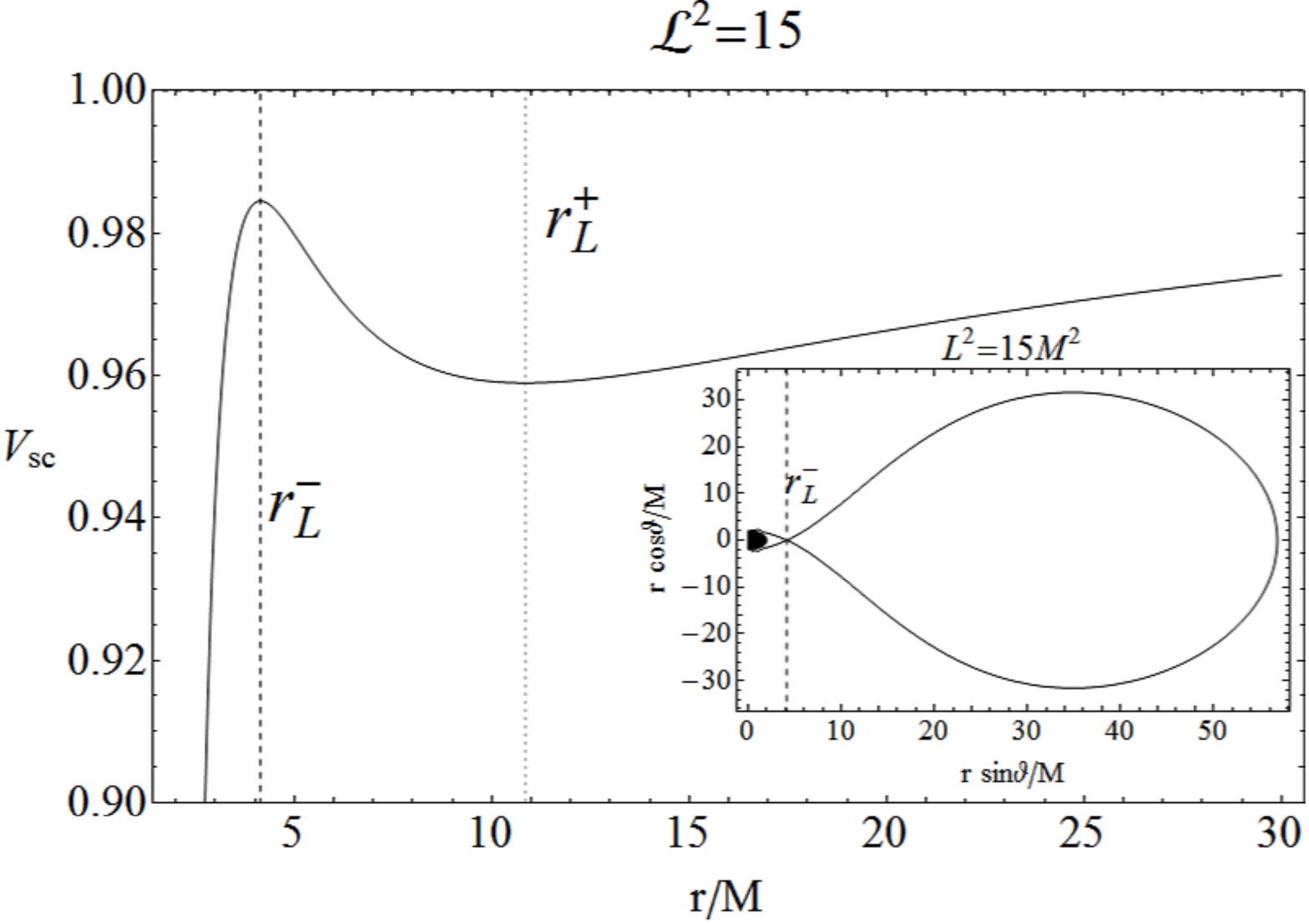}
\includegraphics[width=0.45\hsize,clip]{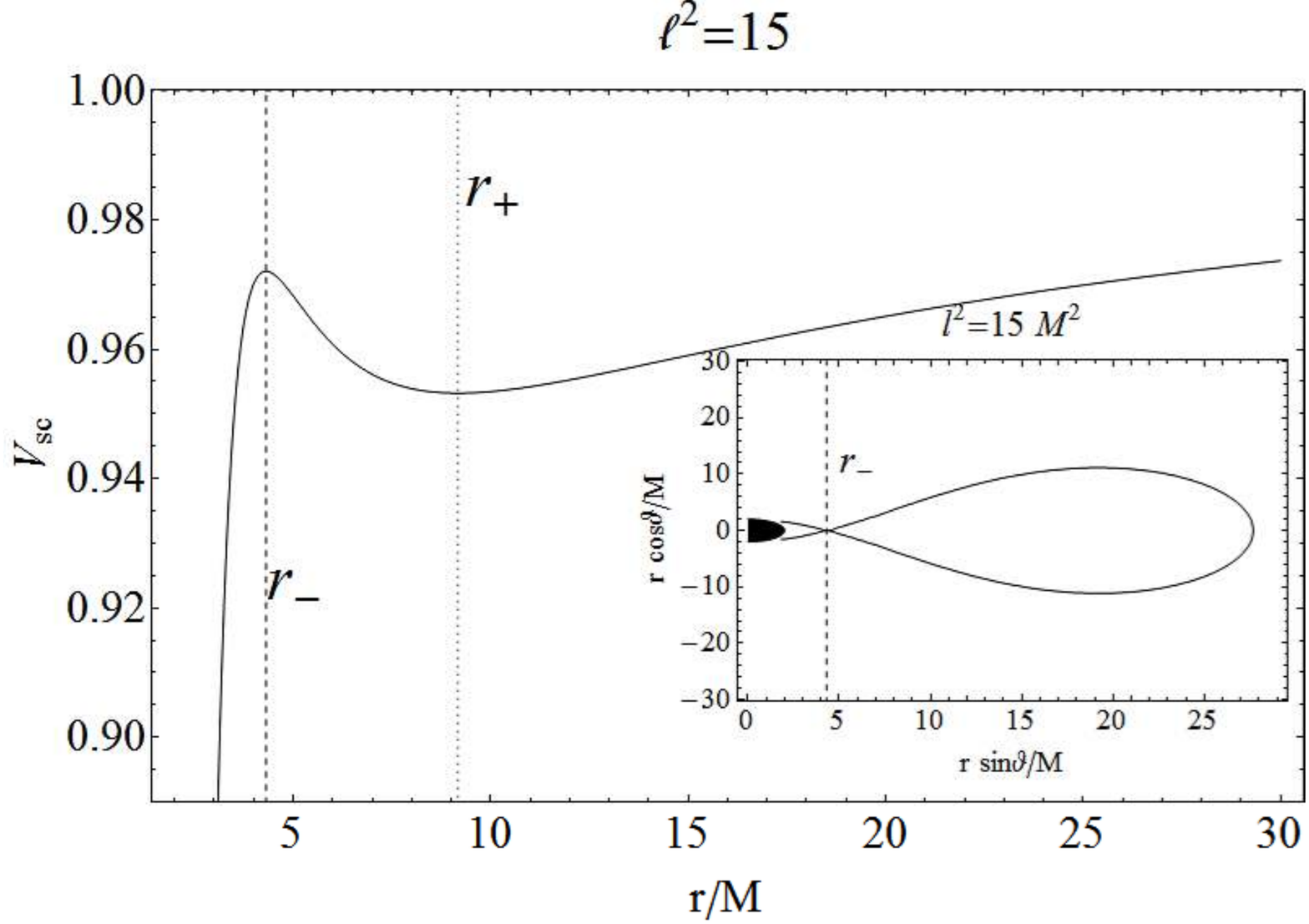}
\includegraphics[width=0.45\hsize,clip]{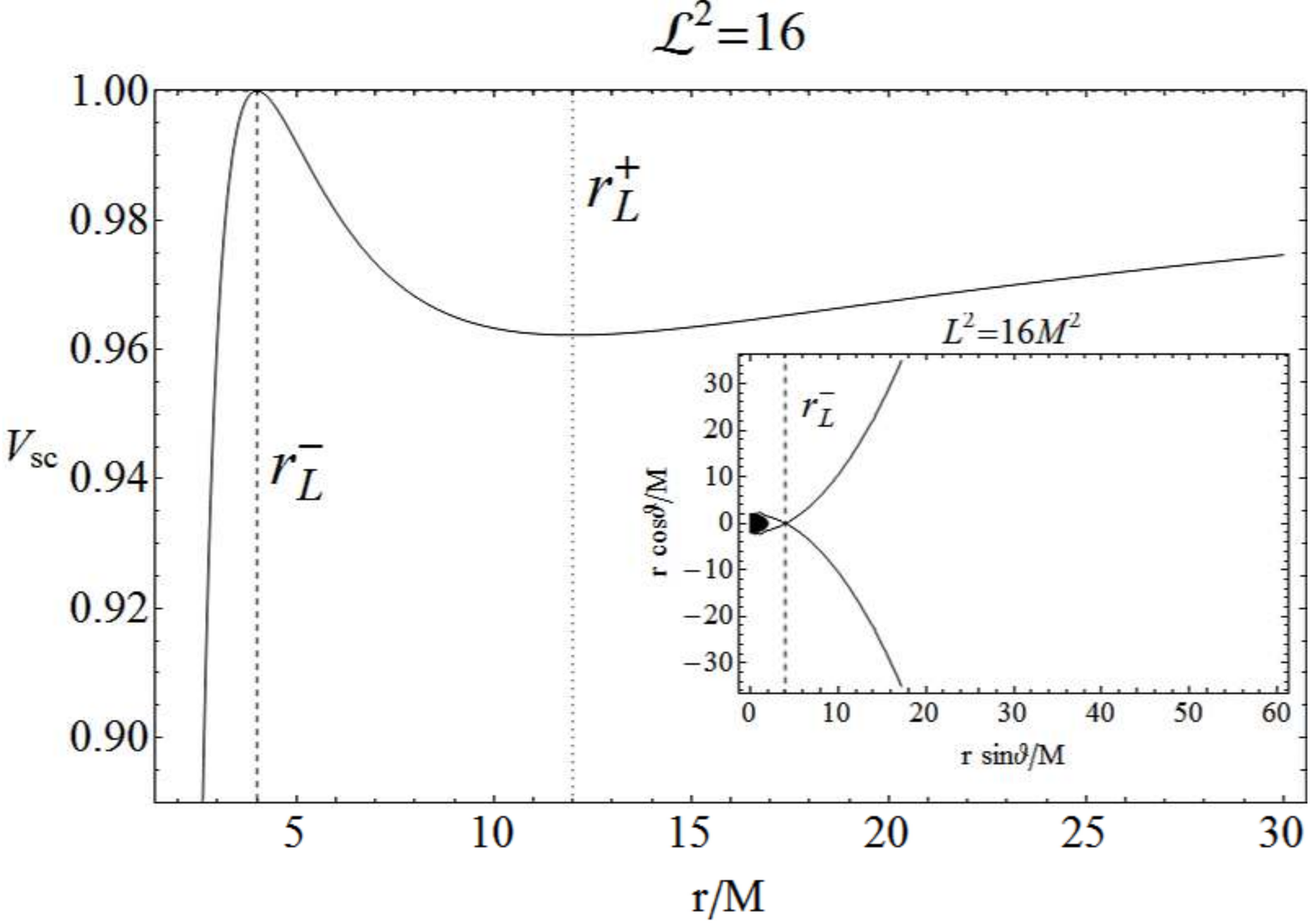}
\includegraphics[width=0.45\hsize,clip]{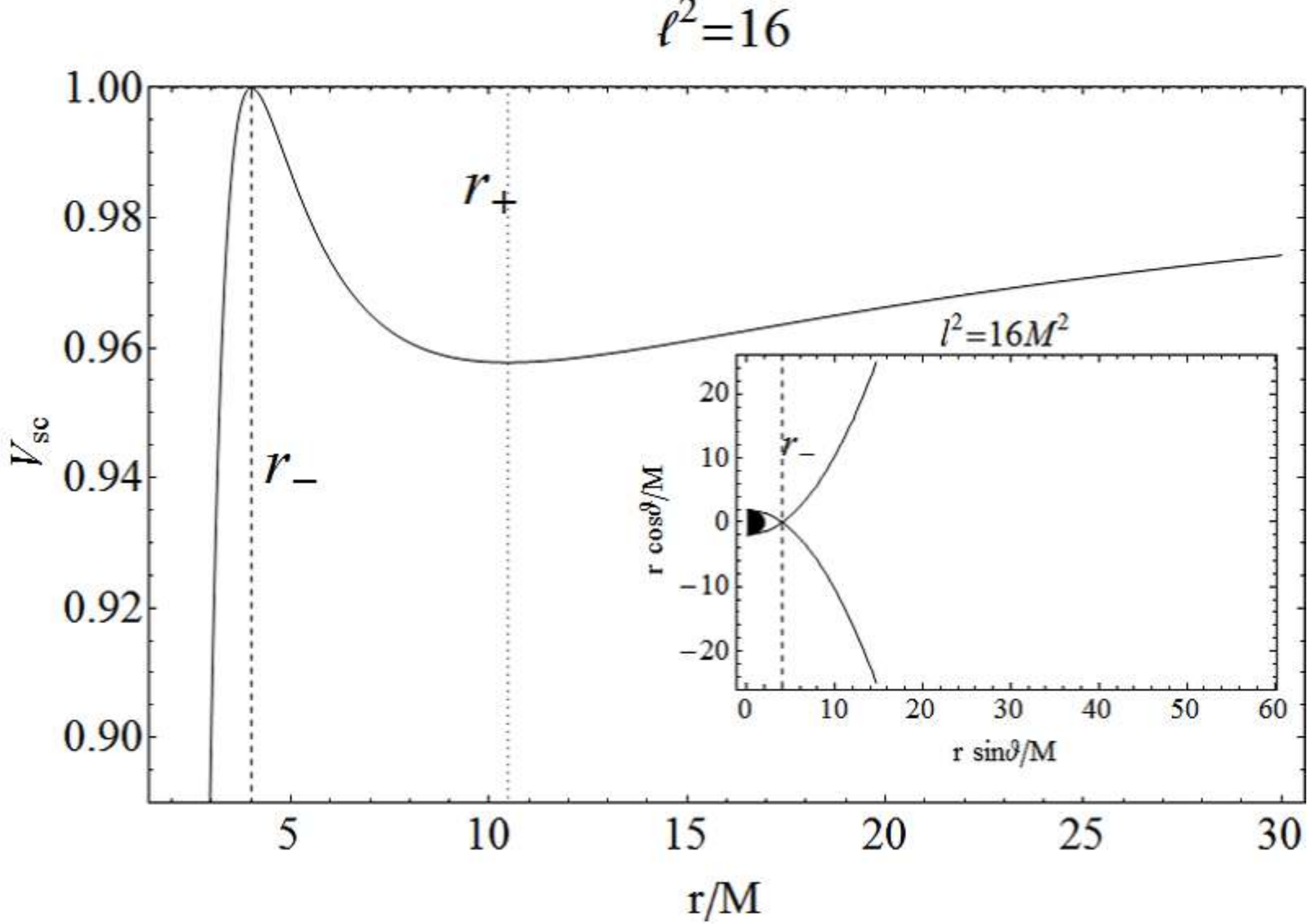}
\includegraphics[width=0.45\hsize,clip]{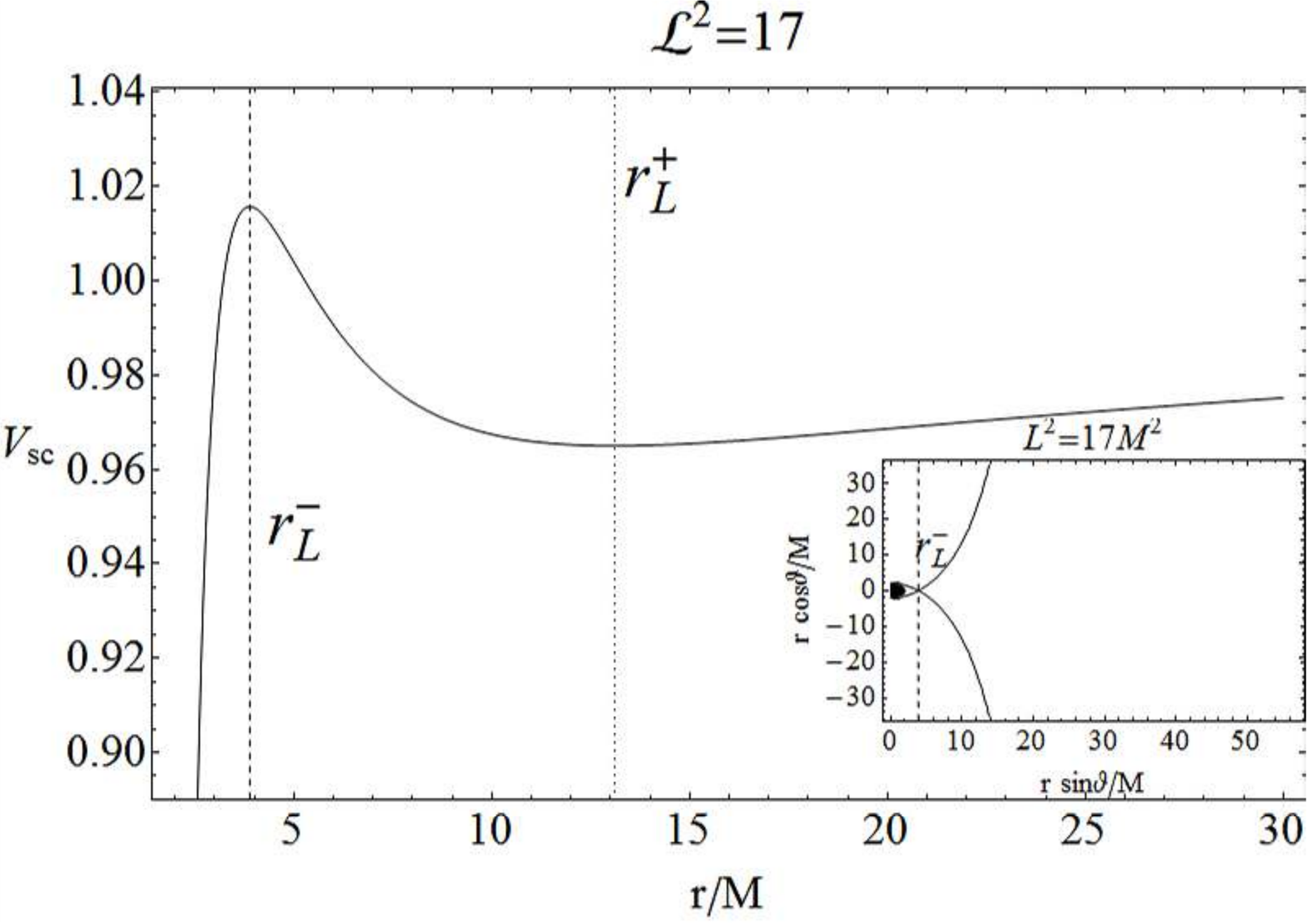}
\includegraphics[width=0.45\hsize,clip]{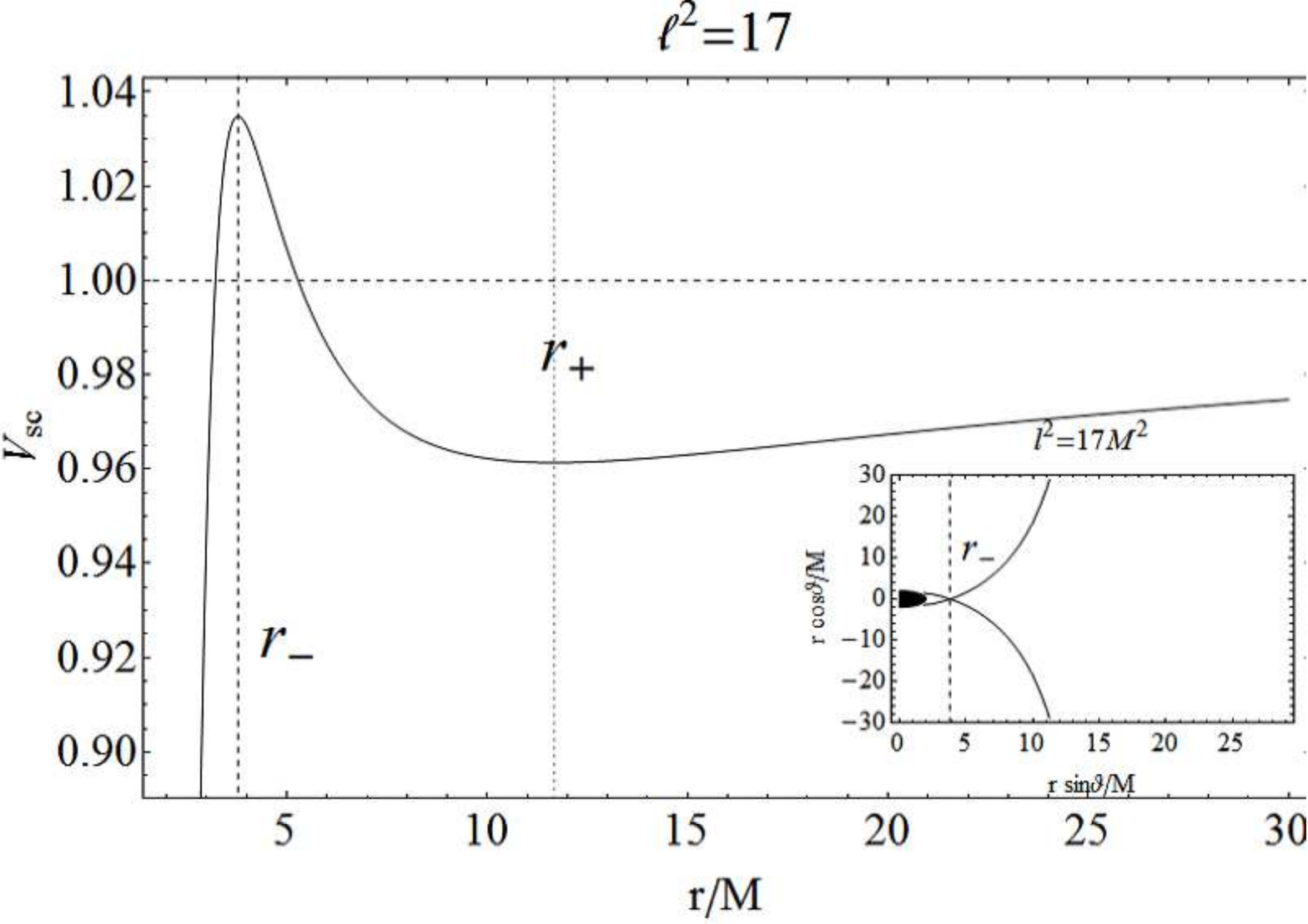}
\caption{{\emph{Left panels}: Effective potential $V_{sc}(L,r)$ as a function of
$r/M$, for selected values of $\lie^2=L^2/M^2\sigma^2$. Dashed lines mark $r_{L}^{\pm}$.
\emph{Inset}: the self--crossing Boyes surfaces corresponding to the maxima of the
effective potential as function of $r$: maximum with energy $\rm{K}>1$ are open surfaces
and  maximum with energy $\rm{K}<1$ are closed surfaces. \emph{Right panels}: Plots of the effective potential $V_{sc}(l,r)$ as
function of $r/M$, for selected values of ${\ell^2}=l^2/M^2\sigma^2$. Dashed lines are
$r_{\pm}$. Correspondingly the self--crossing Boyer surfaces are plotted in the inner
panel. They  correspond to the maxima of the effective potential as function of $r$:
maximum with energy $\rm{K}>1$ are open surfaces and  maximum with energy $\rm{K}<1$ are
closed surfaces.}}
\label{FAI}
\end{figure}

\textbf{Cusps:} the cusps, i.e. the self-crossing surfaces $W=\rm{constant}$, correspond
to the maxima of the effective potential as a function of $r$: open surfaces are maxima
with energy $\rm{K}>1$, and  closed surfaces are maxima with energy $\rm{K}<1$, as
outlined in Fig.\il\ref{FAI} \emph{upper}. It is therefore important to consider  the function $V'$
in the regions of closed and open Boyer surfaces. From Sect.\il\ref{Sec:Neutral-L0} we
know that the solutions of the equations $V'=0$ that correspond to maxima of the
effective potential are located on the radius $r=r_{L}^-$, in particular there are
closed surfaces with a cusps in $r_{L}^->r_{mbo}$, and
$12\leq \lie^2<16$ , viceversa maxima located in $r_{L}^->r_{mbo}$ with $\lie^2\geq16$
are open surfaces with a cusp (see Fig.\il\ref{FAI} \emph{upper}).

\medskip
\noindent
\textbf{Remark:} integrating Eq.\il\ref{equazp} with $V_{sc}$ as function of the
constant of motion $L$, we obtain the following expression for the potential:
\be
{\widetilde{W}\equiv\frac{2 \lie^4 M(r+3M)+\left(4+\lie^4\right) r^2
[\ln(r/M-2)-\ln(r/M)]}{8 r^2}\, .}
\ee

The solutions of $\widetilde{W}=\rm{c}$ ensure the existence of  closed, open and
self--crossing surfaces.

\subsection{Analysis of the Boyer potential vs the angular momentum
$l$}\label{Se:citazione}

\begin{figure}
\centering
\includegraphics[width=0.55\hsize,clip]{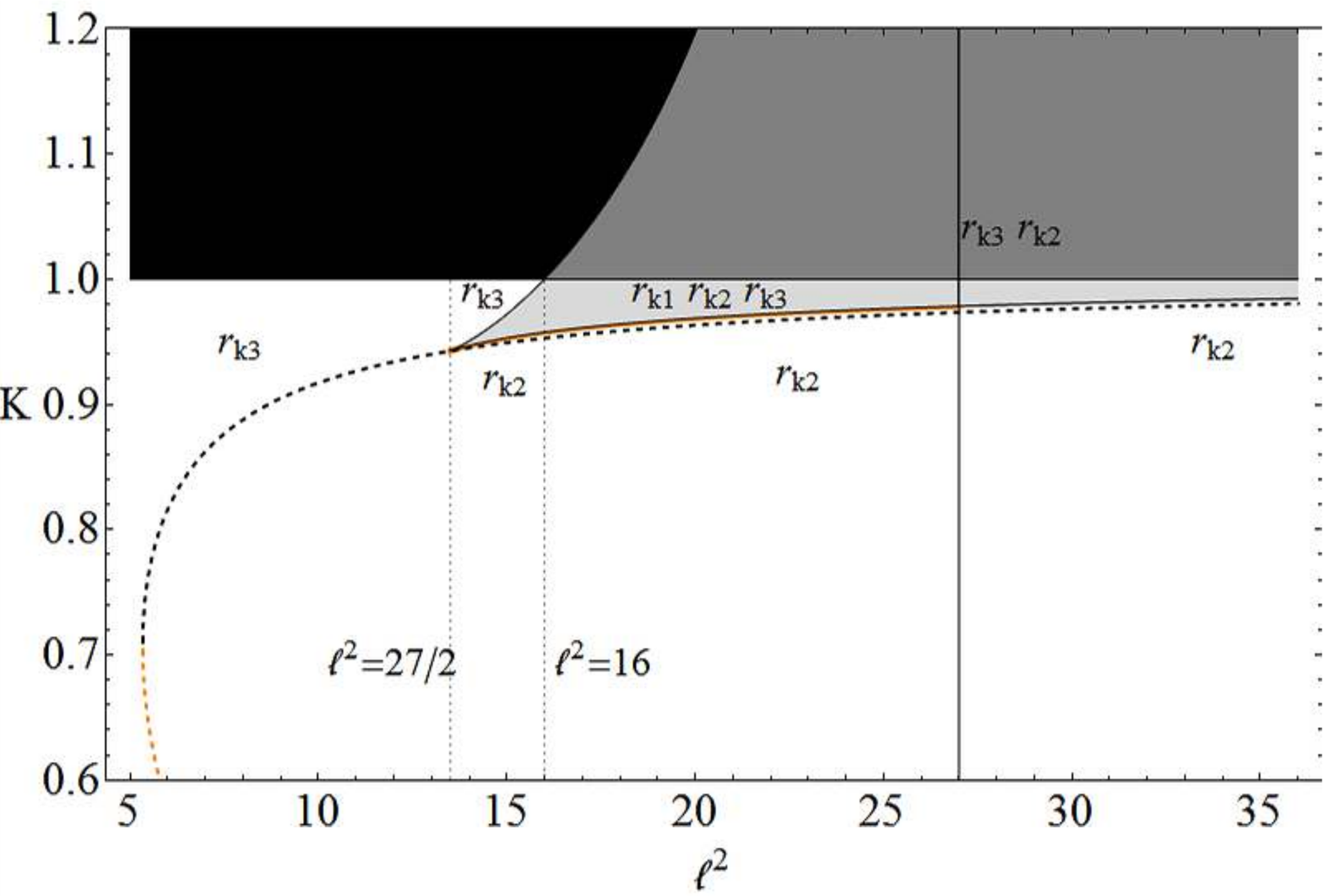}
\includegraphics[width=0.55\hsize,clip]{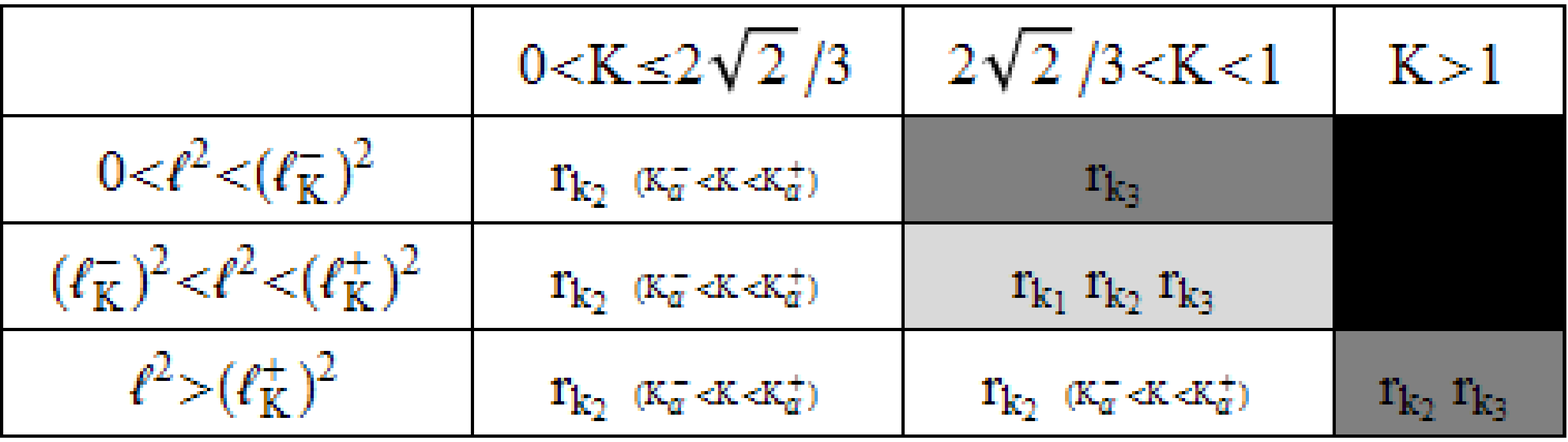}
\caption{(Color online) \emph{Upper panel}: ${\rm{K}}_{a}$ (black curve),
${\rm{K}}_b$ (orange curve), ${\rm{K}}_{\alpha}^{+}$  (dashed thick  black curve) and
${\rm{K}}_{\alpha}^{-}$  (dashed thick  orange curve) as a function of ${\ell^2}$. The
colored regions mark the existence for the radii ${r}_{k_1}, {r}_{k_2}, {r}_{k_3}$. The
black region is forbidden. Dotted lines are $\ell^2=16$ and $\ell^2=27/2$. \emph{Lower
panel}: table summarizing  the  regions of existence for the radii ${r}_{k_1},
{r}_{k_2}, {r}_{k_3}$. Black boxes are forbidden.}
\label{restylingl}
\end{figure}

We face now the analysis  of the Boyer potential in terms of the fluid angular momentum
$l=\mbox{constant}$. Firstly, we note that the Boyer surface in Eq.\il\ref{Edverdi} is
not defined in the region $[r_l^-,r_l^+]$. We detail the study of the sign of
$V_{sc}(l,r)$ in Figs.\il\ref{restylingl}  (see also Fig.\il\ref{restyling},
\emph{left panel}). With respect to the case of $L=\rm{constant}$ (see
Figs.\il\ref{Angelica} and Eqs.\il\ref{ene_boyerL}--\ref{lie_boyerL}), here new
definitions have been introduced for the energies:
\bea
\rm{K}_a&\equiv&\sqrt{\frac{\ell^2-36-2 \beta_1 \sin\left(\frac{1}{3}
\arcsin\alpha_1\right)}{3 \ell^2-81}},
\\
\rm{K}_b&\equiv&\sqrt{\frac{\ell^2-36+2 \beta_1 \cos\left(\frac{1}{3}
\arccos\alpha_1\right)}{3 \ell^2-81}},
\\
\rm{K}_{\alpha}^{\pm}&\equiv&\frac{1}{\sqrt{6}}\sqrt{3\pm\frac{\sqrt{3} \sqrt{\ell^2 (3
\ell^2-16)}}{\ell^2}},
\eea
with
\bea
\alpha_1\equiv\frac{\left(2^3 3^9\right)-\left(108 \sqrt{2}\right)^2\ell^2+\left(6
\sqrt{51}\right)^2 {\ell^2}^2-72 {\ell^2}^3+{\ell^2}^4}{\ell^2 \beta_1^3},\quad
\beta_1&\equiv&(\ell^2-27) \sqrt{\frac{72+(\ell^2-24)^2}{(\ell^2-27)^2}},
\eea
angular momenta:
\bea
{\ell}_{\alpha}^2\equiv-\frac{4}{3 \rm{K}^2 \left(\rm{K}^2-1\right)},\quad
({\ell}^{\pm}_{k})^2&\equiv&\frac{1}{2}\left(\pm\sqrt{\frac{\left(9
\rm{K}^2-8\right)^3}{\rm{K}^2 \left(\rm{K}^2-1\right)^2}}+\frac{27 \rm{K}^4-36
\rm{K}^2+8}{\rm{K}^2 \left(\rm{K}^2-1\right)}\right),
\eea
and radii:
\bea
{r}_{k_1}/M&\equiv&-\frac{2 \left[1+\varpi_k\cos\left(\frac{1}{3}
\arccos\psi_k\right)\right]}{3 \left(\rm{K}^2-1\right)},
\\
{r}_{k_2}/M&\equiv&\frac{2 \left(-1+\varpi_k\cos\left[\frac{1}{3}
\arccos\left(-\psi_k\right)\right]\right)}{3 \left(\rm{K}^2-1\right)},
\\
{r}_{k_3}/M&\equiv&\frac{2 \left[-1+\varpi_k\sin\left(\frac{1}{3}
\arcsin\psi_k\right)\right]}{3 \left(\rm{K}^2-1\right)},
\eea
where:
\be
\psi_k\equiv\frac{8+9 \rm{K}^2 \left(3 \rm{K}^4-5 \rm{K}^2+2\right)
\ell^2}{\varpi_k^3},\quad
\varpi_k\equiv\left(\rm{K}^2-1\right) \sqrt{\frac{4+3 \rm{K}^2 \left(\rm{K}^2-1\right)
\ell^2}{\left(\rm{K}^2-1\right)^2}}\, .
\ee

The surfaces $W(l,r)=\rm{constant}$ exist in all the spacetime with angular momentum
$\ell^2=\ell^2_{k}$ and energy $\rm{K}=\rm{K}_{k}$, where:
\bea
\ell^2_k&\equiv&\frac{r^2 \left[2M+\left(\rm{K}^2-1\right) r\right]}{\rm{K}^2
M^2(r-2M)}<\frac{r^3}{M^2(r-2M)},
\quad
\rm{K}_k\equiv r \sqrt{\frac{r-2M}{r^3-\ell^2 M^2 (r-2M)}}.
\eea
In particular in the case $0<\rm{K}<1$ it is  $2M<r\leq 2M/(1-\rm{K}^2)$.

\begin{figure}
\centering
\includegraphics[width=0.45\hsize,clip]{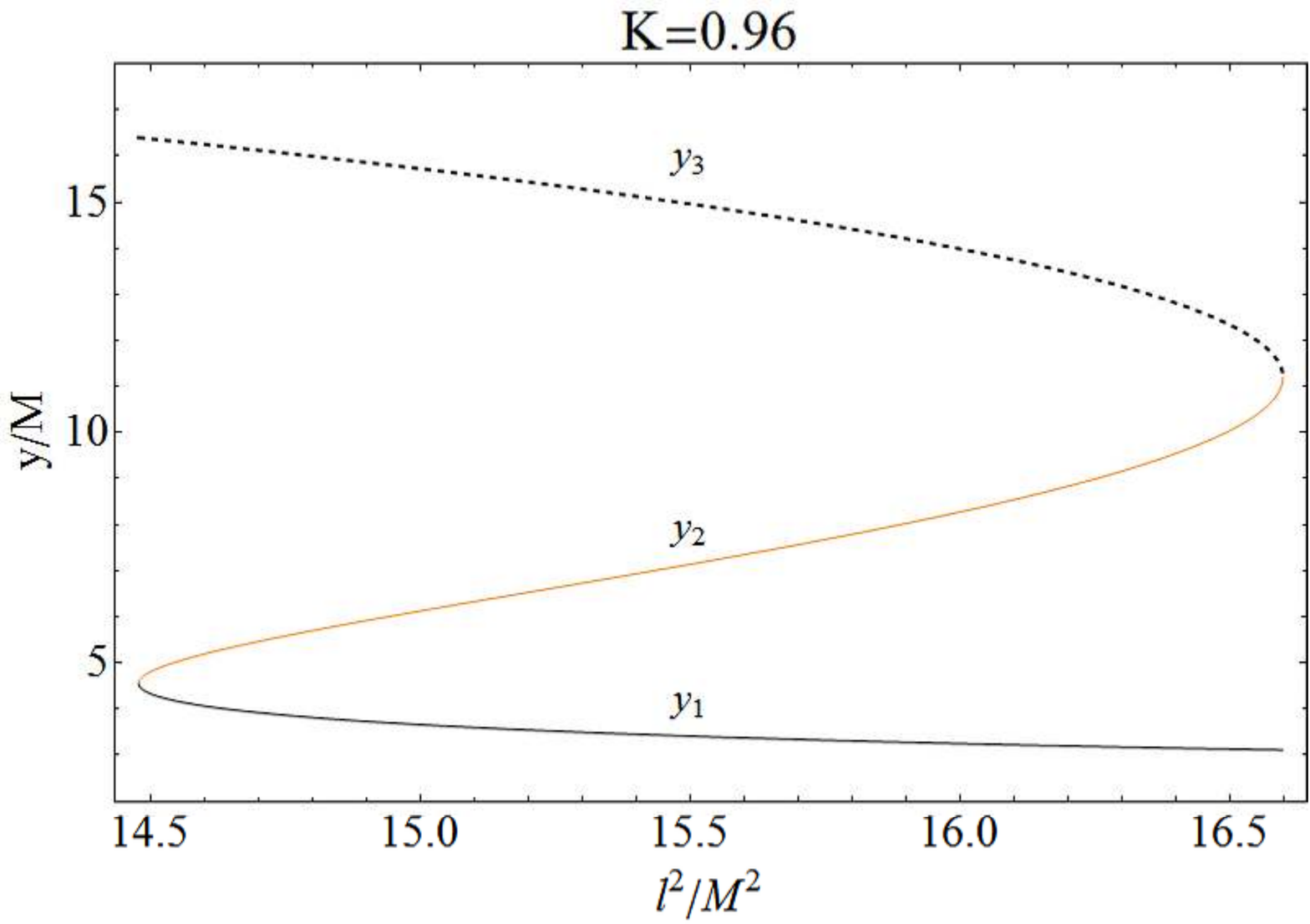}
\includegraphics[width=0.45\hsize,clip]{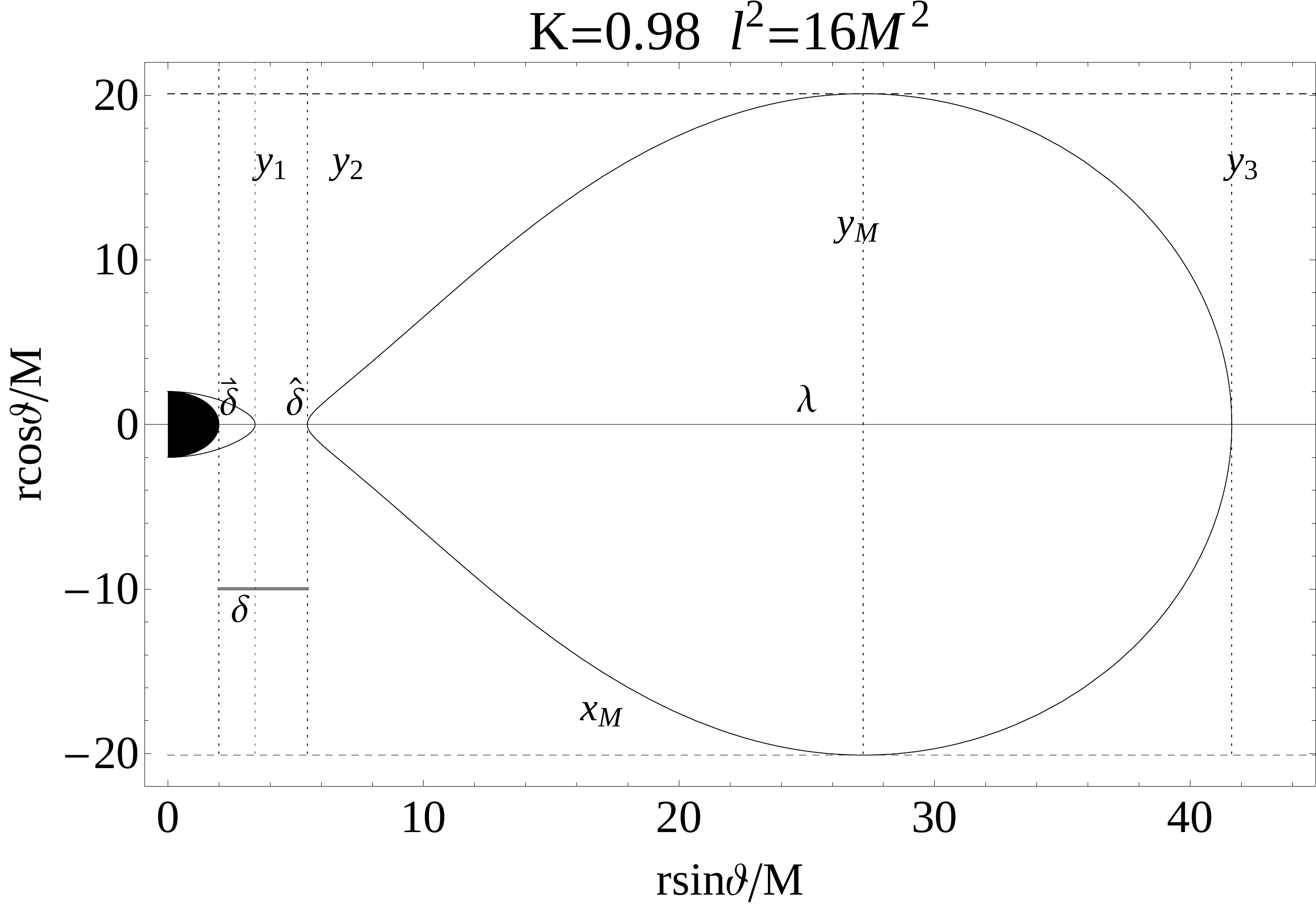}
\caption{(Color online) {Left panel}: $y_1/M$ (black curve), $y_2/M$ (orange
curve)  and $y_3/M$ (dashed curve) as a function of $l^2/M^2$ for $\rm{K}=0.96$.
\emph{Right panel}: the closed Boyer surface at $K=0.98$ and $l^2=16M^2$. $\lambda\equiv
y_3-y_2$ is the surface maximum diameter, $\delta\equiv y_2-2M$ is the distance from the
source, $\hat{\delta}\equiv y_2-y_1$ is the distance from the inner surface,
$\breve{\delta}\equiv y_1-2M$ is the distance of the inner surface from the horizon,
$h\equiv 2 x_M$ and $X_h\equiv y_M$.}
\label{Mattino}
\end{figure}

\medskip
\noindent
\textbf{Closed surfaces:} following the same procedure outlined in the previous
Subsection for $V_{sc}(L,r)$, we find that closed surfaces of the Boyer potential  in
the cartesian  coordinate $(x,y)$ are in the regions $2\sqrt{2}/3<\rm{K}<1$ and
$({\ell}_k^-)^2<{\ell^2}<({\ell}_k^+)^2$, where ${\ell^2}_k^->27/2$ (see
Figs.\il\ref{restylingl}). The surfaces are therefore:
\be
x=\pm\sqrt{\left[\frac{2M \left(\rm{K}^2 l^2+y^2\right)}{\rm{K}^2
\left(l^2-y^2\right)+y^2}\right]^2-y^2}
\ee
(see Fig.\il\ref{Mattino}, \emph{right panel}).

The maximum diameter $(x=0)$, is defined by the points $y={y}_2$ and $y={y}_3$, where
\bea
{y}_{1}/M&\equiv&\sqrt{\frac{1}{9} \left(\varsigma_k-6 \theta_k\sin\left[\frac{1}{6}
(\pi +2 \arccos\varepsilon_k)\right]\right)},
\\
{y}_{2}/M&\equiv&\sqrt{\frac{1}{9} \left(\varsigma_k-6 \theta_k\cos\left[\frac{1}{3}
(\pi +\arccos\varepsilon_k)\right]\right)}
\\
{y}_{3}/M&\equiv&\sqrt{\frac{1}{9} \left(\varsigma_k+6 \theta_k\cos\left[\frac{1}{3}
\arccos\varepsilon_k\right]\right)},
\eea
(see also Eqs.\il\ref{din}--\ref{dan}), being
\bea
\theta_k&\equiv&\sqrt{\frac{16+8 \rm{K}^2 (l/M)^2+\rm{K}^4 [(l/M)^2-32] (l/M)^2-2
\rm{K}^6 [(l/M)^2-12] (l/M)^2+\rm{K}^8 (l/M)^4}{\left(\rm{K}^2-1\right)^4}},
\\
\varepsilon_k&\equiv&\frac{64+K^2 (l/M)^2 \left(48+K^2 \left[48 \left(3 K^2-4\right)+6
\left(K^2-1\right)^2 \left(9 K^4-6 K^2+2\right) (l/M)^2-K^2 \left(K^2-1\right)^3
(l/M)^4\right]\right)}{\left(\rm{K}^2-1\right)^6 \theta_k^{3}},
\\\label{roiso}
\varsigma_k&\equiv&\frac{6 \left[\rm{K}^4 (l/M)^2-\rm{K}^2
(l/M)^2+2\right]}{\left(\rm{K}^2-1\right)^2}\, ,
\eea
(see Fig.\il\ref{Mattino}, \emph{left panel}).

\begin{figure}
\centering
\includegraphics[width=0.43\hsize,clip]{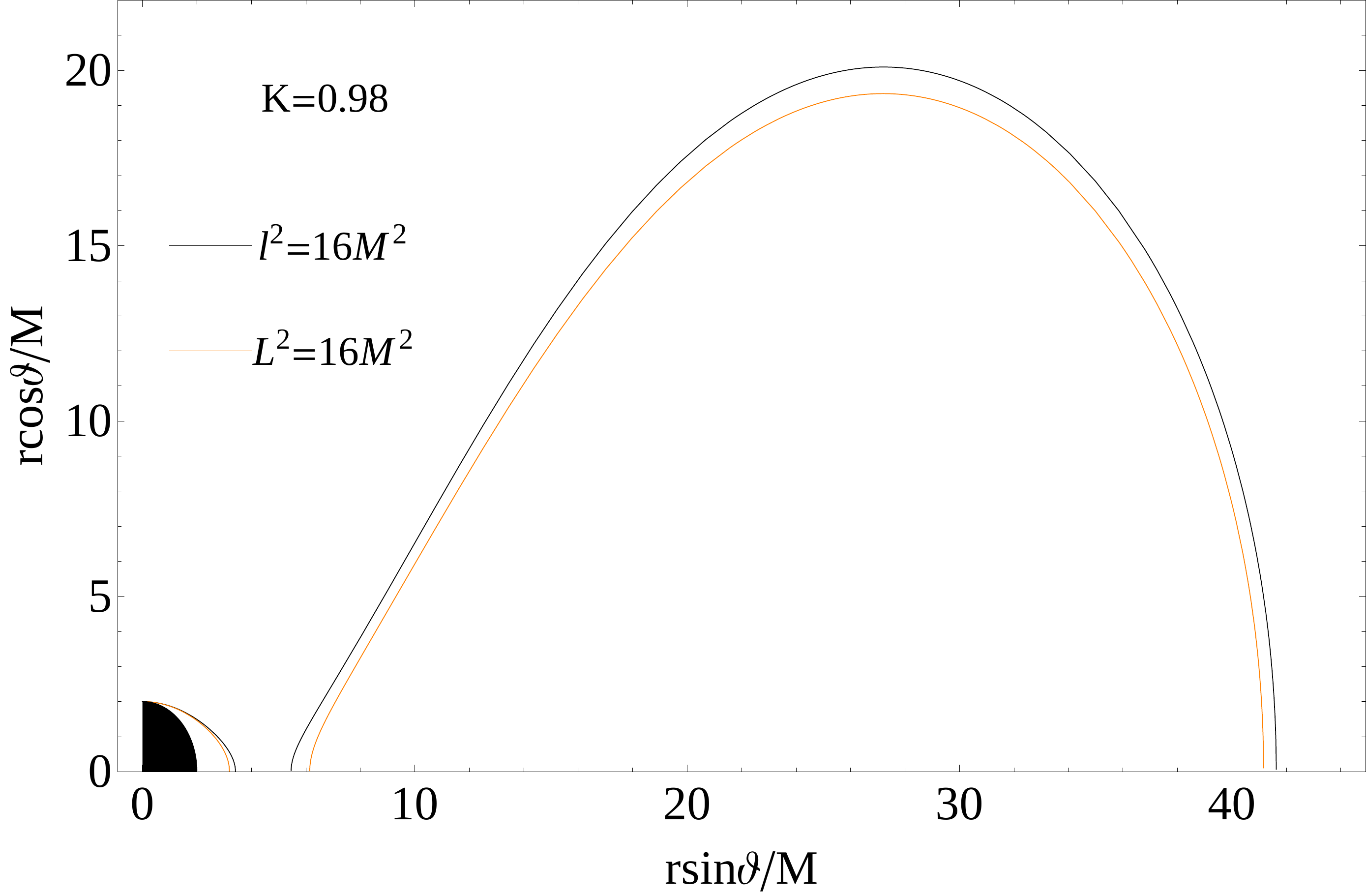}
\includegraphics[width=0.45\hsize,clip]{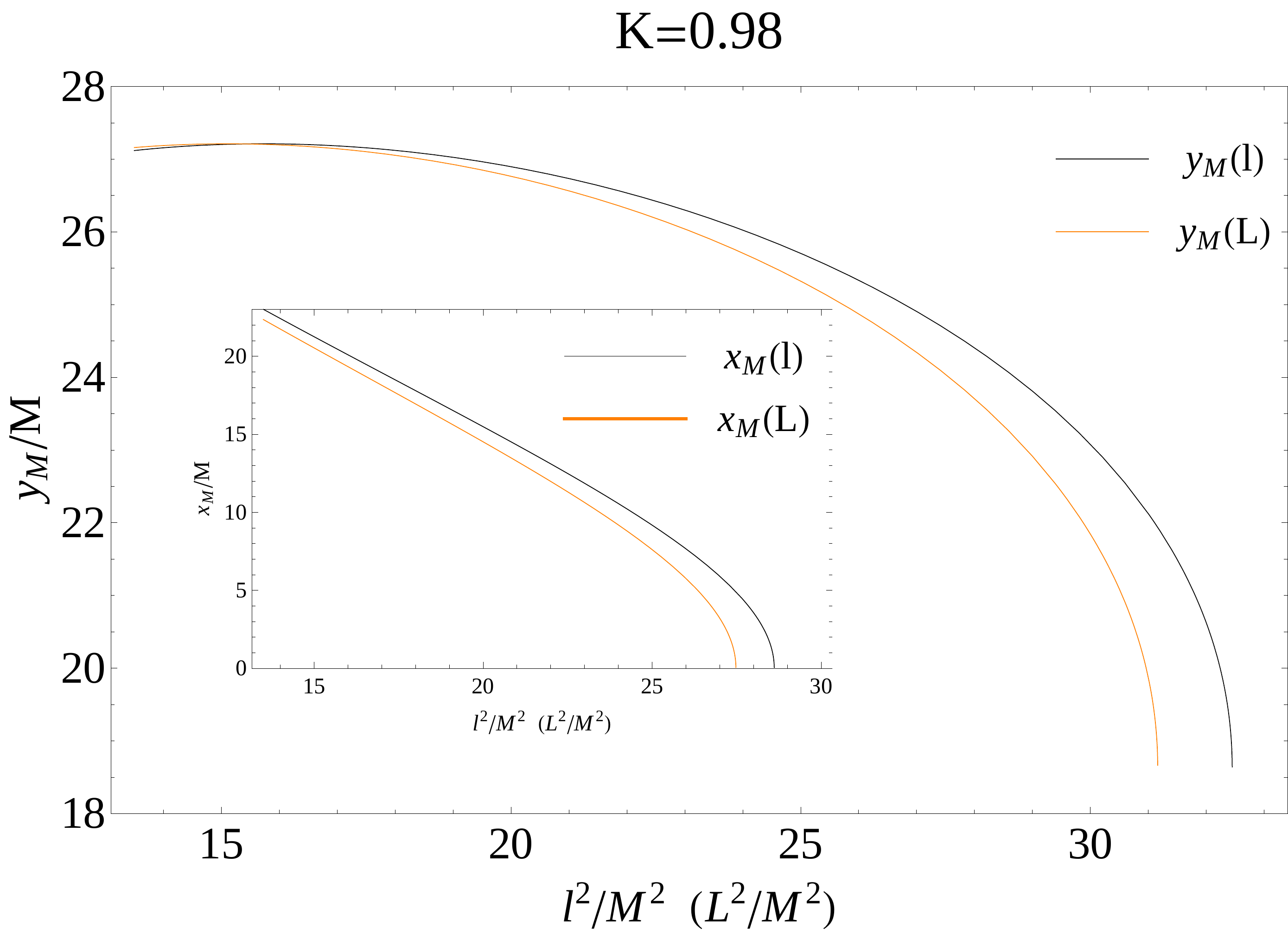}\\
\includegraphics[width=0.45\hsize,clip]{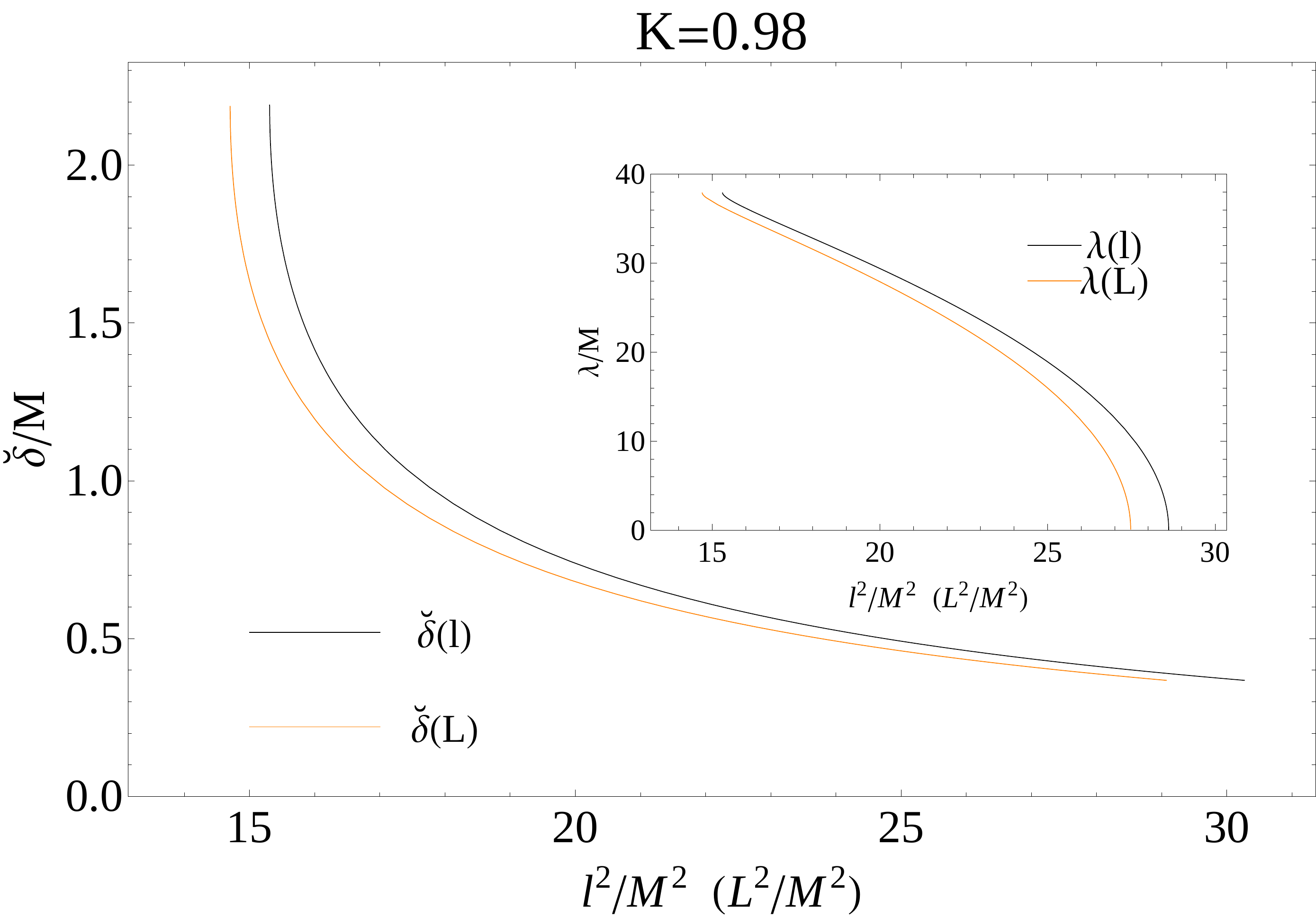}
\includegraphics[width=0.45\hsize,clip]{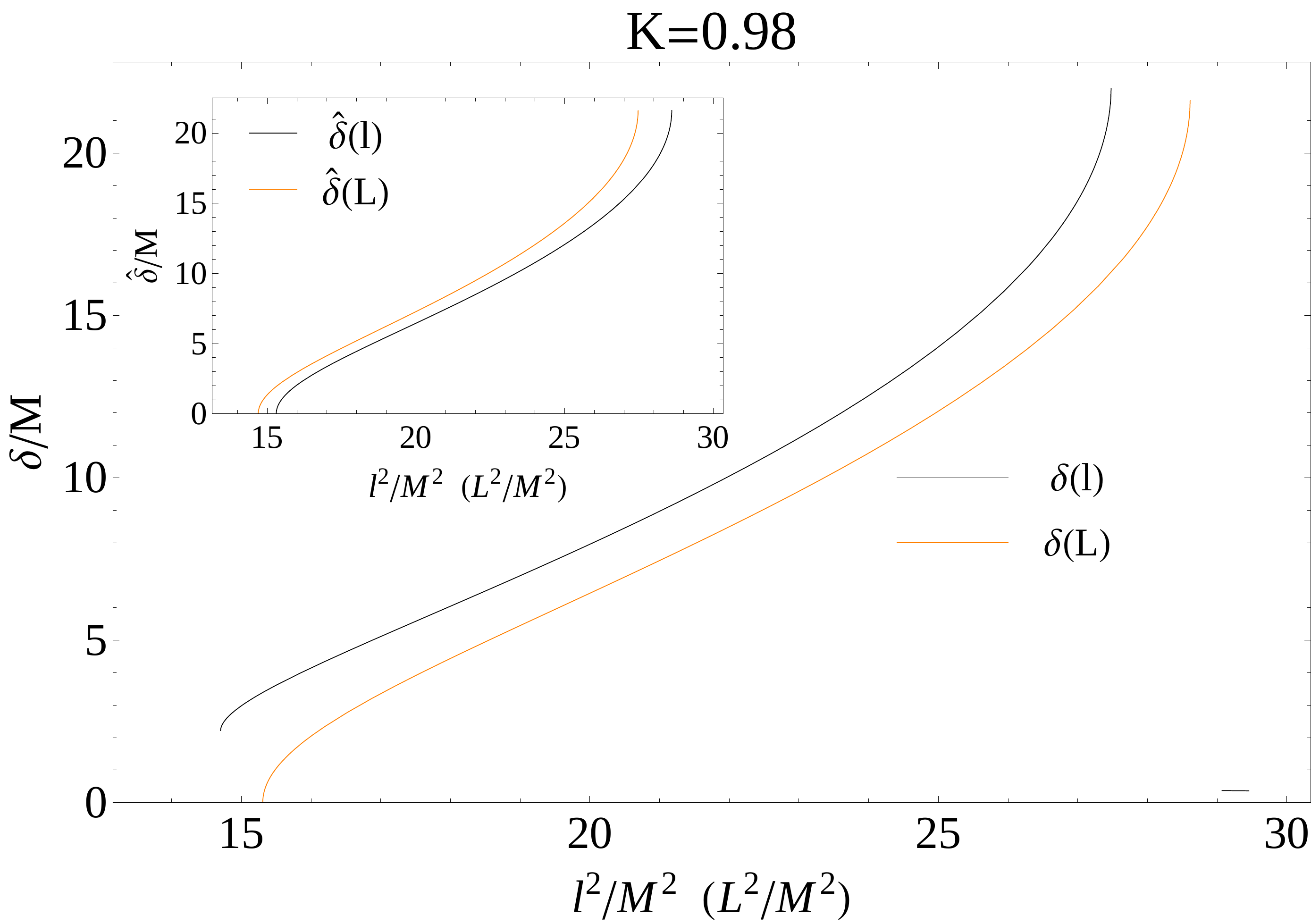}
\caption{(Color online) \emph{Upper left panel}: comparison of the closed Boyer
surfaces with energy $\rm{K}=0.98$ and angular momentum $l^2=16M^2$ (black curves) and
$L^2=16M^2$ (orange curve). It results $x(l^2=16M^2)>x(L^2=16M^2)$. \emph{Upper right
panel}: comparison of the surface maximum point $y_M/M$.  \emph{Inset}: comparison of
the surface maximum point $x_M/M$. \emph{Lower left panel}: comparison of the distance
of the inner surface from the horizon $\breve{\delta}\equiv y_1-2M$. \emph{Inset}:
comparison of the surface maximum diameter $\lambda\equiv y_3-y_2$. \emph{Lower right
panel}: comparison of the surface  distance from the horizon $\delta\equiv y_2-2M$.
\emph{Inset}: comparison of the distance from the inner surface $\hat{\delta}\equiv
y_2-y_1$.}
\label{tethe}
\end{figure}

The  maximum height for the surface is:
\bea
{y}_{M}/M&\equiv&\sqrt{\frac{3 \rm{K}^2 (l/M)^2+4 \sqrt{6} \left(\rm{K}^2-1\right)
\sqrt{-\frac{\rm{K}^4 (l/M)^2}{\left(\rm{K}^2-1\right)^3}} \cos\left[\frac{1}{3}
\arccos\left[-\frac{3}{4} \sqrt{\frac{3}{2}} \left(\rm{K}^2-1\right)^2
\sqrt{-\frac{\rm{K}^4 (l/M)^2}{\left(\rm{K}^2-1\right)^3}}\right]\right]}{3
\left(\rm{K}^2-1\right)}}.
\eea
The surfaces for $l=\rm{constant}$ are larger then those for $L=\rm{constant}$. The two
cases are compared in  Figs.\il\ref{tethe}: the  maximum diameter
\(\lambda(l)>\lambda(L)\); the distance from the source $\delta(l)>\delta(L)$; the
distance of the inner surface from the horizon $\breve{\delta}(l)>\breve{\delta}(L)$;
its maximum height $ h(l)>h(L)$ and finally the quantity $x_h(l)>x_h(L)$; the distance
from the inner surface $\hat{\delta}(l)<\hat{\delta}(L)$.

\noindent

\textbf{Cusps:} the closed (open) self-crossing surfaces $W=\rm{constant}$ are located
on the  maxima of the effective potential with energy $\rm{K}<1$ ($\rm{K}\geq1$) as
function of $r$ (see Figs.\il\ref{FAI} \emph{bottom}). {The critical points are located on
$r_{\pm}$: the maximum is $r_-$ when $3\sqrt{3/2}<\ell<3\sqrt{3}$.
The energy $V_{sc}(r_-)$ at the maximum $r_-$  and  at $\ell=4$   is $V_{sc}(r_-)=1$ and for $\ell<4$ it is
$V_{sc}(r_-)<1$.}

\section{The polytropic equation of state}\label{Sec:polu}

We consider the particular case of a polytropic equation of state
$p(r)=k\rho(r)^{\gamma}$, where the constant $\gamma$ is the polytropic index and $k>0$
is a constant. Using this relation in Eq.\il\ref{boyereq} we have:
\be\label{Volunia}
k\gamma\int^{\rho_{out}}_{\rho_{in}}\frac{
\rho^{(\gamma-1)}d\rho}{\rho(1+k\rho^{\gamma})}=-(W(r_{out})-W(r_{in})).
\ee
Integrating Eq.\il\ref{Volunia}, we obtain:
\be \label{sirena}
\ln\left[\left(\rho \left(k+\rho ^{1-\gamma }\right)^{\frac{1}{-1+\gamma
}}\right)^{\gamma }\right]=-W,\quad\mbox{for}\quad\gamma\neq1,
\ee
and
\be\label{sirenaa}
\ln\left[((1+k) \rho )^{\frac{k}{k+1}}\right]=-W,\quad\mbox{for}\quad \gamma=1,
\ee
(isothermal case).

Solving Eq.\il\ref{sirena} and \ref{sirenaa}  for $\rho$ and using
Eq.\il\ref{Edverdi}, we find respectively:
\be\label{peter}
\bar{\rho}_{\gamma}\equiv\left[\frac{1}{k}\left(V_{sc}^{-\frac{-1+\gamma }{\gamma
}}-1\right)\right]^{\frac{1}{(-1+\gamma)}},\quad\mbox{for}\quad\gamma\neq1
\ee
and
\be\label{peter1}
\rho_k\equiv V_{sc}^{-\frac{1+k}{ k}}\frac{1}{1+k},\quad\mbox{for}\quad \gamma=1.
\ee
In the following we adopt the normalization: $\rho_{\gamma}\equiv
k^{1/(\gamma-1)}\bar{\rho}_{\gamma}$, which is independent from $k$. The following
limits are satisfied:
\be
\lim_{r\rightarrow\infty}\rho_{\gamma}=0,\quad\lim_{r\rightarrow2M}\rho_{\gamma}=\infty,
\ee
and
\be
\lim_{r\rightarrow\infty}{\rho}_{k}=1,\quad\lim_{r\rightarrow2M}{\rho}_{k}=\infty.
\ee

We underline that for $\gamma=1$ it is
\(
\rho^{out}_{k}=\rho^{in}_{k}\left({V^{out}_{sc}/V^{in}_{sc}}\right)^{-\frac{1+k}{k}},
\)
and for $\gamma\neq1$ it is
\(
k (\bar{\rho})_{out}^{\gamma-1}=(k (\bar{\rho})_{in}^{\gamma-1}+1)
\left({V^{out}_{sc}/V^{in}_{sc}}\right)^{\frac{1-\gamma}{\gamma}}-1.
\)
In the polytropic case it is $\rho' =p'/(k\gamma\rho^{\gamma-1})$, thus it is $\rho'=0$
when $p'=0$ and, being $\gamma>0$, the maxima (minima) of $p$ correspond to maxima
(minima) of $\rho$.

\subsection{The case: $\gamma\neq1$}

If the polytropic index is $\gamma\neq1$, the density $\rho=\rho_{\gamma}$ is:
\be
{\rho}_{\gamma}\equiv C^{1/(-1+\gamma )},
\ee
with $C\equiv (V_{sc}^{-2})^{\frac{-1+\gamma }{2 \gamma }}-1$ and
$(V_{sc}^{-2})\equiv\left(\frac{r}{r-2M}-\frac{\ell^2M^2}{r^2}\right)>0$.

\begin{figure}
\centering
\includegraphics[width=0.45\hsize,clip]{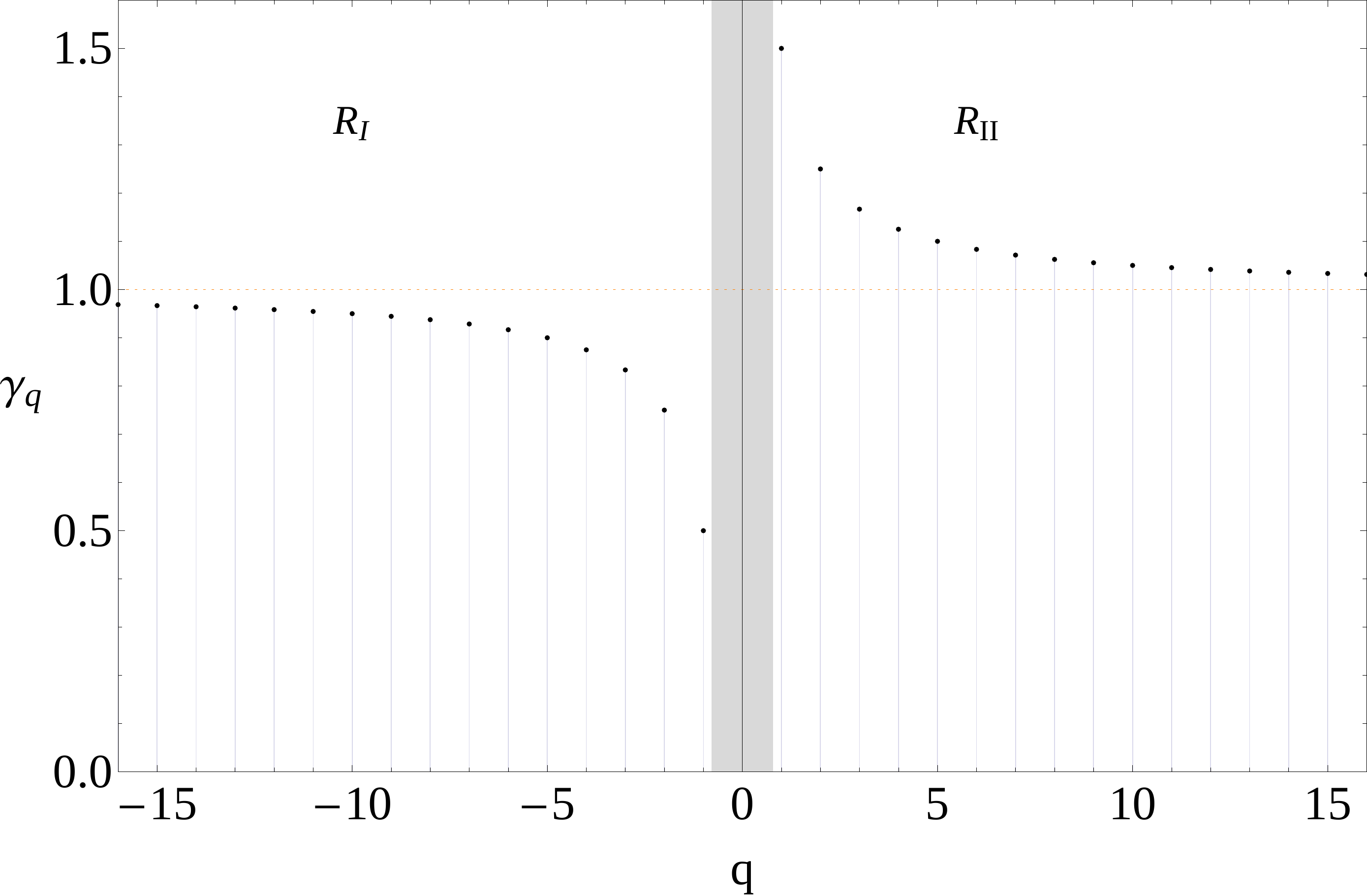}
\includegraphics[width=0.45\hsize,clip]{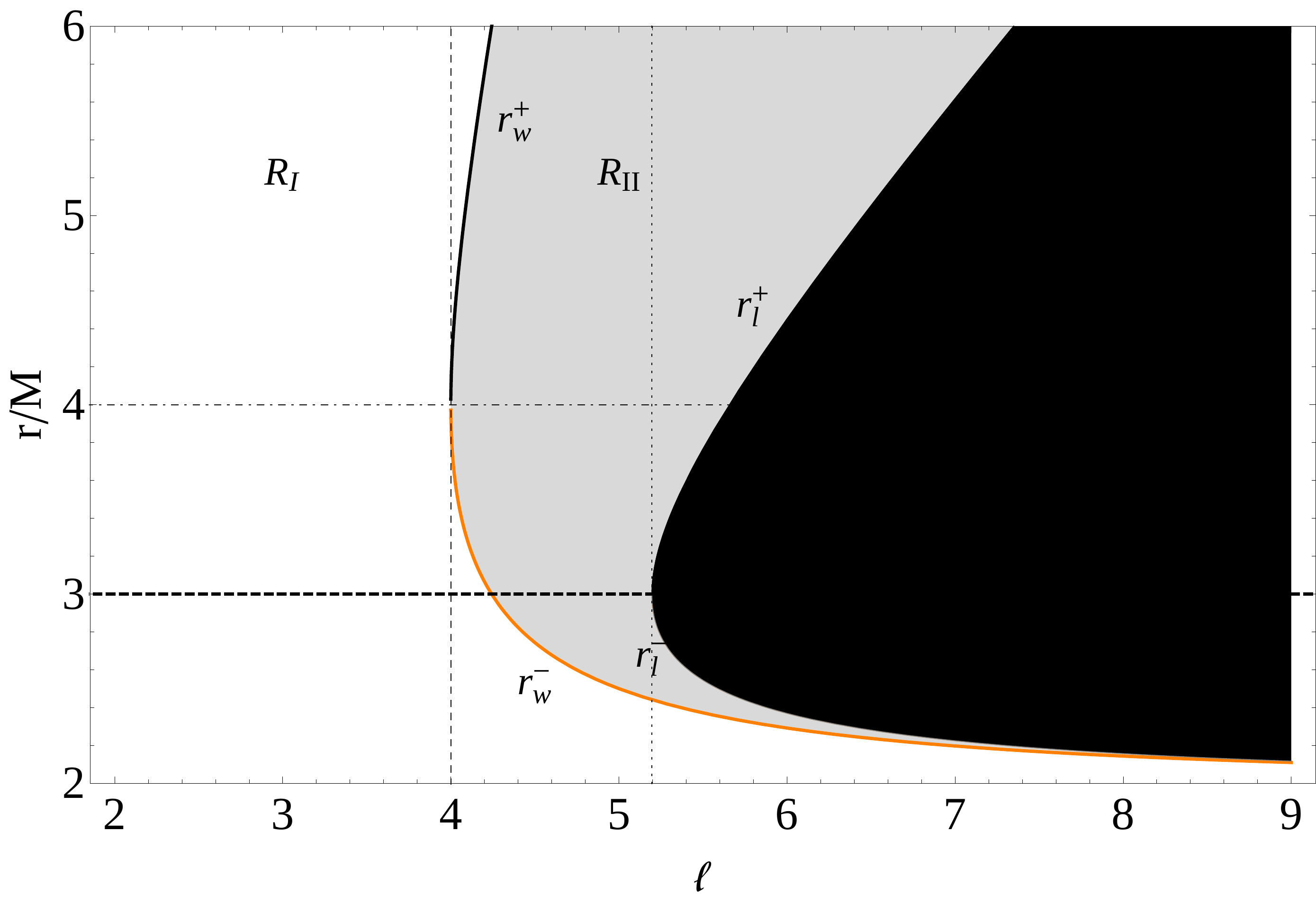}
\caption{(Color online) {\emph{Left panel}: $\gamma_q$ as a
function of $q$, where $|q|\geq 1$ are integers. For large values of $|q|$, the index
$\gamma$ tends asymptotically to $\gamma=1$. Gray region is forbidden. The maximum value
of $\gamma_q$ is $\gamma_q=3/2$, the minimum is $\gamma_9=1/2$. $C<0$ for  $V_{sc}^2<1$
with $0<\gamma_q <1$ in $R_{\ti{I}}$ and $V_{sc}^2>1$ with $\gamma_q >1$ in
$R_{\ti{II}}$.  \emph{Right panel}: $r_l^+$ (black line), $r_l^-$ (orange line), $r_W^+$
(thick  black line) and $r_W^-$ (thick  orange line) as a function of $\ell$. Dashed line
marks $\ell=4$, dotted line $\ell=3\sqrt{3}$, dotted--dashed line $r=r_{mbo}$,
thick--dashed line $r=r_{lco}$.  White region correspond to the range  $R_{\ti{I}}$,
gray region to $R_{\ti{II}}$.}} \label{alba}
\end{figure}

We distinguish between two cases:
\begin{enumerate}
  \item $C>0$ and the density $\rho_{\gamma}$ is defined for all $\gamma$;
  \item $C<0$ and the density $\rho_{\gamma}$ is defined for
      $\gamma=\gamma_q\equiv1+\frac{1}{2q}$, where $|q|\geq 1$ are integers (see
      Fig.\il\ref{alba}, \emph{left panel}).
\end{enumerate}

The condition $C>0$ is satisfied in two cases: where $V_{sc}^2<1$ and $\gamma >1$, in
the ranges:
\be\label{R1}
R_{\ti{I}}\equiv\left\{
\begin{array}{lcl}
0<\ell<4& \mbox{in}& r>2M \\
\ell\geq 4 &\mbox{in} & 2M<r<r_W^- \quad r>r_W^+
\end{array}
\right.
\ee
and where $V_{sc}^2>1$ and $0<\gamma <1$, in the ranges:
\be\label{RII}
R_{\ti{II}}\equiv\left\{
\begin{array}{lcl}
4<\ell<3 \sqrt{3}& \mbox{in}& r_W^-<r<r_W^+\\
\ell=3 \sqrt{3}&\mbox{in} & r_W^-<r<r_W^+\quad r\neq r_{lco}\\
\ell>3 \sqrt{3}&\mbox{in} &r_W^-<r<r_l^-\quad r^+_l<r<r_W^+
\end{array}
\right.
\ee
with
\be
r_W^{\pm}/M\equiv \ell^2\pm\frac{\ell}{4} \sqrt{\ell^2-16 },
\ee
(see Fig.\il\ref{alba}/ \emph{right panel}).

We can summarize as follows: when the polytropic index $\gamma=\gamma_{q}$, the fluid
density $\rho$ is defined for the conditions $R_{\ti{I}}\cup R_{\ti{II}}$, when
$\gamma\neq\gamma_{q}$ it is defined only for the conditions $R_{\ti{I}}$. In the
following subsections we will discuss an example of $\gamma=5/3\neq\gamma_q$ and the
particular case  $\gamma_q(q=1)=3/2$.

\subsubsection{The adiabatic case: $\gamma=5/3$}
We consider now the particular case $\gamma=5/3$. This polytropic index is adopted to
describe a large variety of matter models, as the generic degenerate matter like  star
cores of white dwarfs \citep[see e.g.][]{q}.

\begin{figure}
\centering
\includegraphics[width=0.45\hsize,clip]{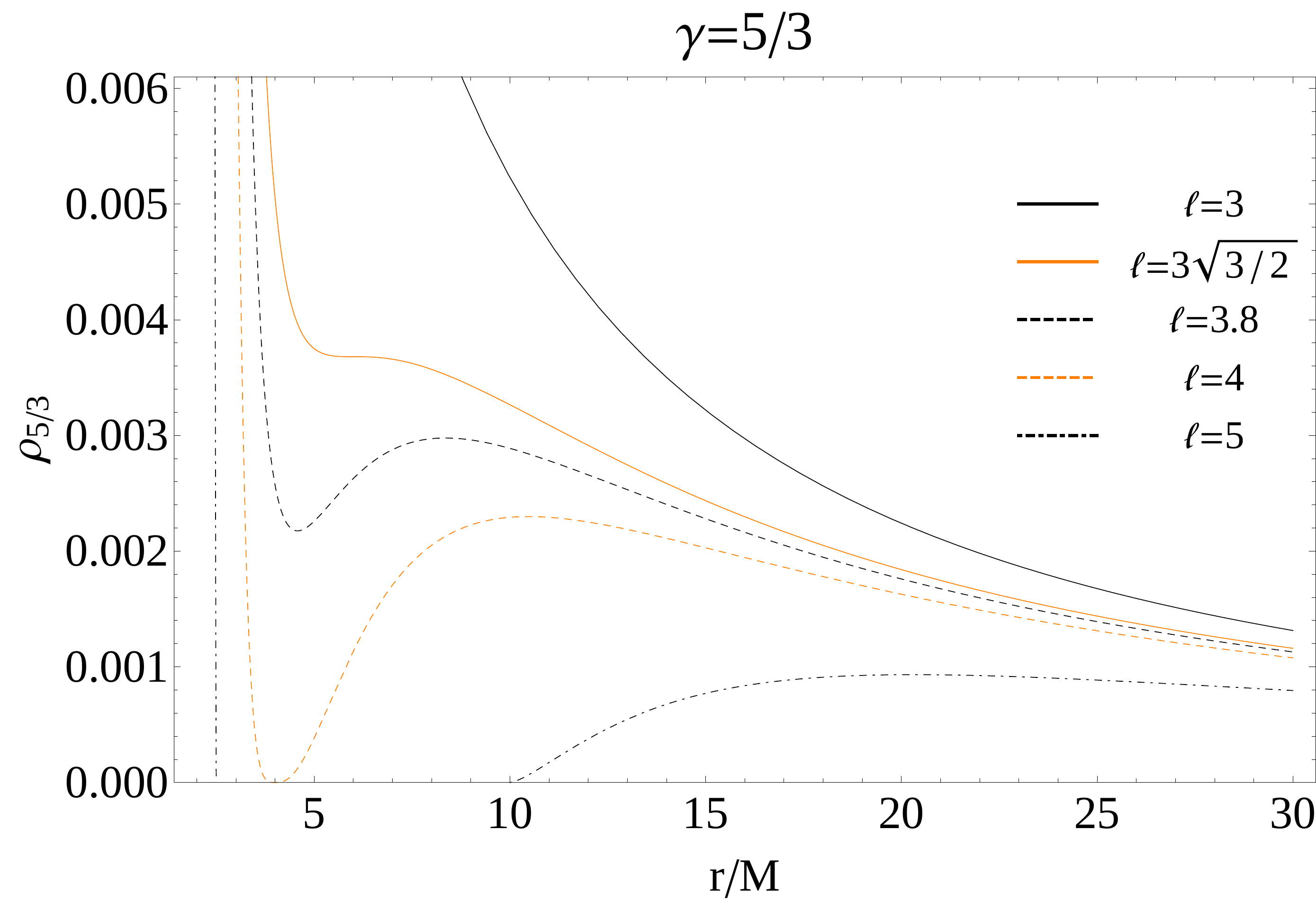}
\includegraphics[width=0.45\hsize,clip]{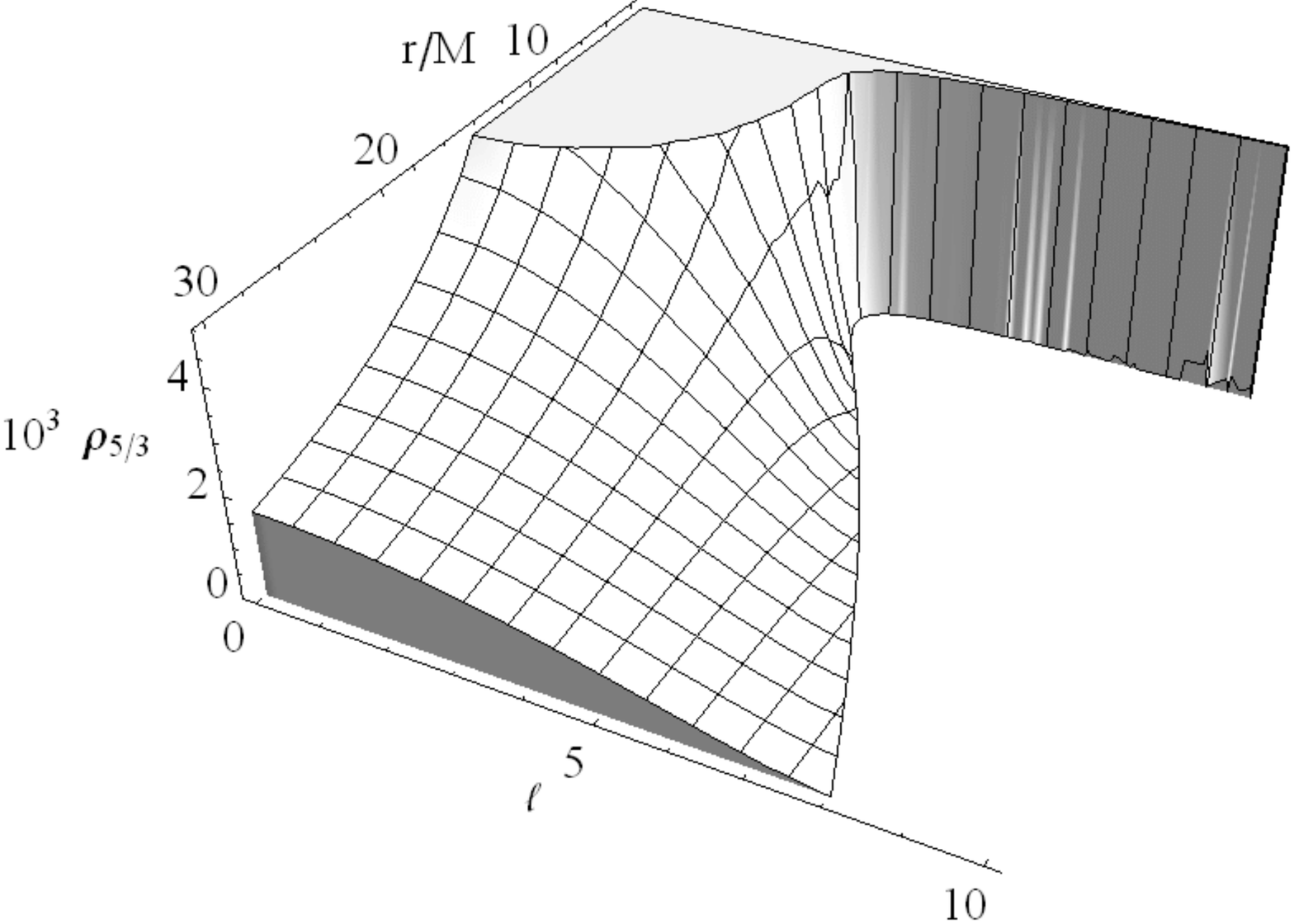}
\caption{{(Color online) {\emph{Left panel}: $\rho_{5/3}$ as a
function of $r/M$ for different values of $\ell$.
 \emph{Right panel}:
$\rho_{5/3}$ as a function of $r/M$ and $\ell$. $\rho_{5/3}$ is not defined in
$[r^-_W,r^+_W]$. }}}
\label{ricPOIp}
\end{figure}

The density  $\rho_{\gamma}$ is then:
\be
\rho_{5/3}=\left[\left(\frac{r}{r-2M}-\frac{M^2\ell^2}{r^2}\right)^{1/5}-1\right]^{3/2},
\ee
 defined in the range $R_{\ti{I}}$ as in Eq.\il\ref{R1}:
\bea
0<\ell<4&\quad\mbox{in}\quad& r>2M,
\\
\ell\geq 4&\quad\mbox{in}\quad&2M<r<r^-_W\;, \;\;r>r^+_W,
\eea
where
\(
\lim_{r\rightarrow r_W^{\pm}}\rho_{5/3}=0,
\)
(see Figs.\il\ref{ricPOIp}).

\begin{figure}
\centering
\includegraphics[width=0.3\hsize,clip]{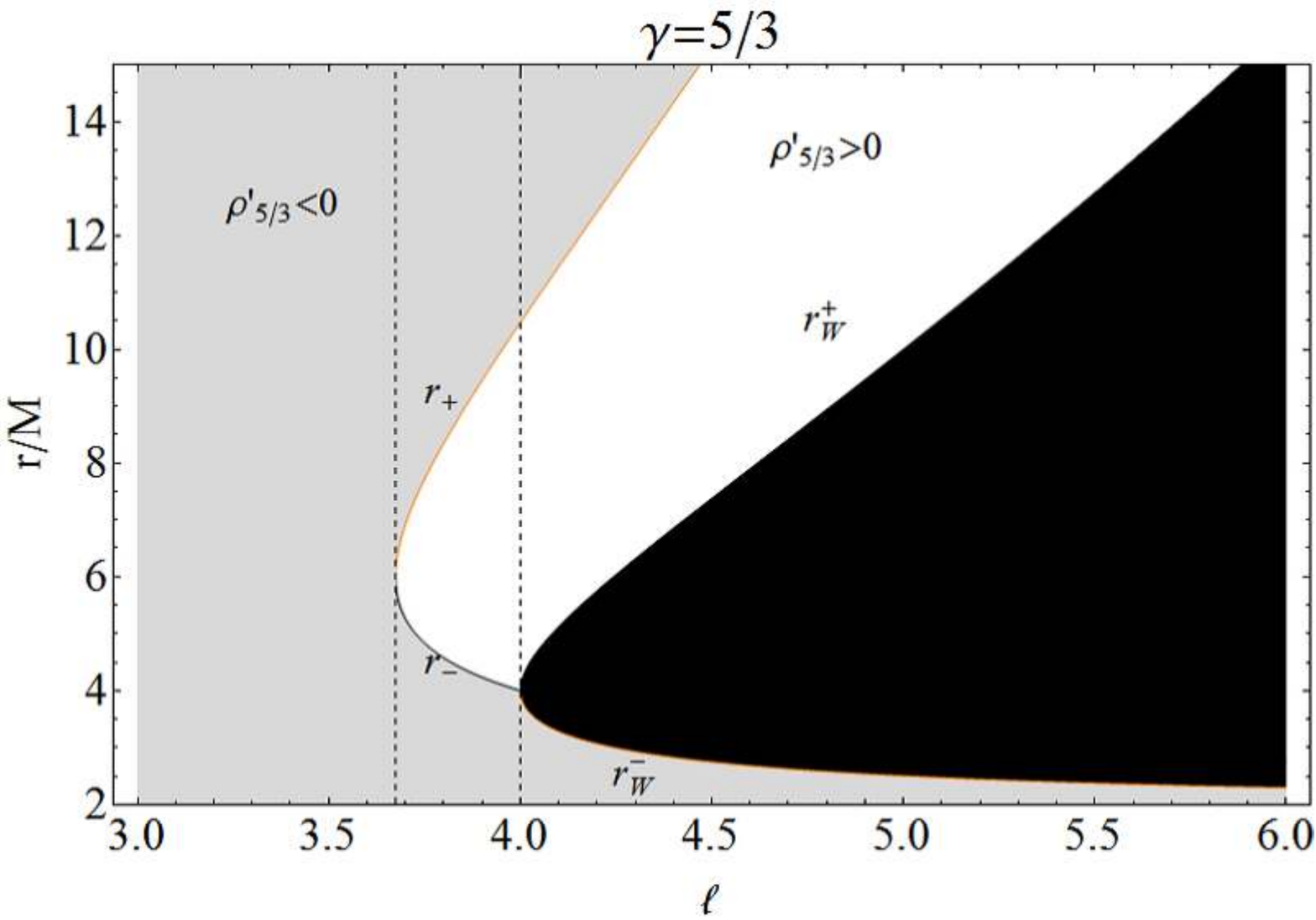}
\includegraphics[width=0.3\hsize,clip]{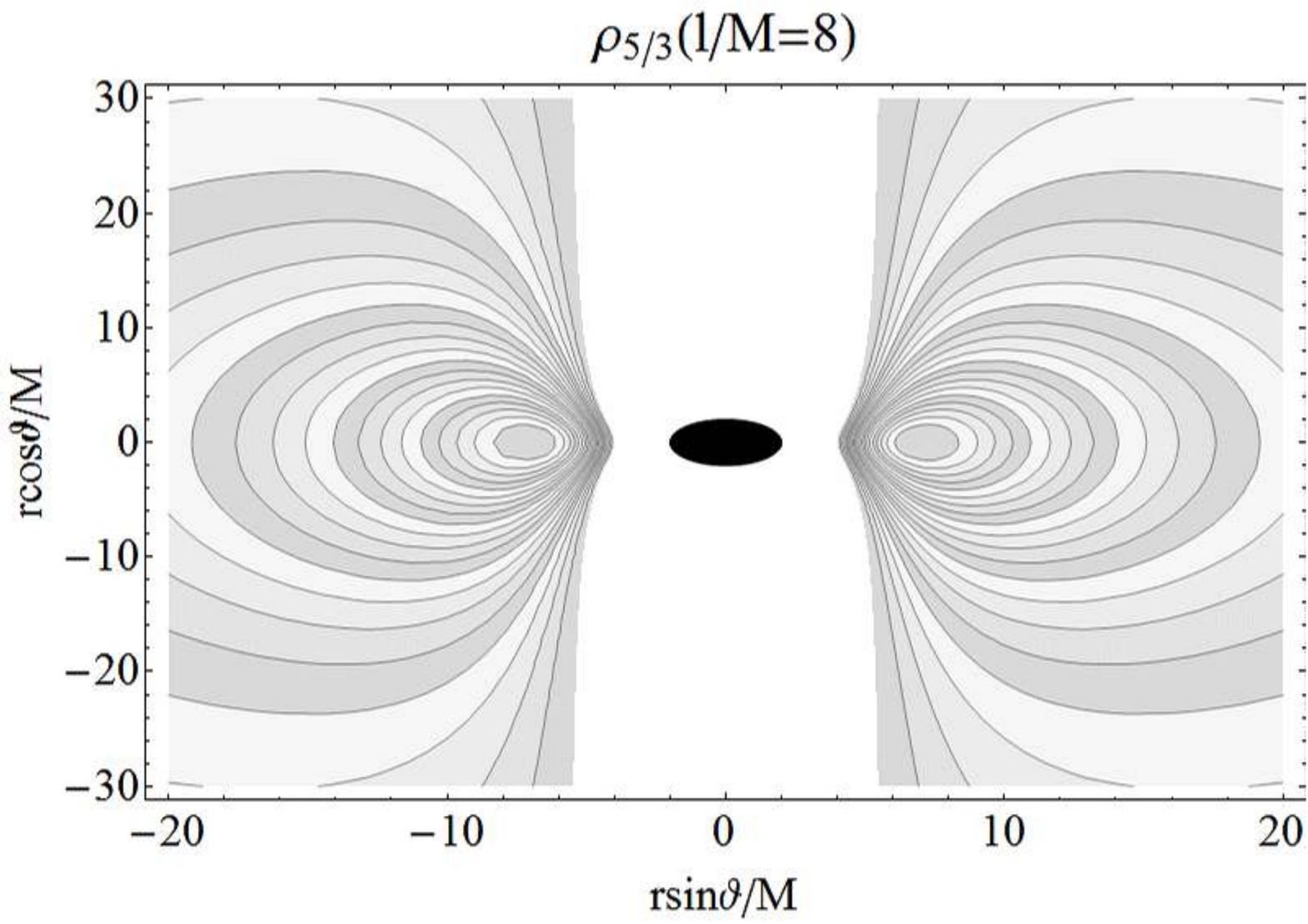}
\includegraphics[width=0.3\hsize,clip]{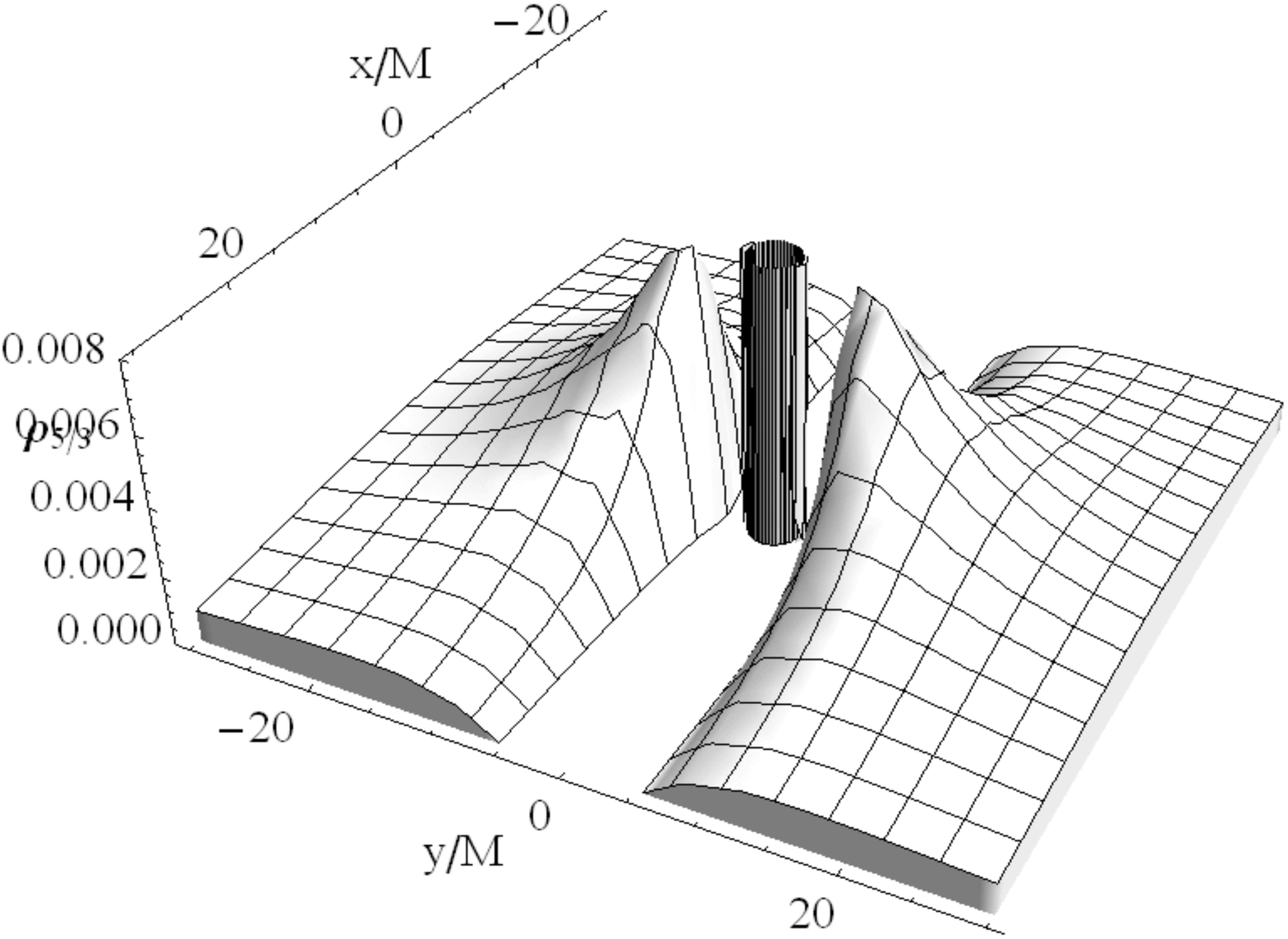}
\caption{(Color online) {\emph{Left panel}: $r_W^{\pm}$ and
$r_{\pm}$ as a function of $\ell$. In the gray (white) region $\rho'_{5/3}<0$
($\rho'_{5/3}>0$), in the black region $\rho'_{5/3}$ is not defined. \emph{Central
panel}: $\rho_{5/3}=\mbox{constant}$ in the plane $(x,y)$. \emph{Right panel}:
$\rho_{5/3}$ as a function of $(x,y)$  for $l=8M$. The black surface is
$\sqrt{x^2+y^2}=2M$.}} \label{ricPOIm}
\end{figure}

The critical points of $\rho_{5/3}$ can be found as solutions of  $\rho'_{5/3}=0$:
\bea
\ell=3\sqrt{{3}/{2}}\quad\mbox{in}\quad r=r_{lsco},
\quad
3 \sqrt{{3}/{2}}<\ell<4\quad\mbox{in}\quad r=r_{\pm},
\quad
\ell\geq 4\quad\mbox{in}\quad r=r_+,
\eea
and it is $\rho'_{5/3}>0$, (density increasing with the orbital radius) for:
\bea
3 \sqrt{{3}/{2}}<\ell\leq 4\quad\mbox{in}\quad r_-<r<r_+,
\quad
\ell>4\quad\mbox{in}\quad  r_W^+<r<r_+.
\eea
We thus conclude that $r_-$ is a minimum and  $r_+$ is a maximum of $\rho_{5/3}$ (see
Figs.\il\ref{ricPOIm}).

\subsubsection{The case: $\gamma=3/2$}

We consider  the particular case  $\gamma_q(q=1)=3/2$, which is in the  extreme cases
$1<\gamma<5/3$.

\begin{figure}
\centering
\includegraphics[width=0.3\hsize,clip]{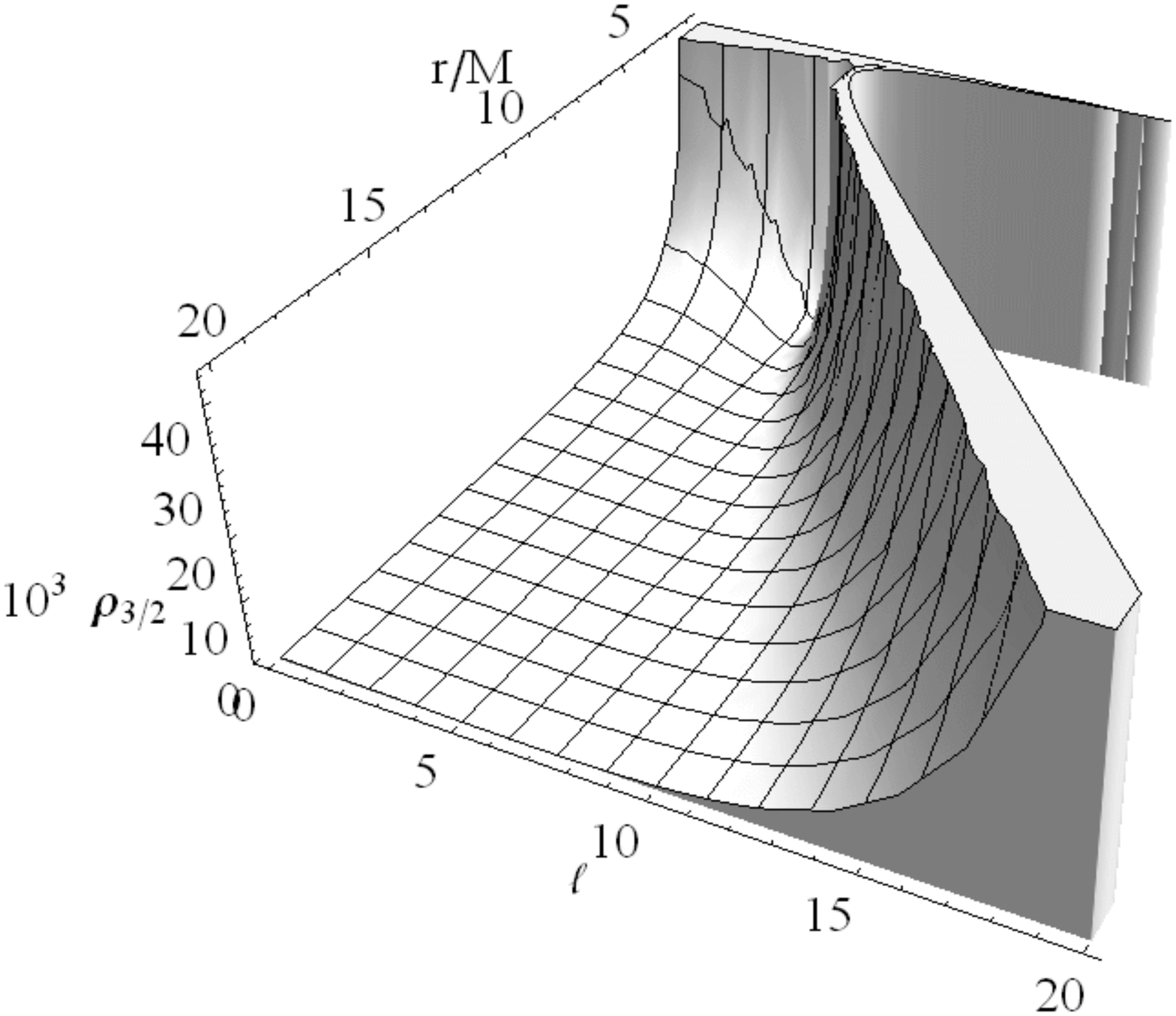}
\includegraphics[width=0.3\hsize,clip]{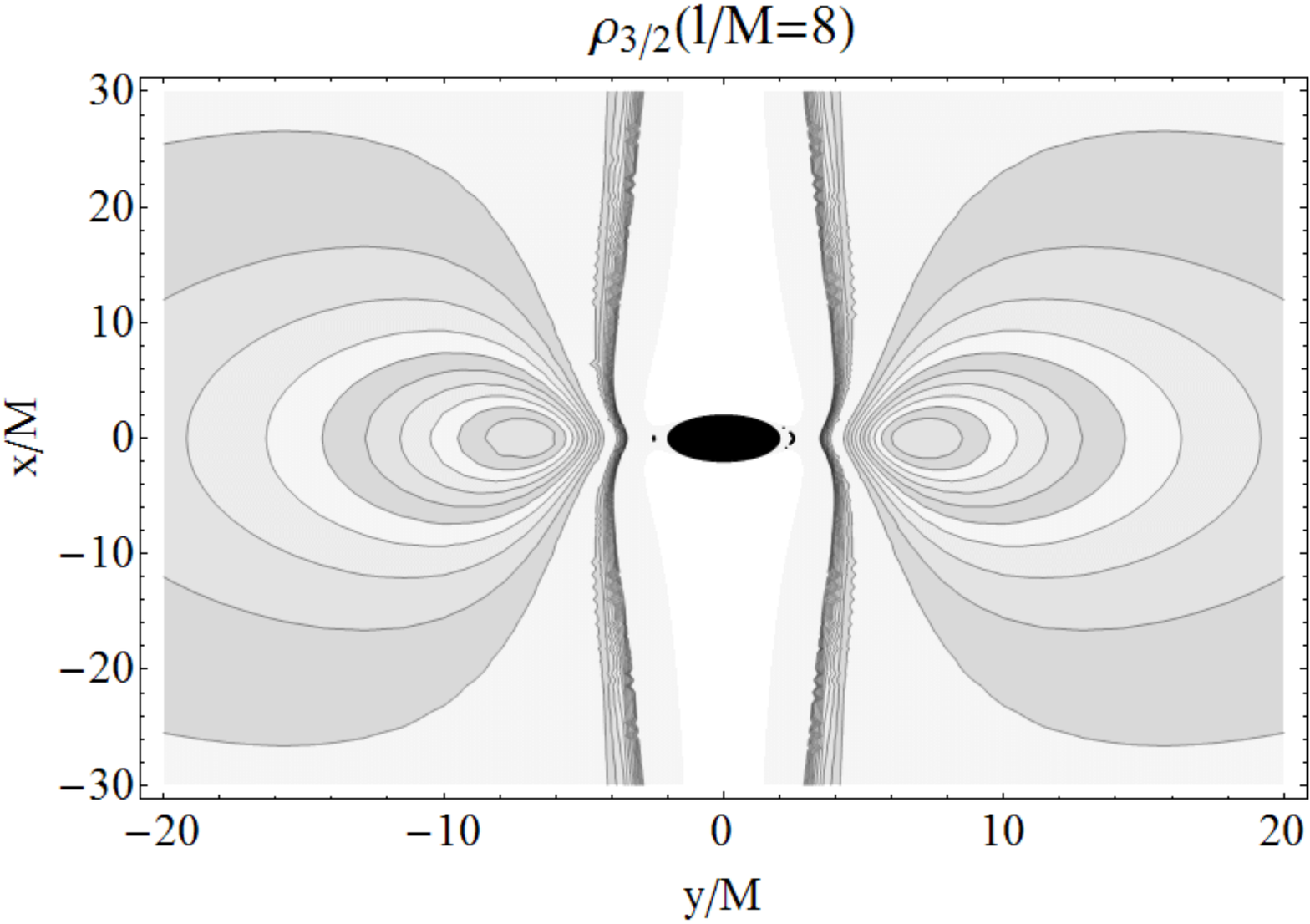}
\includegraphics[width=0.3\hsize,clip]{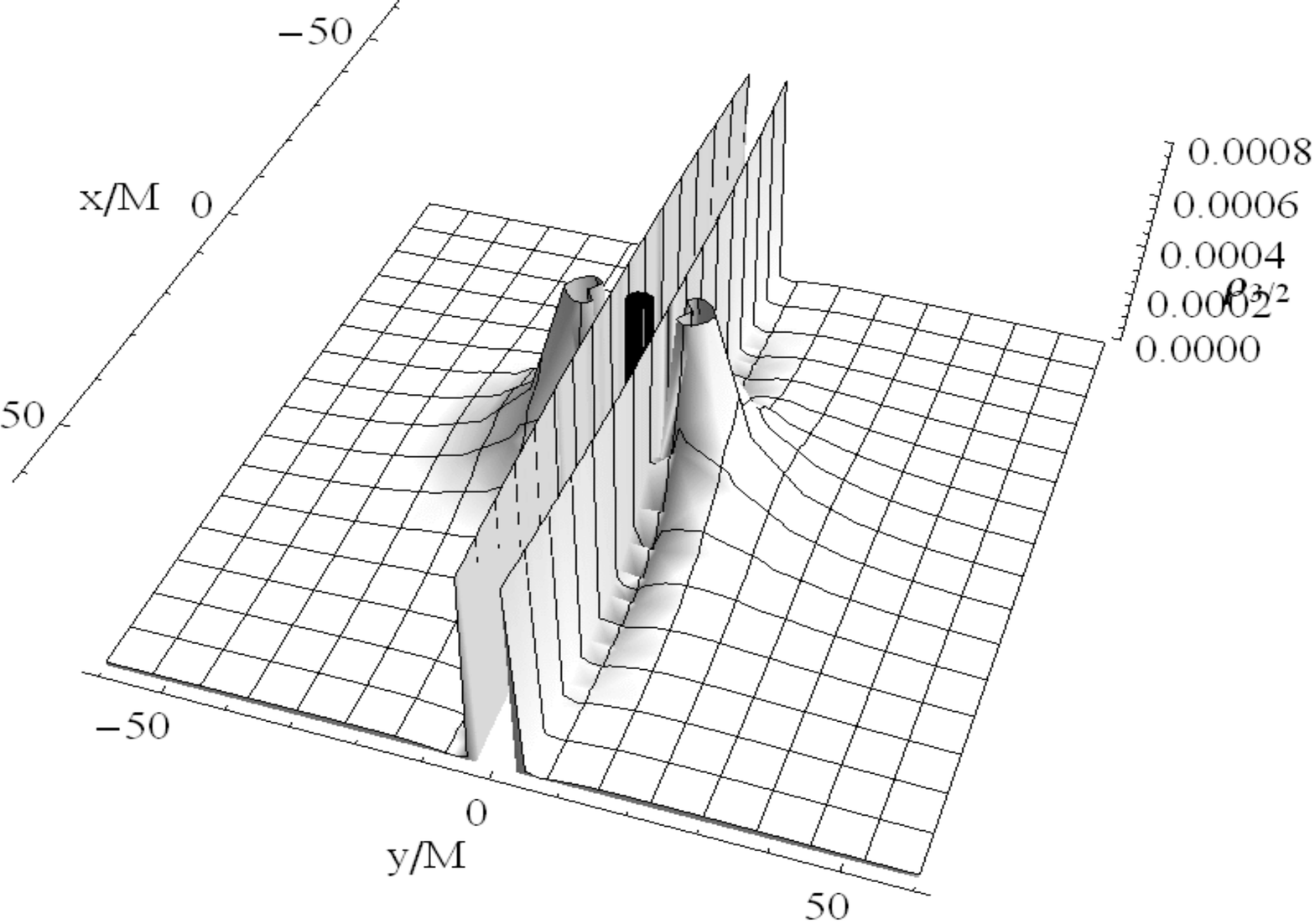}
\caption{{\emph{Left panel}:$\rho_{3/2}$ as a
function of $r/M$ and $\ell$. $\rho_{3/2}$ is not defined in $[r^-_l,r^+_l]$.
\emph{Central panel}: $\rho_{5/3}=\mbox{constant}$ in the plane $(x,y)$. \emph{Right
panel}: $\rho_{3/2}$ as a function of $(x,y)$  for $l=4M$. The black surface is
$\sqrt{x^2+y^2}=2M$.}} \label{ricPOImico}
\end{figure}

The density  $\rho_{3/2}\neq0$ is then:
\be
\rho_{3/2}=\left[\left(\frac{r}{r-2M}-\frac{M^2\ell^2}{r^2}\right)^{1/6}-1\right]^{2},
\ee
 defined in the range  $R_{\ti{I}}\cup R_{\ti{II}}$ as in Eqs.\il\ref{R1}-\ref{RII}:
\bea
0<\ell<4&\quad\mbox{in}\quad& r>2M,
\\
4\leq\ell<3\sqrt{3}&\quad\mbox{in}\quad&r>2M, \quad r\neq r_W^{\pm}
\\
\ell\geq3\sqrt{3}&\quad\mbox{in}\quad&2M<r<r_l^-,\quad r>r_l^+ \quad r\neq r_W^{\pm},
\eea
where
\(
\lim_{r\rightarrow r_W^{\pm}}\rho_{3/2}=0,
\)
(see Figs.\il\ref{ricPOImico}).

The critical points of $\rho_{3/2}$ can be found as solutions of  $\rho'_{3/2}=0$, or
noting that $p'=\frac{3}{2}\rho'\sqrt{\rho}$. We summarize these results concluding that
$r_-$ is a minimum and  $r_+$ is a maximum of $\rho_{3/2}$.

\subsection{The isothermal case: $\gamma=1$}

For  an isothermal  equation of state $(\gamma=1)$, the solution of
Eq.\il\ref{sirenaa} is $\rho=\rho_k$  (see Eq.\il\ref{peter}). This function is not
defined in the range $r_l^-\leq r\leq r_l^+$ and the fluid angular momentum is
$0<\ell<\ell_r$.

In order to describe the regions of maximum and minimum density, we study the function
$\rho_k'$. Being $p=k\rho$, it is clearly $p'=k\rho'$. Thus it is $\rho'=0$ when $p'=0$.
Moreover, being $k>0$, the maxima (minima) of $p$ correspond to maxima (minima) of
$\rho$. The existence of critical points for the isothermal case is therefore studied in
Sect.\il\ref{Sec:neutrall} in terms of the critical points of $G_r$:
\bea\label{1}
\ell=3 \sqrt{{3}/{2}}\quad\mbox{in}\quad r=r_{lsco},
\quad
3 \sqrt{{3}/{2}}<\ell<3 \sqrt{3}\quad\mbox{in}\quad r=r_{\pm},
\quad
\ell\geq 3 \sqrt{3}\quad\mbox{in}\quad r=r_+,
\eea
or also
\(
r>r_{lco} \quad \ell=\ell_{\rm{\rm{K}}}.
\)

\section{The  fluid  proper angular velocity}\label{Sec:proper}
{This Section concerns with the analysis of the  fluid velocity field, in
particular  we are  interested in assessing  the  orbits and the  plans  where the fluid proper
velocity is maximum or minimum. }

\begin{figure}
\centering
\includegraphics[width=0.45\hsize,clip]{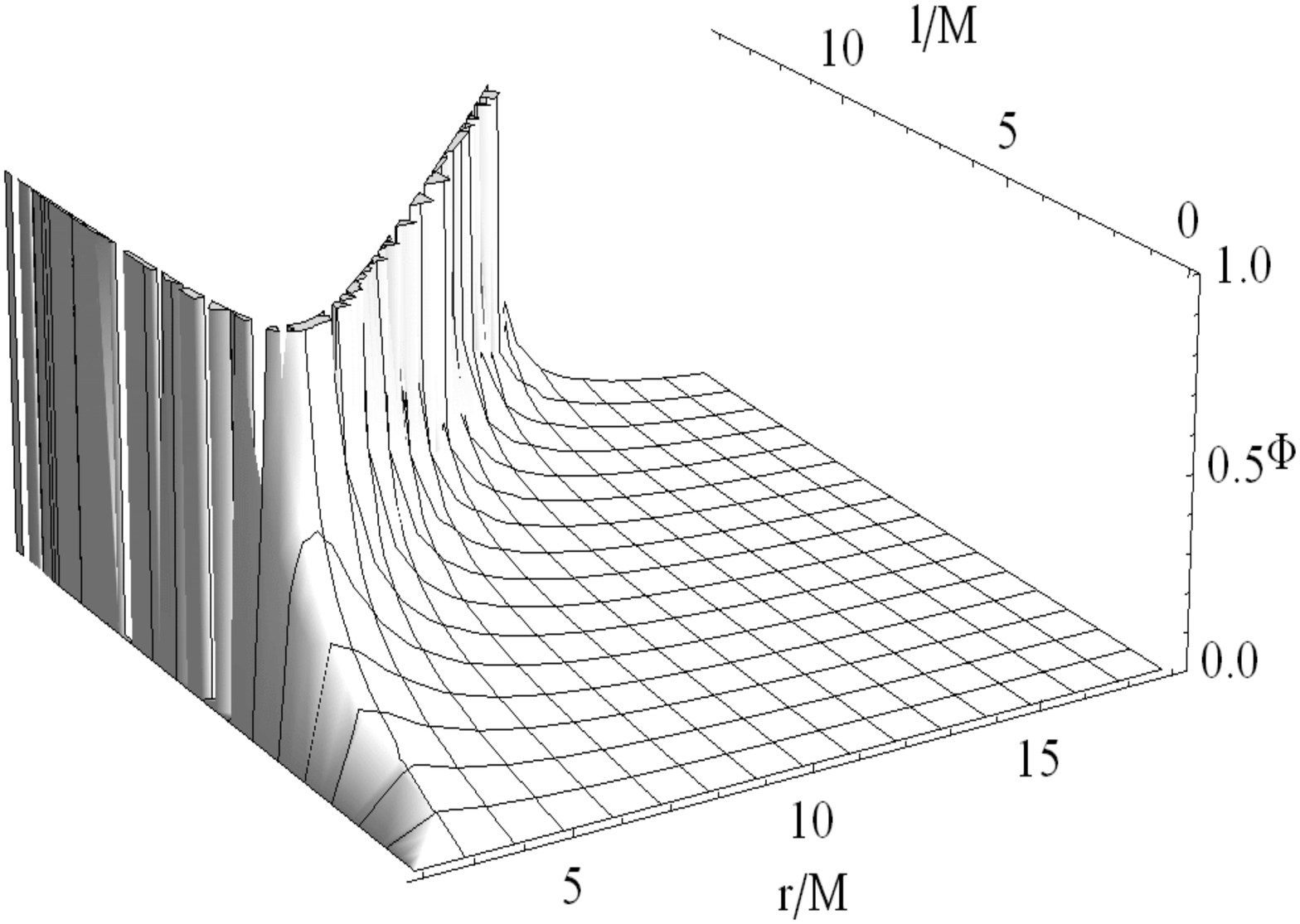}
\includegraphics[width=0.45\hsize,clip]{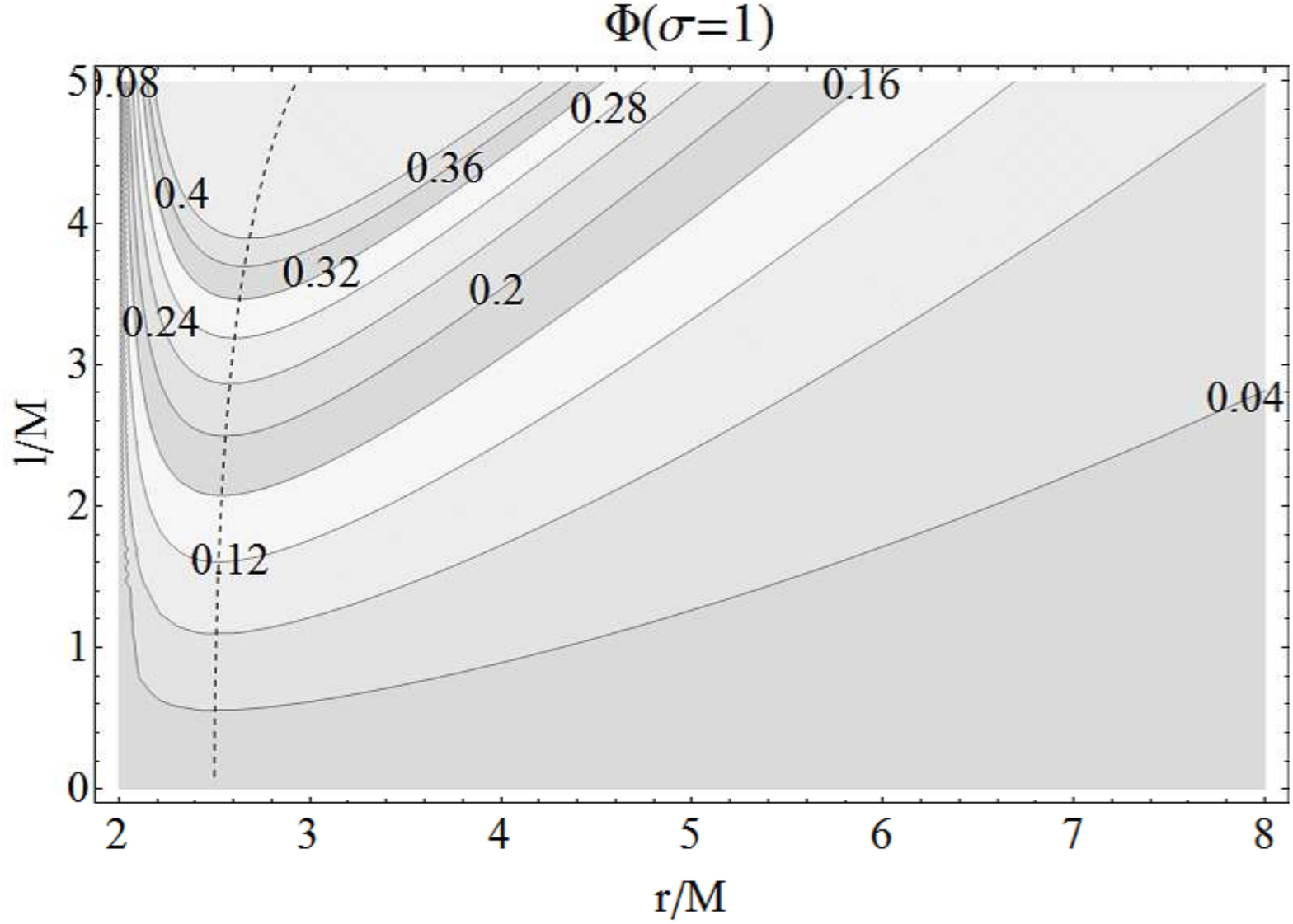}\\
\includegraphics[width=0.45\hsize,clip]{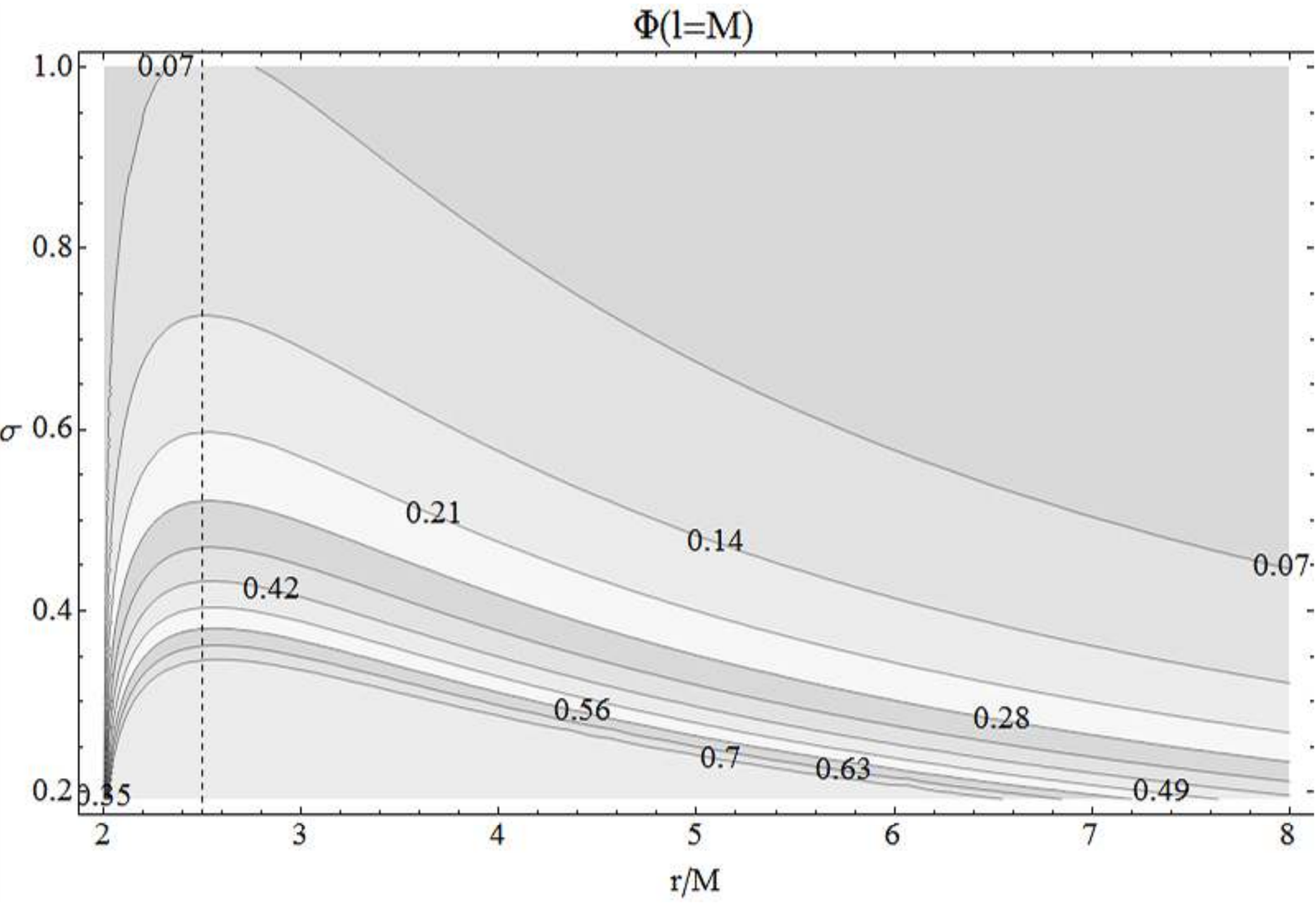}
\includegraphics[width=0.45\hsize,clip]{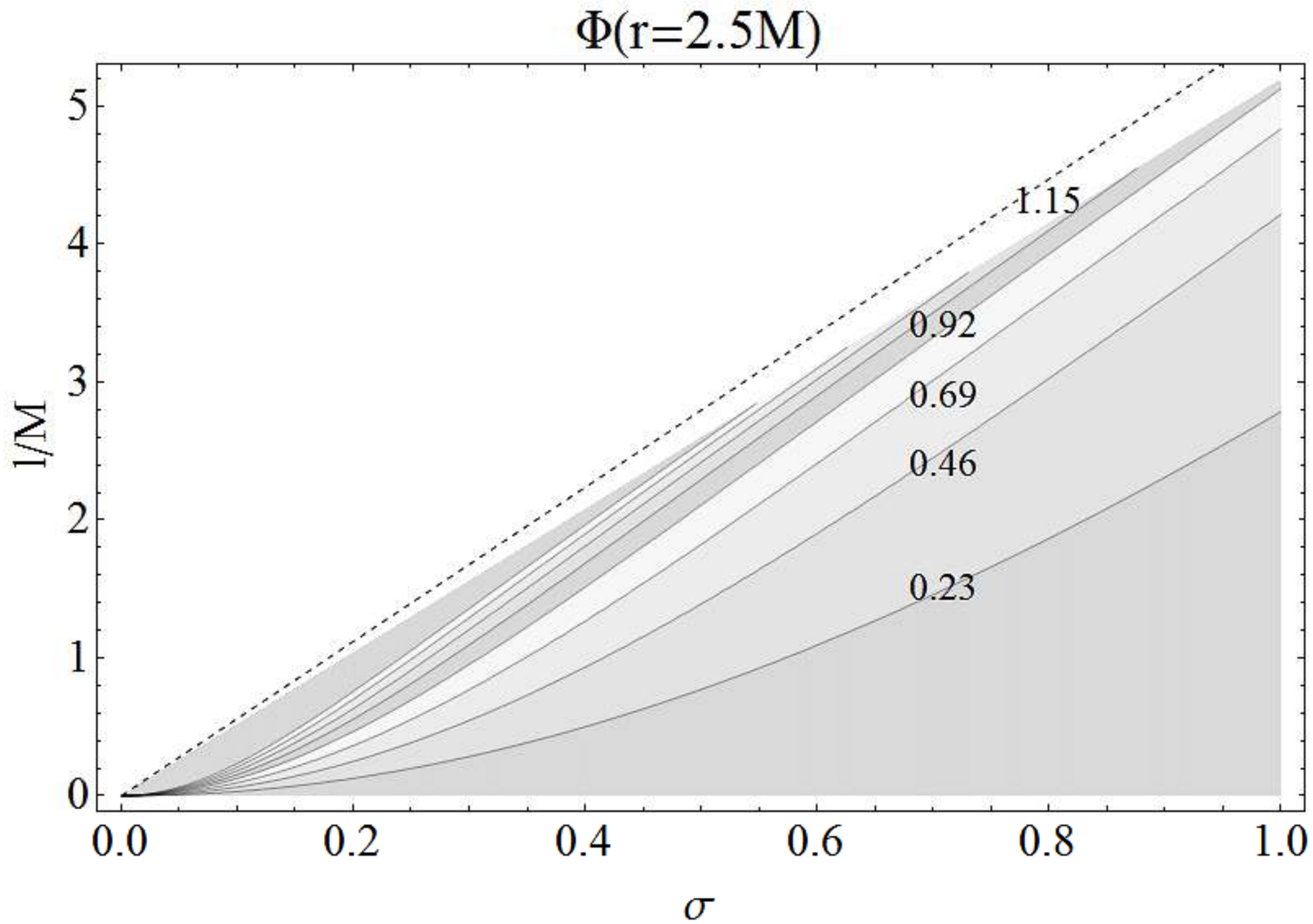}
\caption{\emph{Upper left panel}: $\Phi=a U^{\varphi}/M$ on the
equatorial plane $\sigma=1$ as a function of $r/M$ and $l/M$. The function is  defined for
$l\in[0,l_r]$. \emph{Upper right panel}: $\Phi=\mbox{constant}$ as function of $l/M$ and
$r/M$ on the equatorial plane $\sigma=1$. The curves have minima in $r=r_{\chi}$.
\emph{Lower left panel}: $\Phi=\mbox{constant}$ in the plane  $\sigma$ and $r/M$  for
$l/M=1$. The curves have maxima  in $r=5/2 M$ (dashed line). \emph{Lower right panel}:
$\Phi=\mbox{constant}$ in the plane  $\sigma$ and $\ell$  for  $r/M=5/2$. No motion is
allowed for $\ell>\ell_r$ ($\ell=\ell_r$, dashed line), and it is $\Phi'=0$ on
$r=r_{\chi}^-$, $l=l_{\chi}$ and $\sigma=\sigma_{\chi}$.}
\label{cern2011}
\end{figure}

The fluid four velocity along the $\varphi$ angular direction is:
\be\label{E:lampada}
\Phi=\frac{L}{r^2\sigma^2}=\frac{1}{\sigma^2 r^2}\sqrt{\frac{\sigma^2 l^2 (r-2M) r^2}{\sigma^2
r^3-l^2 (r-2M)}}\, ,
\ee
where  we are always considering $L>0$ and $\Phi>0$\footnote{The convention adopted is
that the matter rotates  in the positive direction of the azimuthal coordinate
$\varphi$.}. We redefine $\Phi$ as a dimensionless quantity:
\be\label{E:phigiro}
\frac{\Phi \sigma^2}{M}=\frac{M^2}{ r^2}\sqrt{\frac{\ell^2 (r-2M) r^2}{r^3-\ell^2 M^2
(r-2M)}}\, .
\ee
Clearly it is $\Phi\gtreqless0$ when $L\gtreqless0$, and they have the same existence
conditions ($0< l <l_r$, $2M< r<r_l^-$ and $r>r_l^+$). The behavior of the angular
velocity as a function of $\sigma$, $l$, and $r$ are portrayed in Figs.\il\ref{cern2011}.

\begin{figure}
\centering
\includegraphics[width=0.45\hsize,clip]{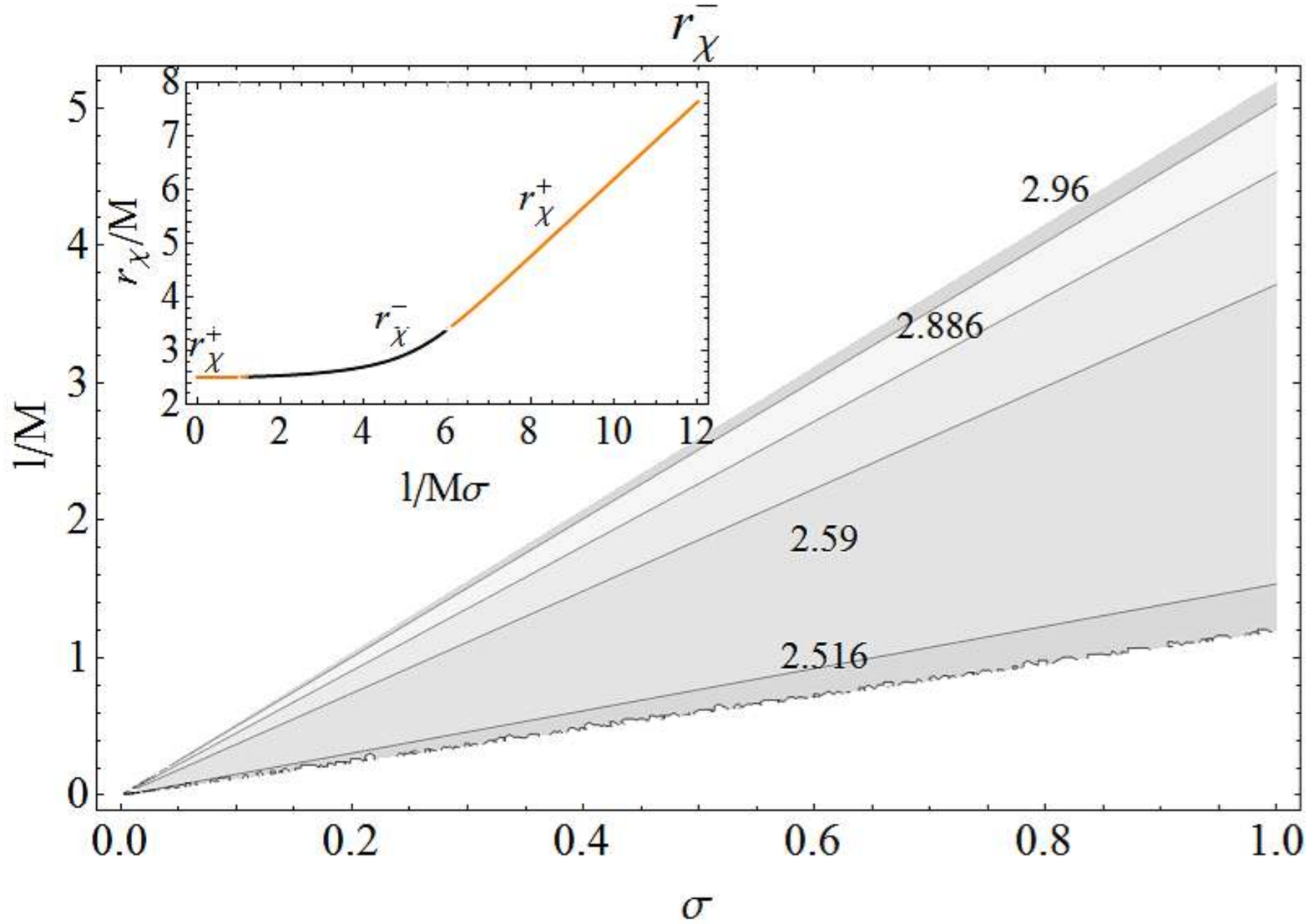}
\includegraphics[width=0.43\hsize,clip]{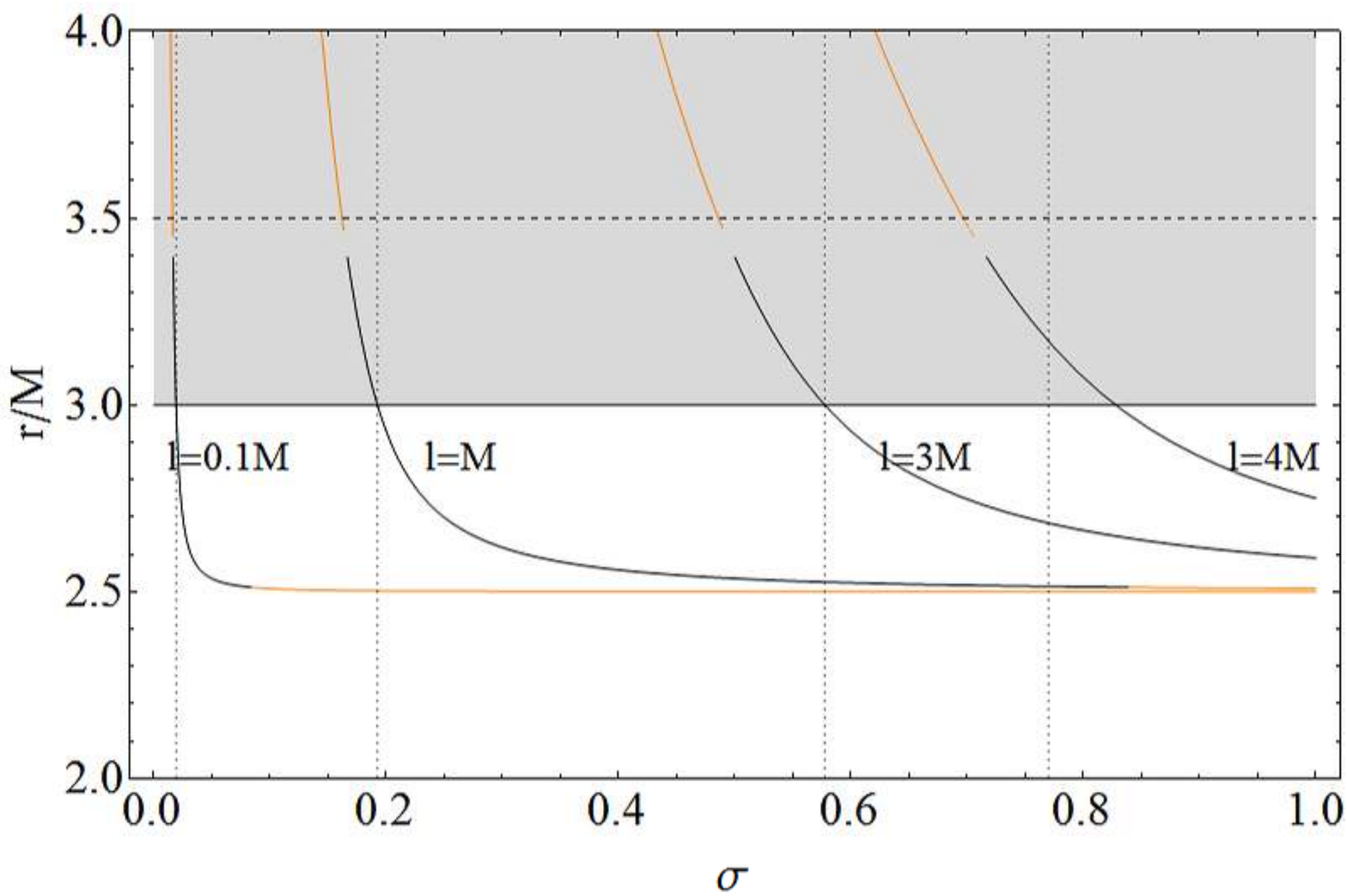}
\caption{(Color online)\emph{Left panel}: $r^{-}_{\chi}=$constant as a
function of $0<\sigma<1$ and $0<l/M<3 \sqrt{3}$. \emph{Inset} $r^{-}_{\chi}$ (black line) and
$r^{+}_{\chi}$ (orange line) as increasing functions of $\ell=l/(\sigma M)$. \emph{Right panel}:
$r^+_{\chi}/M$ (orange line) and  $r^-_{\chi}/M$ (black line) as a function of $\sigma$. Gray
region corresponds to $r\in[r_{lco},\infty]$. Dotted lines mark $\sigma=\sigma_l$ for the selected
$l/M$.}
\label{basica}
\end{figure}

In order to derive the critical points of the proper angular velocity as a function of
$r$, we consider the solutions of the equation $\Phi'=0$. Apart from the  trivial case
$l=0$ and  $\Phi=0$, the proper angular velocity is constant with respect to the radial
coordinate when:
\bea
l=l_{\chi}\equiv\sqrt{\frac{\sigma^2 r^3 (2 r-5M)}{(r-2M)^2}}\quad\mbox{for}\quad {5}/{2}
M<r<3 M,
\eea
where
$0<l_{\chi}<3 \sqrt{3}M$. In terms of the angular momentum and the radius, the orbits of
$\Phi$ belong to the planes:
\bea
\sigma=\sigma_{\chi}\equiv l\sqrt{\frac{(r-2M)^2}{r^3 (2 r-5M)}}\quad\mbox{for}\quad
r_{\chi}<r<r_{lco}
\quad(\sigma=1\quad\mbox{when}\quad r=r_{\chi})\, ,
\eea
where:
\bea\label{rchi}
r_{\chi}=r^{\pm}_{\chi}/M&\equiv&\frac{5}{8}+\frac{1}{2} \sqrt{\frac{25}{8}+\frac{2
\ell^2}{3}-\frac{\ell^2 \left(\ell^2-36\right)}{6
\alpha}-\frac{\alpha}{6}\mp\frac{\sqrt{3} \left(\frac{125}{8}-11
\ell^2\right)}{\beta}}\mp\frac{\beta}{8 \sqrt{3}},
\eea
with
\bea
\alpha&\equiv&\left[-\ell^2 \left(\ell^4-54 \ell^2+1350-6 \sqrt{48 \ell^4-2754
\ell^2+50625}\right)\right]^{1/3},
\\
\beta&\equiv&\sqrt{16 \ell^2+75+\frac{8 \ell^2 \left(\ell^2-36\right)}{\alpha}+8\alpha}\
.
\eea
As shown in Figs.\il\ref{basica},  $2.5M<r_{\chi}<3M$ and the orbital radius
$r_{\chi}$  increases with {${\ell=l/(M\sigma)}$}. An alternative analysis of the proper fluid
angular velocity, as function of { ${(l,r, \sigma)}$}, is presented in Appendix
\il\ref{Sec:appalph}.

\section{Comparing the fluid relativistic angular velocity and the Keplerian angular
momentum}\label{Sec:KeplerANDangular}

\begin{figure}
\centering
\includegraphics[width=0.43\hsize,clip]{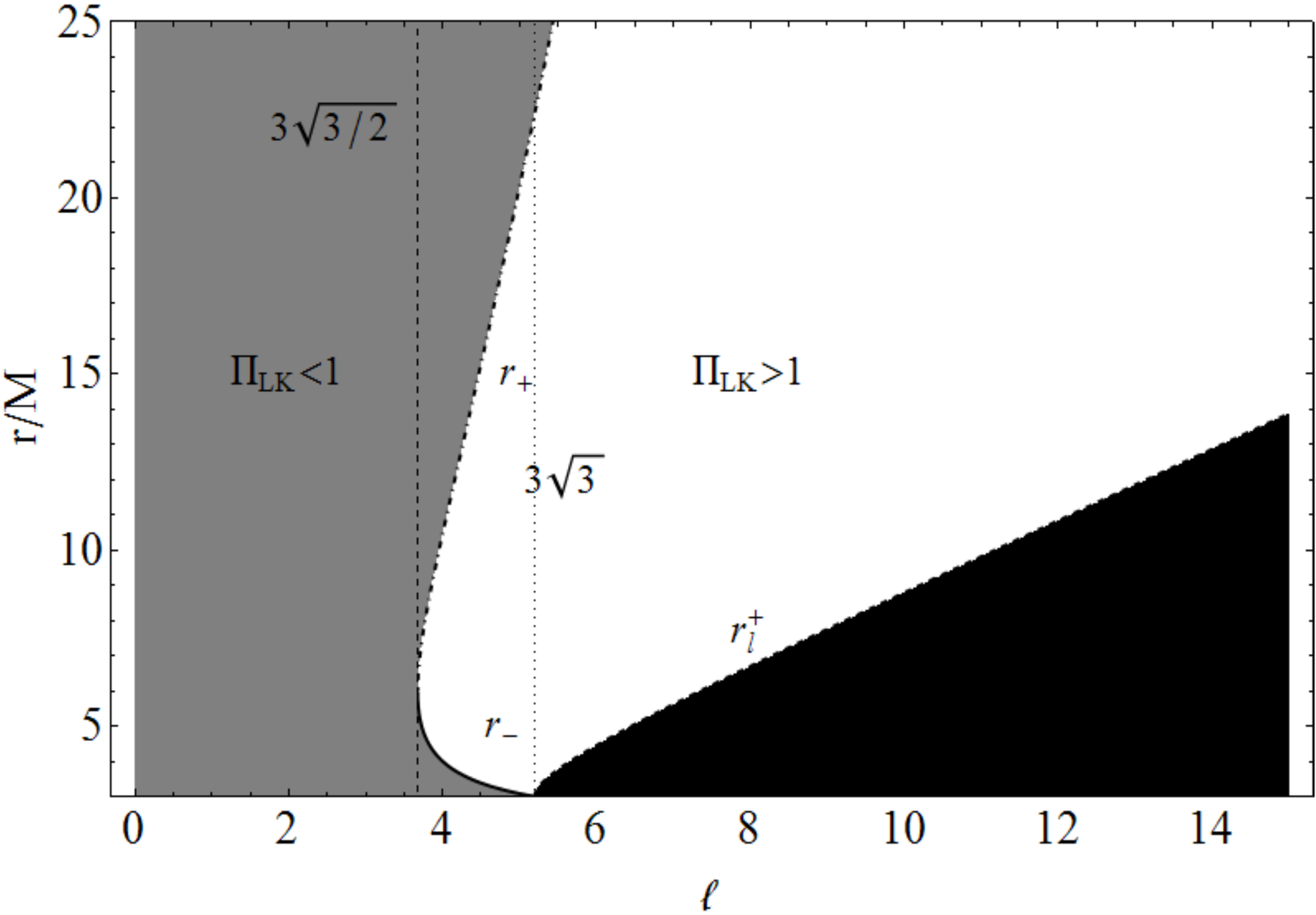}
\includegraphics[width=0.43\hsize,clip]{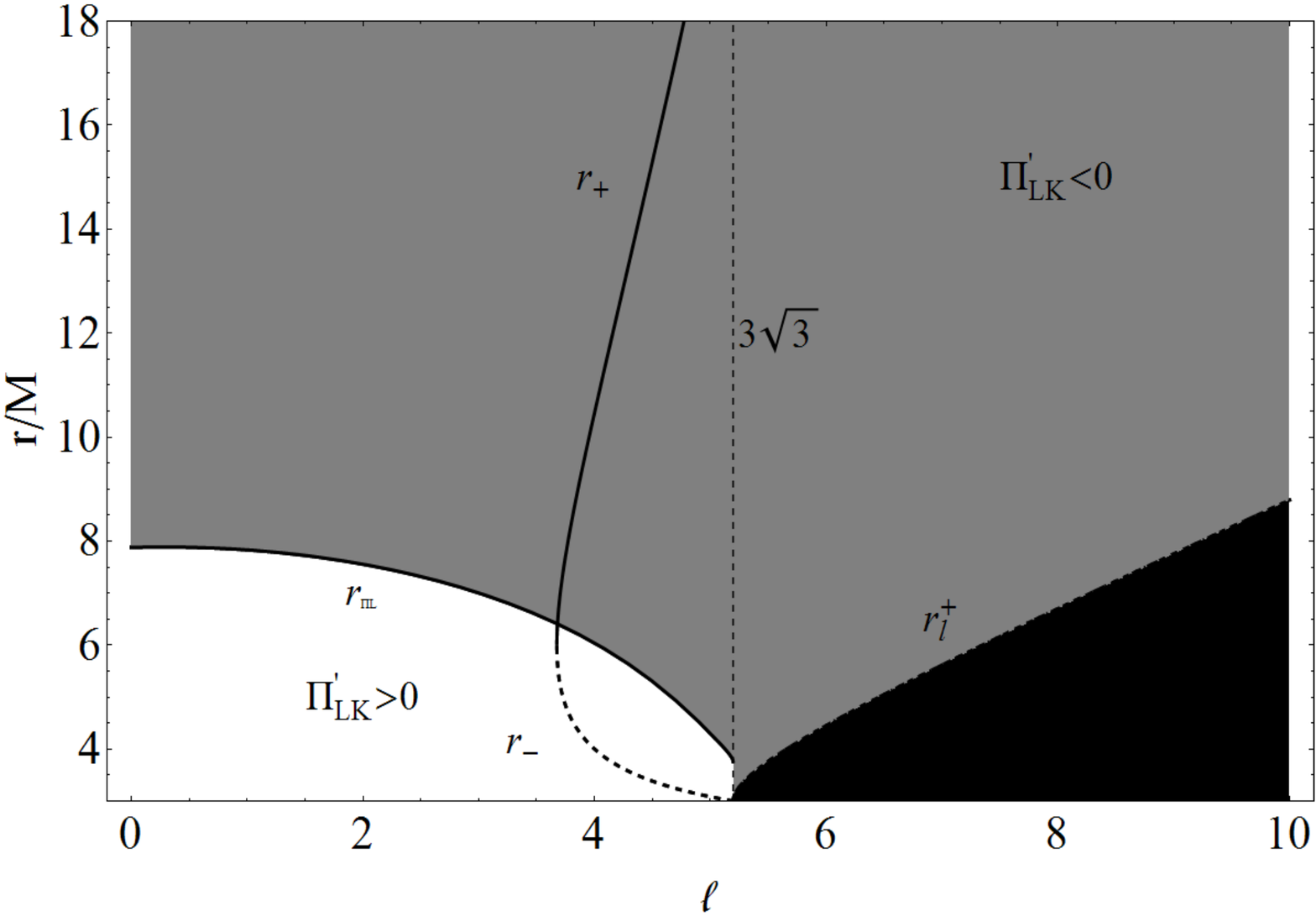}
\\
\includegraphics[width=0.45\hsize,clip]{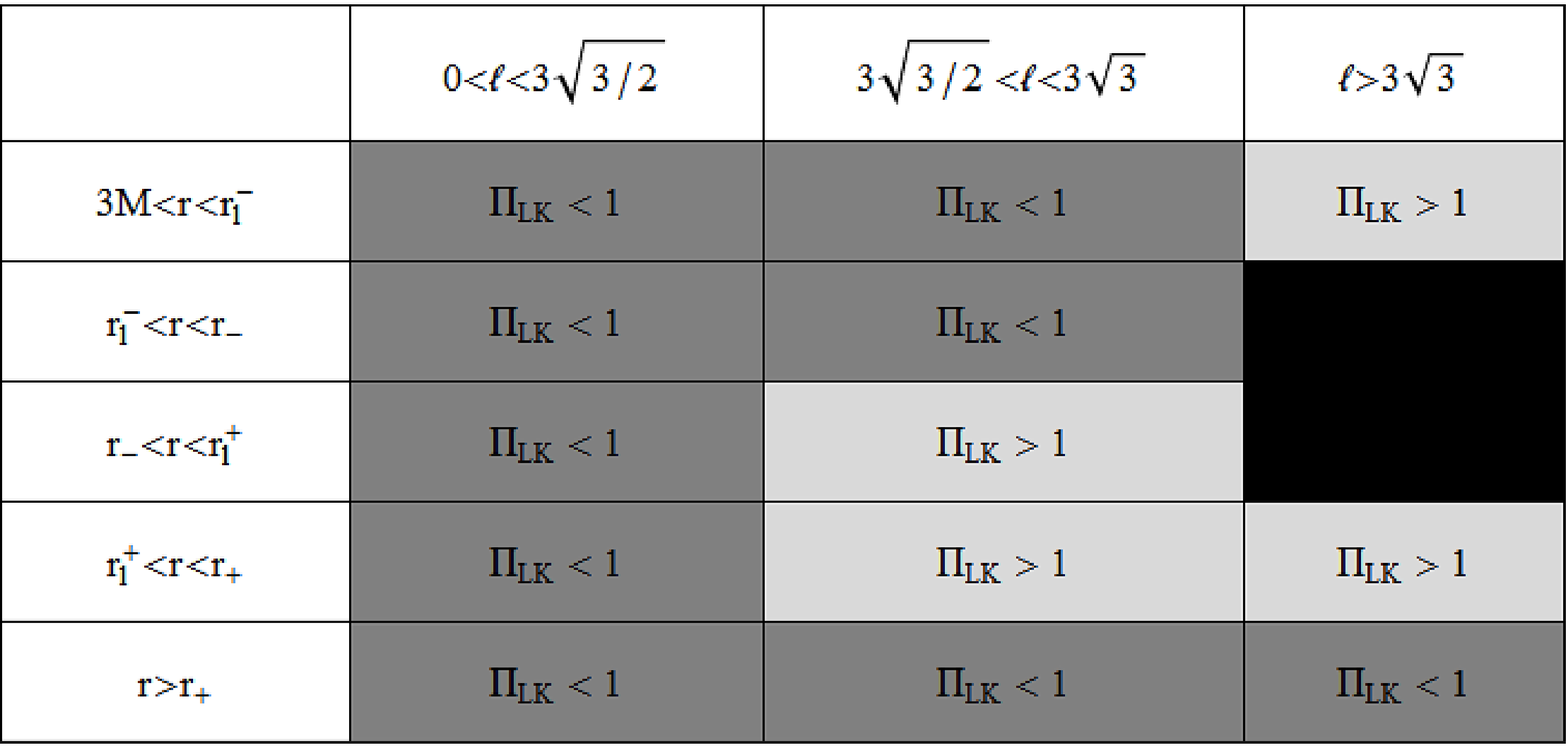}
\includegraphics[width=0.4\hsize,clip]{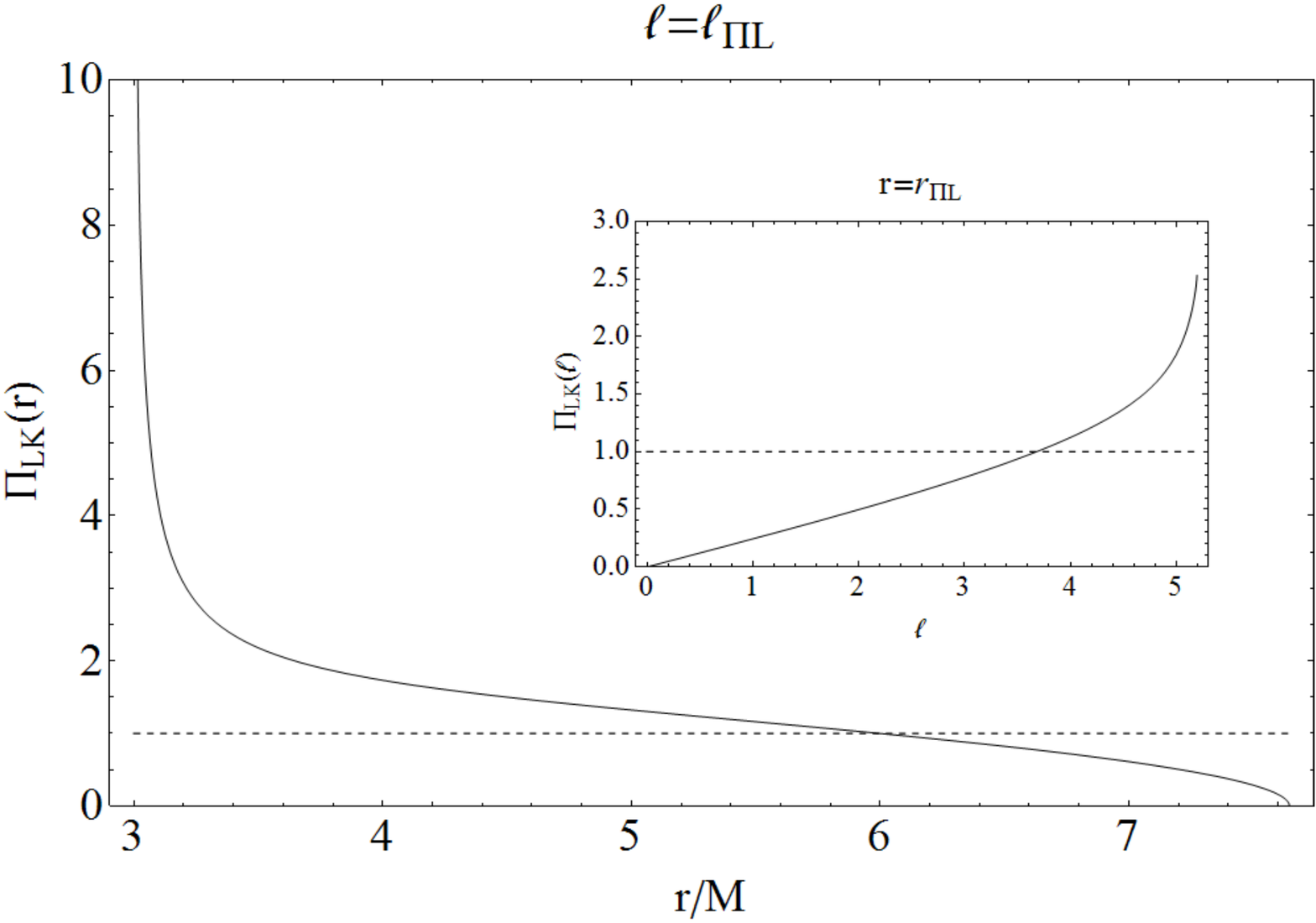}
\caption{{\emph{Upper left panel}:  $r_{\pm}$ and $r^+_l$ as a
function of $\ell$. $\Pi_{\ti{LK}}$ is not defined in the black region.
$\Pi_{\ti{LK}}>1$ in the white region and $\Pi_{\ti{LK}}<1$ in the gray region. Dashed
line marks $\ell=3\sqrt{3/2}$, dotted line $\ell=3\sqrt{3}$. \emph{Upper right panel}: $r_{\Pi_{LK}}$, $r^+_l$, $r_{\pm}$ as a function
of $\ell$. $\Pi_{\ti{LK}}$ is not defined in $r<r^+_l$ (black region). For
$r<r_{\Pi_{LK}}$ it is $\Pi'_{\ti{LK}}>0$ (white region). The gray colored region
corresponds to $\Pi'_{\ti{LK}}<0$. When $r=r_{\pm}$  it is $\Pi_{\ti{LK}}=1$. Dashed
thin line marks $\ell=3\sqrt{3}$. \emph{Lower  left panel}: table
summarizing the regions $\Pi_{\ti{LK}}>1$ (light-gray regions) and $\Pi_{\ti{LK}}<1$
(gray regions) for different values of $r/M$ and $\ell$. $\Pi_{\ti{LK}}$ is not defined
in the black boxes. \emph{Lower right panel }: $\Pi_{\ti{LK}}(\ell_{\Pi_{LK}})$ as
a function of $r/M\in[3.76,7.88]$. Dashed line marks $\Pi_{\ti{LK}}=1$. \emph{Inset}:
$\Pi_{\ti{LK}}(r_{\Pi_{LK}})$ as a function of $\ell\in[0,3\sqrt{3}]$. Dashed line marks
$\Pi_{\ti{LK}}=1$. }}
\label{SPAlla}
\end{figure}

In this Section we compare the fluid configurations with $\partial_rp\neq0$ with the
model $p=$constant where the disk is geodetic by studying the relativistic angular
velocity and comparing our results with the Keplerian definitions. In
Sect.\il\ref{Sec:NeutraLl} we already explored the fluid behavior with respect to
$l_{\ti{K}}$. We summarize here the results considering the ratio:
\be
\Pi_{\ti{LK}}\equiv L/L_{\ti{K}}=\Phi/\Phi_{\ti{K}}=\sqrt{\frac{\ell^2 M (r-3M)
(r-2M)}{r^3-\ell^2 M^2(r-2M)}}\, ,
\ee
which is defined for  $0<\ell\leq 3 \sqrt{3}$  in $r>r_{lco}$, and for $\ell>3 \sqrt{3}$
in $r>r^+_l$. The regions where $L(l)$ is larger or smaller than
$L_{\ti{K}}=L(l_{\ti{K}})$ for different values of the fluid constant of motion $l$ are
portrayed in Figs.\il\ref{SPAlla} \emph{upper left panel}. We have
$\lim_{r\rightarrow\infty}\Pi_{\ti{LK}}=\lim_{r\rightarrow r_{lco}}\Pi_{\ti{LK}}=0$. In
general, $L>L_{\ti{K}}$ for fluid angular momentum $l$ sufficiently large, for $l$
sufficiently low  or far from the source it is  $L<L_{\ti{K}}$ (see
Figs.\il\ref{SPAlla} \emph{upper left panel}).

We are now interested to define the critical points of the ratio $\Pi_{\ti{LK}}$. The
solutions of the equation $\Pi'_{\ti{LK}}=0$ are, in terms of the angular momentum:
\be
\ell_{\Pi_{LK}}\equiv\frac{r \sqrt{(10M-r) r-18M^2}}{M(r-2M)}\, ,
\ee
and in terms of the orbital radius:
\be
r_{\Pi_{LK}}/M=\frac{1}{6} \left[15+\sqrt{3} \sqrt{78-4
\ell^2-\frac{\left(\ell^2-18\right)^2}{\kappa }-\kappa ^{1/3}-\frac{6 \sqrt{3}
\left(\ell^2-35\right)}{\bar{\kappa }}}+\sqrt{3} \bar{\kappa }\right]\, ,
\ee
with
\bea
\kappa &\equiv&\left[54\left( 108+10 \ell^2- \ell^4\right)+\ell^6+12 \sqrt{6}
\sqrt{-\ell^2 \left(\ell^2-27\right) \left(\ell^4-54 \ell^2+756\right)}\right]^{1/3}\,
,
\\
\bar{\kappa }&\equiv&\sqrt{\frac{\ell^4-2 \ell^2 (18+\kappa)+(12+\kappa)
(27+\kappa)}{\kappa }}\, .
\eea
These are defined for $r_{lco}<r_{\Pi_{LK}}<(5+\sqrt{7})$ and
$0<l_{\Pi_{LK}}<3\sqrt{3}$. The sign of $\Pi'_{\ti{LK}}$ as a function of $r$ and $l$ is
summarized in Fig.\il\ref{SPAlla}, \emph{upper right panel}. In particular, it is manifest
that $r_{\Pi_{LK}}$ is a maximum of $\Pi_{\ti{LK}}$. The values of this maximum,
$\Pi_{\Pi_{LK}}(\ell)=\Pi_{\ti{LK}}(\ell=\ell_{\Pi_{LK}})$ and
$\Pi_{\Pi_{LK}}(r)=\Pi_{\ti{LK}}(r=r_{\Pi_{LK}})$, are portrayed in
Fig.\il\ref{SPAlla}, \emph{lower right panel}. We note in particular that $\Pi_{LK}$ is an
increasing function of $l$: this implies that the larger the angular momentum of fluid,
the larger is the ratio between $L$ and the angular momentum of the free particle
$L_K$.

\subsection{The fluid relativistic angular velocity  $\Omega$: the Von Zeipel surfaces}

The fluid relativistic angular velocity $\Omega$ is:
\be
\Omega\equiv\frac{\Phi}{\Sigma}\equiv -
l\frac{g_{tt}}{g_{\varphi\varphi}}=\frac{l(r-2M)}{r^3 \sigma^2}=\frac{l}{l_r^2},
\ee

\begin{figure}
\centering
\includegraphics[width=0.45\hsize,clip]{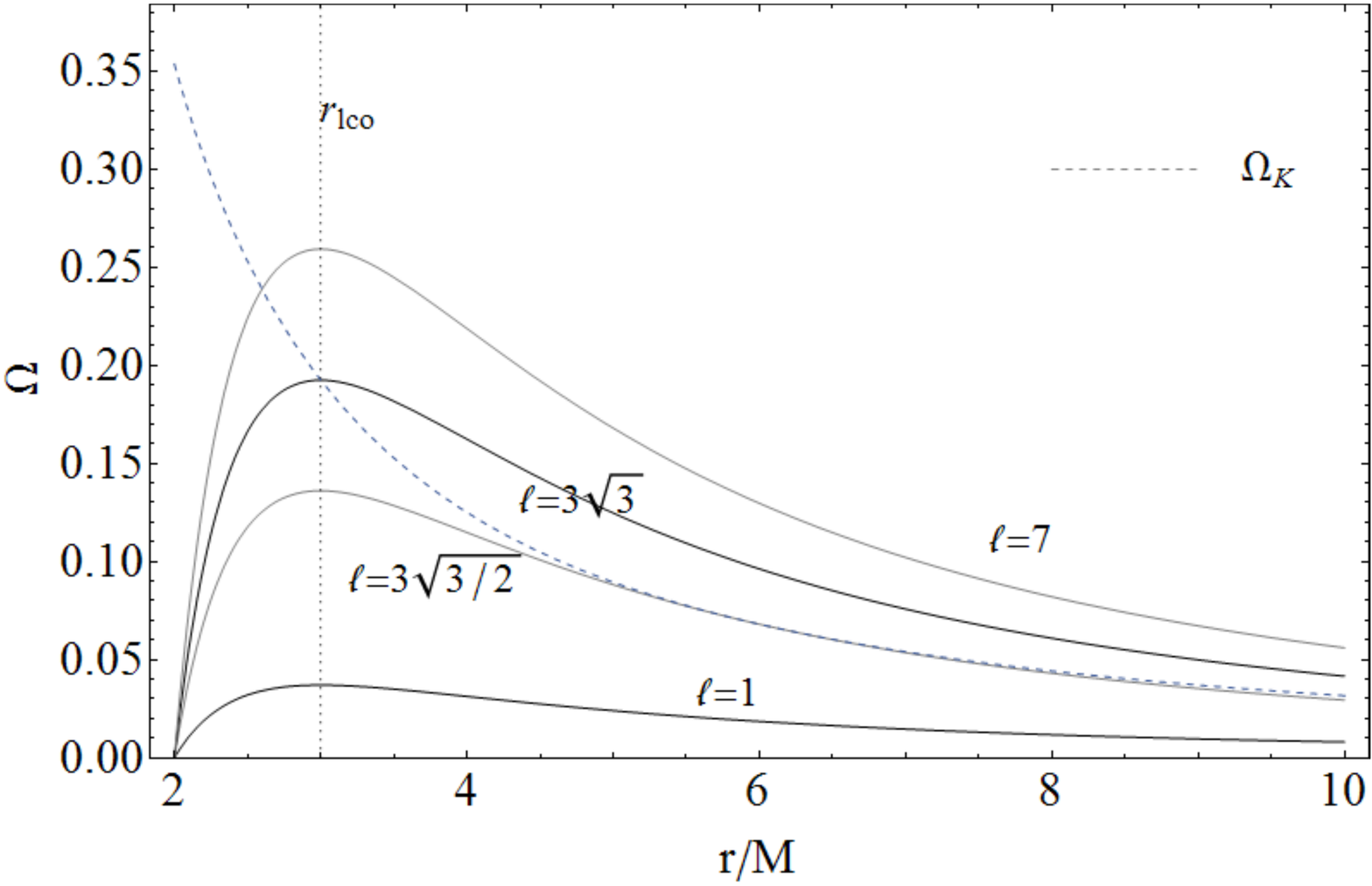}
\includegraphics[width=0.45\hsize,clip]{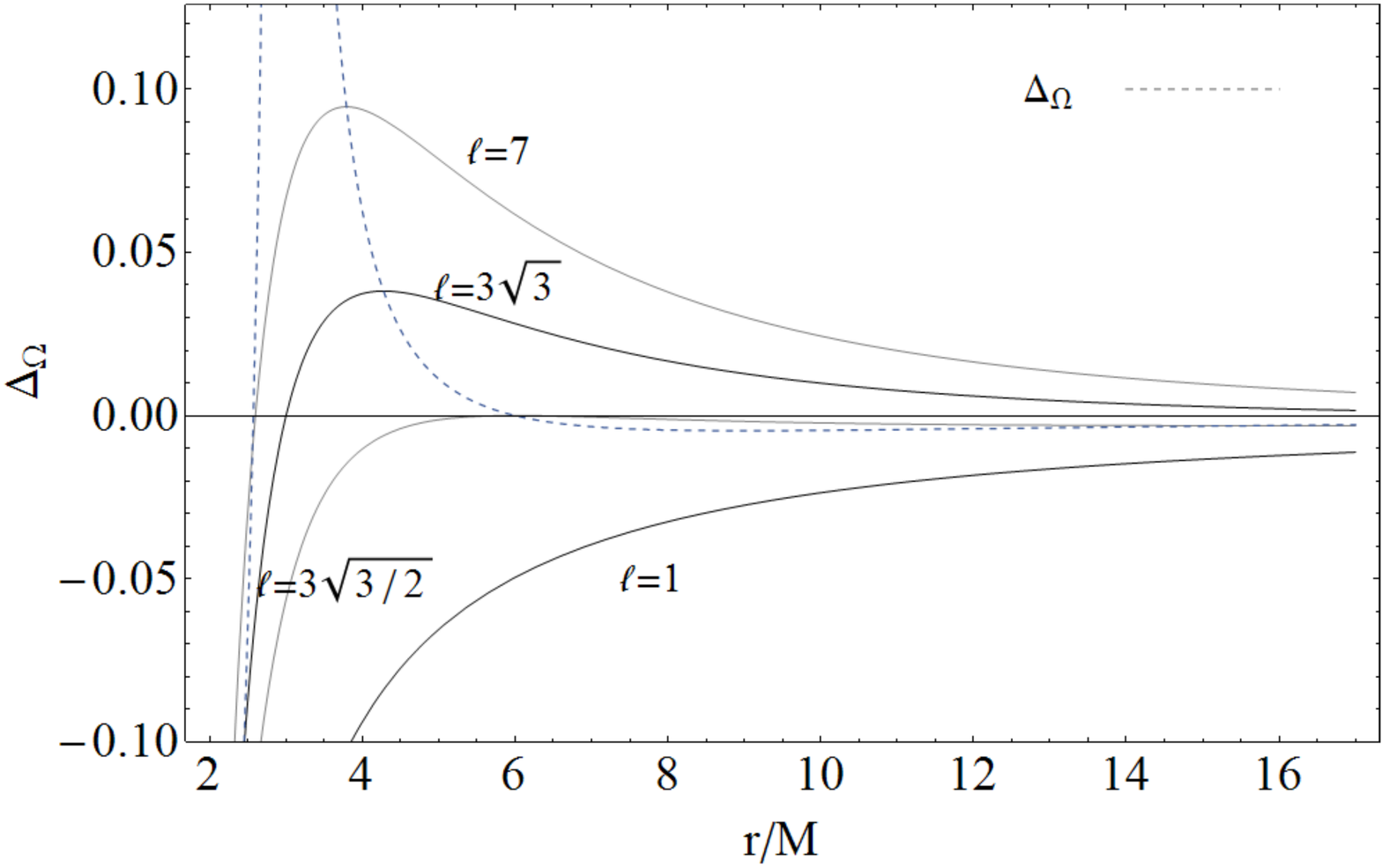}\\
\includegraphics[width=0.45\hsize,clip]{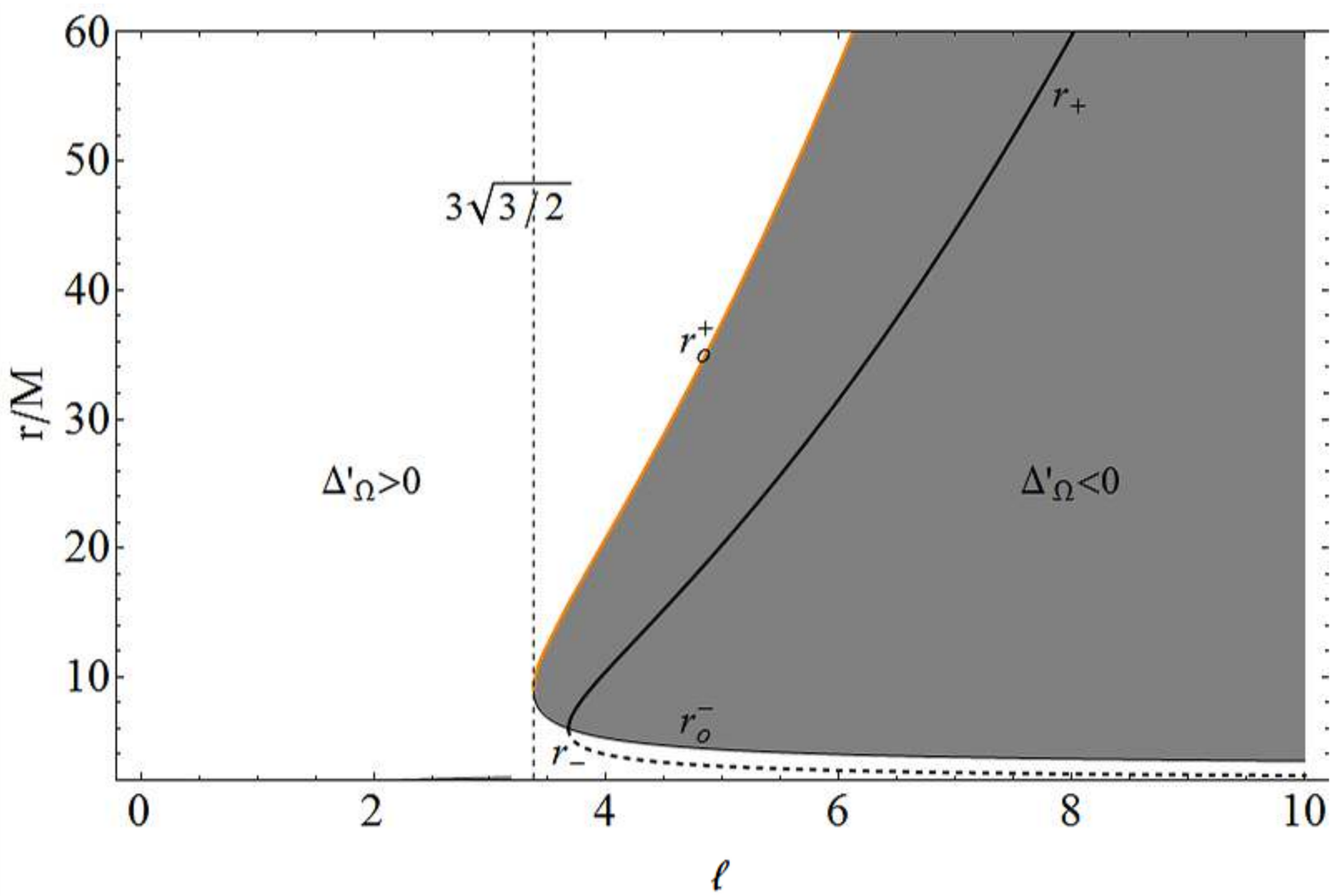}
\includegraphics[width=0.45\hsize,clip]{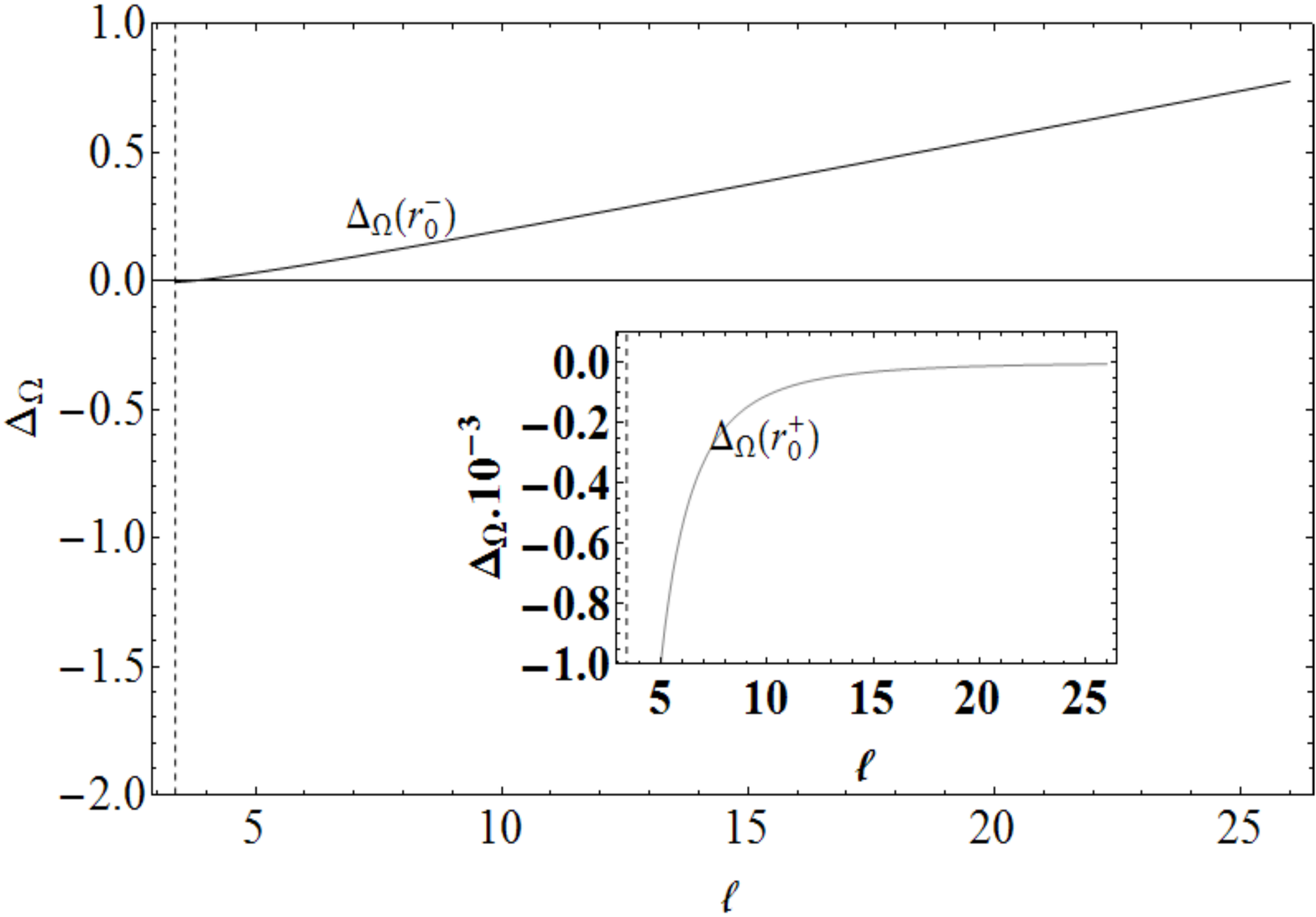}
\caption{(Color online) \emph{Upper left panel}: the relativistic angular
velocity $\Omega$  in units of $1/M$, for $\sigma=1$ as a function of $r/M$,  for different
values of $\ell$. Dashed line marks the Keplerian velocity $\Omega_{\ti{K}}$. Dotted
line marks $r=r_{lco}=3M$, a maximum of $\Omega$. For $\ell=3 \sqrt{3/2}$, it is
$\Omega=\Omega_{\ti{K}}$ in $r=r_{lco}=3M$; for $\ell=3 \sqrt{3}$, it is
$\Omega=\Omega_{\ti{K}}$ in $r=r_{lsco}=6M$ and in $r=6(2+ \sqrt{3})M$. \emph{Upper
right panel}: $\Delta_{\Omega}$ as a function of $r/M$, for different values of $\ell$.
Dashed line marks the maximum of $\Delta_{\Omega}$. \emph{Lower left panel}:
$r_{o}^{\pm}$ and $r_{\pm}$ as a function of $\ell$. In the white (gray) regions it is
$\Delta'_{\Omega}>0$  ($\Delta'_{\Omega}<0$). $\Delta_{\Omega}=0$ when $r=r_{\pm}$, and
$\Delta_{\Omega}$ is maximum in $r={r}_o^{-}$ and minimum in $r={r}_o^{+}$. \emph{Lower
right panel}: $\Delta_{\Omega}(r_o^-)$ as a function of $\ell$. \emph{Inset}:
$\Delta_{\Omega}(r_o^+)$  as a function of $\ell$.}
\label{Arlette}
\end{figure}

The  function  {${\Omega=\sigma\Omega(\ell)}$}:
\be
\Omega=\frac{\ell M(r-2M)}{r^3}
\ee
is defined in $r>2M$, and has a maximum in $r=r_{lco}$ (see Fig.\il\ref{Arlette},
\emph{left panel}).

The surfaces known as the  \emph{von Zeipel's cylinders}, are defined by the conditions:
$l=\mbox{constant}$ and $\Omega=\mbox{constant}$ \citep[see for example][]{M.A.
Abramowicz,Chakrabarti,Chakrabarti0}. In the static spacetimes the family of von
Zeipel's surfaces  does not depend on the particular rotation law of the fluid,
$\Omega=\Omega(l)$, in the sense that it does not depend on nothing but the background
spacetime. In the case of a barotropic fluid,  the von Zeipel's theorem guarantees that
the surfaces $\Omega=\mbox{constant}$ coincide with the surface with constant angular
momentum. More precisely, the von Zeipel condition states: the surfaces at constant
pressure coincide with the surfaces of  constant density  (i.e. the isobar surfaces  are
also isocore) \emph{if and only if} the surfaces with the  angular momentum $l =
\mbox{constant}$ coincide with the surfaces with constant angular relativistic velocity
\citep{Koz-Jar-Abr:1978:ASTRA:,Jaroszynski(1980),M.A. Abramowicz,Chakrabarti,
Chakrabarti0}.

\begin{figure}
\centering
\includegraphics[width=0.43\hsize,clip]{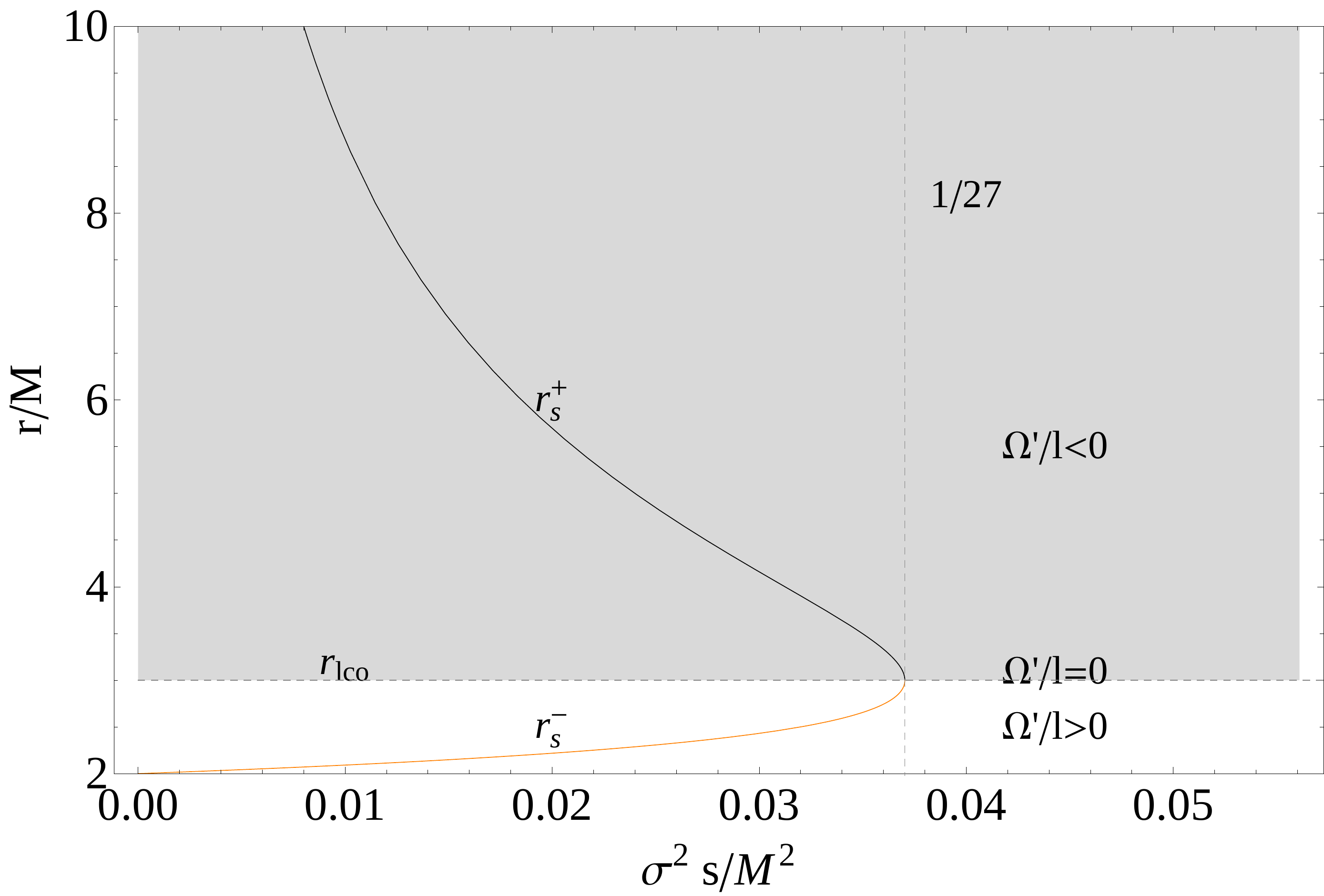}
\includegraphics[width=0.45\hsize,clip]{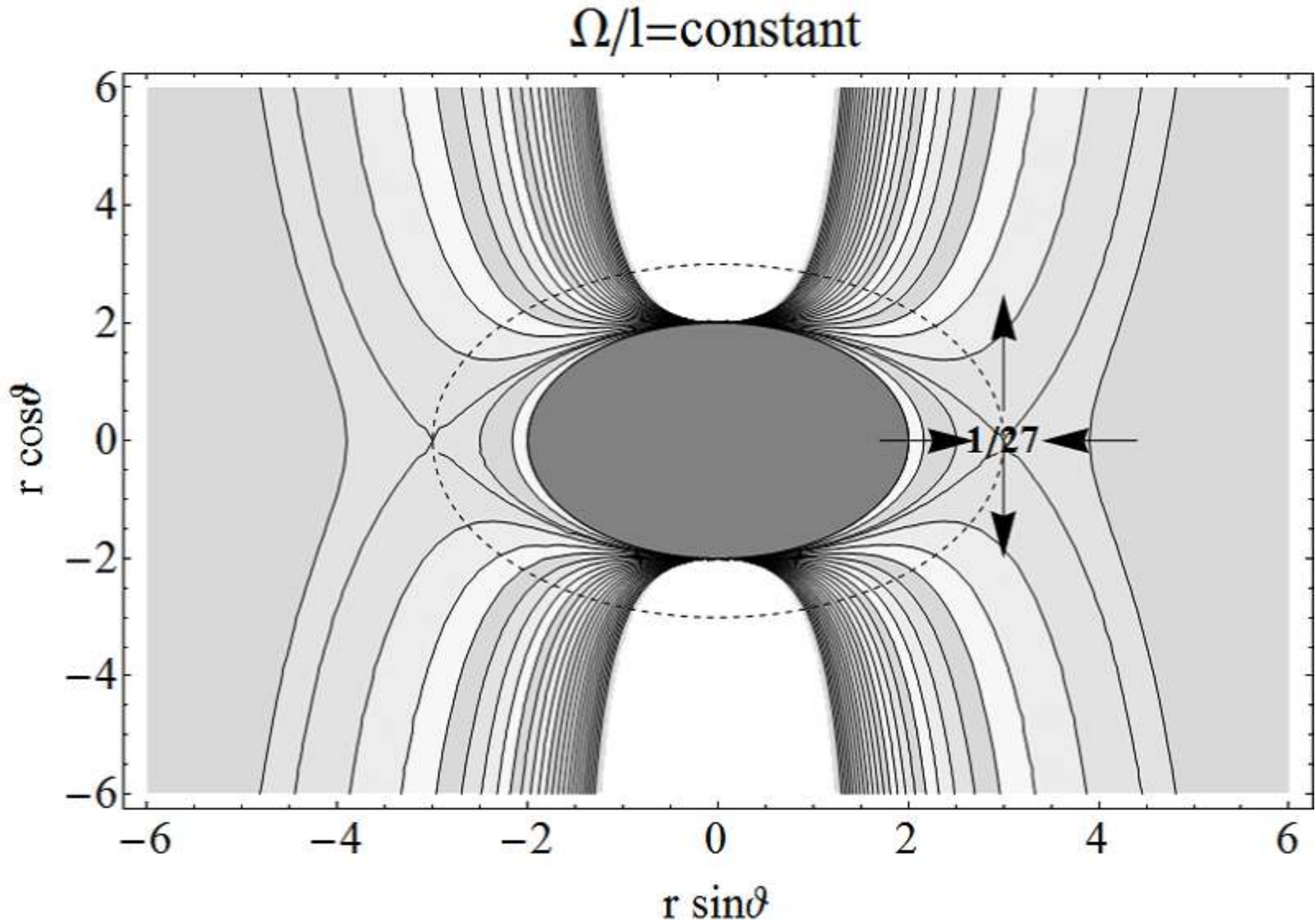}\\
\includegraphics[width=0.3\hsize,clip]{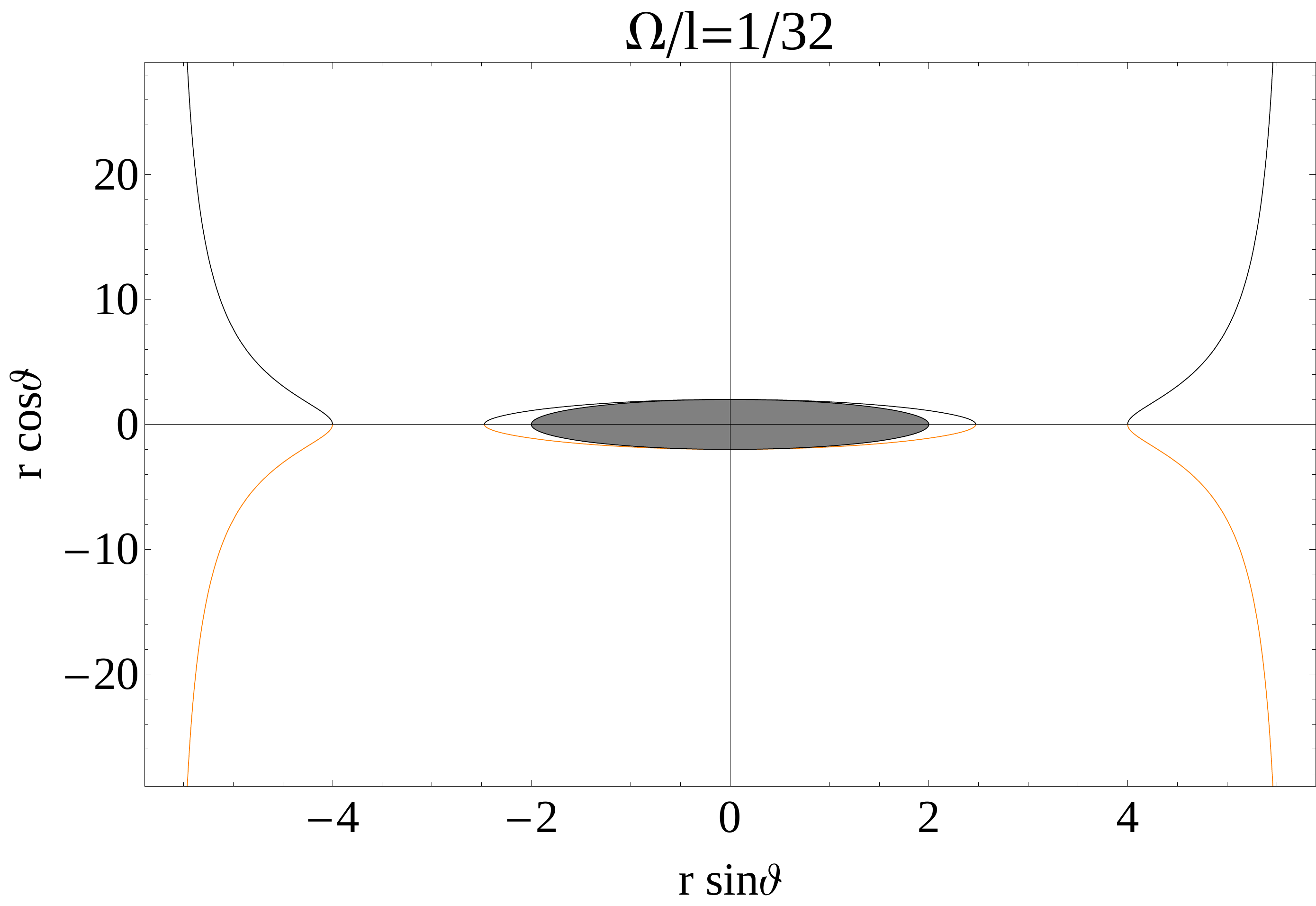}
\includegraphics[width=0.3\hsize,clip]{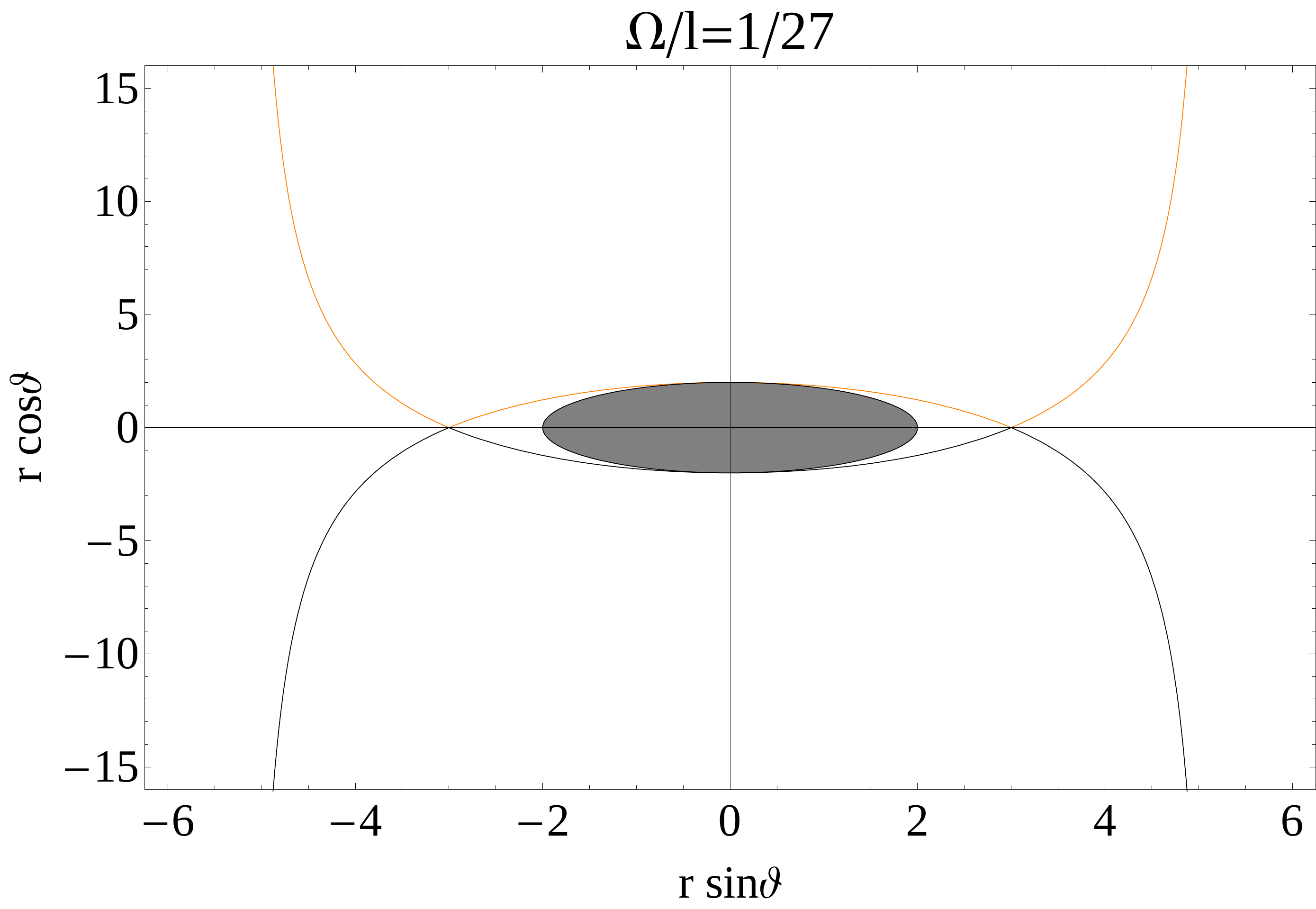}
\includegraphics[width=0.3\hsize,clip]{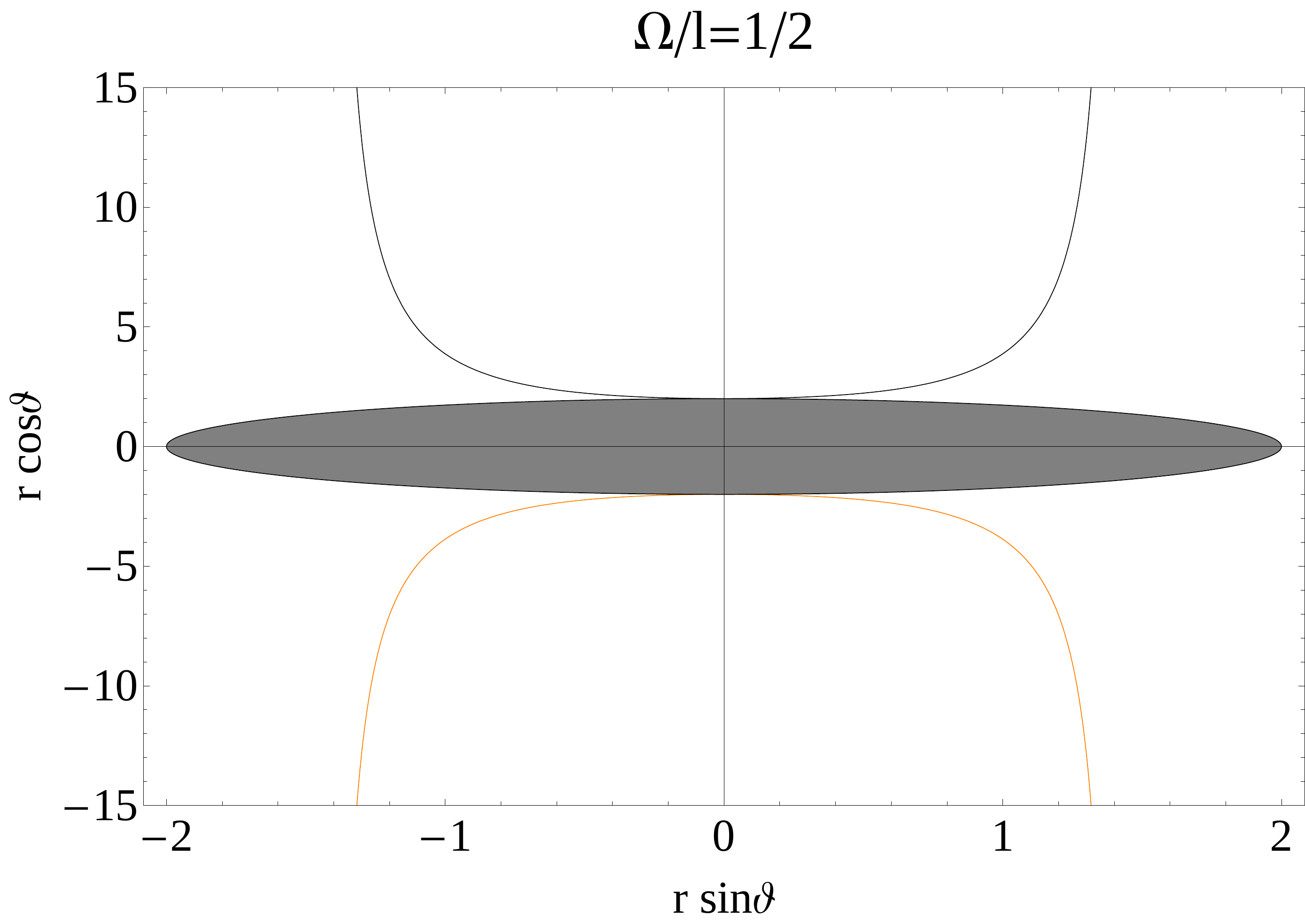}
\caption{(Color online) \emph{Upper left panel}: $r_{s}^{\pm}/M$ as a function
of $(\sigma^2 s)$. Dashed lines mark the conditions $r=r_{lco}$, $\sigma^2  s=(1/27)M^2$. The
white region $(r<r_{lco})$ corresponds to $\Omega'/l>0$, the light-gray region
$(r>r_{lco})$ to $\Omega'/l<0$. $\Omega'/l=0$ in $r=r_{lco}$. \emph{Upper right panel}:
surfaces $\Omega /l=s$ with $r/M=\mbox{constant}$ in coordinate $x=r\cos{\vartheta}$,
$y=r\sin{\vartheta}$. The surface $s=(1/27)M^2$ is marked with a number. The gray region
marks the circle $r\leq2M$.  {Arrows follow the increasing (decreasing) direction respect
of $s$. }\emph{Lower panels}: surfaces $\Omega /l=s$ in units of
$M^2$ as a function of $x=r\cos{\vartheta}$, $y=r\sin{\vartheta}$ in units of $M$, for
different values of $s$. The gray region underlines the circle $r\leq2M$. Dashed circle
corresponds to $r=r_{lco}$.}
\label{milano}
\end{figure}

The surfaces  $\Omega/l=s>0$, being $s$ a constant, are defined by:
\bea
r_{s}^-/M&\equiv&-\frac{2 \sqrt{\frac{M^2}{\sigma^2 s}} \cos\left[\frac{1}{3} \left(\pi
+\arccos\left[-\frac{3 \sqrt{3}}{\sqrt{\frac{M^2}{\sigma^2
s}}}\right]\right)\right]}{\sqrt{3}},
\quad
r_{s}^+/M\equiv\frac{2 \sqrt{\frac{M^2}{\sigma^2  s}} \cos\left[\frac{1}{3}
\arccos\left(-\frac{3 \sqrt{3}}{\sqrt{\frac{M^2}{\sigma^2  s}}}\right)\right]}{\sqrt{3}}.
\eea
These are a function of the product $\sigma^2  s$, and they exist for $0<\sigma^2 s\leq(1/27)M^2$.
In particular, $\sigma^2 s=(1/27) M^2$ corresponds to $r_s^{\pm}=r_{lco}$ (see
Fig.\il\ref{milano}, \emph{upper left panel}). In  the $(x,y)$ coordinates, the
$\Omega/l=s>0$ surfaces read:
\be
x_s^{\pm}\equiv\pm\sqrt{\left(\frac{2M^5}{M^4-s y^2}\right)^2-y^2}
\ee
(see Figs.\il\ref{milano}).

\begin{figure}
\centering
\includegraphics[width=0.55\hsize,clip]{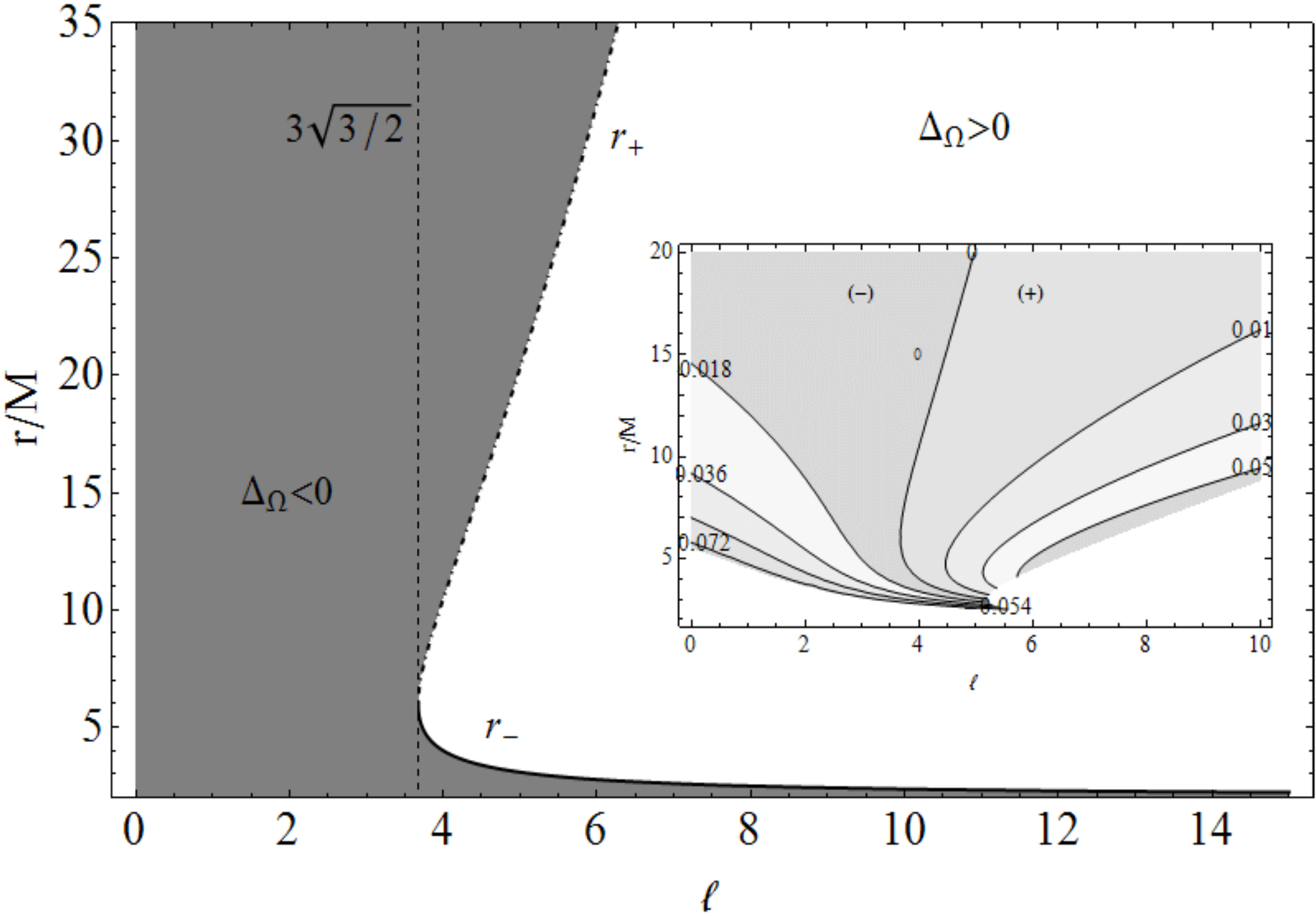}
\includegraphics[width=0.45\hsize,clip]{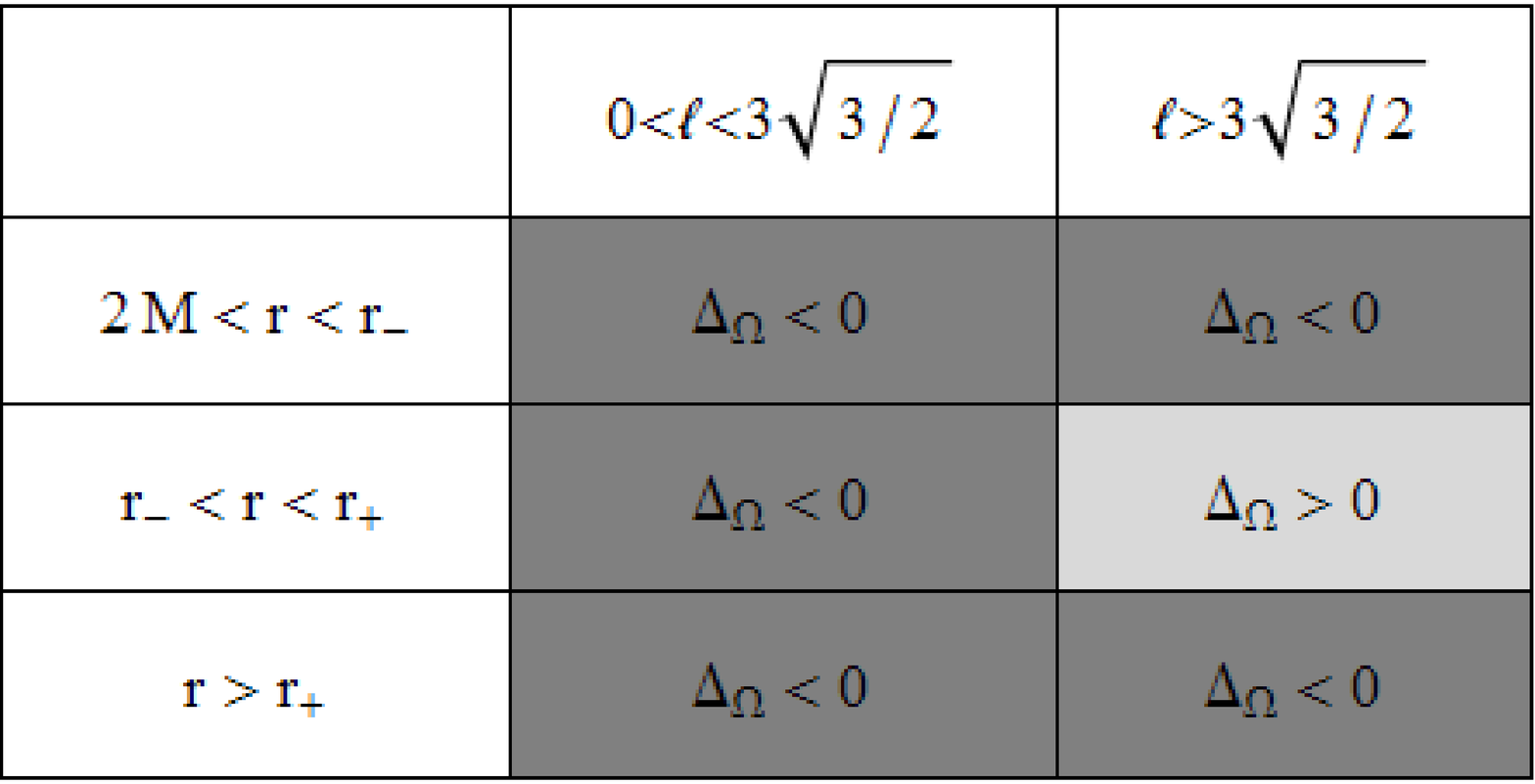}
\caption{\emph{Upper panel}: $r_{\pm}$ as a function of $\ell$.
Gray region corresponds to $\Delta_{\Omega}<0$, white region to  $\Delta_{\Omega}>0$.
$\Delta_{\Omega}=0$ in $r=r_{\pm}$. \emph{Inset}: curves
$\Delta_{\Omega}=\rm{constant}$. The regions of positive and negative $\Delta_{\Omega}$
are underlined. \emph{Lower panel}: table summarizing the regions $\Delta_{\Omega}>0$
(light-gray) and $\Delta_{\Omega}<0$ (gray).}
\label{SPAlla2}
\end{figure}

We are now interested in characterizing the angular velocity with respect to the
Keplerian velocity $\Omega_{\ti{K}}\equiv\frac{\sqrt{M}}{\sigma r^{3/2}}$. We therefore
consider the dimensionless difference
{$\mathbf{\Delta_{\Omega}\equiv(\Omega-\Omega_{\ti{K}})M\sigma}$}:
{
\be
\mathbf{\Delta_{\Omega}=M\sigma\frac{\ell M (r-2M)-\sqrt{r^{3}M}}{\sigma r^3}.
}\ee}
$\Delta_{\Omega}=0$ for $r\rightarrow \infty$, $\Delta_{\Omega}(r=2M)=-1/2 \sqrt{2}$,
$\Delta_{\Omega}(r_{lco})= \left(\ell-3 \sqrt{3}\right)/27$, and
$\Delta_{\Omega}(r_{lsco})=\left(2\ell-3 \sqrt{6}\right)/108$ (see
Fig.\il\ref{Arlette}, \emph{right panel}). $\Delta_{\Omega}(r=2M)$ is negative
irrespective of the value of $\ell$. The sign of  $\Delta_{\Omega}(r_{lco})$ and
$\Delta_{\Omega}(r_{lsco})$, on the contrary, depends on the fluid angular momentum. The
sign of $\Delta_{\Omega}$ as a function of $r$ and $\ell$ is summarized in
Figs.\il\ref{SPAlla2}.

The difference $\Delta_{\Omega}$ is maximum in $r>r_{lco}$ when $\ell=(3/4)
\ell_{\ti{K}}$. Notably, $\ell_{\ti{K}}$ is not a critical point of the angular velocity
difference $\Delta_{\Omega}$. $\Delta_{\Omega}$ always increases in  $2M<r\leq r_{lco}$
for all $\ell$, and for $0<\ell<(3/4)\ell_{\ti{K}}$  in the region $r>r_{lco}$. Other
critical points are for $\ell\geq27/8$ in $r=r^{\pm}_{o}$:
\bea
r_o^-/M&\equiv&\frac{8}{27} \left(2 \ell^2-\sqrt{2} \sqrt{\ell^2 \left(8
\ell^2-81\right)} \cos\left[\frac{1}{3} \left(\pi +\arccos\hat{o}\right)\right]\right),
\\
r_o^+/M&\equiv&\frac{8}{27} \left(2 \ell^2+\sqrt{2} \sqrt{\ell^2 \left(8
\ell^2-81\right)} \cos\left[\frac{1}{3} \arccos\hat{o}\right]\right),
\eea
with
\be
\hat{o}\equiv\frac{\ell^2 \left(512 \ell^4-7776 \ell^2+19683\right)}{16 \sqrt{2}
\left[\ell^2 \left(-8 \ell^2-81\right)\right]^{3/2}},
\ee
where $r_o^-=r_o^+=9M$ for $\ell=27/8$. $\Delta_{\Omega}$ always increases in
$0<\ell<27/8$, and for $\ell\geq27/8$  in the regions $(2M, r_o^-)$ and  $r>r_o^+$,
while it decreases in  $\ell\geq27/8$  in the region $(r_o^-,r_o^+)$. The radius
$r_o^{+}$ is a minimum and $r_o^{-}$ a  maximum point of $\Delta_{\Omega}$. The critical
points values $\Delta_{\Omega}(r_o^{\pm})$ are portrayed in Figs.\il\ref{Arlette},
\emph{lower panels}.

{ We analyze  the profile of the proper angular velocity, $\Phi$, and the relativistic
angular velocity, $\Omega$, respect to the corresponding Keplerian quantities for the
case of a isothermal matter. Clearly, with respect to the general analysis, here we
should take into account the specific density and pressure profile  emerging  by the
choice of the  particular equation of state, $p=k\rho^{\gamma}$.

\section{Summary and conclusions}

{The analysis of a stationary axisymmetric configuration of material, in equilibrium in a Schwarzschild spacetime, as emerging in the Polish doughnut framework constitutes a timely question in view of the interest in astrophysical sources, possibly resulting from super-Eddington accretion onto very compact objects, like Gamma-Ray Bursts, Active Galactic Nuclei, binary systems and Ultraluminous X-ray sources \citep[see e.g.][]{FeB04,So07}. The most characterizing features of the Polish doughnut approach is the thickness of the matter distribution across the equilibrium and the existence of a region enveloping the horizon surface, where the fluid can, in principle, infall onto the black hole. The first property is of impact for a comparison with the wide spectrum of numerical simulation of a thick disk (see \citealt{Straub, Fon03, Abramowicz:2011xu} for recent examples). The non-negligible depth of the accreting profile is typical of the regime where the gravitational effects are strong and it takes a relevant role in all those extreme phenomena associated with the gravitational collapse, characterized by a violent energy-matter release from the central compact object. The second aspect is relevant because the Polish doughnut model can account for a non-zero accretion rate of the torus even when the dissipative effects are negligible. This is in contrast to the original idea by \citet{Shakura1973} that the angular momentum transport is always allowed by the shear viscosity of the accreting material. Indeed, the accreting plasma is in general quasi-ideal and the emergence of dissipative effects as those required to match the observations requires the appearance in the dynamics of a strong turbulent regime, restated as a laminar one in the presence of a significant shear viscosity. Since this picture is not yet settled down \citep[see][]{Balbus2011}, it is very important that an ideal hydrodynamical scheme on a Schwarzschild background like the one offered by the Polish doughnut is able to account for a material infalling onto a black hole.}

{In this work we revisited  the  Polish doughnut model of accretion disks for a perfect fluid circularly orbiting around a Schwarzschild black hole with the effective potential approach for the exact gravitational and centrifugal contributions. We take advantage of the formal analogy between the fluid when the pressure vanishes and the test particle orbiting in the same background to get a comparison between the Polish doughnut, which is supported by the pressure, and the geodetic disk. Our analysis provides a revisited theoretical framework to characterize the accretion processes in presence of general relativistic effects. Indeed we formulate the Polish doughnut model in such a way that the fluid dynamics can be interpreted in terms of the fundamental stability properties of the circular orbits in Schwarzschild background.}

{We  extensively analyzed the Polish doughnut configurations for the fluid $l$ and the particle $L$ angular momenta, taken respectively constant throughout the entire toroidal surface. Then we propose a reinterpretation of torus physics, with respect to the its shape and equilibrium dynamics  in terms of the parameter $l$ (and $L$),  and in terms of the parameter $K$ that was naturally established by introducing  the effective potential for  the fluid motion. This new parameter $K$ has been derived by exploiting the methodological and formal analogy  with the effective potential approach to  the  test particles motion. Note that ($l$, $K$) fully describe the toroidal fluid configuration. This procedure made it possible to emphasize the pressure influence   in the  equilibrium dynamics with respect to the  case of dust, the latter being  treated as a set of test particles not subjected to pressure. This dual aspect, methodological and procedural,   we thought it required a complete and deep analysis of the behavior  of the momenta $l$ and $L$ on the plans and disk orbits, and of $L$ as a function of $l$, and of the surface characteristic parameters  as a function of $K$.}


{The main steps of this analysis are:}
\begin{description}
\item[\emph{Radial pressure gradient vs angular momentum $\lie$}] {We studied the pressure radial gradient as a function of $r$ and of the angular momentum $\lie$. The pressure is a decreasing function of  $r$ in $r<r_{lco}$. and for  $\lie<\lie_K$ it is $p'<0$, this means  that $\lie_K$ identifies the pressure minimum points  located in $[r_{lco}, r_{lsco}]$,  and pressure maximum points in $r> r_{lsco}$ (see Fig. \ref{recognize}, \emph{right panel}). However, at fixed orbit $r$,  $\lie_K$ is always a minimum point of $r$ ie at fixed $r$ the pressure decreases until the angular  momentum  reaches the values $\lie=\lie_K$, and then increases with $\lie$ (note that $ r_{lsco}$ is a minimum point of $\lie_K$).}
\\
\item[\emph{Angular momentum $\lie$ vs. fluid angular momentum $\ell$}] {We found that at fixed $\ell$ the angular momentum $\lie(r,l)$ has a maximum for  $ r=r_-\in [r_{lco}, r_{lsco}]$, and a minimum in $ r_ +> r_{lsco}$, and $\lie(r)$ increases  for  $r<r_-$  and $ r>r_+$ (Fig. \ref{tempo}). We identify three possibilities: $ \ell <3 \sqrt {3/2}$, where the momentum $\lie (r)$ increases with $r$; $ 3 \sqrt {3/2} <\ell <3 \sqrt {3} $, where $\lie$ increases with $r$ up to the maximum point $r_-$, decreases up to the minimum $r_+$; $\ell>3 \sqrt {3} $, where $\lie$ increases with $r$.}
\\
\item[\emph{Pressure gradients vs fluid angular momentum $\ell$}] {The fluid pressure decreases for  $r<r_l^-$ and $r>r_+$ as well as for  $\ell<3\sqrt{3/2}$ (Fig. \ref{filodiff}). The situation is much more complicated for fluid with higher angular momenta  ($\ell>3\sqrt{3/2}$). It appears necessary to consider  fluids with  $3\sqrt{3/2}<\ell<3\sqrt{3}$ and with $\ell>3\sqrt{3}$ separately.  In the first case we observe the presence of a  ring  $[r_-, r_+] $  where the fluid  pressure increases with the orbital radius, $ r_-$ being a pressure minimum and  $r_+$ a pressure maximum. In the case $\ell>3\sqrt{3}$ we find  two rings,  $[r_l^- , r_- ]$ and $ [r_l^+ , r_+]$, where the fluid pressure is an increasing function of $r$. Figure \ref{filodiff} \emph{upper right}  describes this situation from a different point of view: for what concerns the variation of its hydrodynamic pressure with the orbits and  the angular momentum, the fluid dynamics is basically split into two zones,  $r<r_{lco}$ and $r>r_{lco}$, respectively: in the first region, the fluid pressure decreases with increasing  angular momentum up to $\ell=\ell_r$ (minimum),  then it increases with $\ell$ up to $\ell_K $ (maximum), finally it decreases with $\ell$. The trend is precisely the same in the region $r>r_{lco}$, but   $\ell_k$  and  $\ell_r$ are now the minimum and maxima momentum respectively,  i.e. the fluid pressure decreases with $ \ell <\ell_k$, grows in the range  $[\ell_k ,\ell_r ]$  and then decreases with $\ell$. The angular pressure gradient has been studied and compared to the radial and angular gradient in Sec. \ref {Sec:ella}.}
\\
\item[\emph{The Boyer surfaces and polytropic equation of state}] {In Sec.\il\ref{Sec:Poli} we drew a complete and analytic description of the toroidal surface of the disk. A key role of this  analysis was played by the effective potential approach: the toroidal disk can be described once one gives the effective potential and the fluid angular momentum. Our results concerning the disk shape and structure can be summarized as follows:}
\begin{enumerate}
\item the distance from the source of the torus inner surface  ($\delta=y_2-2M$),
    increases with increasing angular momentum of the fluid but decreases with
    increasing energy function defined as the value of the effective potential for
    that momentum;
\item the surface {maximum height} (torus thickness - $h$), increases with the
    energy and decreases with the angular momentum:the torus becomes thinner for
    high angular momenta, and  thicker  for high energies;
\item the location of maximum thickness of the torus moves  towards the external
    regions  with increasing angular momentum and energy, until it reaches a maximum
    an then decreases;
\item the {surface maximum diameter} $(\lambda)$ increases with the energy, but
    decreases with the fluid angular momentum;
\item the {distance of the torus inner surface from the structure inner surface}
    $(\hat{\delta}= y_2-y_1)$ increases with the angular momentum and decreases with
    the energy;
\item the {distance of the structure inner surface from the horizon}
    ($\breve{\delta}\equiv y_1-2M$) increases with the energy and decreases with the
    fluid angular momentum.
\end{enumerate}
{The accreting fluids with a polytropic equation of state were studied in the Section \ref{Sec:polu}, divided into two classes identified by  their polytropic index.}
\\
\item[\emph{The  fluid  angular velocity}] {In Section \ref{Sec:proper} we analyzed the fluid \emph{proper angular velocity} $\Phi$ and we have compared the proper velocity with the Keplerian one $\Phi_K$  in Section \ref{Sec:KeplerANDangular}: the velocity  $\Phi<\Phi_K$ in $ r>r_+$ and  for all $r$  when $\ell<3\sqrt{3/2}$ (see Fig. \ref{SPAlla}). As for  the analysis of the  angular momentum  $L(r,l)$, a distinction is made between fluids with  $\ell\in[3\sqrt{3/2}, 3\sqrt{3}]$, characterized by a ring $[r_-,r_ +]$ where the proper velocity  is higher then the kepler one,  and fluids with $ \ell > 3 \sqrt {3} $, where the proper fluid velocity  is greater  than the Keplerian one in the inner regions, $ r <r_- $. Finally, we have investigated the regions and momenta where  the difference $\Phi-\Phi_K$ is maximum (see Fig. \ref{SPAlla}). We concluded the study of velocity fields by analyzing the \emph{fluid relativistic velocity} $\Omega$ and the Von Zeipel surfaces in Section \ref{Sec:KeplerANDangular}. The velocity $\Omega$ has a maximum at $r_{lco}$ and increases with the  increasing of $\ell$. We also studied the differences $\Delta_{\Omega}\equiv
(\Omega-\Omega_{\ti{K}})M\sigma$ between the relativistic fluid  velocity and the Keplerian  one $\Omega_{\ti{K}}$ (see Fig. \ref{Arlette}). At fixed angular momentum $\ell>3\sqrt{3/2}$, this difference has a maximum and a minimum, while for $ \ell<3\sqrt{3/2}$ it is always decreasing. The fluid velocity $\Omega$  is lower then the Keplerian  $\Omega_K$ for  $r<r_-$ and $ r>r_+$, and for any $r$ when  $\ell<3\sqrt{3/2}$, instead for fluid with with angular momentum greater then $\ell=3\sqrt{3/2}$ there is a ring $[r_-, r_+]$ where the fluid velocity $\Omega$  is larger then the Keplerian one (see Fig. \ref{SPAlla2}).}
\end{description}

{The fundamental merit of the present work is that the analysis of the Polish doughnut features maintain the explicit presence of the effective potential in all the basic expressions describing the matter distribution. In fact, this allows to keep in direct contact our study with the behavior of free test particles moving the gravitational field of the central object and following circular orbit (the dust, pressure-free limit of the present analysis).  What is significant here is the possibility to compare configuration of the considered fluid, as described by certain values of the parameters $K$ and $l$, with the behavior of the test particle system, characterized by the same values of the corresponding parameters $K$ and $L$. These latter quantities have a precise meaning for the  particle (energy and angular momentum as viewed at infinite distance), while the corresponding parameters for the fluid must, on this level, regarded as a classification criterion for the torus morphology. Thus retaining the effective potential in the Polish doughnut treatment allows to identify the Boyer surfaces in terms of parameters that have a precise meaning for a different, but closely related, context, the pressure-free fluid, i.e. the fundamental features of the Schwarzschild spacetime. If this study can not yet directly offer a paradigm for the comparison with the observations, nonetheless, it makes a concrete step in this direction. We are now able to constrain the morphology of the equilibrium configuration by fixing the two parameters $K$ and $l$, i.e. specifying their value, we can define the basic features of the torus shape and of the velocity field in different space regions (this is synthetically sketched in the ten final remarks, achieved by our systematic analysis) and the comparison with the corresponding profile of the particle motion, where the physical comprehension is settled down, offer a valuable tool to interpret what the observations trace out. Our model will be completed including other relevant ingredients, like dissipative effects and the magnetic field and this is probably the natural development of the conceptual paradigm we fixed here.}

\section*{Acknowledgment}

This work has been developed in the framework of the CGW Collaboration
(www.cgwcollaboration.it). DP gratefully acknowledges financial support from the Angelo
Della Riccia Foundation and  wishes to thank the Blanceflor Boncompagni-Ludovisi  Foundation (2012).

 
\appendix

\section{The radial gradient $G_r$ as function of $\sigma$.}\label{Sec:app}

In Section\il\ref{Sec:Neutral-L0} and \il\ref{Sec:neutrall} we detailed the study of the radial gradient $G_r$ as a function of the angular momentum $L$ and of the fluid angular momentum $l$ with a re-parametrization that is independent from the equatorial plane $\sigma$. However, an explicit study of $G_r$ as a function of $\sigma$ is useful for two main reasons: first, to have a direct comparison when the re-parametrization is not possible (as for the polar gradient $G_\vartheta$). Secondly, the angular dependence is necessary to build a three-dimensional characterization of the torus.

In Section\il\ref{Sec:Gr} we showed that the critical points of the pressure are defined by the condition Eq.\il\ref{Eq:wEo}. In terms of the polar coordinate, it reads:
\be
\sigma_{\ti{K}}^{\ti{L}}\equiv \frac{L}{M}\sqrt{\frac{M(r-3M)}{r^2}}\, .
\ee
The condition of existence of $\sigma_{\ti{K}}^{\ti{L}}$ are $r>r_{lco}$ for $0<L\leq2\sqrt{3}M$, and $r_{lco}<r\leq r_{\ti{L}}^-(\sigma=1)$ and $r\geq r_{\ti{L}}^+(\sigma=1)$ for $L>2\sqrt{3}M$, where $r_{\ti{L}}^+(\sigma=1)$ are the circular orbit radii for a test particle in Schwarzschild spacetime defined by Eq.\il\ref{testpartradii}, evaluated at $\sigma=1$. The point $r_{lsco}$ is a maximum for $\sigma_{\ti{K}}^{\ti{L}}$: the condition $r_{\ti{L}}^+=r_{\ti{L}}^-=r_{lsco}$ is fulfilled when $\sigma_{\ti{K}}^{\ti{L}}=\sigma_{\ti{K}}^{\ti{L*}}\equiv{\sqrt{L^2}}/{2 \sqrt{3}M}$. When $L=2\sqrt{3}M$, it is $\sigma_{\ti{K}}^{\ti{L*}}=1$. The sign of $G_r$ is summarized in Figs.\il\ref{I}.

\begin{figure}
\centering
\includegraphics[width=0.45\hsize,clip]{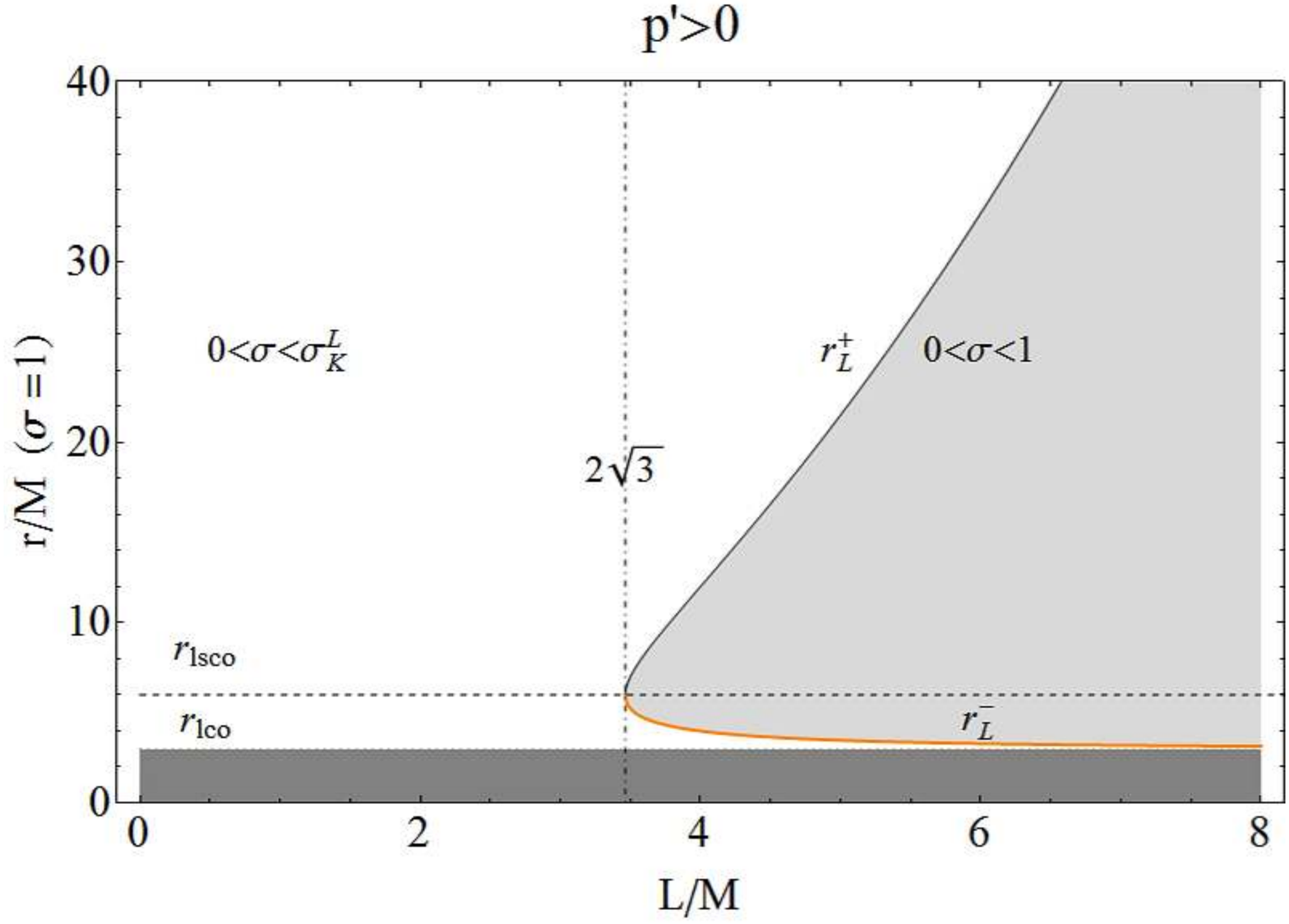}
\includegraphics[width=0.45\hsize,clip]{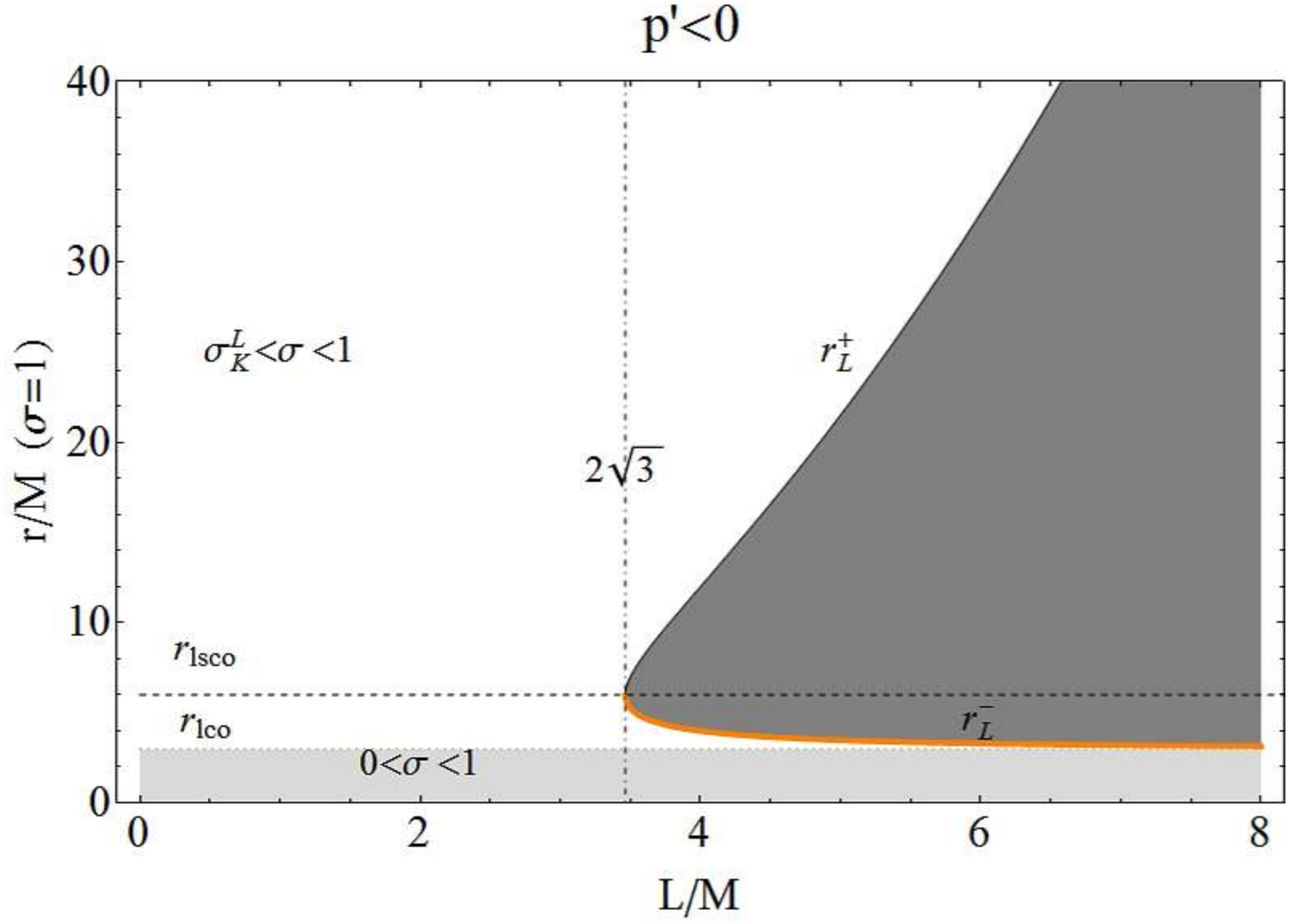}
\caption[font={footnotesize,it}]{(Color online) \footnotesize{\emph{Left panel}: conditions for  \ttb{$G_r>0$}: $0\leq \sigma< \sigma_{\ti{K}}^{\ti{L}}$ when $0<L<2\sqrt{3}M$ and $r>r_{lco}$, and when $L>2\sqrt{3}M$ and $r_{lco}<r<r_{\ti{L}}^-(\sigma=1)$ and $r> r_{\ti{L}}^+(\sigma=1)$.  \ttb{$G_r>0$} for all $0<\sigma<1$ when $L>2\sqrt{3}M$ and $r_{\ti{L}}^-(\sigma=1)< r< r_{\ti{L}}^+(\sigma=1)$, while it is always positive when $r<r_{lco}$. \emph{Right panel}: conditions for \ttb{$G_r<0$}: $\sigma_{\ti{K}}^{\ti{L}}< \sigma<1$ when $0<L<2\sqrt{3}M$ and $r>r_{lco}$, and when $L>2\sqrt{3}M$ and $r_{lco}<r<r_{\ti{L}}^-(\sigma=1)$ and $r> r_{\ti{L}}^+(\sigma=1)$.  \ttb{$G_r<0$} for all $0<\sigma<1$ when $r<r_{lco}$, while it is always negative when $L>2\sqrt{3}M$ and $r_{\ti{L}}^-(\sigma=1)< r< r_{\ti{L}}^+(\sigma=1)$. Dotted lines mark $r=r_{lco}$, dashed lines $r=r_{lsco}$, dot-dashed lines $L=2\sqrt{3}M$.} }\label{I}
\end{figure}

We characterize now the angular momentum $L$ as an explicit function of $l$ and $\sigma$. We know from Section\il\ref{Sec:NeutraLl} that $L$ is not defined in the interval $[r^-_l,r^+_l]$, where $r_l^\pm$ are introduced in Eq.\il\ref{testpartradiil}. In terms of $\sigma$ this region corresponds to $0\leq a \leq \sigma_{l_r}$, where $\sigma_{l_r}\equiv\sqrt{\frac{l^2 (r-2M)}{r^3}}$. $\sigma_{l_r}$ is defined for $r>2M$ when $0<l\leq3 \sqrt{3}M$, and for $2M<r<r_l^-(\sigma=1)$ and $r>r_l^+(\sigma=1)$ when $l>3 \sqrt{3}M$. $r=r_{lco}$ is a maximum for $\sigma_{l_r}$, where $\sigma_{l_r}=\sigma_{l_r}^*\equiv\frac{\sqrt{l^2}}{3 \sqrt{3}M}$. When $l=3\sqrt{3}M$, it is $\sigma_{l_r}^*=1$.

\begin{table*}
 \centering
 \begin{minipage}{140mm}
\caption[font={footnotesize,it}]{\footnotesize{Regions where $\partial_rL=0$.}}\label{16Set}
\begin{tabular}{ll|ll|ll}
\hline
$0<l/M\leq3\sqrt{3/2}$& &$3\sqrt{3/2}<l/M<3\sqrt{3}$&&$l/M\geq3\sqrt{3}$
\\
\hline
$\sigma$&$r/M$&$\sigma$&$r/M$&$\sigma$&$r/M$
\\
\hline
$(0,\sigma_{l_r}^*]$&$r_+$&$(0,\sigma_{l_r}^*]$& $r_+$ &$(0,1]$&$r_+$
\\
$(\sigma_{l_r}^*,\sqrt{2}\sigma_{l_r}^*)$&$r_+$, $r_-$&$(\sigma_{l_r}^*,1]$& $r_+$, $r_-$ &$ $&$ $
\\
$\sqrt{2}\sigma_{l_r}^*$&$r_+=r_-=r_{lsco}$&$ $& $ $ &$ $&$ $\\
\hline
\end{tabular}
\end{minipage}
\end{table*}

The critical points of $L$ are determined by the condition:
\be
\sigma_{\ti{K}}^{\ti{l}}\equiv\sqrt{\frac{l^2 (r-2M)^2}{M r^3}}\,.
\ee
(see Eq.\il\ref{lkkeple}). $\sigma_{\ti{K}}^{\ti{l}}$ is defined for $r>r_{lco}$ when $0<l\leq 3\sqrt{3/2}M$, and for $r_{lco}<r\leq r_-(\sigma=1)$ and $r\geq r_+(\sigma=1)$ when $ 3\sqrt{3/2}M<l<3\sqrt{3}M$, and also for $r\geq r_+(\sigma=1)$  for $l\geq3\sqrt{3}M$, where $r_\pm(\sigma=1)$ are the radii defined by Eq.\il\ref{critrad}, evaluated at $\sigma=1$. $\sigma_{\ti{K}}^{\ti{l}}$ is maximum for $r=r_{lsco}$, where $\sigma_{\ti{K}}^{\ti{l}}=\sigma_{\ti{K}}^{\ti{l}*}=\sqrt{2}\sigma_{l_r}^*$. These results are summarized in Table\il\ref{16Set}.

Finally the study of $G_r$ as a function of $(r,l, \sigma)$ is illustrated in Figs.\il\ref{seenvisto}.

\begin{figure}
\centering
\includegraphics[width=0.45\hsize,clip]{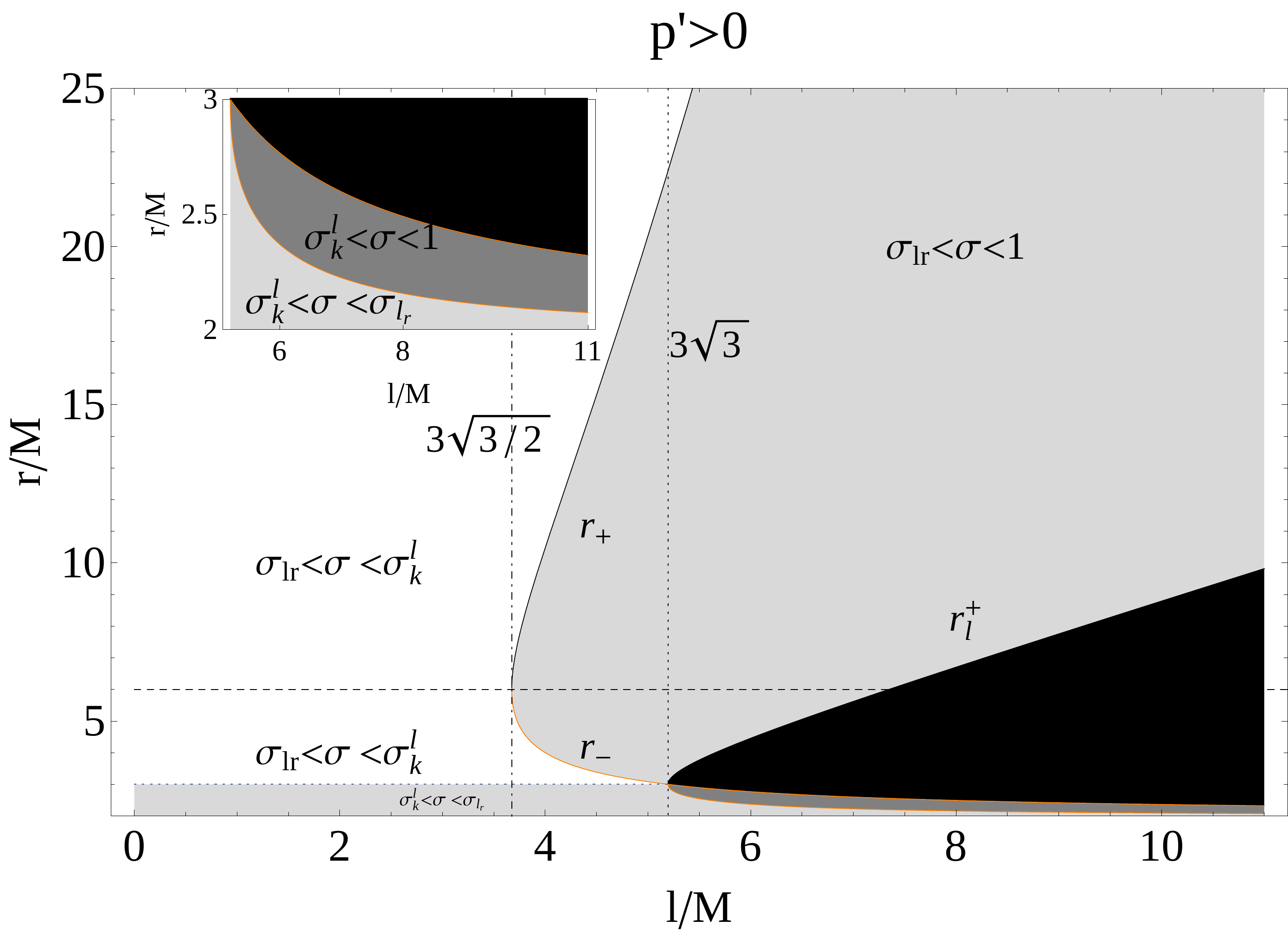}
\includegraphics[width=0.45\hsize,clip]{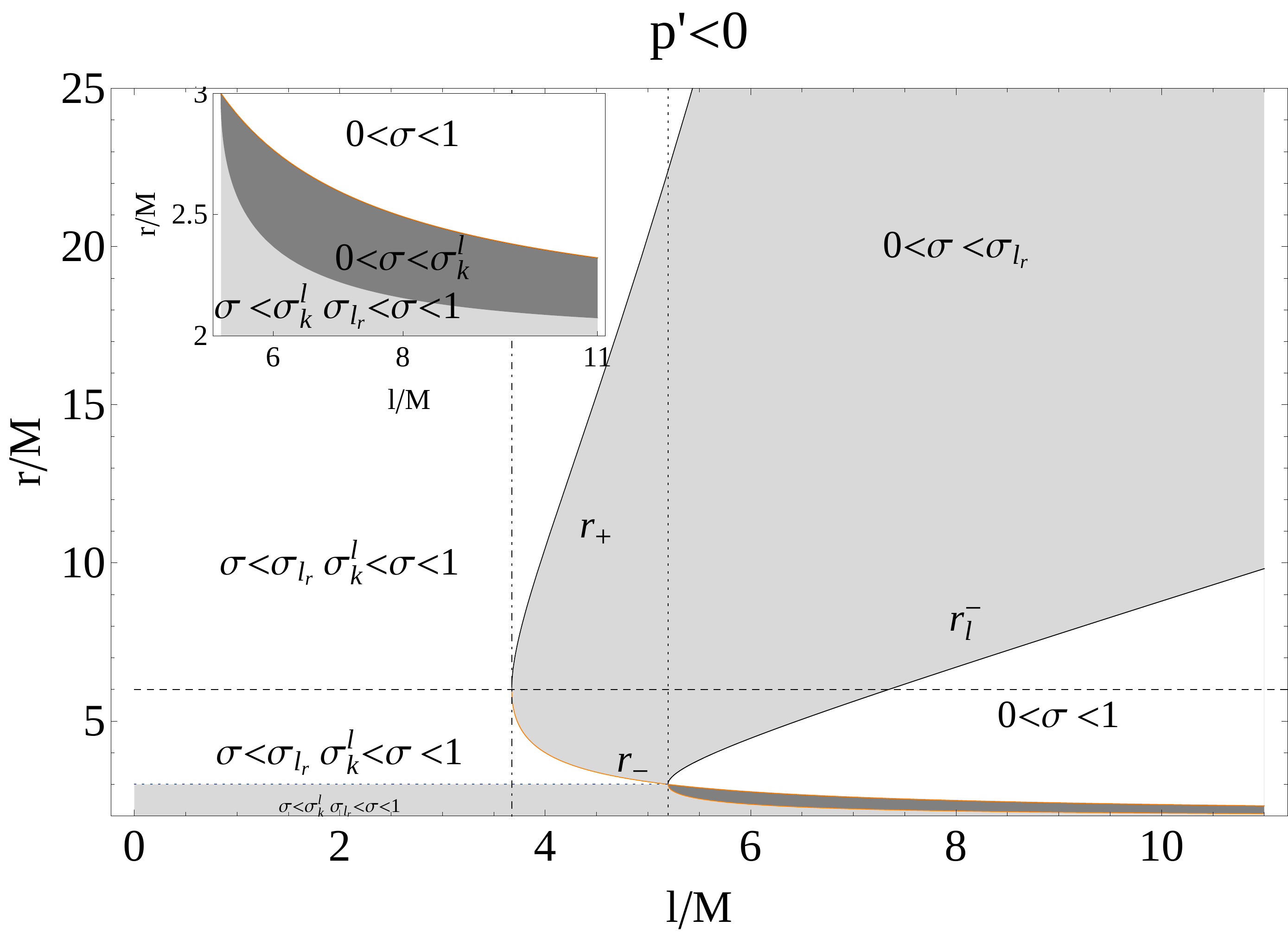}
\caption[font={footnotesize,it}]{(Color online) \emph{Left panel}: conditions for \ttb{$G_r>0$}: $\sigma_{l_r}< \sigma< \sigma_{\ti{K}}^{\ti{l}}$ when $0<l<3\sqrt{3/2}M$ and $r>r_{lco}$, and when $3\sqrt{3/2}M<l<3\sqrt{3}M$ and $r_{lco}<r<r_-(\sigma=1)$ and $r> r _+(\sigma=1)$, and when $l>3\sqrt{3}M$ and $r> r _+(\sigma=1)$; $\sigma_{l_r}< \sigma< 1$ when $3\sqrt{3/2}M<l<3\sqrt{3}M$ and $r_-(\sigma=1)<r< _+(\sigma=1)$, and when $l>3\sqrt{3}M$ and $r_l^+(\sigma=1)<r<r _+(\sigma=1)$; $\sigma_{\ti{K}}^{\ti{l}}< \sigma< 1$ when $l>3\sqrt{3}M$ and $r_l^-(\sigma=1)<r< r_-(\sigma=1)$; $\sigma_{\ti{K}}^{\ti{l}}<\sigma<\sigma_{l_r}$ when $l<3\sqrt{3}M$ and $2M<r<r_{lco}$, and when $l>3\sqrt{3}M$ and $2M<r< r_l^-(\sigma=1)$. \emph{Right panel}: conditions for $G_r>0$: $0<\sigma<\sigma_{l_r}$ and $\sigma_{\ti{K}}^{\ti{l}}<\sigma<1$ when $0<l<3\sqrt{3/2}M$ and $r>r_{lco}$, and when $3\sqrt{3/2}M<l<3\sqrt{3}M$ and $r_{lco}<r<r_-(\sigma=1)$ and $r> r _+(\sigma=1)$, and when $l>3\sqrt{3}M$ and $r> r _+(\sigma=1)$; $)<\sigma<\sigma_{l_r}$ when $3\sqrt{3/2}M<l<3\sqrt{3}M$ and $r_-(\sigma=1)<r< _+(\sigma=1)$, and when $l>3\sqrt{3}M$ and $r_l^+(\sigma=1)<r<r _+(\sigma=1)$; $0<\sigma<\sigma_{\ti{K}}^{\ti{l}}$ when $l>3\sqrt{3}M$ and $r_l^-(\sigma=1)<r< r_-(\sigma=1)$; $0<\sigma<\sigma_{\ti{K}}^{\ti{l}}$ and $\sigma_{l_r}<\sigma<1$ when $l<3\sqrt{3}M$ and $2M<r<r_{lco}$, and when $l>3\sqrt{3}M$ and $2M<r< r_l^-(\sigma=1)$.  \ttb{$G_r<0$} for all $0<\sigma<1$ when $L>3\sqrt{3}M$ and $r_-(\sigma=1)< r< r_l^+(\sigma=1)$. Dotted lines mark $l=3\sqrt{3}M$, dot-dashed lines $l=3\sqrt{3/2}M$, dashed lines $r=r_{lsco}$ and dotted gray lines $r=r_{lco}$. Black regions are forbidden. \emph{Insets}: zoom of the region $l/M\in[6,11]$ and $r/M\in[2,r_{lco}]$.}
\label{seenvisto}
\end{figure}

\section{Analysis of the  proper velocity profile}\label{Sec:appalph}

We redefine the radius in Eq.\il\ref{rchi} in terms of $\sigma$  and $l$:
\be
r_{\chi}/M=\frac{1}{4}\left(\frac{5}{2}-\frac{\beta }{2 \sqrt{3}}+\frac{\check{\beta}}{ \sqrt{6}}\right)\, ,
\ee
where
\ttb{\bea
\check{\beta}&\equiv&\sqrt{\frac{264 \sqrt{3} (l/M)^2 \alpha -4 \left[(l/M)^4-4 (l/M)^2 \alpha +\alpha ^2\right] \beta +\sigma^2 \left[144 (l/M)^2 \beta +75 \alpha  \left(\beta -5 \sqrt{3}\right)\right]}{\sigma^2 \alpha  \beta }},
\\
\alpha &\equiv&\left[-(l/M)^2 \left(1350 \sigma^4+(l/M)^4-6 \sigma^2 \left[9 (l/M)^2+\sqrt{48 (l/M)^4-2754 \sigma^2 (l/M)^2+50625 \sigma^4}\right]\right)\right]^{1/3},
\\
\beta &\equiv&\sqrt{\frac{8 \left[(l/M)^2+\alpha \right]^2+\sigma^2 \left[75 \alpha -288 (l/M)^2\right]}{\sigma^2 \alpha }},
\eea}
for $\sigma\neq0$ (for $\sigma=0$ it is $r^{\pm}_{\chi}=2M$).

The fluid proper angular  velocity  increases with $r/M$ ($\Phi'>0$) when:
\ttb{
\bea
0<l<3 \sqrt{3}M\quad\mbox{for}\quad0<\sigma\leq \sigma_{l_r}\quad\mbox{and}\quad 2M<r<r_l^-\, ,
\\
\sigma_{l_r}<\sigma\leq 1\quad\mbox{and}\quad 2M<r<r_{\chi}\, ;
\\
 l\geq 3 \sqrt{3}M\quad\mbox{for}\quad 0<\sigma\leq 1\quad\mbox{and}\quad2M<r<r_l^-\, .
\eea}
It is $\Phi'<0$ when:
\ttb{
\bea
0<l<3 \sqrt{3}M\quad\mbox{for}\quad 0<\sigma<\sigma_{l_r}\quad\mbox{and}\quad r>r_l^+\, ,
\\
\sigma=\sigma_{l_r}\quad\mbox{and}\quad r>r_l^-\, ,
\\
\sigma_{l_r}<\sigma\leq 1\quad\mbox{and}\quad r>r_{\chi}\, ;
\\
l=3 \sqrt{3}M\quad\mbox{for}\quad 0<\sigma<1\quad\mbox{and}\quad r>r_l^+\, ,
\\
\sigma=1\quad\mbox{and}\quad r>r_l^-\,;
\\
l>3 \sqrt{3}M\quad\mbox{for}\quad 0<\sigma\leq 1\quad\mbox{and}\quad r>r_l^+
\eea}
(see Fig.\il\ref{acceento}). It follows that the angular velocity   increases in $2M<r<r_{\chi}<3M$  until it reaches the point $r=r_{\chi}$, that is a maximum, where $2.5M<r_{\chi}<3M$ with angular momentum $l_{\chi}$ on the equatorial plane $\sigma_{\chi}$, then it decreases with the radius (see Fig.\il\ref{acceento}).

\begin{figure}
\centering
\includegraphics[width=0.45\hsize,clip]{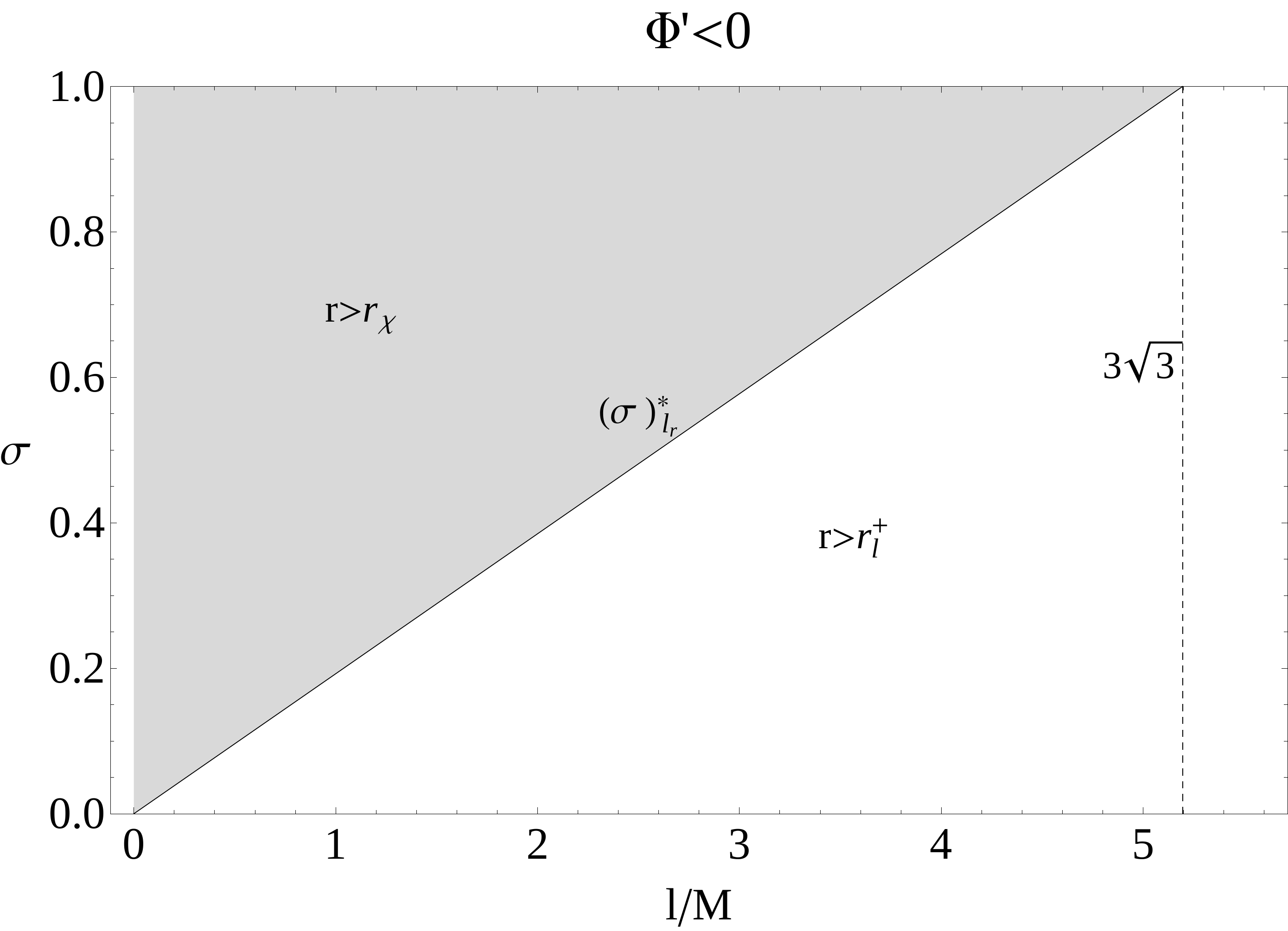}
\includegraphics[width=0.45\hsize,clip]{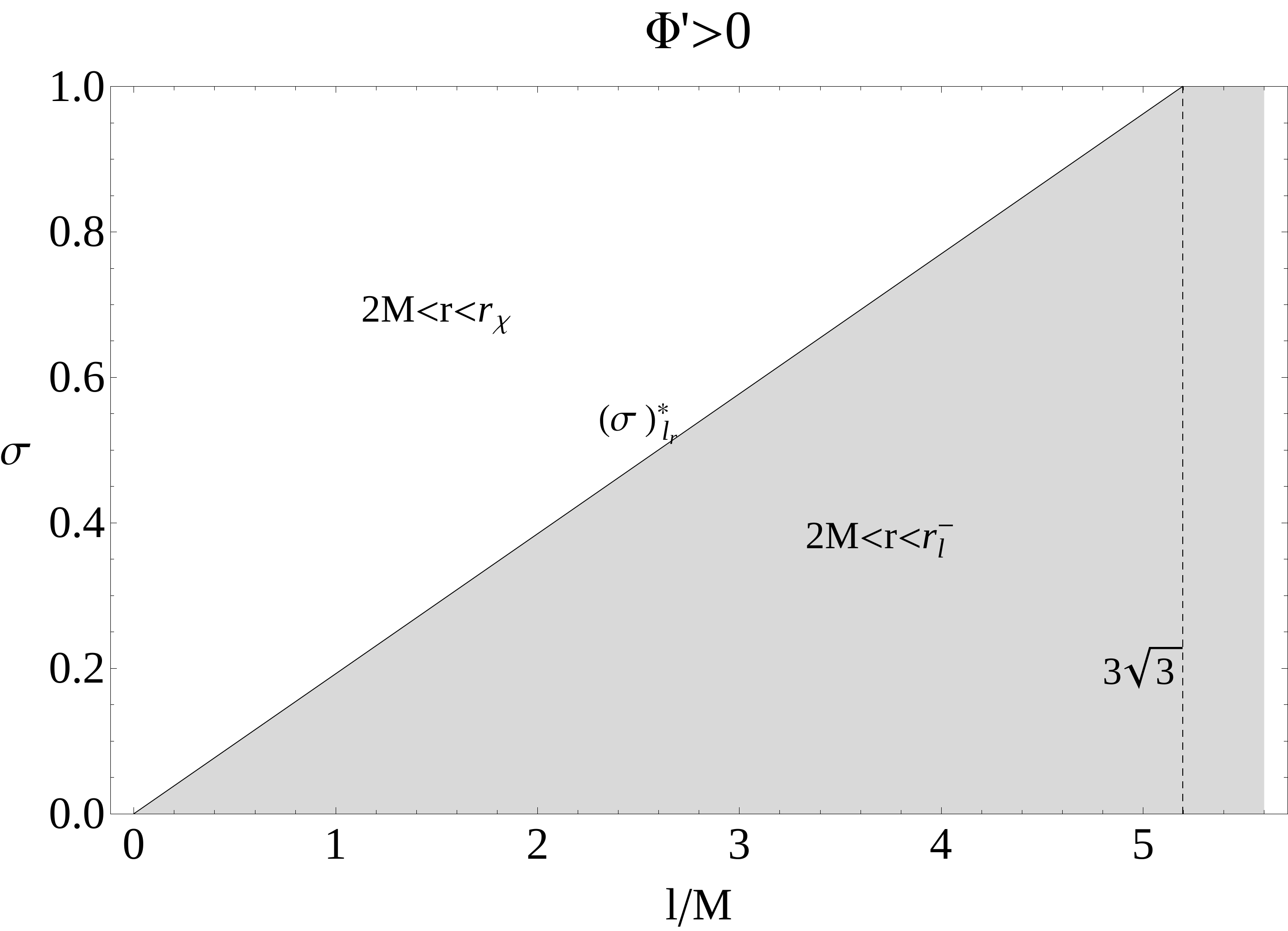}
\caption[font={footnotesize,it}]{\emph{Left panel}: conditions for $\Phi'<0$. \emph{Right panel}: conditions for $\Phi'>0$. Solid line marks $\sigma_{l_r}^*$ as a function of $l/M$, dashed line $l/M=3\sqrt{3}$.}
\label{acceento}
\end{figure}


\end{document}